%% file: main.tex
\documentclass[12pt,a4paper,twoside]{book}

\usepackage[a4paper,inner=3cm,outer=2cm,top=3cm,bottom=3cm]{geometry} 
\usepackage[utf8]{inputenc} 
\usepackage{newlfont}
\usepackage{fancyhdr}
\usepackage[T1]{fontenc} 
\usepackage{times}

\usepackage{amsmath,amssymb,amsthm,amscd,amsfonts}
\usepackage{mathtools}
\mathtoolsset{showonlyrefs=true} 
\usepackage{array}
\usepackage{makecell} 

\usepackage[usenames,dvipsnames]{color}
\usepackage{xcolor,graphicx,float} 
\usepackage{subfigure}  
\usepackage{caption}
\usepackage{adjustbox} 
\usepackage{booktabs} 
\graphicspath{{./fig_ch0/}}

\definecolor{uscblue}{RGB}{12,37,119}

\PassOptionsToPackage{hyphens}{url}\usepackage[final,pdftex,bookmarks=true,bookmarksopen=true,breaklinks,colorlinks,bookmarksnumbered=true,pdfsubject={ },pdfauthor={Fernando Castro-Prado},urlcolor=black,citecolor=black,linkcolor=black]{hyperref} 

\usepackage[nottoc,numbib]{tocbibind} 
\usepackage[authoryear]{natbib}

\usepackage{setspace,titlesec}
\usepackage{enumerate}
\usepackage[shortlabels]{enumitem} 

\newcommand{\headerright}{Nonparametric independence tests in high-dimensional settings, with applications to the genetics of complex disease}

\usepackage{etoolbox}

\fancypagestyle{head-normal}{%
\fancyhf{}
\fancyhead[LO]{\rightmark} 
\fancyhead[RE]{\leftmark}  
\fancyhead[RO,LE]{\thepage} 
}
\let\origdoublepage\cleardoublepage
\newcommand{\clearemptydoublepage}{\clearpage {\pagestyle{empty}\origdoublepage}}
\newcommand{\com}[1]{} 

\titleformat{\chapter}[display]{\Huge}{\vspace{-1.5cm}\titlerule\begin{flushright}
		{\LARGE \chaptername~\thechapter}
\end{flushright}}{-1.5cm}
{\filleft}[\vspace{-0.5cm}\rule{\textwidth}{1.5pt}\vspace{-1cm} ]

\theoremstyle{plain}
\newtheorem{theorem}{Theorem}[chapter]
\newtheorem{lemma}{Lemma}[chapter]
\newtheorem{proposition}{Proposition}[chapter]
\newtheorem{corollary}{Corollary}[chapter]
\theoremstyle{definition}

\newtheorem{remark}{Remark}[chapter]

\title{\huge{\textbf{PhD thesis}}}
\author{Fernando Castro Prado}
\date{July 2024}

\DeclareMathOperator{\dcov}{dcov}
\DeclareMathOperator{\dvar}{dvar}
\DeclareMathOperator{\dcor}{dcor}
\DeclareMathOperator{\dCov}{dCov}
\DeclareMathOperator{\dMv}{dMvar}

\DeclareMathOperator{\dCor}{dCor}
\newcommand{\dcovh}{\widehat{\dcov}}

\DeclareMathOperator{\E}{E}
\DeclareMathOperator{\Prob}{P}
\DeclareMathOperator{\Var}{Var}
\DeclareMathOperator{\Cov}{Cov}
\DeclareMathOperator{\Cor}{Cor}
\DeclareMathOperator{\Normal}{\mathcal{N}}

\DeclareMathOperator{\tr}{tr}

\DeclareMathOperator*{\argmin}{arg\,min}
\DeclareMathOperator{\M}{\mathcal{M}}
\newcommand{\implica}{\Rightarrow} 
\newcommand{\eqv}{\Leftrightarrow}
\newcommand{\flecha}{\longrightarrow} 
\newcommand{\flechita}{\mapsto} 
\newcommand{\inner}[2]{\left\langle{#1},{#2}\right\rangle}
\newcommand{\norm}[1]{\left\lVert#1\right\rVert}
\newcommand{\llaves}[1]{\left\lbrace{#1}\right\rbrace} 
\newcommand{\detlim}{\underset{n\to\infty}{\longrightarrow}}

\newcommand{\distrilim}{\overset{\mathcal{D}}{\detlim}}

\newcommand{\Dlim}{\stackrel{\mathcal{D}}{\longrightarrow} }

\newcommand{\aslim}{\overset{a.s.}{\detlim}}
\newcommand{\iid}{\text{ IID }}
\newcommand{\boxx}{\framebox[1.1\width]}

\newcommand{\overeq}[1]{\overset{\text{#1}}{=}}
\newcommand{\overdef}[1]{\overset{\text{#1}}{:=}}
\newcommand{\overleq}[1]{\overset{\text{#1}}{\leq}}
\newcommand{\overless}[1]{\overset{\text{#1}}{<}}

\newcommand{\overthen}[1]{\overset{\text{#1}}{\implica}}

\newcommand{\overarrow}[1]{\overset{\text{#1}}{\flecha}}
\newcommand{\comb}[2]{\binom{#1}{#2}}
\newcommand{\disjoint}{\sqcup}
\newcommand{\comp}{\circ}
\newcommand{\card}{\#}
\newcommand{\transp}{^\text{t}}

\newcommand{\wrt}{\,\mathrm{d}}
\newcommand{\dmu}{\,\mathrm{d}\mu}
\newcommand{\dnu}{\,\mathrm{d}\nu}
\newcommand{\damu}{\,\mathrm{d}|\mu|}
\newcommand{\amu}{a_\mu}
\newcommand{\anu}{a_\nu}
\newcommand{\dtheta}{\,\mathrm{d}\theta}
\newcommand{\spx}{\mathcal{X}}
\newcommand{\spy}{\mathcal{Y}}
\newcommand{\spz}{\mathcal{Z}}
\newcommand{\xy}{\mathcal{X}\times\mathcal{Y}}
\newcommand{\xd}{\left(\mathcal{X},d_{\mathcal{X}}\right)}
\newcommand{\yd}{\left(\mathcal{Y},d_{\mathcal{Y}}\right)}
\newcommand{\xsd}{\left(\mathcal{X},\sqrt{d_{\mathcal{X}}}\right)}
\newcommand{\ysd}{\left(\mathcal{Y},\sqrt{d_{\mathcal{Y}}}\right)}
\newcommand{\intx}{\int_{\mathcal{X}}}
\newcommand{\h}{\mathcal{H}} 
\newcommand{\Borel}[1]{\mathcal{B}\left({#1}\right)}
\newcommand{\R}{\mathbb{R}}
\newcommand{\Rplus}{\mathbb{R}^{+}}
\newcommand{\Q}{\mathbb{Q}}
\newcommand{\Z}{\mathbb{Z}}
\newcommand{\N}{\mathbb{N}}
\newcommand{\Zplus}{\mathbb{Z}^{+}}
\newcommand{\Nstar}{\Zplus}
\newcommand{\bA}{{\mathbf{A}}}
\newcommand{\ba}{{\mathbf{a}}}
\newcommand{\bB}{{\mathbf{B}}}
\newcommand{\bb}{{\mathbf{b}}}
\newcommand{\bc}{\mathbf{c}}
\newcommand{\bC}{\mathbf{C}}
\newcommand{\bD}{{\mathbf{D}}}
\newcommand{\bE}{\mathbf{E}}
\newcommand{\bG}{\mathbf{G}}
\newcommand{\bH}{\mathbf{H}}
\newcommand{\bI}{\mathbf{I}}

\newcommand{\bK}{{\mathbf{K}}}

\newcommand{\bL}{{\mathbf{L}}}
\newcommand{\bM}{\mathbf{M}}

\newcommand{\bO}{\mathbf{O}}
\newcommand{\bo}{{\mathbf{o}}}
\newcommand{\bp}{\mathbf{p}}
\newcommand{\bq}{\mathbf{q}}
\newcommand{\bQ}{\mathbf{Q}}
\newcommand{\br}{\mathbf{r}}
\newcommand{\bs}{{\mathbf{s}}}
\newcommand{\bt}{{\mathbf{t}}}
\newcommand{\bU}{{\mathbf{U}}}
\newcommand{\bV}{\mathbf{V}}
\newcommand{\bv}{{\mathbf{v}}}
\newcommand{\bw}{{\mathbf{w}}}
\newcommand{\bX}{{\mathbf{X}}}

\newcommand{\bY}{{\mathbf{Y}}}

\newcommand{\bZ}{{\mathbf{Z}}}
\newcommand{\bz}{\mathbf{z}}
\newcommand{\bzero}{\boldsymbol{0}}
\newcommand{\bone}{\boldsymbol{1}}
\newcommand{\balpha}{\boldsymbol{\alpha}}
\newcommand{\bgamma}{\boldsymbol{\gamma}}
\newcommand{\bGamma}{\boldsymbol{\Gamma}}
\newcommand{\bLambda}{\boldsymbol{\Lambda}}
\newcommand{\bmu}{\boldsymbol{\mu}}
\newcommand{\bPhi}{\boldsymbol{\Phi}}

\newcommand{\V}{{\mathcal V}}

\newcommand{\mX}{{\mathcal{X}}}
\newcommand{\mY}{{\mathcal{Y}}}
\newcommand{\mZ}{{\mathcal{Z}}}

\newcommand{\BState}{\State\hskip-\ALG@thistlm}

\newcommand{\dx}{d_{\spx}}
\newcommand{\dy}{d_{\spy}}
\newcommand{\dz}{d_{\spz}}
\newcommand{\dCovh}{\widehat{\dCov}}
\DeclareMathOperator{\Multin}{Multinomial}
\DeclareMathOperator{\Multib}{Multi-Bernoulli}
\DeclareMathOperator{\rank}{rank}
\DeclareMathOperator{\vecop}{vec}
\DeclareMathOperator{\HSIC}{HSIC}
\newcommand{\HSICh}{\widehat{\HSIC}}
\DeclareMathOperator{\GT}{GT}
\newcommand{\GTh}{\widehat{\GT}}

\newcommand{\ceil}[1]{\left\lceil{#1}\right\rceil}
\newcommand{\brc}[1]{\left\lbrace{#1}\right\rbrace}
\newcommand{\tensor}{\otimes}

\newcommand{\eps}{\varepsilon}

\addtocontents{toc}{\protect\thispagestyle{empty}}

\begin{document}

\begin{titlepage}
\includegraphics[scale=1.3]{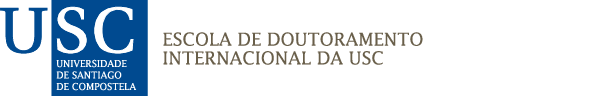}
    \begin{center}
        \vspace*{.5cm}
        \LARGE
        {PhD thesis}
       
        \vspace*{.3cm}

        {\Huge \color{uscblue} 
        \textbf{ \textsc{
       Nonparametric Independence  \\[.3cm]
       Tests in High-Dimensional  \\[.3cm]
       Settings, with Applications to \\[.3cm]
       the Genetics of Complex Disease  }}} 
            
        \vspace*{1cm}         
            
     {\fontfamily{ppl}   \textbf
     	{Fernando Castro Prado}
     }
     
      \vspace*{3.5cm}
            
            {\large Supervised by:}
            {\Large\fontfamily{ppl}   
            	{Wenceslao González Manteiga}\\[-.1cm]
            	\hspace*{1.2cm}{Javier Costas Costas}
            }
            
        \vspace*{1.7cm}
        
            {\small
        
{ \sf
	PROGRAMA DE DOUTORAMENTO\\[-.5cm]
EN ESTAT\'ISTICA E INVESTIGACI\'ON OPERATIVA}
}  
        \vspace*{.7cm}
            
        \large
        {\sf SANTIAGO DE COMPOSTELA\\
        2024}
    \end{center}
\end{titlepage}


\clearemptydoublepage
\include{ack_v14.tex}

\clearemptydoublepage
\thispagestyle{empty}
\newpage
\vspace*{\fill}

\subsection*{Funding and academic support}
\pagestyle{empty}

This work has been funded by projects \mbox{PID2020-116587GB-I00} ({MICIU / AEI / 10.13039 / 501100011033}; Spanish Ministry of Science, Innovation and Universities) and \mbox{ED431C 2021/24} (Department Culture, Education and Universities; Government of Galicia), as well as by the \mbox{FPU19/04091} grant of the Spanish Ministry of Science, Innovation and Universities. We also want to show gratitude to the USC Institute of Mathematics (IMAT) and the German Cancer Research Centre (DKFZ) for their support.

We thank the Galician Supercomputing Centre (CESGA) for the access to their facilities, in order to carry out the most computer-intensive experiments in this dissertation. The schizophrenia data that is analysed in this dissertation was generated under support of the Instituto de Salud Carlos III (grant number ISCIII/PI14/01020) to Javier Costas, co-founded by European Regional Development Fund (ERDF).

Regarding the dataset for the study of liver enzymes, we thank the participants of the Trinity Student Study (dbGaP accession \href{https://www.ncbi.nlm.nih.gov/projects/gap/cgi-bin/study.cgi?study_id=phs000789.v1.p1}{\textit{phs000789.v1.p1}}), first published by \citet{Mills:Molloy}. Their research was supported by the Intramural Research Programs of the National Institutes of Health, the National Human Genome Research Institute, and the Eunice Kennedy Shriver National Institute of Child Health and Development.

We also thank Dr Dominic Russ (University of Birmingham) and Prof Thomas Berrett (University of Warwick) for help in reproducing their research, and Prof Rosa Crujeiras (University of Santiago de Compostela and CITMAga) for her involvement in the application for dbGaP data.

\clearemptydoublepage

\frontmatter

\pagestyle{fancy}
\renewcommand{\headerright}{Contents}
\fancyhead[LO]{{Contents}} 
\fancyhead[RE]{{Contents}}
\hypersetup{linkcolor=black}
\tableofcontents

\hypersetup{urlcolor=gray,linkcolor=black,citecolor=gray}
\clearemptydoublepage
\pagestyle{head-normal}

\mainmatter

\captionsetup[table]{labelfont=it,textfont={it}}
\captionsetup[figure]{labelfont=it,textfont={it}}

\clearemptydoublepage
\include{ch1_v14.tex}
\clearemptydoublepage
\include{ch2_v14.tex}
\clearemptydoublepage
\include{ch3_v14.tex}
\clearemptydoublepage
\include{ch4_v14.tex}
\clearemptydoublepage
\include{ch5_v14.tex}

\clearemptydoublepage
\include{ch6_v14.tex}
%
\renewcommand{\chaptername}{Appendix}
\appendix
\clearemptydoublepage
\include{apA_v14.tex}
\clearemptydoublepage
\include{apB_v14.tex}

\clearemptydoublepage
\backmatter

\include{res_gl_v14.tex}
\clearemptydoublepage
\include{fi_v14.tex}
\clearemptydoublepage
\include{bib_v14.tex}

\clearemptydoublepage

\end{document}

%% file: ack_v14.tex
\chapter*{Acknowledgements}
\thispagestyle{empty}%
The last few years have been the best of my life, and many of the good things that have happened have been (in)direct consequences of doing this PhD. When looking back, it is with a smile, and it is with the gratitude to a number of people who have played a role in this story, which is somewhere in between a \textit{Bildungsroman} and my \textit{Wanderjahre}. It is the story of the learning of the crafts of a noble profession, of the transition to (scientific) adulthood.

I cannot start without thanking everybody who has put direct effort into producing the content of this dissertation --- first and foremost, to my supervisors, Wences and Xabi. When writing these lines, I feel that you have done your job --- you have taught me all you had to, we are presenting a dissertation that we are proud of, and I am ready for the world out there. Xabi, thank you for being such an inspiration as a scientist and a person, for always listening to me, and for always knowing what to say. Wences, thank you for having believed in me back in 2016, for showing me with your example how I want to be when I `grow up', and for having made me learn to believe in myself and my work.

Thanks to my other co-authors of manuscripts. David, Jelle, Fer --- it is a pleasure to work with you. The same holds for Dominic, who deserves a special mention for having hosted me at the DKFZ in Heidelberg in 2021. That research stay was the light at the end of the pandemic, and I will never forget how satisfactory it was, both personally and scientifically.

Thanks to my \textit{tribunal de seguimento} ---to Rosa, Ricardo and Antonio (with Carmen in our memory)---, who have been providing their feedback all along the way. Thanks, too, to the CAPD (co-ordination of our PhD programme, especially to Alberto and Rosa), and to the two international referees of the dissertation. Thanks to each of you, for your work and help.

I am also honoured of having worked with Laila, Pablo J, Fer F \textit{et al.} during my time in Lab 15 of the IDIS. Thanks to the MODESTYA group and its members for their support of all sorts. Thanks to Diego B, Alejandra L, Arís and Dani C for all your advice and help. Thanks to the many nice people I have studied with. To name one, thanks to Xabi L, who deserves extra credit for having endured me so much.

Thanks to all the good professors and researchers that have ever inspired or helped me, in particular to Fernando A and Elena VC for their continuous support. Thanks to the USC PTXAS (i.e., the non-scientific staff), who are always there for us. Starting with Julia, Edi and the rest of the administration; and continuing with janitors, librarians, cleaners, the SNL --- thank you.

Thanks to two further extraordinary teachers, who have not qualified to these acknowledgements because of their job, but because of being my parents. My PhD would not have been possible without all your efforts of all kinds, nor without all the efforts that your parents made. Thanks to so many dear relatives for being in my life, in particular to Paquita and Antonio. Thanks, too, to everyone who was there during my childhood in the town of Bertamiráns.

There is not a font type large enough to highlight how important having supportive friends around me has been. Thanks to Javi MC, Lu V, Elizabeth C, Robert K, Olli B and Pablo G for being excellent human beings. Thanks, too, to everyone who made me feel at home during my time in southern Germany, like Jannik, Khwab, Chantal, Paul R, Dylan, Berkay, the MVD, the SV Nikar, the HLFF, the Sprachschule, the `Escándalo' gang, and the DKFZ Biostatistics Unit.

Thanks to many others who helped shape the good memories of these years --- Michelle A, Christian G, Alberto H, Chip, Nieve, Iza, Pablo P, Rodri, David K, David O, Ris, Gabi F, Alexey, Borja B, Tito S, Sebastian H, James, \textit{et al.} In a mix of the scientific and personal part, I am glad of having the friendship of members of QuinteScience like Paula C, Gemma ML, Marta D and Javi MF, to name a few; and of those who are still running the association or did so with me in the past. Thanks to the organisers and participants of science Olympiads, language exchanges, ESTALMAT, the `Jóvenes Investigadores' and `Arquímedes' research contests, the `Eladio Viñuela' summer school, and of every conference that has motivated and inspired me.

Thanks to all the medical professionals who helped me overcome many small health issues, such as Puri, María, Antonio, Elena, Mercedes, Rosa and Will. Thanks to all the people related to swimming and athletics, which have kept me healthy between my two ears. Thanks to those who made my time in TV sets and its aftermath better --- Ana, Rosana, Laura, Cris A, Moisés, Romay, Gabi R, Carlos A, Jero, Nacho M, Lucía M. And thanks to the Spanish TV for having provided me with some economic stability, something very unusual when doing a PhD in Spain.

I am aware that naming names (and having limited space available) may make any not named names feel excluded. Therefore, I just want to say thanks to all the good people who have played a role in the beautiful journey towards this point of time and space. We are now at the end of a stage, but I am confident that the best is yet to come. Thanks to all for so much.

\begin{flushright}
	\textit{Fer}
\end{flushright}

%% file: ch1_v14.tex
\fancyhead[LO]{\rightmark}
\fancyhead[RE]{\leftmark}
\renewcommand{\headerright}{\thechapter}

\setcounter{chapter}{-1}

\chapter{Introduction}
\thispagestyle{empty}
\label{ch0}
\graphicspath{{./fig_ch1/}}

In this chapter, we will be providing a general overview of the topic of the dissertation. It begins with some motivation and context of its broader area of knowledge (\S~\ref{BMDS}), to then get into more specific basic concepts (\S~\ref{gwas:snp}).

\section{Statistics, genomics and biomedical data science}\label{BMDS}

The past few years have been witness to unprecedented developments in the ways we produce, store and process information; much in the same fashion as the first industrial revolution was essentially a transformation of how energy was produced, stored and processed \citep{Schoelkopf}. This revolution, like the one in the 18th century, has only been possible with an enormous amount of progress in the science and technology related to the resource at the core of the revolution --- in today's world, data.

We are speaking about the science of data, or \emph{data science}. This discipline corresponds to the enlargement of statistics that John Tukey foresaw 60 years ago \citep{FoDA}, which ---he claimed--- is an empirical science (unlike e.g., mathematics), in the sense of having:
\begin{enumerate}[(a)]
	\item intellectual content,
	\item organization in an understandable form, and
	\item reliance upon the test of experience as the ultimate standard	of validity.
\end{enumerate}

This `new' science of data, as we see it in this century \citep{Donoho}, has as its core mathematical statistics, but it is also being driven by advances in computing (both in hardware and software), data visualisation, the spread of larger and more heterogeneous data, the growing interest in quantification across all fields of knowledge, and so forth. Regardless of how we call it, the \emph{science of those who learn from data} lies in the intersection between statistical formalism, computing skills, and knowledge of the domain of application. It is also the place where the `two cultures' of \citet{Breiman} meet --- where it may sometimes be useful to drop the assumption of any data-generating model and go for algorithmics, or where on the contrary it may be the goal to perform inference taking into account the data mechanism.

In parallel to the data revolution, the field of (human) biology has undergone its own transformation, evolving from a discipline that used to yield few observations of a small number of variables of similar nature, to a true high-throughput science that produces extremely large, often heterogeneous datasets, with the advent of the `omic' era \citep{MSMB}. This is part of the more general phenomenon of transitioning from \emph{data} to \emph{big data} --- we are producing, collecting and processing information at higher volume, velocity and variety than ever before \citep{Galeano}.

Nowadays, genetics scales up to studying the whole hereditary information in an individual, and we talk about the science of \emph{genomics}. What is more, it scales up to studying the whole hereditary information in a cohort of hundreds of thousands or even millions of individuals --- we are now in the biobank era. All this new information available to scientists at an ever-increasing pace has been possible due to the equally rapid advances in technology, with the cost of sequencing a human genome decreasing at an even higher rate than Moore's law, currently around $100\,000$ times cheaper than in the early times of the field, two decades ago \citep{genome:cost}.

Such efforts at the largest scale were pioneered by the United Kingdom, beginning two decades ago, and have produced hundreds of research articles with insight on a large variety of human traits \citep{Bahcall}. To name another example, in Galicia, where this PhD dissertation has been written, an ambitious project to sequence $400\,000$ of its inhabitants (roughly 15 \% of the population), for the advancement of precision healthcare, has been announced very recently.\footnote{ The Galician Genome Project was inaugurated on January 26th, 2024. See, for example, \url{https://web.archive.org/web/20240127173320/https://www.gciencia.com/saude/angel-carracedo-proxecto-xenoma-galicia-mais-ambicioso-mundo} or \url{https://yewtu.be/watch?v=DIx8w8bEKVM} .}

Even in 2024, with millions of sampled individuals across thousands of studies, there is still a large margin for progress \citep{Tam}, with more basic scientific discoveries to be made, and many lives to be improved through better healthcare, by means of precision medicine \citep{Korosok,Denny}. Genetics ---and more generally, all biomedical science--- still has much work ahead, and this poses the challenge of understanding the very high-dimensional and heterogeneous datasets that are produced in this field of knowledge. The challenge is complexity, of both the data and the science questions. In today's world, the best science that will come out of biomedical data will combine statistical methodology, computational skills, and sound knowledge of the domain of application. This is why we talk about a science of biomedical data, about \emph{biomedical data science} \citep{Levitt}.

In this dissertation, we present a dialogue, back and forth, between the statistical contributions and the applications to genetics, with due prominence given to computing and algorithmics too. After the previous general introduction to the broad field of knowledge, we now present our research goals and, with them, the contents and structure of the upcoming chapters of the thesis.

In the next section, we present more specific concepts of the kind of genetic studies and data we will be dealing with.

\section{Genome-wide association studies and single-nucleotide\\polymorphisms}\label{gwas:snp}

Genetic studies have given profound insight into the variability among individuals, for any imaginable trait of interest. Although some features of humans vary almost exclusively because of the environment, and others are inherited in a simple Mendelian fashion (i.e., the phenotype is linked to one or a few genes, each with a very high effect), the truth is that a vast majority of the variables one can measure or observe in a human being are complex traits (let them be risk of schizophrenia, height or blood levels of metabolites). The hereditary component of complex traits is highly polygenic --- it lays on a large number of variants along the genome, each of them with a small marginal effect \citep{Brandes}.

Today, it is widely accepted that complex human traits are mostly influenced by common genetic variants \citep{Lander2019,Chatterjee}, which altogether have turned out to polygenically explain a considerable proportion of the overall trait heritability \citep{10y}. That said, even today, when genetic data for millions of individuals across thousands of studies is available, there is still much progress to be done, with new variants to identify, heritability estimates to be refined, or predictions of phenotypes to be made. To give an idea of the extremely large sample sizes that are required, a recent work that found almost all the genetic component of human height based on common variants \citep{height:all:g} used an $n$ greater than 5 million.

The role of heredity in psychiatry has been studied for more than a century, since the times of Francis Galton, with \citet{Pearson:psych} not having ``the least hesitation'' in asserting its relevance. Today it is known that a majority of psychiatric disorders are multifactorial, complex traits. They occur as a result of a combination of genetic and environmental factors, none of which are necessary or sufficient on their own. Furthermore, the individual effect of each of them is generally trifling. More precisely, the genome can explain up to 80 \% of the susceptibility to suffer some of these diseases, like schizophrenia \citep{Sullivan:PGC}.

Over the last 15 years, genome-wide association studies (acronymically known as GWA studies or GWASs) have evolved from a promising, incipient idea to a reality that has revolutionised the way research in human trait genetics is conducted \citep{15y}. GWA studies involve genotyping many (human) individuals to perform tests of statistical hypotheses, estimations, predictions, and so forth; with the goal of advancing in the knowledge of the relationship between phenotype and genotype in human complex traits --- in fact the name derives from them originally being aimed exclusively at detecting \emph{associations} between phenotypes and genetic variants.

In these studies, the response variable measures a phenotypic characteristic of interest, which can be binary (typically, the indicator of presence/absence of a common disease) or continuous (e.g., physical measures of the human body, concentration of certain molecules in the blood, cardiological parameters, age at which a body development hallmark is achieved, and so forth). Whereas the binary scenario requires two groups (called \emph{cases} and \emph{controls}), for quantitative outcomes, a single large cohort of individuals will be enough \citep{Zhang,Cardon}. Chapter~\ref{ch3} will focus on the former of those two settings and Chapter~\ref{ch4}, on the latter. In Chapter~\ref{ch6}, on the other hand, we will diverge slightly from these settings, with the aim of studying categorical phenotypes.

From the beginning of the Human Genome Project in the 1990s, it was already a goal to sequence large cohorts of individuals to unravel the molecular causes of human trait variation \citep{Lander1996}. After all, since the days of Gregor Mendel, a key motivation of genetic research, if not the most relevant one, has been to understand the link between genotype and phenotype \citep{Zschocke,Brandes}.

Despite the diversity of existing technologies to analyse the human genome, GWAS databases often focus on \emph{single-nucleotide polymorphisms} (SNPs), which are the most simple and common form genetic variation among humans \citep{Tam}. Each SNP represents a change in one of the 3 billion letters \citep{HG} that form the ``book of life'', that is, the alternation between the reference nucleotide for that specific position of the human genome, and another nucleotide that can be observed in a proportion of people that is over a certain threshold (which traditionally used to be set as 1~\%, but that nowadays varies across different authors). This restriction on frequency means that only a few of the positions contain a SNP for a given population.

For instance, let us assume that a certain SNP can manifest as two possible nucleotides (i.e., it is \emph{biallelic}; as it is almost always the case) and that those are $A$ and $G$. In phylogenetic terms, it is common to refer to one of them as \emph{ancestral} and the other one as \emph{derived}, based on the evolutionary history of that position of the genome (or \emph{locus}). Regardless of that, each individual will have one of the three following genotypes in their (diploid) genome:$$\{AA,AG,GG\}.$$

In the following, we will also be referring to the alleles ($A$ or $G$) as \emph{major} or \emph{minor} depending on which of them is found at a greater frequency in humans. The acronym \emph{MAF} will also be seen more than once in this dissertation and it stands for \emph{minor allele frequency}, that is, $\textrm{MAF}:=f(A)$ whenever $f(A)\leq f(G)$, where $f$'s are for geneticist what statisticians call \emph{proportions} in a population.

Another important concept when working with GWASs and SNPs is the \emph{Hardy--Weinberg equilibrium} (HWE), a phenomenon that consists in the stability of the frequencies of both alleles ($A$ and $G$) and of each possible genotype ($AA$, $AG$ and $GG$) from generation to generation, under panmixia and in the absence of evolutionary influences \citep{Hardy,Weinberg}. Namely, if we denote $\theta:=f(A)$, the HW proportions for the genotypes are:
$$
f(AA)=\theta^2;\;\;\; f(AG)=2\theta(1-\theta);\;\;\; f(AA)=(1-\theta)^2.
$$
The concept of HWE will be very relevant during the sections devoted to the genetic motivation and applications in the upcoming chapters, and so will be that of \emph{linkage disequilibrium} (LD). Let us first consider two SNPs, and denote their alleles by $A/G$ and $C/T$, respectively. LD is defined as the phenomenon by which the joint distribution of both SNPs differs from the product of the marginals, as a result of a low recombination rate between the two loci, which in turn is almost invariably due to physical proximity within the same chromosome (with the exception being the so-called \emph{long-range LD}). In our example, under LD, we would have that the probability of observing the first SNP being an $A$ and the second, a $C$, differs from $f(A)$ times $f(C)$.

In Chapters~\ref{ch3}, \ref{ch4} and \ref{ch6} we will be reiterating these biological concepts as they arise, giving additional detail. They will feature most prominently in the introductory sections of those chapters, as well in the passages devoted to real data applications.

\clearemptydoublepage

\chapter{Research goals and techniques}
\thispagestyle{empty}
\label{ch1}

Once we have set the general context for our research, it is due time to present our objectives and methodology for the whole dissertation. Section~\ref{Obj} presents our main goals and research hypotheses, giving an overview of the structure of this document. On the other hand, in Section~\ref{Meth} we introduce the reader to the most important methodology that we will be using in the remaining chapters.

\section{Objectives}\label{Obj}


Statistical independence is a kind of relation between two traits of the units that are being studied, which corresponds to the informal concept of them not being associated in any way. Totally deterministic dependence is the opposite of statistical independence, with a continuum of intensity of association between those two extreme cases. Mathematically, two random variables are independent if, and only if, their joint probability distribution is the product of the marginals.

The main aim of this dissertation is to use nonparametric methods to derive new procedures for independence tests in general metric, semimetric and premetric spaces; in different high-dimensional settings that are of interest in complex disease genomics. This will turn out to lead to several meaningful applications, since most of the problems of interest in genetics (and in most empirical sciences) boil down to looking for associations between variables. All our biological research goals have to do with understanding the relationship between genes and the variability in phenotypic features (i.e., traits that are observable or, at least, measurable at the protein level).

In the genetic literature, it is almost universally assumed that genetic variants act in a linear and additive fashion, a simplification that does not necessarily hold. We will therefore focus on state-of-the-art methodology that is able to capture associations of any kind ---not only linear ones--- and that works in a large variety of marginal support spaces.

There will be four fundamental lines of work:

\begin{enumerate}[(i)]
\item Nonparametric independence tests between ternary random variables in a context of large-scale multiple tests in metric spaces.

\item Nonparametric independence tests between a continuous random variable, and a random element in a 3-point premetric space; and interpretations related to linear regression in a transformed space.

\item Extension of the methods based in distances and kernels of bullet points \textit{(i)} and \textit{(ii)} to the testing for association between discrete random variables with supports of arbitrary cardinality, and for the goodness of fit of a discrete random variable to a given distribution; and comparison with classical methodology for categorical data.

\item Computational implementation of the algorithms developed in \textit{(i)}--\textit{(iii)} and application to real datasets related to the genetics of complex disease, with emphasis on psychiatry.
\end{enumerate}
We now provide further detail about our goals and sketch the contents of the remainder of the dissertation, chapter by chapter.

\subsection*{Chapter~\ref{ch2}. Testing for statistical dependence in metric spaces and beyond}

When two random variables (or vectors) $X$ and $Y$ take values in Euclidean spaces, it is possible to define a measure that characterises their independence, called \emph{distance covariance} \citep{SRB}, defined as a weighted $L^2$ norm of the difference of the joint characteristic function and the product of the marginals. Distance covariance features a property that other, more conventional, population parameters do not --- it vanishes if \emph{and only if}, there is independence:
$$
\dCov(X,Y)=0\iff X, Y \;\text{independent.}
$$
This theory is part of a broader field of research known as the \emph{energy of data}, which one can extend to settings where the marginal supports are metric, semimetric or premetric spaces \citep{Jakobsen,Lyons}. The fundamental idea can be metaphorically described as considering data as celestial bodies that gravitate governed by statistical forces and energies \citep{TEOD}.

The previous tradition of testing comes from the more inference-based of the two cultures described by \citet{Breiman}. The more algorithmic scientists who learn from data, however, have had as one of their `hot topics' for the past two decades the \emph{learning with kernels}. Instead of transforming the complex, big and heterogeneous data with a distance, they resort to functions called \emph{kernels} with seemingly different properties to distances, but dual to them \citep{Sejdinovic}. Throughout the dissertation, we will use the word `distance' not only to refer to a metric, but to also encompass any premetric (including semimetrics). Finally, not only do these two traditions of independence testing converge, but they also do so with a third school --- the so-called Global Tests \citep{goeman2006testing,Jelle:Biomet}, which are locally most powerful tests in certain Gaussian regression models.

The goal of Chapter~\ref{ch2} is to provide the theoretical framework for the aforementioned methodology, to review literature in the topic and to serve as a gentle introduction to the statistical machinery used in the rest of the thesis. This chapter features most of the contents of \citet{F:review}, with some parts being based on \citet{F:gwas} and \citet{F:epistasis}. Some of it is not to be seen in any preprint nor work in the editorial process by us, and we are presenting it for the first time with this dissertation.

\subsection*{Chapter~\ref{ch3}. Testing for gene-gene interaction in complex disease}

Despite many research efforts of the scientific community since the beginning of the 21st century, the heritability of common human diseases is not yet fully understood at the molecular level \citep{Brandes}. One of the missing pieces of the puzzle is the lack of insight into genetic interactions, which are considered by biologists as one of the most relevant unknown `parameters' in the human complex disease `equation' \citep{Manolio}.

A limitation of the existing methodology for detecting such gene-gene interactions is that it generally assumes linearity in the effects. There is no biological reason for doing so, whence we argue that distance covariance (which characterises general statistical independence, not only the linear one) can be an interesting approach to this problem.

The large size of genomic datasets is going to make it computationally unwieldy to perform the distance-based hypothesis testing as it is usually found in the literature, that is, with permutations. For this reason, we will work out the explicit asymptotic null distribution of the empirical distance covariance in our case.

As it is always the case with novel statistical methodology, we will use simulated examples to check that we control the type I error, and that we have reasonable power. We will also compare our results with the well-known competing method BOOST \citep{BOOST}. Finally, we will apply all of the above to a genomic dataset generated by us in the context of a study of schizophrenia, a disorder with a high socioeconomic burden and therefore of strategic research interest.

The contents of this chapter are collected in \citet{F:epistasis}.

\subsection*{Chapter~\ref{ch4}. Testing for gene-phenotype associations in human complex traits}

A main goal of genomic studies is to detect variants in the human DNA that are significantly associated with the variability of a quantitative (phenotypic) trait of interest \citep{Tam}. Much in the same way that the previous chapter aimed at detecting genotype-genotype interactions, this one focuses on phenotype-genotype associations.

Once again arguing that genetic variants do not necessarily act in an additive manner, we want to develop statistical testing procedures based on distance covariance in premetric spaces. This opens the door to selecting a priori the kind of genetic model that it is desired to test for, by simply choosing a distance that reflects it; which is of high interest from a biological perspective.

Further interpretations would be possible from the point of view of mathematical statistics, by exploring the dualities introduced in Chapter~\ref{ch2} --- testing with a distance is equivalent to testing with the kernel induced by that distance, which in turn is equivalent to to testing in a linear Gaussian regression model in the space of the so-called \emph{features} of that kernel.

A relevant research question will be to identify all distances that make sense for the purposes of this chapter, and to see how the testing procedure and its interpretation vary depending on the choice of the premetric (within the family of all feasible ones).

For the sake of computational efficiency, given the size of genomic datasets, we will aim at not approximating the null distribution of our test statistics with permutations, instead exploiting the simplicity of the marginal spaces and their geometries to derive closed-form formulae that can be quickly evaluated in practice.

Once that adequate statistical methodology has been developed, and assuming that it controls type I error and shows reasonable power in simulations, the idea will be to analyse a relevant biological dataset. To stay on-topic with psychiatric genetics, we will study continuous biomarkers of disorders related to alcoholism, namely the serum levels of liver enzymes (which are biomarkers of cirrhosis). We will compare the results of our approach with those provided by one of the most commonly applied tests, namely the linear one in PLINK \citep{PLINK}.

The contents of this chapter are collected in \citet{F:gwas}.

\subsection*{Chapter~\ref{ch6}. Comparison of distance-based tests with classical methodology for categorical data}

Categorical data is ubiquitous in biomedical research, so it is very relevant to wonder what happens to the methodology of Chapter~\ref{ch3} when the marginal spaces have an arbitrary (finite) number of categories. It will be interesting to compare the form of the resulting test statistic with well-known classics such as Pearson's and the likelihood ratio (i.e., the test statistic for the $G$-test).

On the other hand, another common hypothesis regarding arbitrary categorical variables that one may want to test for, is the goodness of fit to a discrete distribution. We will also aim at testing for this hypothesis with distances, using results by \citet{ED}.

In both settings, we would once more like to compute explicit (asymptotic) null distributions, in order to avoid the time-consuming permutations. Moreover, we want to see if our tests perform well in simulations, in terms of significance and power, both in absolute terms and relatively to competing methods.

Finally, we want to study relevant problems in psychiatric genetics with the aforementioned techniques. For the independence testing, we will try to see if the genome has significant predictive ability of the severity of schizophrenia. On the other hand, when it comes to applying the goodness-of-fit test, we will check if a cohort of schizophrenia patients shows deviations from the genotype frequencies that are expected in the general population, and see if the genetic variants that present such deviations are associated with this psychiatric disorder (otherwise, any positives would tend to indicate putative problems in genotyping). 

The contents of this chapter are collected in \citet{F:categorical}.

\vspace*{1cm}

Moreover, the dissertation includes the discussion and conclusions of our research, in \textbf{Chapter~\ref{ch7}}, which also features an overview of open problems and promising lines of future work. \textbf{Appendix~\ref{apA}} contains technical information, consisting in mathematical proofs and other theoretical remarks. \textbf{Appendix~\ref{apB}} gives the necessary information for reproducing our simulation studies and real data experiments. Additionally, we provide a `\textbf{Further information}' section, which outlines the research output of this dissertation, and a \textbf{summary} of the thesis in Galician, which is the official language of the university where we have been conducting the doctoral studies. The thesis concludes by listing its references, in the \textbf{Bibliography} section.

\section{Methodology}\label{Meth}

We now present the methodology of our research. For a more organised structure, we have opted for creating three subsections, devoted to statistics (\S~\ref{stat:meth}), genetics (\S~\ref{gen:meth}) and computing (\S~\ref{comp:meth}). That subdivision should merely be seen as a convenient way of arranging the text, rather than as a rigorous taxonomy, since there is non-empty overlap between each pair of those three categories.

\subsection{Statistical methodology}\label{stat:meth}

The general guiding principle of each of our research problems follows the same pattern:
\begin{enumerate}
	\item developing novel statistical methodology and proving desirable theoretical properties;
	\item confirming the performance of our proposal with simulated data;
	\item analysing real data to address a biological research question;
	\item discussing the results and drawing some conclusions.
\end{enumerate}

The main problem we are interested in is, as previously indicated, the testing for general statistical independence in a nonparametric setting. All the testing procedures that we will use have in common that there is some form of statistical distance between distributions underneath, as the ones defined by \citet{Lindsay} and reviewed in detail by \citet{Markatou}.

First and foremost, we will make use of distance covariance \citep{SRB}, which is defined as a weighted $L^2$ norm of the difference between the product of the marginal characteristic functions and the joint one. It vanishes if \emph{and only if} there is statistical independence, whence we say that it captures all kinds of associations. This theory was initially conceived for random variables with support in Euclidean spaces, but it we will be of greater interest to us in its extension to metric, semimetric, and premetric spaces.

Let us first clarify what we mean with those terms. Given a set $\mZ\neq\emptyset$, we say that a function $$\rho: \mZ \times \mZ \to [0,+\infty[$$ is a \emph{premetric} or \emph{distance} if it is symmetric in its arguments and satisfies $\rho(z,z)=0$ for all $z\in\mZ$. Then $(\mZ,\rho)$ is called a \emph{premetric space} or \emph{distance space} \citep[][\S~3.1]{Deza:Laurent}. If $\rho$ satisfies, in addition, the `identity of indiscernibles', that is,
$$\rho(z,z')=0\Rightarrow z=z'$$
for all $z,z'\in\mZ$, we speak of \emph{semimetric} $\rho$ and \emph{semimetric space} $(\mZ,\rho)$, as defined by \citet{Sejdinovic}. Moreover, if the following inequality (known as the \emph{triangle inequality}) also holds for every $z_1,z_2,z_3\in\mZ$:
$$\rho(z_1,z_3)\leq\rho(z_1,z_2)+\rho(z_2,z_3),$$
$\rho$ is called a \emph{metric} and $(\mZ,\rho)$, a \emph{metric space}. We would like to highlight that, although the words `premetric' and `semimetric' are widely used to refer to functions with less properties than a metric, their exact meaning varies across different bibliographical sources. This is why we introduce these definitions and nomenclature now, and we will be consistent with them throughout the dissertation.

Distance covariance is dual to the \emph{Hilbert--Schmidt independence} criterion (HSIC), which instead of being based on premetrics, has kernels as its backbone. These functions ---often defined as being symmetric and positive definite--- have widely been used in statistical learning theory since the inception of the discipline \citep{Genton}. The kernel approach leads to a distance between distributions (known as \emph{maximum mean discrepancy} or MMD) that is dual to the one associated with distance covariance (known as $\mathcal E$-distance or \emph{energy distance}), with both of them being examples of the well-known class of conditionally negative distances between probability distributions \citep{Markatou}.

Those two philosophies to independence testing are, in turn, equivalent to certain locally most powerful tests in Gaussian regression models, known as Global Tests \citep{DJ}. These three testing schools will be presented in detail in Chapter~\ref{ch2}, together with some important concepts for their understanding.

It is also of interest to note that the test statistics are, in all these cases, either $U$- or $V$-statistics. Each of them has as its asymptotic null distribution a quadratic form of standard Gaussian variables, with coefficients given by the eigenvalues of a certain operator. In the literature, this distribution is seldom used for any practical purpose, and permutation testing \citep{Goeman:2021} is used instead. This has the advantage of not having to deal with the highly non-trivial estimation of the coefficients, but it is extremely computationally-intensive and barely feasible in the ultra-high-dimensional setting of genomics (as it will be illustrated in the upcoming chapters).

\subsection{Genetic methodology}\label{gen:meth}

We now present the genetic methodology of the dissertation, namely the specific tools we use for working with GWAS data, both for retrieving information and for conducting analyses.

We resorted to the well-known genetic software package PLINK v1.9 \citep{PLINK}, calling it from R \citep{R}, for a number of tasks:
\begin{itemize}
	\item testing for genetic interaction with preexisting methodology \citep{BOOST};
	\item classical linear association testing between a genetic variant and a continuous phenotype;
	\item reading, writing and transforming file formats in which genotype data are usually stored and shared;
	\item principal component analysis of genotype matrices;
	\item quality controls of GWAS data, based on the proportion of missing data and divergence from the HWE;
	\item management of genetic variants in LD.
\end{itemize}

Additionally, we greatly benefited from web apps and online resources made available by the biomedical community and public institutions from around the globe, including:

\begin{itemize}
	\item UCSC Genome Browser <\url{https://genome.ucsc.edu/}>, to visualise the human genome in high-resolution, exploring the genetic variants in a specific region;
	\item ENSEMBL <\url{https://www.ensembl.org}> and its Biomart <\url{https://mart.ensembl.org}>, for the retrieval of functional annotations of specific genetic variants;
	\item NHGRI--EBI GWAS Catalog <\url{https://www.ebi.ac.uk/gwas/}>, which is a comprehensive collection of GWAS summary statistics for human complex traits;
	\item GTEx portal <\url{https://www.gtexportal.org}>, for the determination of which tissue type each genetic variant is expressed in;
	\item SynGO portal <\url{https://www.syngoportal.org}>, to annotate which genes are related to synaptic transmission, a brain function of interest in psychiatry;
	\item NCBI dbGaP <\url{https://www.ncbi.nlm.nih.gov/gap/}>, through which we obtained individual-level SNP data for some of our applications;
	\item Genome Aggregation Database (gnomAD) <\url{https://gnomad.broadinstitute.org/}>, which aggregates the available information for the interpretation of individual human genetic variants, and that we used mainly for comparing allele frequencies across different ancestries;
\end{itemize}

Our data came either from work of our group in psychiatric genetics \citep{Galicia,Facal:Scand} or from the aforementioned public repository dbGaP \citep{dbGaP}, of the USA National Institutes of Health. And we will be introducing the details on each dataset and the relevant information on the studies that produced them in the central chapters of this document.

\subsection{Computational methodology}\label{comp:meth}

To achieve our research goals, not only should the statistical methods be powerful and efficient, but the same holds for the computational resources.

When attempting to approach the testing in Chapter~\ref{ch3} with resampling strategies (which is common practice in distance covariance literature), we encountered that conventional computers were insufficient to perform this task. We therefore needed to resort to supercomputer \emph{Finisterrae II} in the Galician Supercomputing Centre (CESGA) and use very fast implementations of low-level computations, and to do this in a parallel architecture. However, all this turned out to be more an illustration of what one should not do when developing statistical tools that are user-friendly to practitioners, rather than a useful approach per se. As part of those first attempts, we also wrote some code in the programming language C and in Matlab, but none of them lead to any meaningful insight.

The results that are featured in this dissertation were produced using mostly R \citep{R} as the programming language, with a few lines of code written in Python. Some functions of R's \texttt{tidyverse} \citep{Hadley} were used for graphics and some of \texttt{data.table} were helpful when dealing with large datasets \citep{data.table}. We also used R to call PLINK \citep{PLINK} from it and that way have a cleaner data analysis pipeline.

Our implementation of distance covariance methods (which is available in Appendix~\ref{apB}) is self-contained, in the sense that it does not include nor depend on preexisting code for distance covariance. However, we would like to provide a brief overview on what software is available for energy statistics. The first implementation was made by the authors of the original articles on these techniques, as the R package \texttt{energy}, and it has recently been updated \citep{energy:package}. This was already quite computationally efficient, by doing most of the numerical crunching in C and leaving R as a wrapper. The algorithm by \citet{Huo} provided further advance in speed. \citet{dcortools} developed a comprehensive collection of functions for distance covariance estimation and testing for R; whereas \citet{Torrecilla} did the same for Python.

In Chapters~\ref{ch3}, \ref{ch4} and \ref{ch6} we will be providing an overview of the computational methodology used, with further details for reproducibility in Appendix~\ref{apB}.


%% file: ch2_v14.tex
\fancyhead[LO]{\rightmark}
\fancyhead[RE]{\leftmark}
\renewcommand{\headerright}{\thechapter}

\chapter{Testing for statistical dependence in metric spaces and beyond}
\thispagestyle{empty}
\label{ch2}
\graphicspath{{./fig_ch2/}}

The \emph{energy of data} \citep{SR:book} is a branch of mathematical statistics that has been recently developed and it includes the characterisation of statistical independence in Euclidean spaces via an association measure called \emph{distance correlation}. In Section~\ref{SRB} we will introduce those concepts in the Euclidean setting, to then extend the paradigm to metric spaces (\S\S~\ref{Intro_Jakobsen}--\ref{Test dcor}). The extension of distance covariance to metric spaces is a non-trivial issue, to which we will devote a few pages, in an effort to provide the readership with a gentle introduction to the abstract mathematical concepts that this theory requires.

We also provide an overview of the duality of this approach and the kernel techniques popular in the machine learning community (namely with the Hilbert--Schmidt independence criterion and associated methodology) and with the theory of locally most powerful tests in Gaussian regression models (the so-called Global Tests). Sections~\ref{hsic} and~\ref{gt} are devoted to such topics; as well as to remarking other important ideas, such as the extension from metric to semi- and premetric spaces, and the concept of feature maps (which will be ubiquitous in the upcoming chapters).

An earlier version of the contents of most of this chapter are available as a stand-alone technical report \citep{F:review}, which is a self-contained introduction to distance covariance in metric spaces. Sections~\ref{gdc:semimetric},~\ref{hsic} and~\ref{gt} can also overlap with preprints by us, namely in the introductory sections of \citet{F:gwas} and \citet{F:epistasis}.

\section{Distance covariance in Euclidean spaces}\label{SRB}

When two random elements (vectors) $\mathbf{X}$ and $\mathbf{Y}$ are Euclidean-space-valued (let $\mathbf{X}$ be $L$-dimensional and $\mathbf{Y}$ be $M$-dimensional, for $L,M\in\Zplus$), it is possible to construct an association measure that characterises their independence called \emph{distance correlation} \citep{SRB,Brownian}. In order to be able to define it, we should first introduce distance covariance, which is no more than a weighted $L^2$ norm of the difference of the joint characteristic function and the product of the marginals:
\[\dCov(\mathbf{X},\mathbf{Y}):=\norm{\varphi_{\mathbf{X},\mathbf{Y}}-\varphi_\mathbf{X}\varphi_\mathbf{Y}}_w\equiv\sqrt{\int_{\mathbb{R}^L\times\mathbb{R}^M}|\varphi_{\mathbf{X},\mathbf{Y}}(\bs,\bt)-\varphi_\mathbf{X}(\bs)\varphi_\mathbf{Y}(\bt)|^2w(\bs,\bt)\,\mathrm{d}\bs\,\mathrm{d}\bt}\text{;}\]\setcitestyle{square}%
where $w$ is a weight function which is dependent of the dimension of the Euclidean spaces in which the supports of $\bX$ and $\bY$ are contained (and it has a property of uniqueness \citep{Uniqueness}):\setcitestyle{round}%
\[
w(\bs,\bt):=
\frac{\Gamma\left(\frac{L+1}{2}\right)}{\left(\norm{\bs}\sqrt{\pi}\right)^{L+1}}\:
\frac{\Gamma\left(\frac{M+1}{2}\right)}{\left(\norm{\bt}\sqrt{\pi}\right)^{M+1}}
,\;(\bs,\bt)\in\mathbb{R}^L\times\mathbb{R}^M\text{;}
\]
where $\Gamma(\cdot)$ denotes the complete gamma function and, as usually:
\[
\varphi_{\bX}(\bs):=\E\left[e^{i\inner{\bs}{\bX}}\right],\:\bs\in\mathbb{R}^L; \;\;\;\;\;\;
\varphi_{\bY}(\bt):=\E\left[e^{i\inner{\bt}{\bY}}\right],\:\bt\in\mathbb{R}^M\text{.}
\]

In the two equations above, we are assuming that $\inner{\cdot}{\cdot}$ and $\norm{\cdot}$ denote the standard inner product of a Euclidean space and the norm derived from it.

Analogously to its non-distance counterpart, when calculating the distance covariance of a random variable and itself, one obtains the square of a measure of spread called \emph{distance standard deviation}. Both distance variance and its square root are meaningful measures of dispersion, as studied by \citet{dSD}, which can be applied to random vectors of arbitrary (finite) dimensionality.

Logically, distance correlation is defined as the quotient of distance covariance and the product of distance standard deviations (as long as none of the latter vanish):
\[\dCor(\mathbf{X},\mathbf{Y}):=\frac{\dCov(\mathbf{X},\mathbf{Y})}{\sqrt{\dCov(\mathbf{X},\mathbf{X})\dCov(\mathbf{Y},\mathbf{Y})}}\text{,}\]
and so it has no sign. It is an improved version of the square of Pearson's correlation because:
\begin{itemize}
	\item It has values in [0,1]. This is unsurprising, since $\R$ is totally ordered and, as such, one can only move ``leftwards'' or ``rightwards'' and so the sign of Pearson's correlation expresses this structure. However, this notion is not valid in Euclidean spaces of arbitrary dimensionality.
	\item It is zero if \underline{and only if} $X$ and $Y$ are independent (thus, its interest). This means that, unlike with Pearson's, the nullity of \emph{dCor} ---or of \emph{dCov}--- is equivalent to independence. Therefore, testing for values of distance correlation ---or of distance covariance--- significantly different of zero is the same as searching for dependency, and it is a search for dependencies of all kind (not only linear ones, as with ordinary correlation).
\end{itemize}

It is also frequent to see $\dCov$ and $\dCor$ represented by the calligraphic letters $\V$ and $\mathcal R$ in the literature and we will be using both notations over the forthcoming sections and chapters. When in need of specifying the metrics being used, we will add a subindex to indicate it.

\begin{sloppypar}
	Notwithstanding the convoluted initial definition of \emph{dCov}, its sample version can easily be computed. Given a paired sample $$(\mathbf{X}_1,\mathbf{Y}_1),\ldots,(\mathbf{X}_n,\mathbf{Y}_n)\text{ IID }(\mathbf{X},\mathbf{Y})\text{;}$$ let $a_{ij}:=\dx(\mathbf{X}_i,\mathbf{X}_j)$ be the Euclidean distances in $\spx=\R^L$ between the observed $\mathbf{X}$'s with indices \mbox{$i,j\in\{1,\ldots,n\}$}. Then, the doubly centred distances are:
	\begin{equation} A_{ij}:=a_{ij}-\frac{1}{n}\sum_{k=1}^na_{ik}-\frac{1}{n}\sum_{k=1}^na_{kj}+\frac{1}{n^2}\sum_{k,l=1}^na_{kl} \label{eq:center} \end{equation}
	If $\{b_{ij}\}_{i,j}$ and $\{B_{ij}\}_{i,j}$ are analogously defined for $\{\mathbf{Y}_i\}_i$, the empirical distance covariance is simply the nonnegative real number whose square is:
	\begin{equation} \widehat{\dCov}_n(\mathbf{X},\mathbf{Y})^2:=\frac{1}{n^2}\sum_{i,j=1}^n A_{ij}B_{ij} \label{eq:empdc} \end{equation}
	so that it is, indeed, a covariance of distances. And hence distance correlation is a correlation of distances, with the latter name being found in some the earliest literature in the topic \citep{SRB}.
\end{sloppypar}

The estimator in (\ref{eq:empdc}) is reminiscent of the following alternative representation of $\dCov^2$:
\[\dCov(\mathbf{X},\mathbf{Y})^2\!=\!\E\Big[\Big(\dx(\mathbf{X},\mathbf{X}')-\E\lbrace \dx(\mathbf{X},\mathbf{X}'')\rbrace-\E\lbrace \dx(\mathbf{X}',\mathbf{X}''')\rbrace+\E\lbrace \dx(\mathbf{X}'',\mathbf{X}''')\rbrace\Big)\]
\[\times\Big(\dy(\mathbf{Y},\mathbf{Y}')-\E\lbrace \dy(\mathbf{Y},\mathbf{Y}'''')\rbrace-\E\lbrace \dy(\mathbf{Y}',\mathbf{Y}''''')\rbrace+\E\lbrace \dy(\mathbf{Y}'''',\mathbf{Y}''''')\rbrace\Big)\Big]\text{,}\]
which is valid as long as moments of order $2$ are finite \citep[][Remark 4.6]{Jakobsen}. In the equation above, primed letters refer to IID copies of the corresponding random vector; while $\dx$ and $\dy$ denote the Euclidean metrics in $\spx=\R^L$ and $\spy=\R^M$, respectively.

The previous identity also holds when the double centring is only performed in one of the marginals (e.g., the $\bY$'s):
\[\dCov(\mathbf{X},\mathbf{Y})^2\!=\!\E\Big[\dx(\mathbf{X},\mathbf{X}')\,\Big(\dy(\mathbf{Y},\mathbf{Y}')-\E\lbrace \dy(\mathbf{Y},\mathbf{Y}'')\rbrace-\E\lbrace \dy(\mathbf{Y}',\mathbf{Y}''')\rbrace+\E\lbrace \dy(\mathbf{Y}'',\mathbf{Y}''')\rbrace\Big)\Big]\text{;}\]
which is well-defined as long as all first moments are finite. We will see in the following that simple matrix algebra justifies that for computing the empirical distance covariance, it also suffices to doubly centre one of the marginals.

Whenever $\{\mathbf{X},\mathbf{Y}\}$ are independent and have finite first moments, the asymptotic distribution of a scaled version of the preceding statistic is a linear combination of independent chi-squared variables with one degree of freedom. More precisely:
\[n\:\widehat{\dCov}_n(\mathbf{X},\mathbf{Y})^2\distrilim\sum_{j=1}^\infty \lambda_j Z_j^2\text{,}\] 
where $\{Z_j\}_j$ are IID $\Normal(0,1)$ and where $\{\lambda_j\}_j\subset\mathbb{R}^+$. Such quadratic forms arise often when dealing with $U$- and $V$-statistics.

Unfortunately, knowing the form of the theoretical null distribution is often not helpful in practice. As a result, almost all the distance correlation literature we are aware of resorts to resampling techniques when it comes to approximating the critical values for the independence test. They generally design the resampling scheme based on the information that the null hypothesis provides, which in this setting (i.e., independence) leads to permutation testing. The theoretical quadratic form is not used by most authors due to the difficulty of estimating the $\lambda_j$'s, since one would need to deal with an abstract linear operator and obtain its non-zero eigenvalues (potentially, an infinity of them), all this under no model assumptions (note that our setting is nonparametric). See, for example \citet{Jakobsen} or \citet{SRB}.

A common element to Chapters~\ref{ch3}, \ref{ch4} and \ref{ch6} of this dissertation will be that, starting from certain data types that are of interest in genetics, we will see that the geometries that come up when studying independence with distance-based techniques are such that we are able to explicitly compute a closed form for the asymptotic null distribution and show that it performs well in practice, both in terms of the results obtained but also computationally (which is crucial when working in high-throughput sciences like genomics).

\section{Context and notations}\label{Intro_Jakobsen}
\subsection{General statement of the nonparametric problem of independence}
Let $\xd$ and $\yd$ be two arbitrary separable metric spaces (the need for separability is dealt with in~\ref{Sep}). The random element $Z=(X,Y)$ is defined over a probability space $(\Omega,\mathcal F,\Prob)$ and has values in $\spx\times\spy$, with its distribution being:
\[\theta:\Borel{\spx\times\spy}\longrightarrow[0,1]\text{.}\]
The following notation will be used for the marginal distributions:
\begin{itemize}
	\item $X\sim\mu:=\theta\comp\pi_1^{-1}$, marginal over $\spx$; where $\pi_1:(x,y)\in\spx\times\spy\flechita x\in\spx$.
	\item $Y\sim\nu:=\theta\comp\pi_2^{-1}$, \hspace*{.5mm}marginal over $\spy$; \hspace*{.5mm}where $\pi_2:(x,y)\in\spx\times\spy\flechita y\in\spy$.
\end{itemize}
\begin{sloppypar}
	Thus, the nonparametric test of independence for $X$ and $Y$ consists in testing ${H_0:\theta=\mu\times\nu}$ versus ${H_1:\theta\neq\mu\times\nu}$. For the sake of clarity, it is important to note that the product $\mu\times\nu$ is defined conventionally: it is the only measure in $\Borel{\spx}\tensor\Borel{\spy}$ so that
\end{sloppypar}
\[(\mu\times\nu)(A\times B):=\mu(A)\nu(B);\;A\in\Borel{\spx},\:B\in\Borel{\spy}\text{.}\]

\subsection{Separability of marginal spaces}\label{Sep}
The first perquisite of assuming the separability of $\spx$ and $\spy$ is that, this way, the $\sigma-$algebra that their topological product generates is simply the product $\sigma-$algebra:
\[\Borel{\spx\times\spy}=\Borel{\spx}\tensor\Borel{\spy}:=\sigma\brc{A\times B:A\in\Borel{\spx},B\in\Borel{\spy}}\text{.}\]\setcitestyle{square}%

\begin{sloppypar}
	This equality is useful by itself (e.g., it is crucial to the proof of Lemma 3.10 in \citet{Jakobsen}), but its most important corollary is that it guarantees that the metrics of the marginal spaces are jointly measurable: for $\spz\in\brc{\spx,\spy}$, $\dz$ is {$\Borel{\spz}\tensor\Borel{\spz}/\Borel{\mathbb{R}}-$measurable}. This, in turn, is what ensures that the Lebesgue integrals that appear in the definition of distance covariance (\S~\ref{Def_dcov}) are defined. A counterexample would be $\spx:=\mathbb{R}^{\mathbb{R}}$, equipped with the discrete metric. This is a particular case of \emph{Nedoma's pathology}
	(see \citet[Proposition  21.8]{Schechter} and \citet[Example 6.4.3]{Bogachev} for further details), which states that the diagonal set $\{(x,x):x\in\spx\}$ is not in $\Borel{\spx}\tensor\Borel{\spx}$ when the cardinality of $\spx$ is greater than that of the continuum.
\end{sloppypar}\setcitestyle{round}%

Finally, separability is explicitly used in the proofs of some important properties of distance covariance \citep[Theorem 4.4 and Lemma 5.8]{Jakobsen}, which indicates that it is not an ungodly hypothesis.

The original article that presented distance correlation in metric spaces \citep{Lyons} was oblivious of the crucial role of separability in the theory.

\subsection{Signed measures}\label{Signed}
The map $\mu:\Borel{\spx}\longrightarrow\mathbb{R}$ is said to be a finite signed (Borel) measure, and it is denoted $\mu\in \M(\spx)$, if and only if $|\mu|$ is a finite measure. For each $\mu\in \M(\spx)$, there is a \emph{Hahn--Jordan decomposition} and it is essentially unique \citep[Theorem 3.2.1]{Billingsley} or, in other words, it is possible to find a couple of nonnegative measures $\mu^{\pm}\in \M(\spx)$ so that
\[\mu=\mu^+-\mu^-\]
and a partition of the space $\spx=\spx^+\disjoint\spx^-$ satisfying:
\[\mu^+(\spx^-)=0=\mu^-(\spx^+)\text{;}\]
which is the same as saying that $\mu^+$ and $\mu^-$ are \emph{orthogonal} or \emph{mutually singular}. This allows to naturally define (Lebesgue) integrals with respect to signed measures. For $f:\spx\longrightarrow\mathbb{R}$ measurable,
\[\intx f\dmu:=\intx f\dmu^+-\intx f\dmu^-\text{;}\]
which is well-defined whenever $f$ is integrable with respect to $|\mu|=\mu^++\mu^-$.

On the other hand, it will also be necessary to integrate with respect to product measures. To begin with, consider $\nu\in \M(\spy)$, with Hahn--Jordan decomposition given by $(\spy^{\pm},\nu^{\pm})$. Then:
\begin{itemize}
	\item $\mu^+\times\nu^++\mu^-\times\nu^-$ is a (nonnegative) measure with support $(\spx^+\times\spy^+)\disjoint(\spx^-\times\spy^-)$;
	\item $\mu^+\times\nu^-+\mu^-\times\nu^+$ is a (nonnegative) measure with support $(\spx^+\times\spy^-)\disjoint(\spx^-\times\spy^+)$.
\end{itemize}
Because of their disjoint supports, the aforementioned two measures are mutually singular and, consequently \citep[corollary of Theorem 6.14]{Rudin}, they form the Hahn--Jordan decomposition of $\mu\times\nu$:
\[\mu\times\nu=(\mu^+\times\nu^++\mu^-\times\nu^-)-(\mu^+\times\nu^-+\mu^-\times\nu^+)\text{.}\]
Thus, the integral of a Borel-measurable function $h:\spx\times\spy\longrightarrow\mathbb{R}$ with respect to $\mu\times\nu$ is:
\[\int h\wrt\mu\times\nu=\int h\wrt\mu^+\times\nu^++\int h\wrt\mu^-\times\nu^--\int h\wrt\mu^+\times\nu^--\int h\wrt\mu^-\times\nu^+\text{;}\]
which entails that $\mathcal L^1(\mu\times\nu)$ is the intersection of the four function spaces $\mathcal L^1(\mu^{\pm}\times\nu^{\pm})$.

On the last equation, the integration sets were omitted, as it is superfluous to underscore that it is the largest possible one (in this case, $\spx\times\spy$). This notation abuse, taken from \citet{Lyons}, is among the few ones that will be used on the present chapter, while the ones that caused mistakes and confusion on Lyons' article (and even in its corrigendum \setcitestyle{square}\citep{Oops}) will be avoided. \setcitestyle{round}

The last relevant remark about the integration with respect to the product of signed measures is that they satisfy a generalised Fubini--Tonelli theorem \citep[\S~3.3]{Bogachev}:
\[\forall\:h\in\mathcal L^1(\mu\times\nu),\;\int h\wrt\mu\times\nu=\iint h\dmu\dnu=\iint h\dnu\dmu\text{.}\]

\subsection{Regularity of a measure}\label{Moments}

The following result, known as the $c_r-$\emph{inequality}, will be useful in the upcoming development of this chapter. Its proof can be found in Appendix~\ref{apA}.
\begin{proposition}\label{prop:cr_ineq}
For any $\alpha,\beta,r\in\Rplus$: $(\alpha+\beta)^r\leq c_r(\alpha^r+\beta^r)$, where
\[c_r=\begin{cases}1,&r<1\\2^{r-1},&r\geq1\end{cases}\text{.}\]
\end{proposition}

At this point, we can introduce the concept of regularity of a signed measure. Let $\mu\in \M(\spx)$. Then, $\mu$ is said to have finite moments of order $r$, and it is written as $\mu\in \M^r(\spx)$, if and only if
\[\exists\:o\in\spx,\;\int\dx(o,x)^r\damu(x)<+\infty\text{.}\]
Applying the $c_r-$inequality, it is straightforward to see that when the condition above holds, it does so for any origin:
\[\mu\in M^r(\spx)\eqv\forall o\in\spx,\;\int\dx(o,x)^r\damu(x)<+\infty\text{.}\]
In addition, a signed measure on a product of two spaces $\theta\in \M(\spx\times\spy)$ is said to belong to $\M^{r,r}(\spx\times\spy)$ if both its marginals have finite moments of order $r$. Finally, the subindex $1$ will be used as a notation for probability measures:
\[\M_1(\spx):=\big\lbrace{\mu\in \M(\spx):\:\mu\geq0,\:\mu(\spx)=1}\big\rbrace;\]
\[\M_1^r(\spx):=\M^r(\spx)\cap \M_1(\spx);\;\;\;\;\M_1^{r,r}(\spx\times\spy):=\M^{r,r}(\spx\times\spy)\cap \M_1(\xy)\text{.}\]

\section{Formal definition of \emph{dcov}}\label{Def_dcov}

\com{I am not sure if I am introducing a mess with $\dCov^2=\V^2=\dcov$. See how well this meshes with the following chapters.}

The previous section set the theoretical framework in which speaking of distance covariance makes sense, thus solving some inconsistencies of \citet{Lyons}. This will enable to define the operator \emph{dcov} rigorously, simplifying and illustrating the explanations by \citet{Jakobsen}.

\subsection{Integrability of the metric}\label{Int}
In order to define \emph{dcov}, it is important to keep in mind that:
\begin{equation}\label{eq:integrab:metric}
\forall\:\mu_1,\mu_2\in \M^1(\spx):\;\dx\in\mathcal L^1(\mu_1\times\mu_2)\text{.}
\end{equation}
This is a consequence of Fubini and the triangle inequality:
\[\int\dx\wrt|\mu_1|\times|\mu_2|\leq\int\dx(x,o)\wrt|\mu_1|\times|\mu_2|(x,x')+\int\dx(o,x')\wrt|\mu_1|\times|\mu_2|(x,x')=\]
\[=|\mu_2|(\spx)\int\dx(x,o)\wrt|\mu_1|(x)+|\mu_1|(\spx)\int\dx(x,o)\wrt|\mu_2|(x)<+\infty\text{.}\]

\subsection{Expected distances and some inequalities}\label{aD}
The definition of distance covariance involves doubly centred distances (\S~\ref{dmu}), but first the various expected values that are to appear should be checked to be well-defined.

For $\mu\in \M^1(\spx)$, the following function maps each point $x\in\spx$ to its expected distance to a random element with distribution $\mu$:
\begin{align*}
	\amu:\;&\spx\longrightarrow\mathbb{R}\\
	&x\longmapsto\int\dx(x,x')\dmu(x')
\end{align*}
Obviously, it is well-defined. On top of that, it is $|\mu|(\spx)-$Lipschitzian  (and, therefore, continuous):
\[\forall x,x'\in\spx:\;|\amu(x)-\amu(x')|\leq\int|\dx(x,z)-\dx(x',z)|\damu(z)\leq\]
\[\leq\int\dx(x,x')\damu(z)=|\mu|(\spx)\dx(x,x')\text{.}\]
On the other hand, recalling Equation~\eqref{eq:integrab:metric}, the integral $D(\mu)$ is always a real number:
\[D(\mu):=\int\amu\dmu=\int\dx\dmu\times\mu\text{.}\]
The four inequalities in the following proposition can easily be derived from the previous results (as shown in Appendix~\ref{apA}) and they will be very useful hereinafter.

\begin{proposition}\label{prop:ineq_amu}
For $\mu\in \M_1^1(\spx)$ and $x,y\in\spx$:
\begin{enumerate}
	\item $D(\mu)\leq2\amu(x)$;
	\item $D(\mu)\leq\amu(x)+\amu(y)$;
	\item $\dx(x,y)\leq\amu(x)+\amu(y)$;
	\item $\amu(x)\leq\dx(x,y)+\amu(y)$.
\end{enumerate}
\end{proposition}

\subsection{Doubly centred distances}\label{dmu}
For $\mu\in \M^1(\spx)$, the doubly $\mu-$centred version of $\dx$ is:
\begin{align*}
	d_\mu:\;&\spx\times\spx\longrightarrow\mathbb{R}\\
	&(x_1,x_2)\flechita\dx(x_1,x_2)-\amu(x_1)-\amu(x_2)+D(\mu)
\end{align*}

This modification of $\dx$, in general, is not a metric; although it is always continuous (since $\dx$, $\amu$, $\pi_1$ and $\pi_2$ are) and, in particular, Borel-measurable. Moreover, it is important to note that, when writing $d_\mu$, there is no explicit reference to the metric space over which this map is defined. Such an abuse of notation makes formulae easier to read and write without creating any misunderstanding. That is not the case of some abbreviations by Lyons, such as the usage of $d:=\dx$ and $d:=\dy$, which mistakenly suggests that there is a need for $\spx$ and $\spy$ to share the same metric structure, which is an unnecessary restriction for the theory that would render some interesting applications impossible, like the ones in Chapter~\ref{ch4}.

The last remarkable property of $d_\mu$ is given by the following integrability theorem, which is proven in Appendix~\ref{apA}.
\begin{theorem}\label{th:dmu_L2}
For any $\mu,\mu_1,\mu_2\in \M^1_1(\spx)$, it holds that:
\[\;d_\mu\in\mathcal L^2(\mu_1\times\mu_2)\text{.}\]
\end{theorem}

\subsection{The association measure \emph{dcov}}\label{dcov}
In the context of metric spaces, distance covariance is defined as:
\[\dcov(\theta):=\int_{(\spx\times\spy)^2}d_\mu(x,x')d_\nu(y,y')\dtheta^2\left((x,y),(x',y')\right),\;\theta\in M_1^{1,1}(\spx\times\spy)\text{;}\]
where, once again, $\mu:=\theta\comp\pi_1^{-1}$ and $\nu:=\theta\comp\pi_2^{-1}$.

For the above expression to be finite, it suffices to have finite first moments, as stated in the following theorem, which is proven in Appendix~\ref{apA}.

\begin{theorem}\label{th:dcov:well_def}
	For every $\theta\in\M^{1,1}_1(\spx\times\spy)$, $\dcov(\theta)$ is well-defined.
\end{theorem}

The different integrability checks that have been conducted so far allow to write \emph{dcov} in terms of expected values. Taking $X\sim\mu\in \M_1^1(\spx)$ and $Y\sim\nu\in \M_1^1(\spy)$, with joint distribution $\theta:=\Prob\comp\binom{X}{Y}^{-1}$, their distance covariance is given by:
\[\dcov(X,Y)\overdef{Abuse}\dcov(\theta)=\E[d_\mu(X,X')d_\nu(Y,Y')]=\]
\[=\E\Big\lbrace\Big(\dx(X,X')-\E[\dx(X,X')|X]-\E[\dx(X,X')|X']+\E[\dx(X,X')]\Big)\cdot\]
\[\cdot\Big(\dy(Y,Y')-\E[\dy(Y,Y')|Y]-\E[\dy(Y,Y')|Y']+\E[\dy(Y,Y')]\Big)\Big\rbrace\text{;}\]
where primed letters refer to independent and identically distributed (IID) copies of the corresponding random element.

Finally, note that \emph{dcov} is always an association measure, in the sense that it vanishes under independence:
\[\dcov(\mu\times\nu)=\int d_\mu d_\nu\wrt(\mu\times\nu)^2\overeq{Fubini}\]
\[=\left(\int\dx\dmu^2-2\int\amu\dmu^2+\int D(\mu)\dmu^2\right)\left(\int\dy\dnu^2-2\int\anu\dnu^2+\int D(\nu)\dnu^2\right)=\]
\[=[D(\mu)-2D(\mu)+D(\mu)][D(\nu)-2D(\nu)+D(\nu)]=0\text{.}\]
Moreover, under certain conditions, \emph{dcov} is nonnegative and it can be rescaled into the interval $[0,1]$ (see~\ref{Def_dcor}), becoming a normalised association measure \citep[pages 375--376]{Bishop}.

\section{Distance covariance in negative type spaces}\label{TN}
The fact that:
\[\theta=\mu\times\nu\implica\dcov(\theta)=0\text{,}\]
makes it natural to wonder which spaces ensure that the reciprocal implication also holds. The answer is: \emph{strong negative type} spaces, since in them $\dcov(\theta)$ is an injective function of $\theta-\mu\times\nu$.

In order to explain this, negative type spaces will be firstly introduced (\S~\ref{Def_TN}), as they are the ones in which \emph{dcov} admits the aforementioned representation (although injectivity is not guaranteed). Then the strong version of this condition will be defined (\S~\ref{TNF}) and a pivotal result will be put forward --- strong negative type is not only a necessary condition for \emph{dcov} to characterise independence, but it is also sufficient (with a little exception, by no means restrictive).

\subsection{Metric spaces of negative type}\label{Def_TN}
The concept of negative type is not a recent invention \citep{Wilson} and it has recently been enjoying its ``second youth'': firstly, because of its role in computational algorithmics (\citeauthor{Deza:Laurent}, \citeyear{Deza:Laurent}, \S~6.1.; \citeauthor{Naor}, \citeyear{Naor}) and, more recently, in relation to the \emph{energy of data} \citep{TEOD} and learning theory for reproducing kernel Hilbert spaces (RKHSs), as studied by \citet{gretton2008kernel}, \citet{Sejdinovic} and many others. The concept of RKHS will be studied more in detail in Section~\ref{hsic}.

The metric space $\xd$ is said to be of negative type if and only if:
\[\forall n\in\Z^+;\:\forall \textcolor{blue}{x},\textcolor{red}{y}\in\spx^n:\:2\sum_{i,j=1}^n\dx(\textcolor{blue}{x_i},\textcolor{red}{y_j})\geq\sum_{i,j=1}^n[\dx(\textcolor{blue}{x_i},\textcolor{blue}{x_j})+\dx(\textcolor{red}{y_i},\textcolor{red}{y_j})]\text{.}\]
The analytic expression above has the following geometrical interpretation --- given $n$ red points and as many blue ones, the sum of the distances between the $2n^2$ ordered pairs of the same colour is not greater than the corresponding sum for different colours. Moreover, this condition can be stated in another way, that is apparently more general, which is the
\emph{conditionally negative definiteness} of the metric. However, both are actually equivalent (which can be checked by taking repetitions of the points and recalling that $\Q$ is dense in $\mathbb{R}$):
\[\forall n\in\N;\:\forall x\in\spx^n;\:\forall\alpha\in\mathbb{R}^n,\sum_{i=1}^n\alpha_i=0:\;\sum_{i,j=1}^n\alpha_i\alpha_j\dx(x_i,x_j)\leq0\text{.}\]
\setcitestyle{square}%
This is not to say that negative type metric spaces are the ones in which the metric acts like a negative definite function (such as the ones thoroughly studied by \citet{Klebanov} and \citet{Berg}).\setcitestyle{round}%
However, an equivalent definition in terms of the negative definiteness of a certain kernel exists. Namely, $\xd$ is a negative type space if and only if there is a point $o\in\spx$ so that the \emph{absolute antipodal divergence}
\begin{equation}\label{eq:antipodal}
d_o(x,y):=\dx(x,o)+\dx(y,o)-\dx(x,y),\;(x,y)\in\spx^2
\end{equation}
is positive definite. In the above, the word \emph{kernel} is being used to denote any function on a non-empty Cartesian square which is symmetric in its arguments. This notion will be introduced in more detail in Section~\ref{hsic} and used extensively throughout the dissertation from that point on. In our definition, all kernels will be symmetric and positive definite. An example of this is the function $d_o$ defined above, to which we will give more meaning in Section~\ref{hsic}.

There are many familiar examples of negative type spaces, like the Euclidean ones and, more generally, all Hilbert spaces, as it will be explained next, in Section~\ref{Hilbert}.

\subsection{Representation in Hilbert spaces}\label{Hilbert}
Now some results involving Hilbert spaces are to be presented. For the sake of simplicity, assume that the scalar field is $\mathbb{R}$ in every case, but, as a general rule, every statement that will be made is also true for $\mathbb{C}$, \textit{mutatis mutandi}. This can be proven by realifying or complexifying \citep[pages 132--135 of][]{Jakobsen}, according to the case.

It will be necessary to integrate functions $f:\spx\longrightarrow\h$ which have a Hilbert space as their codomain. Had $\spx$ not been assumed to be separable (see~\S~\ref{Sep}), as in \citet{Lyons}, the spaces $\h$ that arise later on would not necessarily be separable, which would only allow to perform weak integration \citep{Pettis}, and not the strong one \citep{Bochner}. Given $\mu\in \M(\spx)$, if $f$ is a scalarly $\mu-$integrable, then the integral $I\in\h$ of $f$ with respect to $\mu$ exists and is unambiguously defined by its commutativity with respect to every map of the dual space $\h^*$:
\[I=\int_{\spx}f\dmu\eqv\forall h^*:\h\longrightarrow\mathbb{R}\text{ linear and continuous},\;h^*(I)=\int_{\spx}(h^*\comp f)\dmu\text{.}\]
Hereinafter, every Hilbert space that will arise is going to be separable, which means that Pettis integrals are Bochner integrals.

After these technical remarks, \emph{Schoenberg's theorem} (\citeauthor{Schoenberg:1937}, \citeyear{Schoenberg:1937} and \citeyear{Schoenberg:1938}) can be stated. It characterises negative type spaces $\xd$ as those such that $\xsd$ can be isometrically embedded into a Hilbert space:
\[\exists\;\h\text{ Hilbert space};\:\exists\:\varphi:\spx\longrightarrow\h;\:\forall x,y\in\spx:\;\norm{\varphi(x)-\varphi(y)}_{\h}^2=\dx(x,y)\text{.}\]

For a simple proof, using the absolute antipodal divergence (see Equation~\eqref{eq:antipodal}), refer to \citet[Theorem 3.7]{Jakobsen}, which corrects \citet{Lyons}. Regardless of this, Schoenberg's theorem ensures that the separability of the original metric spaces (\S~\ref{Sep}) is inherited by all the Hilbert spaces that arise. Before the Hilbert space representation of \emph{dcov} can be tackled, the \emph{barycentre operator} has to be defined: given an isometric map $\varphi:\xsd\longrightarrow\h_1$ (like the one on the preceding theorem) and $\mu\in \M^1(\spx)$, the following Pettis integral always exists
\[\beta_\varphi(\mu):=\int_\spx\varphi\dmu\in\h_1\]
and it is called \textit{barycentre}, because it is the average of a $\h_1$-field over $\spx$ according to the distribution given by $\mu$ (thus resembling the geometrical idea of a gravity centre). In fact, if $X\sim\mu\in \M_1^1(\spx)$,
\[\beta_\varphi(\mu)=\E[\varphi(X)]\text{.}\]
On the other hand, if $\psi:\ysd\longrightarrow\h_2$ is also isometric, the barycentre of the tensor product $\varphi\otimes\psi$ for $\theta\in \M^{1,1}(\spx\times\spy)$ is defined as:
\[\beta_{\varphi\tensor\psi}(\theta):=\int_{\spx\times\spy}(\varphi\tensor\psi)\dtheta\in\h_1\tensor\h_2\text{.}\]

More importantly, if $(\mu,\nu)$ are the marginals of $\theta\in \M_1^{1,1}(\spx\times\spy)$, the following equality holds:
\[\dcov(\theta)=4\norm{\beta_{\phi\tensor\psi}(\theta-\mu\times\nu)}_{\h_1\tensor\h_2}^2\text{.}\]
In conclusion, \emph{dcov} characterises independence in those spaces in which $\beta_{\phi\tensor\psi}$ is injective, which are going to be dealt with right below.

\subsection{Strong negative type space}\label{TNF}\setcitestyle{square}%
If $\xd$ has negative type, one can derive the following inequality (whose proof is remarkably long \citep[Lemma 3.16]{Jakobsen}):
\[\forall\mu_1,\mu_2\in \M^1_1(\spx):\:D(\mu_1-\mu_2)\leq0\text{.}\]
On top of that, if the operator $D$ separates probability measures (with finite first moments) in $\xd$, that space is said to have \emph{strong} negative type:
\[D(\mu_1-\mu_2)=0\eqv\mu_1=\mu_2\text{.}\]

The extended Schoenberg's theorem shows the equivalence of the strong negative type of $\xd$ and the existence of an isometric map $\varphi:(\spx,\sqrt{\dx})\longrightarrow\h_1$ such that $\beta_\varphi$ is injective. Furthermore, for strong negative type $\spx$ and $\spy$, two isometric maps $\varphi:(\spx,\sqrt{\dx})\longrightarrow\h_1$ and $\psi:(\spy,\sqrt{\dy})\longrightarrow\h_2$ can be found so that $\beta_{\varphi\tensor\psi}:\M^{1,1}(\spx\times\spy)\longrightarrow\h_1\tensor\h_2$ is injective. As a result, whenever $\spx$ and $\spy$ have strong negative type, the equivalence
\[\dcov(X,Y)=0\eqv X,Y\text{ independent}\]
holds for any random element $Z=(X,Y):\Omega\longrightarrow\spx\times\spy$.
\setcitestyle{round}%

Thus, the strong negative type of marginal spaces is a \emph{sufficient} condition for the equivalence above to hold, but is it also \emph{necessary}? The answer is \emph{yes}, but with the exception of a ``pathological'' case.

If $\yd$ was not of strong negative type (symmetrically for $\spx$), it is indeed possible to find $\theta\in \M^{1,1}_1(\spx\times\spy)$ so that:
\[\dcov(\theta)=0\text{ and, at the same time, }\theta\neq(\theta\comp\pi_1^{-1})\times(\theta\comp\pi_2^{-1})\text{;}\]
whenever $\min\brc{\card\spx,\card\spy}>1$. Such $\theta$ can be constructed as follows:
\[\theta:=\frac{\delta_{x_1}\times\nu_1+\delta_{x_2}\times\nu_2}{2}\text{;}\]
where $\nu_1,\nu_2$ are two different measures in $M_1^1(\spy)$ so that $D(\nu_1-\nu_2)=0$, and $x_1,x_2\in\spx$ are two distinct points. For each $x\in\spx$, $\delta_x\in M^1(\spx)$ denotes point mass at $x$.

This way, the aforementioned pathological case consists of one of the marginal spaces being a singleton. Such exception is not a restriction because, whenever $\card\spy=1$ (symmetrically for $\spx$), $\dcov\equiv0$ (since $d_\nu\equiv0$) and every $\theta\in \M_1^{1,1}(\spx\times\spy)$ is the product of its marginals. To see this last part, note that:
\[\spy=\{y\}\implica\Borel{\spy}=\brc{\emptyset,\{y\}}=\brc{\emptyset,\spy}\text{.}\]
And consequently, for $B\in\Borel{\spy}$,
\[\forall A\in\Borel{\spx},\;\theta(A\times B)=\begin{cases}\theta(A\times\emptyset)=\theta(\emptyset)=0=\mu(A)\nu(\emptyset)\\
	\theta(A\times\spy)=\theta\left[\pi_1^{-1}(A)\right]\equiv\mu(A)=\mu(A)\nu(\spy)\end{cases}\text{;}\]
and so $\theta=\mu\times\nu$. 
This analytical result is the formalisation of the intuitive notion that, if a random element $Y$ has constantly a certain value, the observations of any other random $X$ are bound to be independent of those of $Y$.

After the previous theoretical discussion, the interest of identifying practical examples of strong negative type spaces is clear. With regard to this, for most real data applications, it suffices to know that all separable Hilbert spaces have strong negative type. Although this is an unsurprising result, its proof is by no means straightforward \citep[pages 49--60]{Jakobsen}.

\section{Distance correlation in metric spaces}\label{dcor}

\subsection{The association measure \emph{dcor}}\label{Def_dcor}

Like previously, let $(X,Y)\sim\theta\in \M_1^{1,1}(\spx\times\spy)$ have marginals $(\mu,\nu)$, where $\xd$ and $\yd$ are two separable metric spaces. Then, the following inequalities hold:
\[|\dcov(X,Y)|\leq\sqrt{\dvar(X)\dvar(Y)}\leq D(\mu)D(\nu)\text{;}\]
where $\dvar(X):=\dcov(X,X)$. If, in addition, $\xd$ and $\yd$ have negative type:
\[\dcov(X,Y)=4\norm{\beta_{\varphi\times\psi}(\theta-\mu\times\nu)}^2_{\h_1\tensor\h_2}\geq0\text{.}\]
In this context, \emph{distance correlation} (for metric spaces) is defined as:
\[\dcor(X,Y):=\frac{\dcov(X,Y)}{\sqrt{\dvar(X)\dvar(Y)}}\in[0,1]\]
whenever the denominator is non-zero. For nondegenerate cases, this will not be a matter of concern, for $\dvar(X)$ only reaches the extreme values of its range $[0,D(\mu)^2]$ when it is concentrated on one or two points (respectively):
\[\dvar(X)=0\eqv\exists\:x\in\spx,\;\mu=\delta_x\text{ ``$\mu-$almost surely'';}\]
\[\dvar(X)=D(\mu)^2\eqv\exists\:x,x'\in\spx,\;\mu=\frac{\delta_x+\delta_{x'}}{2}\text{ ``$\mu-$almost surely''.}\]
When $\dvar(X)=0$, as in the Euclidean case, $\dcor(X,Y):=0$.

\subsection{\emph{dcor} in Euclidean spaces}\label{dcor=dCor^2}
In has already been shown that \emph{dcor} has range $[0,1]$ and is zero if and only if there is independence, which recapitulates the property for Euclidean spaces (\S~\ref{SRB}). Indeed, it is possible to prove (via the Hilbert space representations introduced in~\ref{Hilbert}) that, when $\xd$ and $\yd$ are (finitely dimensional) Euclidean spaces, the value of distance correlation of \S~\ref{Def_dcor} \citep{Lyons} equals the square of the one in \S~\ref{SRB} \citep{SRB}:
\[\dcov(X,Y)=\dCov(X,Y)^2;\;\dcor(X,Y)=\dCor(X,Y)^2\text{.}\]
For $\theta\in \M_1^{2,2}(\spx\times\spy)$, $\dcov(X,Y)$ becomes a product of expectations. By expanding it and simplifying, one can easily get the generalisation of Remark 3 in \citet{SRB} to general metric spaces:
\[\dcov(X,Y)\!=\!\E[\dx(X,X')\dy(Y,Y')]+\E[\dx(X,X')]\E[\dy(Y,Y')]-\]\[-2\E[\dx(X,X')\dy(Y,Y'')]\text{.}\]
In conclusion, \emph{dcov} satisfactorily extends \emph{dCov} squared.

\section{Nonparametric test of independence in metric spaces}\label{Test dcor}

\subsection{Kernel associated to \emph{dcov}}\label{h}
The following map will be key to the construction of the sample version of \emph{dcov}:
\begin{align*}
	h:\;&(\spx\times\spy)^6\flecha\mathbb{R}\\
	&\big((x_i,y_i)\big)_{i=1}^6\flechita f_{\spx}(x_1,x_2,x_3,x_4)\,f_{\spy}(y_1,y_2,y_5,y_6)\text{;}
\end{align*}
where, for $\spz\in\brc{\spx,\spy}$,
\[f_{\spz}(\bz):=\dz(z_1,z_2)+\dz(z_3,z_4)-\dz(z_1,z_3)-\dz(z_2,z_4),\:\bz\in\spz^4\text{.}\]
\setcitestyle{square}%
The functions $f_{\spz}$ and $h$ are clearly measurable and proving their integrability can be accomplished by sequentially deriving inequalities from the triangle inequality (see pages 148--150 of \citet{Jakobsen} for the correction of the attempt by \citet{Lyons}). Integrating these functions is pretty straightforward. Firstly, for $f_{\spx}$:\setcitestyle{round}%
\[\int_{(\xy)^2}f_{\spx}(x_1,x_2,x_3,x_4)\dtheta^2((x_3,y_3),(x_4,y_4))=
\]
\[=\dx(x_1,x_2)-\amu(x_1)-\amu(x_2)+D(\mu)\equiv d_\mu(x_1,x_2),\;(x_1,x_2)\in\spx^2\text{;}\]
where $\theta\in M_1^{1,1}(\xy)$ has marginals $(\mu,\nu)$. Given that the same (\textit{mutatis mutandi}) holds for $f_{\spy}$,
\[\dcov(\theta)=\int_{(\xy)^2}d_\mu(x_1,x_2)d_\nu(y_1,y_2)\dtheta^2\left((x_1,y_1),(x_2,y_2)\right)=\int_{(\xy)^6}h\dtheta^6\text{.}\]
This means that, if $(X_i,Y_i)_{i=1}^6$ denotes a vector that contains $6$ random elements that are independent and identically distributed to $(X,Y)\sim\theta$,
\[\dcov(\theta)=\E\left[h\left((X_i,Y_i)_{i=1}^6\right)\right]\]
and, consequently, its sample version is a $V$-statistic.

\subsection{Empirical distance covariance}\label{UV}

For $n\in\mathbb{Z}^{+}$, the following notation will be used for the \emph{empirical measure} associated to a certain sample $\brc{(X_i,Y_i)}_{i=1}^n\iid(X,Y)\sim\theta$: 
\[\theta_n:=\frac{1}{n}\sum_{i=1}^n\delta_{(X_i,Y_i)}:\Omega\longrightarrow M_1^{1,1}(\xy)\text{.}\]
A few routine computations yield that the natural estimator
\[\dcovh(\theta):=\dcov(\theta_n)\]
is, unsurprisingly, the $V$-statistic with (nonsymmetric) kernel $h$:
\[\dcov(\theta_n)=\frac{1}{n^6}\sum_{i_1=1}^n\cdots\sum_{i_6=1}^n h\left((X_{i_\lambda},Y_{i_\lambda})_{\lambda=1}^6\right)\equiv V_n^6(h)\text{.}\]
We want to highlight that we are now using the word \emph{kernel} to refer to a function that is used to define a $U$- or $V$-statistic, since it is customary to use that word instead of `function'. Unfortunately, this coincides with the common term for referring to the positive definite functions on which the Hilbert--Schmidt independence criterion (Section~\ref{hsic}) is based. We will be helping the reader tell both apart by denoting with $h$ the former and with $k$ the latter.

Now coming back to $V_n^6(h)$, it is logical to consider the analogous $U$-statistic as an alternative estimator of $\dcov(\theta)$, which will be shown to require less stringent conditions to behave satisfactorily than $\dcov(\theta_n)$. For $n\geq7$, let:
\[\tilde U_n^6(h):=\frac{1}{6!\binom{n}{6}}\sum_{\brc{i_\lambda}_{\lambda}\subset[1,n]\cap\mathbb{Z}\text{ different}}h\left((X_{i_\lambda},Y_{i_\lambda})_{\lambda=1}^6\right)\text{;}\]
where the tilde indicates that this is not a $U$-statistic \textit{sensu stricto}, but rather one built upon a kernel that is nonsymmetric. To correct this, let $\bar h$ be the symmetrisation of $h$:
\[\bar h(z):=\frac{1}{6!}\sum_{\sigma\in S_6}h\left(z_{\sigma(j)}\right)_{j=1}^6\equiv\frac{1}{6!}\sum_{\sigma\in S_6}h(z_\sigma),\:z\in(\xy)^6\text{;}\]
where $S_6:=\brc{\sigma:[1,6]\cap\mathbb{Z}\flecha[1,6]\cap\mathbb{Z}:\;\sigma\text{ bijective}}$ is the symmetric group of order $6$.
So $\tilde U_n^6(h)$ is the $U$-statistic based on $\bar h$:
\[\tilde U_n^6(h)=\frac{1}{\binom{n}{6}}\sum_{i_1<\ldots<i_6}\bar h\left((X_{i_\lambda},Y_{i_\lambda})_{\lambda=1}^6\right)\text{.}\]
The analogous for the $V$-statistic also holds:
\[\forall\:\sigma\in S_6,\;\dcov(\theta_n)\equiv V_n^6(h)=\int_{(\xy)^6}h(z)\dtheta_n^6(z)\overeq{Fubini}\]
\[=\int_{(\xy)^6}h(z)\dtheta_n^6(z_{\sigma^{-1}})\overeq{ACOV}\int_{(\xy)^6}h(z_\sigma)\dtheta_n^6(z)=V_n^6(\bar h)\]
and the same arguments can prove that $\dcov(\theta)=\int\bar h\dtheta^6$.

Now that the usual symmetric kernels can be used, it is possible to resort to the \emph{strong law of large numbers} (SLLN) for $U$-statistics \citep{Hoeffding} to infer that, for $\theta\in \M_1^{1,1}(\xy)$,
\[\tilde U_n^6(h)\aslim\dcov(\theta)\text{;}\]
where ``a.s.'' stands for \emph{almost surely}.
\begin{sloppypar}
	\citet{Lyons} mistook the hypotheses of the aforementioned Hoeffding theorem for the ones of the SLLN for $V$-statistics \citep[page 274]{Gine:Zinn}. The weakest conditions under which the SLLN for $V$-statistics holds in this context are: ${\theta\in \M_1^{5/3,5/3}(\xy)}$ \citep[Theorem 5.5]{Jakobsen}. In other words, the finiteness of moments of order $\frac{5}{3}$ suffices to ensure the asymptotic consistency of $\dcov(\theta_n)$:
\end{sloppypar}
\[V_n^6(h)\aslim\dcov(\theta)\text{.}\]

\subsection{Null distribution of the test statistic}\label{NullDist}
If $\theta\in M_1^{1,1}(\xy)$ is the product of its marginals $(\mu,\nu)$ and these are nondegenerate, the asymptotic distributions of the estimators introduced in~\ref{UV} are:
\begin{align*}
	nV_n^6(h)&\distrilim\sum_{i=1}^\infty\lambda_i(Z_i^2-1)+D(\mu)D(\nu)\text{;}\\
	n\tilde U_n^6(h)&\distrilim\sum_{i=1}^\infty\lambda_i(Z_i^2-1)\text{;}
\end{align*}
where $\brc{Z_i}_{i\in\Nstar}\iid\Normal(0,1)$ and where $\brc{\lambda_i}_{i\in\Nstar}$ are the eigenvalues (with multiplicity) of the linear operator $S:\mathcal L^2(\theta)\flecha\mathcal L^2(\theta)$ that maps $f$ into $S(f):\xy\flecha\mathbb{R}$, which is defined as:
\[S(f)(x,y):=\int_{\xy}d_\mu(x,x')d_\nu(y,y')f(x',y')\dtheta(x',y'),\:(x,y)\in\xy\text{.}\]

The original attempt of proving the result for the $V$-statistic \citep{Lyons} included some incorrect arguments to conclude that $\sum_{i=1}^\infty\lambda_i=D(\mu)D(\nu)$. \citet{Oops} states that the previous identity does hold as long as both marginal spaces have negative type, which leads to the exact same asymptotic distribution that \citet{SRB} had derived.

Anyhow, this cannot be brought to practical usefulness (as in~\ref{SRB}), since the eigenvalues $\brc{\lambda_i}_i$ depend on $\theta$ (unknown) and cannot be easily estimated. The most logical approach to this is, once again as in~\ref{SRB}, a resampling strategy. One way of arguing for this procedure would be to summon the results of \citet{Arcones:Gine}, that ensure that approximating the thresholds for the test statistic via naive bootstrap leads to a consistent resampling technique, as $\bar h$ satisfies the integrability condition required by those authors.

\section{Semimetric spaces and beyond}\label{gdc:semimetric}

We will now once more define distance covariance in this section, but now with the goal of providing a very simple framework ---albeit slightly less intuitive--- that is extendable to semimetric and premetric spaces, and that allows for a direct parallelism with what happens in kernel spaces (Section~\ref{hsic}).

Given random vectors $\bX \in \R^L$ and $\bY \in \R^M$ with finite first moments, their distance covariance can be expressed as:
\begin{equation}
	\label{eq:dcovalt}
	\V^2(X,Y) = \E\Big [\|\bX -\bX'\| \left\{ \|\bY -\bY' \| - \|\bY-\bY'' \|  
	- \|\bY'-\bY''\|+\|\bY''-\bY'''\| \right\} \Big],
\end{equation}
where primed letters denote IID copies of $(\bX,\bY)$ and $\|\cdot\|$ is the Euclidean norm.

In the following, we will work with the \emph{generalised distance covariance} (GDC), in the terminology of \citet{Sejdinovic}, that is, we will extend $\V$ to metric, semimetric and even premetric spaces.

Given a set $\mZ\neq\emptyset$, we say that a function $\rho: \mZ \times \mZ \to [0,+\infty[$ is a premetric if it is symmetric in its arguments and satisfies $\rho(z,z)=0$ for all $z\in\mZ$. Then $(\mZ,\rho)$ is called a premetric space, as already stated in Chapter~\ref{ch1}.

A premetric space $(\mZ,\rho)$ is said to have negative type if, for all $n \geq 2$, $z_1,\ldots,z_n \in \mZ$ and $a_1,\ldots,a_n \in \R$ with $\sum_{i=1}^n a_i = 0$, it holds that
$$
\sum_{i,j=1}^n a_i a_j \rho(z_i,z_j) \leq 0.
$$
Now let $\rho_{\mathcal{X}}$ and $\rho_{\mathcal{Y}}$ denote premetrics of negative type on $\mX$ and $\mY$, which are assumed to be probability spaces for certain $\sigma$-algebras.  Then, the \emph{(generalised) distance covariance} of two random elements $X\in\mX$ and $Y\in\mY$ such that $\E |\rho_\mX(X,X')+\rho_\mY(Y,Y')| < \infty$ is defined as:
\begin{equation}
	\label{eq:gendcov2}
	\V^2_{\rho_\mX,\rho_\mY}(X,Y) = \E\Big[\rho_\mX(X,X') \left\{\rho_\mY(Y,Y')-\rho_\mY(Y,Y'')\\ -\rho_\mY(Y',Y'')+\rho_\mY(Y'',Y''')\right\}\Big],
\end{equation}
which is clearly reminiscent of Equation~\eqref{eq:dcovalt}.

The $\V^2$ statistic is never negative and it vanishes under independence. The converse (i.e., nullity of the GDC implies independence) holds if and only if the premetrics $\rho_\mX$ and $\rho_\mY$ are of \emph{strong} negative type. A premetric $\rho$ of negative type on $\mZ$ is said to be strong if, for every pair of probability measures $P,Q$ on $\mZ$, the following equivalence holds:
$$
P=Q \iff \int_{\mZ\times\mZ}\rho\,\mathrm{d}(P-Q)^2=0.
$$
This is the same as stating that the \emph{energy distance} \citep{Energy:dist} is able to separate probability distributions on $\mZ$.

Consider now IID joint samples $\bX = (X_1,\ldots,X_n)$ and $\bY = (Y_1,\ldots,Y_n)$ of $X$ and $Y$, and define the distance matrices for each sample:
$$\bD^{\bX}:=\left( \rho_\mX(X_i,X_j)\right) _{n\times n} , \quad \quad  \bD^{\bY} := \left( \rho_\mY(Y_i,Y_j)\right)  _{n\times n}.$$
Then their doubly centred versions, $\tilde{\bD}^{\bX}$ and $\tilde{\bD}^{\bY}$, can be computed as follows:
\begin{equation}\label{eq:doublecenter}
\tilde{\bD}^{\bX} = (\bI_n-\bH) \bD^{\bX} (\bI_n-\bH) , \quad \quad  \tilde{\bD}^{\bY} = (\bI_n-\bH) \bD^{\bY} (\bI_n-\bH);
\end{equation}
where $\bH= \frac{1}{n} \boldsymbol{1} \boldsymbol{1}^t\in\mathbb R ^{n\times n}$ and $\boldsymbol{1}$ is an $n$-vector of ones. With this notation, a consistent empirical estimator of~\eqref{eq:gendcov2} is:

\begin{equation}
	\label{eq:dcovemp}
	\widehat{\V}^2_{\rho_\mX,\rho_\mY}  (X, Y) =
	\frac{1}{n^2} \sum_{i,j=1}^{n} \tilde{\bD}^{\bX}_{ij} \tilde{\bD}^{\bY}_{ij}  = 
	\frac{1}{n^2} 
	\tr ({\bD}^{\bX} \tilde{\bD}^{\bY})=
	\frac{1}{n^2} 
	\tr (\tilde{\bD}^{\bX}{\bD}^{\bY}).
\end{equation}

The last two versions of the formula in the equation above, which are due to $\bI-\bH$ being idempotent and matrix products commuting inside the trace operator, were not featured in the earliest distance covariance literature \citep{SRB}. Nevertheless, they are quite interesting, since they allow for a very large gain in computation speed in practice, by reducing the number of times the most time-consuming steps have to be performed. This is specially true when one thinks not only about estimating $\V^2$, but to then test for independence with permutation testing --- the compact formula means that one only has to doubly centre once, instead of $1+B$ times (where $B$ denotes the number of resamples).

\section{Hilbert--Schmidt independence criterion}\label{hsic}

The {\it Hilbert-Schmidt independence criterion} (HSIC) is an association measure that was proposed as recently as the GDC \citep{gretton2005measuring,gretton2008kernel}, whose popularity is more biased towards the machine learning community. Let $\mZ\neq\emptyset$, as in Section~\ref{gdc:semimetric}. For the purposes of this dissertation, we will say that a \emph{kernel} is a function $k: \mZ \times \mZ \to \R$ which is symmetric in its arguments and positive definite. We define the latter condition as $k$ satisfying:
$$\sum_{i,j=1}^n a_i a_j k(z_i,z_j) \geq 0$$
for all $n \geq 1$, $z_1,\ldots,z_n \in \mZ$ and $a_1,\ldots,a_n \in \R$. In that case, we call $(\mZ,k)$ a \emph{kernel space}.

We remark that some authors do not assume that positiveness is part of the definition of a kernel \citep{Genton}, but in any case, a kernel satisfying that property is under the hypotheses of Mercer's theorem \citep{Mercer}. This result ensures the existence of an embedding (called \emph{feature map}) of $\mZ$ into an inner product space (known as the \emph{feature space}), in a way that using the kernel in $\mZ\times\mZ$ is identified with evaluating the inner product in the feature space. This concept will be formally introduced at the beginning of Section~\ref{gt} and extensively used throughout the following chapters.

Given kernels $k_\mX : \mathcal{X} \times \mathcal{X} \to \R$ and $k_\mY : \mY \times \mY \to \R$, the HSIC statistic of random elements $X\in\mX$ and $Y\in\mY$ is \citep{Sejdinovic}:
\begin{equation}
	\label{eq:hsic}
	\mbox{HSIC}_{k_\mX,k_\mY}(X,Y) = \E \Big[ k_\mX(X,X') \big\{k_\mY(Y,Y')-k_\mY(Y,Y'')- k_\mY(Y',Y'')+k_\mY(Y'',Y''') \big\} \Big];
\end{equation}
whenever  $\E\left[  |k_\mX(X,X')|+|k_\mY(Y,Y')| \right] < \infty$.

The $\mbox{HSIC}$ is always nonnegative and it vanishes under independence. The converse (i.e., nullity of the HSIC implies independence) holds if and only if the kernels $k_\mX$ and $k_\mY$ are \emph{characteristic}. A kernel $k$ on $\mZ$ is characteristic if, for every pair of probability measures $P,Q$ on $\mZ$, the following equivalence holds:
$$
P=Q \iff \int_{\mZ\times\mZ}k\,\mathrm{d}(P-Q)^2=0.
$$
This is the same as saying that the \emph{maximum mean discrepancy} \citep[MMD;][]{MMD} separates probability distributions on $\mZ$. We can see the first instance of the distance-kernel duality at this level, with the energy distance corresponding to twice the squared MMD \citep{Sejdinovic}. This and the similarity of Equations \eqref{eq:gendcov2} and \eqref{eq:hsic} are the basis of the GDC-HSIC equivalence.

If we again consider samples $\bX$ and $\bY$, we can construct kernel matrices

$$\bK^{\bX}=\left( k_\mX(X_i,X_j)\right)_{n\times n} \text{ and } \bK^{\bY}=\left( k_\mY(Y_i,Y_j)\right)_{n\times n}$$

and doubly centre them as we did for GDC, to consistently estimate \eqref{eq:hsic} as:
\begin{equation}
	\label{eq:HSICemp}
	\HSICh_{k_\mX,k_\mY} (X, Y) = \frac{1}{n^2} \sum_{i,j=1}^{n} \tilde{\bK}^{\bX}_{ij} \tilde{\bK}^{\bY}_{ij} = \frac{1}{n^2} \tr ({\bK}^{\bX} \tilde{\bK}^{\bY})= \frac{1}{n^2} \tr (\tilde{\bK}^{\bX} {\bK}^{\bY}).
\end{equation}
Unlike what happened with distance covariance, the kernel literature did very explicitly point out to the simpler ways of computing the empirical HSIC (i.e., the last two formulae in the equation above) from the very beginning, as in Equation 9 of \citet{gretton2005measuring}.

We finally summarise the equivalence of GDC and the HSIC derived by \citet{Sejdinovic}. On the one hand, let $k:\mZ\times\mZ\to\mathbb R$ be a kernel. Then the following function is a semimetric on $\mZ$, and it is said to be the semimetric induced by $k$:
\begin{equation}
	\label{eq:kinducedrho}
	\rho_k (z,z') = \frac{k(z,z) + k(z',z')}{2}  - k(z,z').
\end{equation}
The squared GDC on the semimetric marginal spaces induced by two kernels equals the HSIC on those marginal kernel spaces:
\begin{equation}
	\label{equiv:hsic:dc}
	\mbox{HSIC}_{k_{\mX},k_{\mY}}(X,Y)=
	\V^2_{\rho_{k_{\mX}},
		\rho_{k_{\mY}}
	}(X,Y)
\end{equation}
for any $X\in(\mX,k_\mX)$ and $Y\in(\mY,k_\mY)$.

On the other hand, let $\rho:\mZ\times\mZ\to\mathbb R$ be a premetric. Then the following function is a kernel on $\mZ$ for any point $z_0\in\mZ$, and it is said to be the kernel induced by $\rho$ with centre $z_0$:
\begin{equation}
	\label{eq:smindkernel}
	k_{\rho,z_0}(z,z') = \rho(z,z_0) + \rho(z',z_0) -  \rho(z,z').
\end{equation}
The expression above coincides with the absolute antipodal divergence, introduced in Equation~\eqref{eq:antipodal}.

The HSIC on the kernel marginal spaces induced by two premetrics equals the squared GDC on those marginal premetric spaces:
$$
\V^2_{\rho_{k_\mX,x_0},\rho_{k_\mY,y_0}}(X,Y)=\mbox{HSIC}_{k_{\mX},k_\mY}(X,Y)
$$
for any $X\in(\mX,k_\mX)$ and $Y\in(\mY,k_\mY)$, regardless of the choice of $(x_0,y_0)\in\mX\times\mY$.

It follows quite trivially that the same equivalence holds, \textit{mutatis mutandi}, for the empirical versions of GDC and the HSIC.

As a last remark, we want to stress that the Moore--Aronszajn theorem ensures that each of our (symmetric, positive definite) kernels induces a unique \emph{reproducing kernel Hilbert space} (RKHS). These structures arise very often in the literature of the field, so we will define the concept of RKHS for the sake of completion of the current section.

Firstly, let $S\neq\emptyset$ be an arbitrary non-empty set and assume that $\h\subset\R^S$ is a Hilbert space whose inner product we will denote by $\inner{\cdot}{\cdot}_{\h}$. Then, $\h$ is said to be an RKHS if, and only if, the evaluation functional
\begin{align}
	\mathcal L_x:\;\h &\longrightarrow \mathbb{R}\\
	f &\longmapsto f(x)
\end{align}
is continuous for each $x\in S$. This is the same as saying that $\mathcal L_x$ is a bounded operator on $\h$:
$$
\forall x\in S,\:\exists\,M_x\in\Rplus,\:\forall f\in\h:\; |\mathcal L_x(f)|\equiv|f(x)|\leq M_x\norm{f}_{\h}\; ;
$$
where $\norm{\cdot}_{\h}$ is the norm induced by $\inner{\cdot}{\cdot}_{\h}$.

The name of \emph{reproducing kernel Hilbert space} is due to the fact that, by the Riesz representation theorem, the condition above ensures that:
$$
\forall x\in S,\:\exists^{\cdot} K_x\in\h,\:\forall f\in\h:\:f(x)\equiv\mathcal L_x(f)=\inner{f}{K_x}_{\h} .
$$
This means that $K_x(y)=K_y(x)=:K(x,y)$ is a (symmetric, positive definite) kernel on $S$ and that it uniquely characterises the RKHS.

\section{The Global Test}\label{gt}

A kernel $k: \mZ \times \mZ \to \R$ (according to our definition, which includes positive definiteness) can always be decomposed into \emph{features}, that is, one can embed the abstract domain of $k$ into a Euclidean space, in which one can apply more conventional classical statistical techniques. In that linear world, we will focus on Gaussian regression, to show how performing the Global Test by \citet{goeman2006testing} on the data transformed by the feature map is equivalent to testing for independence both with the GDC and the HSIC.

We say that $\boldsymbol{\Phi}: \mZ \to \R^d$ is a feature map of $k$ whenever
$$k(z,z') = \langle \boldsymbol{\Phi}(z), \boldsymbol{\Phi}(z') \rangle$$
for all $z,z'\in\mZ$. We use the bracket notation $\langle \cdot,\cdot \rangle$ for the ordinary inner product in $\R^d$, and allow $d$ to be in $\mathbb Z^+\cup\{ \infty\}$. The existence of such  $\bPhi$ is ensured for any of our kernels, which are positive definite by definition \citep{Mercer}. Throughout this dissertation, we will abuse nomenclature by saying ``feature map of a (pre)metric $\rho$'' when referring to a feature map of a kernel $k_{\rho,z_0}$ which is induced by a (pre)metric $\rho$.

For a first illustration of feature maps, let us assume that for certain $r_{\mX},r_{\mY}\in\mathbb Z^+\cup\{\infty\}$ there are feature maps $\bPhi^{\mX}:\mX\to\R^{r_\mX}$ and $\bPhi^{\mY}:\mY\to\R^{r_\mY}$ for kernels $k_\mX$ and $k_\mY$, respectively. Then, the HSIC is a linear combination of squared ordinary (product-moment) covariances in the linear world of the feature marginal spaces:
\begin{equation}
	\label{eq:hsic:feat:cov}
	\mbox{HSIC}_{k_{\mX},k_{\mY}}(X,Y)=
	\sum_{l=1}^{r_{\mX}} \sum_{m=1}^{r_{\mY}}
	\Cov^2\left( \bPhi^{\mX}_l(X), \bPhi^{\mY}_m(Y)\right) ;
\end{equation}
where $X\in(\mX,k_\mX)$ and $Y\in(\mY,k_\mY)$ are such that all moments involved in the equation above exist.

Now, consider an empirical Bayes linear model, for the regression of univariate $y$ against a $p$-dimensional random vector $\bX$, with intercept $\mu \in \R$ and error variance $\sigma^2\in\Rplus$:
\begin{equation} \label{eq:linmod}
	y\mid\boldsymbol{\beta} \sim \mathcal{N}(\mu + \boldsymbol{\beta}^t \bX,\sigma^2);
\end{equation}
where $\boldsymbol{\beta}\in\R^p$ is a random vector given by $\boldsymbol{\beta} = \tau \bb$. Here, $\tau \in \R$ is an unknown parameter, with $\bb$ capturing all the randomness of $\boldsymbol{\beta}$. We assume that $\E[\bb] = \bzero$ and $\E [\bb \bb^t] = \bI_p$.

In this model, it is natural to test:
$$
H_0: \tau^2 = 0\;\;\text{ against }\;\;H_1: \tau^2 > 0;
$$
which amounts to wondering whether $\boldsymbol{\beta}$ is significantly different from the null $p$-vector.

Following the developments by \citet{Jelle:Biomet}, we derive the pivot:
\begin{equation}  \label{eq:globaltestuniv}
	\GT(\bX,y) = \frac{1}{n^2} \, \sum_{i,j = 1}^n \langle \bX_i, \bX_j \rangle \,  (y_i - \widehat{\mu}) (y_j - \widehat{\mu});
\end{equation}
where $\hat{\mu} = \frac{1}{n} \sum_{i=1}^n y_i$ is the maximum-likelihood estimator of $\mu=\E[y\mid\boldsymbol{\beta}=\bzero]$ derived from the joint sample $$(\bX_1,y_1),\ldots,(\bX_n,y_n)\iid(\bX,y).$$ $\widehat{GT}$ is what \citet{Jelle:Biomet} introduced as the Global Test (GT) statistic for the Gaussian linear model. The nomenclature ``Global Test'' means, in this context, the locally most powerful test for the regression model that one is considering in each case. We will capitalise those words when referring to this concept, in order to avoid any confusion with any other hypothesis tests that are global in some way.

\citet{chaturvedi2017global} showed that the GT for a multivariate response $\bY \in \R^q$ can simply be written as:
\begin{equation}  \label{eq:globaltest}
	\GTh(\bX,\bY) = \frac{1}{n^2} \sum_{i,j = 1}^n \langle \bX_i, \bX_j \rangle \,  \langle \bY_i - \widehat{\boldsymbol{\mu}} ,  \bY_j - \widehat{\boldsymbol{\mu}} \rangle.
\end{equation} 

Then, one can see \citep{DJ} that it is equivalent to perform the GT on our data transformed by their feature maps, and to test independence on the original spaces (both with GDC and the HSIC):
$$
\HSICh_{k_\mX,k_\mY}  (X, Y) =
\widehat{\V}^2_{\rho(\cdot,\cdot;k_{\mX}),\rho(\cdot,\cdot;k_{\mY})}  (X, Y) =
\GTh\left( {\bPhi}^{\mX}(X) ,{\bPhi}^{\mY}(Y) \right) ,
$$
for any random elements $X$ and $Y$ with supports in some arbitrary kernel spaces $(\mX,k_\mX)$ and $(\mY,k_\mY)$.

\vspace*{1cm}

We now have defined and explored all the mathematical machinery that will be used in the following chapters of the dissertation, where we present our main contributions, in line with the research goals outlined in Section~\ref{Obj}. We now proceed to Chapter~\ref{ch3}, where distance covariance and associated techniques will first be used.


%% file: ch3_v14.tex
\fancyhead[LO]{\rightmark}
\fancyhead[RE]{\leftmark}
\renewcommand{\headerright}{\thechapter}

\chapter{Testing for gene-gene interactions in complex disease}
\thispagestyle{empty}
\label{ch3}
\graphicspath{{./fig_ch3/}}

Understanding epistasis (genetic interaction) may shed some light on the genomic basis of common diseases, including disorders of maximum interest due to their high socioeconomic burden, like schizophrenia. In this chapter, we propose distance correlation as a novel tool for the detection of epistasis from case-control data of SNPs.

On the methodological side, we highlight the derivation of the explicit asymptotic null distribution of the test statistic. We show that this is the only way to obtain enough computational speed for the method to be used in practice, in a scenario where the resampling techniques found in the literature are impractical. Our simulations demonstrate satisfactory calibration of significance, as well as comparable or better power than preexisting methodology. We conclude with the application of our technique to a schizophrenia genetics dataset, obtaining biologically sound insights.

This chapter is organised as follows. Section~\ref{miss:her} introduces some biological context about genetic interaction and the relevance of this problem. Section~\ref{epist:tools} reviews the many solutions that the scientific community has tried to offer in recent years. Section~\ref{LCT} serves as an overview of the large-correlation tests (LCTs), a family of methods that inspired our methodology. Section~\ref{epistasis:test} presents a novel testing procedure for association between SNPs, based on the techniques explained in detail in Chapter~\ref{ch2}. Some results of our simulation study are reported in Section \ref{simu:epistasis}. In Section \ref{SCZ}, we apply the method to a genomic dataset of schizophrenia, to finally discuss the results and draw some conclusions in Section~\ref{epist:discu}.

An earlier version of the contents that we now present can be found in \citet{F:epistasis}.

\section{Missing heritability in complex disease}\label{miss:her}
As indicated in Chapter~\ref{ch1}, there is strong evidence of the relevance of genetics in psychiatry, and this field of study has more than a century of history. Today there is no doubt that most psychiatric disorders are multifactorial, complex traits. To give a more precise idea, nowadays it is estimated that the genome can explain up to 80 \% of the susceptibility to suffer some of these diseases, like schizophrenia \citep{Sullivan:PGC}.

The genetic susceptibility to a psychiatric disorder lies on a large number of variants along the genome, none of which are necessary or sufficient on their own. Although the specialised literature usually focuses simply on additive models \citep{Purcell}, biological knowledge suggests that gene-gene interactions (or \emph{epistasis}) could be one of the factors that explain the phenomenon of  \emph{missing heritability}, which contributes to the inefficiency of genome-wide association studies when it comes to explaining causality of complex diseases \citep{Manolio,Brandes}. Evidence from studies on model organisms also support the importance of genetic interactions in the understanding of complex traits \citep{Mackay:Moore}.

We are interested in datasets of case-control GWASs (i.e., a collection of genotypes of ``healthy'' individuals and ``patients'') for schizophrenia. The statistical challenge hinges on using this data to detect pairs of genetic variants that significantly increase or decrease the susceptibility to develop schizophrenia, which further research can confirm with biological criteria.

This data corresponds to SNPs, which are variants on one of the ``letters'' of the DNA (i.e., each of them occurs at one specific point of the genome). Given that we will only consider autosomal variants, each individual can carry 0, 1 or 2 copies of the \emph{minor} allele (the least frequent of the two variants) on their diploid genome. The aforementioned setting requires performing statistical inference in a context of high dimension and low sample size, where the covariates are \emph{ternary} (discrete with support of cardinality $3$). The scope of this chapter will be how to do so, using the distance-based techniques that we introduced in Chapter~\ref{ch2}.

In the next section, we provide a review of the extremely wide variety of approaches to the detection of epistasis that can be found in the biostatistical literature. The main conclusion of that study effort is that there is no clear winner among the different available techniques, which justifies the maintained interest in this problem over the past few years.

\section{Statistical approaches for epistasis detection}\label{epist:tools}
The recent development of the ``-omic'' disciplines has been parallel to the creation of bioinformatic tools to process the vast amount of data that these experimental sciences produce. The diversity of the available ``-omic'' software is so large that it has even been necessary to develop meta-tools to index the existing techniques. For example, the directory of one of them \citep{OMICtools} contains more than $20\,000$, 900 of which are designed for GWAS data analyses, which in turn contain a subset of 100 that are suitable for epistasis detection.

The existence of such a wide spectrum of proposed solutions for such a specific task owes to the surprisingly high diversity of statistical methods that are valid for it---e.g. linear models (standard and generalised), logistic regression, tests on Pearson's correlations, permutation tests, Bayesian nonparametric statistical inference, random forests, Markov chains, co-information indices, graph theory, or maximal entropy probability models.

Another cause of that diversity of alternatives is the fact that some of the available techniques only focus on a specific subproblem (pairwise gene-gene interactions versus higher orders, binary versus continuous response variable, pedigrees, stratified populations and so forth) and on the different computing strategies that they use in order to obtain results within reasonable amounts of time (for instance, initial filters based on biological knowledge, code parallelisation, graphical processing units, Boolean operations, machine learning approaches, or ant colony optimisation algorithms).

\begin{table}[!htbp]
\centering
\caption{Some remarkable epistasis detection tools for GWAS data analysis.}
\begin{adjustbox}{width=.95\textwidth}
	\begin{tabular}{@{}llll@{}}
		\toprule
		Tool & Statistical techniques & Computational tricks & Reference \\
		\cmidrule(l{-1pt}r{-1pt}){1-4}
		2S-LRM  & Logistic regression & Pre-filtering & \citet{Vaart} \\
		AntEpiSeeker & $\chi^2$ tests & Ant colony optimisation & \citet{AntEpiSeeker} \\
		BEAM  & Bayesian MCMC & None & \citet{BEAM} \\
		BOOST & Logistic regression & Boolean operations, parallelisation & \citet{BOOST} \\
		BiForce & Linear regression & Boolean operations, parallelisation & \citet{BiForce} \\
		CES   & Evolutionary algorithms & Artificial intelligence & \citet{CES} \\
		CINOEDV & Information theory & Swarm intelligence on hypergraphs & \citet{CINOEDV} \\
		EpiGPU & Linear regression & GPU architectures & \citet{EpiGPU} \\
		EpiACO & Information theory & Ant colony optimisation & \citet{EpiACO} \\
		EpiBlaster & Pearson's correlations & GPU architectures & \citet{EpiBlaster} \\
		Fiúncho & Information theory & Parallelisation & \citet{Fiuncho} \\
		GLIDE & Linear regression & GPU architectures & \citet{GLIDE} \\
		GWIS  & ROC curve analysis & GPU architectures & \citet{GWIS} \\
		IndOR & Logistic regression & Pre-filtering & \citet{IndOR} \\
		MDR   & Combinatorics, resampling & Pre-filtering & \citet{MDR} \\
		Random Jungle & Random forests & Parallelisation & \citet{Random Jungle} \\
		SNPruler & Information theory & Branch and bound algorithms & \citet{SNPruler} \\
		Stage-wise LRT & GLMs, closed testing & Hierarchical testing & \citet{Franberg} \\
		Wtest & Logistic regression & None & \citet{wtest} \\
		\bottomrule
	\end{tabular}
\end{adjustbox}
\label{HumGenet}
\end{table}

Table~\ref{HumGenet} summarises some of the existing methods, including the ones reviewed by \citet{Gusareva}, \citet{Niel} and \citet{Russ} and some other that we consider representative. Some are very widely used, like BOOST, due to it being implemented in the popular genetics toolset PLINK \citep{PLINK}; whereas other of the methods on the table have not been used much in practice.

\section{Large-scale correlation tests (LCTs)}\label{LCT}

In Table~\ref{HumGenet}, it is shown that one conspicuous epistasis detector \citep{EpiBlaster} is based on scanning for differential behaviours of (Pearson's) correlations between cases and controls. This is unsurprising, since several authors \citep{De la Fuente,Camacho,Haeseleer} support the idea of correlation tests in this context when the data is continuous (gene expression, metabolomics and so forth), which however is not the case of SNPs (ternary variables).

Moreover, such techniques usually rely on the normality of the covariates, a hypothesis that turns out to be excessively restrictive in most cases. Therefore, the procedure by \citet{Cai:Liu} contains an interesting approach, as they manage to establish a rigorous theoretical framework for the kind of correlation tests that are convenient for epistasis detection. This technique is part of the hot topic of hypothesis testing on high-dimensional covariance structure, that has been developed almost from scratch during the past few years \citep{AnnuRev:Cai}.

We now will be presenting some aspects related to the LCTs by \citet{Cai:Liu}. Given $L\in\mathbb{Z}^{+}$ SNPs, let $\mathbf{X}=\left(X_j\right)_{j=1}^L$ and $\mathbf{Y}=\left(Y_j\right)_{j=1}^L$ be the corresponding random vectors of 0's, 1's and 2's for case and control individuals, respectively. If their correlation matrices are: $(\rho_{ij1})_{i,j}\in\mathbb{R}^{L\times L}\;\;\;\;\text{and}\;\;\;\;
(\rho_{ij2})_{i,j}\in\mathbb{R}^{L\times L}$, the aim is testing:
\[\begin{cases}{H_{0ij}}:\rho_{ij1}=\rho_{ij2}\\H_{1ij}:\rho_{ij1}\neq\rho_{ij2}\end{cases}\]
for each pair $(i,j)\in([1,L]\cap\mathbb{Z})^2$ so that $i<j$; using samples $\{\mathbf{X}_k\}_{k=1}^{n_1}\text{ IID }\mathbf{X}$ and $\{\mathbf{Y}_k\}_{k=1}^{n_2}\text{ IID }\mathbf{Y}$, which are assumed to be independent of each other.

\subsection{LCT: classical approach}

A scarcely innovative approach would be to stabilise the variance of the sample correlation coefficients via Fisher's $Z$ transformation (\textit{atanh}). One could think of combining this strategy with a procedure that controls the false discovery rate (FDR), such as the ones by \citet{BH} or \citet{BY}, thus establishing the desired large-scale correlation test (LCT). The main drawback to this idea is that, when normality is not ensured, the behaviour of the test statistic differs from the well-known asymptotic distribution of the Gaussian case. Simulation studies \citep{Cai:Liu} show that this method performs very poorly (both with Benjamini--Hochberg and Benjamini--Yekutieli), especially when compared to the LCT that will be introduced next.

\subsection{LCT with with bootstrap}
\citet{Cai:Liu} devised an LCT with bootstrap (the LCT-B), which is based on the test statistic
\[T_{ij}:=\frac{\hat\rho_{ij1}-\hat\rho_{ij2}}
{\sqrt{\frac{\hat\kappa_1}{3n_1}\left(1-\tilde\rho_{ij}^2\right)^2+\frac{\hat\kappa_2}{3n_2}\left(1-\tilde\rho_{ij}^2\right)^2}};\]
where $\hat\kappa_1$ and $\hat\kappa_2$ are the respective sample kurtoses of $\mathbf{X}$ and $\mathbf{Y}$, and $\tilde\rho_{ijl}$ is a thresholded version of $\hat\rho_{ijl}$, for $l\in\{1,2\}$; with $\tilde{\rho}_{ij}^2:=\max\{\tilde{\rho}_{ij1}^2,\tilde{\rho}_{ij2}^2\}$.

$H_{0ij}$ will be rejected when $|T_{ij}|$ is greater than a certain threshold $\hat t_{\alpha}\in\Rplus$, which depends on the nominal value $\alpha\in]0,1[$ under which one wants to maintain the FDR. If the distributions of $\mathbf{X}$ and $\mathbf{Y}$ are totally unknown, it is reasonable to use resampling techniques in order to approximate the tail of the distribution of $T_{ij}$, which determines $\hat t_{\alpha}$. The bootstrap scheme that \citet{Cai:Liu} built to this purpose is consistent and leads to a threshold $\hat t_{\alpha}^*$, which defines the LCT with bootstrap (LCT-B). This test is supported by strong theoretical results, which were proven in the original \citeyear{Cai:Liu} article.

\subsection{Unsuitability of the LCT for SNP data}

\begin{figure}
	\centering\includegraphics[width=0.8\textwidth]{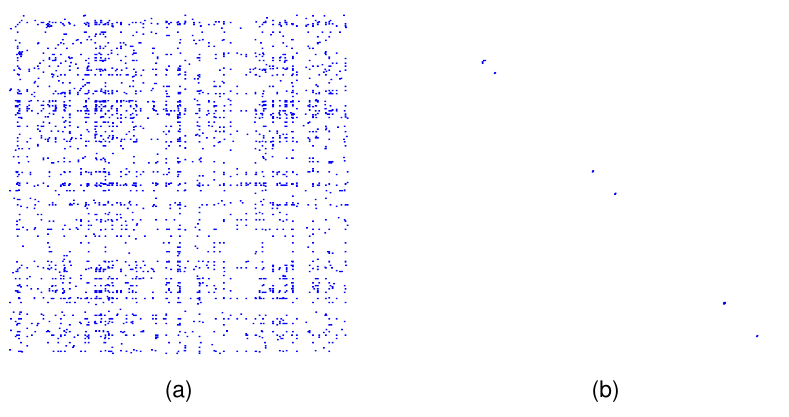}
	\caption{Adjacency matrix of the putative epistatic network detected by the LCT-B, for (a) gene expression data for prostate cancer (Broad Institute) and (b) SNP data for schizophrenia (Health Research Institute, Santiago de Compostela)}
	\label{Mcoexpr_LCT-B}
\end{figure}

We have implemented the LCTs of \citet{Cai:Liu} in the \emph{R} programming language \citep{R} to further illustrate the motivation of our the present chapter. We firstly reproduced the real-data example on the original LCT article, obtaining the adjacency matrix in Fig.~\ref{Mcoexpr_LCT-B}a. To accomplish that, we applied the method known as LCT-B to the data by \citet{Singh}, in which dimensionality was trimmed down to 500 using the Welch--Satterthwaite test (Behrens--Fisher problem). Since the variables involved are assumed to be continuous, the LCT-B yields believable results; in the sense that the resulting matrix is sparse, but not too much. However, a biological validation of all those results would be extremely difficult to accomplish.

On the other hand, when the schizophrenia SNP data (remarkably discrete) are analysed, the adjacency matrix looks very differently (Fig.~\ref{Mcoexpr_LCT-B}b) to the previous one (Fig.~\ref{Mcoexpr_LCT-B}a). The only nonzero elements are close to the diagonal, owing to the fact that the only pairs that are being detected are in linkage disequilibrium (i.e., the frequency of such SNP pairs is significantly different from the product of the marginal frequencies, due to their physical proximity within a certain chromosome). Such findings are useless from the point of view of psychiatric genetics because they do not show an association that is related to schizophrenia, but rather one that is independent of this disease.

Some authors, like \citet{EpiBlaster}, argue that treating clearly discrete SNP data as continuous is an acceptable simplification. Nevertheless, even if that could be anecdotally true in some specific setting, this is clearly not the case, as Fig.~\ref{Mcoexpr_LCT-B}b clearly displays.

The unsatisfactory behaviour of the LCT-B (when applied to a different setting from the one it was originally intended to) is the main motivation of the present chapter. In this context, it is justified to wonder which association measures characterise the independence of ternary variables, as well as how to extend the LCTs by \citet{Cai:Liu} to less stringent conditions so that they become applicable to SNP data.

\section{A distance-based test for epistasis}\label{epistasis:test}

We now present the particularisation of the theoretical framework in Sections~\ref{SRB} to \ref{gdc:semimetric} to spaces of cardinality $3$ (Section~\ref{sec:space3}), to then build upon it our own proposal of a testing procedure for epistasis detection (Section~\ref{sec:test}).

\subsection{Distance correlation in spaces of cardinality $3$}\label{sec:space3}
Clearly, in a finite space, the finiteness of moments (of any order) and separability are not an issue. Alternatively, one can resort to brute-force and solve the system of inequations that are derived from simply using the definitions \citep{Klebanov,Lyons}, obtaining a direct \textemdash albeit cumbersome\textemdash~proof of the fact that any 3-point space $\xd$ is necessarily of strong negative type. Such proof is, in principle, superfluous, as long as one wants to make use of strong theorems, such as Schoenberg's: $\xsd$ can clearly be embedded into a Hilbert space, isometric to the vertices of a triangle in $\mathbb{R}^2$ (note that the square root transformation preserves the triangle inequality of the metric). Nevertheless, it is interesting to check that, when the metric structure becomes so simple, abstract arguments (like the ones that arise in the proof of Schoenberg's theorem) become unnecessary.

Let $\spx:=\{0,1,2\}$ be the set of the three possible genotypes for each SNP. There is no biological reason to assume that $2\in\spx$ copies of the minor allele affect twice as much as one \citep{Bush}, neither when it comes to increasing the susceptibility to a psychiatric disorder nor to decreasing it. As a matter of fact, in some cases this susceptibility is maximal under heterozygosis \citep{Costas:Heteroz_opt}, which is coded by $1\in\spx$.

Therefore, there is no rationale for prioritising the Euclidean distance:
\begin{equation}\label{Euc:metric}
d(0,2)=2d(0,1)=2d(1,2)\text{,}
\end{equation}
instead of more general (non-``linear'') metric spaces. And this is why distance correlation turns out to be a way to extend the ideas of \citet{Cai:Liu}. As previously commented, the marked discreteness of SNP data provides another incentive for transcending the idea of linear correlation.

No specific type of interaction is being looked for --- our aim is to simply detect epistasis. For this reason, we will henceforward focus on the \emph{discrete metric}:
\begin{equation}\label{disc:metric}
d(0,1)=d(1,2)=d(0,2)=1\text{;}
\end{equation}
which conveys agnosticism on the underlying genetic model.

Other distances with straightforward genetic interpretation can be defined. For instance, by dropping the identity of indiscernibles, one can reflect the following inheritance models with very simple premetrics:
\begin{itemize}
	\item Recessive: $d(0,1)=0;\;\;d(0,2)=d(1,2)=1$.
	\item Heterozygous: $d(0,2)=0;\;\;d(0,1)=d(2,1)=1$.
	\item Dominant: $d(1,2)=0;\;\;d(1,0)=d(2,0)=1$.
\end{itemize}

All the aforementioned geometries allow for distance covariance to work, as stated in Chapter~\ref{ch2}. It is possible to define even more premetrics with a genetic interpretation, which we will explore in more detail in the next chapter (Section~\ref{tailoring}). However, for the current problem of interest (i.e., testing for epistasis) we will restrict ourselves to the discrete metric, as previously mentioned, in order not to complicate the interpretation even more, in the context of a very elusive genetic concept as epistasis \citep{Russ}.

\subsection{Proposal of a hypothesis test}\label{sec:test}
Searching for epistasis consists in looking for significantly different dependence structures between the case and control groups, as previously discussed. Let us focus on a pair of indices $(i,j)$ in $\{1,\ldots,L\}$ such as $i<j$. To simplify notation, let $Z_i$ and $Z_j$ be random variables with support $\spz\in\llaves{\spx,\spy}$, corresponding to two different SNPs, for which a joint sample of size $n\in\mathbb{Z}^{+}$ is available:
\[(Z_{i,1},Z_{j,1}),\ldots,(Z_{i,n},Z_{j,n})\text{ IID }(Z_i,Z_j)\text{.}\]
The aim is testing the independence of $\{Z_i,Z_j\}$ or, equivalently,
\[\begin{cases}{H_{0ij}}:\dCov(Z_i,Z_j)=0\\H_{1ij}:\dCov(Z_i,Z_j)\neq0\end{cases}\]
with the philosophy of the large-scale multiple tests by \citet{AnnuRev:Cai}.

Our test statistic will be $\widehat{\dCov}(Z_i,Z_j),$ as defined in Equation~\eqref{eq:dcovemp}. When it comes to approximating its null distribution, one can take advantage of the finiteness of the marginal spaces --- in this setting, only a few of the coefficients of the quadratic form that gives the asymptotic null distribution of distance covariance will be non-null. Namely, we present two theorems for such distributions, both for the geometry of maximum interest to us (i.e., the discrete metric) and for the Euclidean distance (i.e., for classical distance covariance). Proofs can be found in Section~\ref{proofs:ch3} of the appendix.

\begin{theorem} \label{th:discrete}
	Let $(X_1,\ldots,X_n)$ and $(Y_1,\ldots,Y_n)$ be IID samples of jointly distributed random variables $(X,Y) \in \{0,1,2\} \times \{0,1,2\}$, with marginal probabilities $p_j = \Pr(X=j) q_j$ and $\Pr(Y=j)$, for $j=0,1,2$.
	
	Consider $\mathcal X = \mathcal Y =\{0,1,2\}$ equipped with the discrete metric, as in Equation~\eqref{disc:metric}.
	
	Then, whenever $X$ and $Y$ are independent, for $n \to \infty$,
	$$
	n \, \widehat{\dCov}_{\mathrm{discrete}}^2(X,Y) \Dlim
	\lambda_1 \, \mu_1 Z_{11}^2 + \lambda_1 \mu_2 Z_{12}^2 + \lambda_2 \mu_1 Z_{21}^2 + \lambda_1 \mu_2 Z_{22}^2;
	$$
	
	where $Z_{11}^2, Z_{12}^2, Z_{21}^2, Z_{22}^2$ are independently chi-squared distributed with one degree of freedom. $\lambda_1$ and $\lambda_2$ are given by:
	$$
	\frac{1-\sum_{j=0}^2 p_j^2}{2} \pm \sqrt{\frac{(1-\sum_{j=0}^2 p_j^2)^2}{4} - 3 \prod_{j=0}^2 p_j}. 
	$$
	Similarly $\mu_1$ and $\mu_2$ are given by
	$$
	\frac{1-\sum_{j=0}^2 q_j^2}{2} \pm \sqrt{\frac{(1-\sum_{j=0}^2 q_j^2)^2}{4} - 3 \prod_{j=0}^2 q_j}.
	$$\qed
\end{theorem}

\begin{theorem}\label{th:Euclidean}
	Let $(X_1,\ldots,X_n)$ and $(Y_1,\ldots,Y_n)$ be IID samples of jointly distributed random variables $(X,Y) \in \{0,1,2\} \times \{0,1,2\}$ with $p_j = P(X=j)$ and $q_j = P(Y=j), j=0,1,2$.
	
	Consider $\mathcal X = \mathcal Y =\{0,1,2\}$ equipped with the Euclidean metric, as in Equation~\eqref{Euc:metric}.
	
	Then, whenever $X$ and $Y$ are independent, for $n \to \infty$,
	$$
	n \, \widehat{\dCov}_{\mathrm{Euclidean}}^2(X,Y) \Dlim  
	\lambda_1 \, \mu_1 Z_{11}^2 + \lambda_1 \mu_2 Z_{12}^2 + \lambda_2 \mu_1 Z_{21}^2 + \lambda_1 \mu_2 Z_{22}^2;
	$$
	
	where $Z_{11}^2, Z_{12}^2, Z_{21}^2, Z_{22}^2$ are independently chi-squared distributed with one degree of freedom, and ``$\Dlim$'' denotes convergence in distribution. $\lambda_1$ and $\lambda_2$ are given by
	$$
	p_0(1-p_0)+p_2(1-p_2) \pm \sqrt{\Big(p_0(1-p_0)+p_2(1-p_2)\Big)^2 - 4 \prod p_j}.
	$$
	Similarly $\mu_1$ and $\mu_2$ are given by
	$$
	q_0(1-q_0)+q_2(1-q_2) \pm \sqrt{\Big(q_0(1-q_0)+q_2(1-q_2)\Big)^2 - 4 \prod q_j}.
	$$\qed
\end{theorem}

It is crucial to note that we do not want to directly test for the equality of distance correlations, as \citet{Cai:Liu} did when looking for differential gene co-expression, following the rationale by \citet{De la Fuente} and others. In our search for epistasis, however, we are just interested in finding SNP pairs that are dependent for the cases and independent for the controls. When, for some SNP pair, independence is rejected for healthy individuals and not for patients, it will attributed to a spurious interaction resulting from \emph{population substructure} \citep{Brandes}, i.e., from the effect of unmeasured (and often unmeasurable) covariates.

\subsection{Extensions to interactions among more than two SNPs}

A limitation of the procedure described above is that it is restricted to testing interactions of two SNPs. There are at least two straightforward possibilities for extending Theorems \ref{th:discrete} and \ref{th:Euclidean} to settings involving more than two SNPs. The first one would be to resort to the concept of distance multivariance \citep{Boettcher1, Boettcher2}, a natural generalisation of distance covariance for testing the independence of more than two random vectors. In particular, given a sample

$$(\mathbf{X}_1,\mathbf{Y}_1, \mathbf{Z}_1),\ldots,(\mathbf{X}_n,\mathbf{Y}_n, \mathbf{Z}_n)\text{ IID }(\mathbf{X},\mathbf{Y}, \mathbf{Z})\text{;}$$

let $a_{ij}:=d(\mathbf{X}_i,\mathbf{X}_j), \, b_{ij}:=d(\mathbf{Y}_i,\mathbf{Y}_j), \, c_{ij}:=d(\mathbf{Z}_i,\mathbf{Z}_j)$ and define the centred distances $A_{ij} , \, B_{ij}, \, C_{ij}$ as in Equation \eqref{eq:center}. Then the sample version of the distance multivariance  between the three vectors $\mathbf{X},\mathbf{Y}, \mathbf{Z}$ is:
$$
\widehat{\dMv}_n(\mathbf{X},\mathbf{Y},\mathbf{Z}  )^2:= - \frac{1}{n^2}\sum_{i,j=1}^n A_{ij}B_{ij} C_{ij},
$$
as in \citet[Theorem 4.1]{Boettcher1}, which extends Equation~\eqref{eq:empdc} for sample distance covariance.

For the case where $\mathbf{X}$, $\mathbf{Y}$ and $\mathbf{Z}$ are discrete-valued with support $\{0,1,2\}$, we can then prove extensions of Theorems~\ref{th:discrete} and~\ref{th:Euclidean} for testing interactions of three SNPs (see Section~\ref{ap:mvar} in the appendix). Results for the distance multivariance for SNP interactions of order 4 and higher can be derived analogously. 

A second generalisation of our methodology to testing for interactions between SNP sets arises from considering the generalised distance covariance between SNP vectors $\mathbf{X} \in \{0,1,2\}^L$ and $\mathbf{Y} \in \{0,1,2\}^M$. However, it appears to be challenging to derive product metrics $d_L:  \{0,1,2\}^L \times \{0,1,2\}^L \to [0,+\infty[$ leading to meaningful dependence tests for SNP sets.

An analogue of Theorem~\ref{th:discrete} for generalised distance covariance based on the discrete distances $d_L$ and $d_M$, respectively, can be easily obtained using Lemma~\ref{lem:discrete:dist}. The resulting asymptotic null distribution is a weighted sum of $(2^L-1) \times (2^M-1)$ independent chi-squared variables, with one degree of freedom each.

For this problem, when considering a large number of SNPs, the usefulness of the discrete metric appears questionable, since it does not take into account for how many components two SNP vectors differ. More useful product metrics may possibly be defined in an ``$L^p$ fashion'':
$$
\rho_L(\mathbf{x}, \mathbf{x}') = \left(  \sum_{j=1}^L d (\mathbf{x}_{j}, \mathbf{x}'_{j})^L\right) ^{1/L}
$$
where $d:  \{0,1,2\} \times \{0,1,2\} \to [0,+\infty[$ is a metric defined at the single-SNP level, as described in Section \ref{sec:space3}. The detailed study of generalised distance covariances based on this type of metrics may lead to powerful testing procedures for dependence and interaction between SNP sets.

\subsection{Naive resampling strategies and computational challenges}\label{resampling:scheme}
We initially attempted to approximate the $p$-values of our test by using a permutation-based approach, which is the gold standard in the distance covariance literature \citep{TEOD}. We briefly discuss this brute-force approach, as a ``negative result'' that can be of utility to other scientists. Although theoretically sound, this strategy leads to such high computation times that it is not suitable for big data, as illustrated in Section~\ref{comput:challenge}. This is particularly true for genomics, which is so data-intensive that not even the high-performance computing (HPC) described in the next section make resampling feasible. Nonetheless, for other scenarios (of lower dimensionality, or in which no theoretical derivation of the asymptotic null distribution is possible) the following approach might be of interest.

In order to approximate the null distribution of the test statistic $\widehat{\dCov}(Z_i,Z_j)$, it is possible to devise a resampling scheme according to the relevant information that is available under the null hypothesis, which in this case is the independence of $Z_i$ and $Z_j$. As a result, the reasonable thing to do is not to resample from $\{(Z_{i,k},Z_{j,k})\}_k$, but to do it separately from $\mathcal Z_i:=\{Z_{i,k}\}_k$ and $\mathcal Z_j:=\{Z_{j,k}\}_k$ (permutation tests). Thus, it suffices to compute $B\in\mathbb{Z}^{+}$ statistics of the form $$\dCovh(\mathcal Z_i^{*(b)},\mathcal Z_j^{*(b)})$$ to obtain a Monte--Carlo approximation of the sampling distribution of the empirical distance covariance under $H_{0ij}$.

The usage of permutation tests in this context of metric spaces was inspired by the excellent performance of the same scheme in Euclidean spaces \citep{SRB,TEOD}. It has the drawback that there is not the same kind of fully-fledged formal justification of consistency (as the one by \setcitestyle{square}\citet{Arcones:Gine} for the naive bootstrap that \citet[page 100]{Jakobsen} outlined),\setcitestyle{round}%
which should not be a source of concern in practice, like in the Euclidean case.

It should also be clarified that authors such as \citet{Cai:Liu} and \citet{SRB} argue that the number of resamples $B$ is relatively unimportant for their methods to work, as long as it is not extremely small. With this in mind, and also taking into account that the running time is $O(B)$, we decided to use a moderate value for $B$ in the present chapter, namely the one devised by \citet{SRB} as a function of sample size $n$:
\[B(n)=200+\lfloor{5000}/{n}\rfloor\text{,}\]
where $\lfloor\cdot\rfloor:\;\mathbb{R}\to\mathbb{Z}$ is the floor function. Some empirical checks confirm that increasing~$B$ with respect to the value above causes barely noticeable improvements (if any) both in terms of the calibration of significance levels (as long as the nominal value is not extremely small) and of power.

\subsection{Computational challenge of the resampling approach}\label{comput:challenge}

The implementation of the test, as presented in Section~\ref{resampling:scheme}, was an extremely challenging issue from the computational point of view, given the high dimensionality of the data, the amount of samples, and the high number of hypothesis tests resulting from the combinatorial explosion. Thus, a quite sophisticated set of computer technologies and strategies was required to obtain results within somewhat manageable computational times. 

As a general rule, any statistical technique based on GWAS data will suffer from the issues that are inherent to such input (high dimension and low sample size). To illustrate this point, Table~\ref{Tempos} compares the running times of the original \emph{R} code with another one, whose core is implemented in the compiled language \emph{C}, this way making the numerical crunching far swifter. This second code ---labelled ``R \& C'' on the table--- also includes some high-performance computing (HPC) improvements and, what is more, it can be executed in sequential or in parallel mode (i.e., the workload can be distributed among different processors, decreasing the execution time by a factor that is approximately equal to the number of available processors). For a comparison of performance like the one on Table~\ref{Tempos}, it is crucial to carry on the experiments in the same environment---in our case, the supercomputer \textit{Finisterrae II} (Galician Supercomputing Centre, CESGA).

\begin{table}[H] 
	\centering
	\caption{Comparison of running times for the different versions of the code, all of them referring to the permutation testing approach. It should be noted that the times for runs of the the sequential code on the largest GWAS are estimations.}
		\begin{tabular}{@{}lrrr@{}}
			\cmidrule(l{-7pt}r{-7pt}){1-4}
			& Simulation, & GWAS, & GWAS,\\
			Code version & $R=10^3$ & $L=1000$ & $L=4000$\\
			\cmidrule(l{-7pt}r{-7pt}){1-4}
			R sequential & 12 h 10 min & 42 days 1 h & 2 years \\
			R \& C sequential & 3 h 59 min & 2 days 1 h & 30 days  \\
			R \& C parallel & 50 min & 2 h 41 min & 2 days  \\
			\cmidrule(l{-7pt}r{-7pt}){1-4}
		\end{tabular}
	\vspace*{.2cm}
	\label{Tempos}
\end{table}

Hence, in light of the order of magnitude of these times (the \emph{R} version would need up to two years in large-scale settings, while the \emph{R} \& \emph{C} parallel implementation only requires ten hours), it is fully justified to resort to HPC strategies in a compiled language, especially if one takes into account that a GWAS can easily involve millions SNPs, with the running time being a linear and monotonically increasing function of $\binom{L}{2}$ and, consequently, $O(L^2)$; as illustrated by the ratio between the GWAS columns of Table~\ref{Tempos}.

For the parallel version of the \emph{R \& C} code, in each case, the lowest amount of hardware that yielded results within a reasonable amount of time was used: 12 cores for simulations, and 48 processors for real data analyses. To reduce the times by a factor of $f$, it would suffice to increase the number of processors $f$ times, as long as economic and logistic constraints make it possible.

Moreover, the algorithm was parallelised in two alternative ways:
\begin{enumerate}
	\item using a shared-memory paradigm via the OpenMP library, distributing the computational effort among the different cores that exist within a processor;
	\item applying a distributed-memory strategy, where different computational nodes ---that belong to various machines--- are able to share their workload via a message protocol, which in this case is the MPI library.
\end{enumerate}

The first parallelisation (which is very easy to implement in the main loop of the algorithm) was useful to apply the test in simulated data, where the dimensionality was not too problematic. However, the number of parallel execution threads one can add is limited by the number of cores available on a CPU processor chip, which is not enough to address real data examples. For this reason, a distributed-memory parallelisation was developed,  with a classical Master/slave paradigm, where hundreds of processors can work together to reduce complexity. It consists in:
\begin{enumerate}
	\item A processor (\emph{Master}) calls \emph{R} routines that load the matrices that contain the input, split it and distribute it among several processors (\emph{slaves}).
	\item Each processor works with one fragment of the matrix, running the iterations that have been assigned to it (i.e., performing independence tests for a fraction of the total of SNP pairs).
	\item Once each slave finishes its part, it sends the results to its Master.
	\item Finally, the Master builds the final \textit{p}-value matrix, which is later used to wrap up the results in \emph{R}.
\end{enumerate}

The \emph{R \& C} version combines an interface in the programming language \emph{R} with a core in \emph{C}, with the latter being devoted to perform low-level computations in a time-efficient manner. Another important factor that helps decrease the computational time in our implementation is the use of specific libraries to codify low-level operations that involve large vector and matrices. Namely, the well-known Intel MKL libraries and SIMD (Single Instruction, Multiple Data) techniques have been applied to exploit data-level parallelism --- using an extension in the registers and the arithmetic and logic instructions present in modern microprocessors, they can process the same operation simultaneously on the elements of an array through a single instruction. In the present case, it was particularly useful to implement matrix operations.

It was not possible to resort to preexisting software because the most efficient distance-correlation-related algorithms \citep[like the one by][]{Chaudhuri} are only designed for the Euclidean case and, therefore, not adaptable to the structure of the 3-point spaces that are the scope of the present chapter.

\section{Simulation study}\label{simu:epistasis}
In order to validate our testing procedure, we have designed some population models in which the intensity of dependence can be adjusted by tuning a parameter. We firstly introduce those models, to then use them to compare the performance of our method with that of BOOST \citep{BOOST}, one of the most popular epistasis detectors within the genomics community. For distance covariance, we will consider the discrete metric in every scenario because it reflects our agnosticism on the underlying genetic model.

\subsection{Design of population models for the validation of our methodology}
The theoretical models that are about to be defined refer to the interaction between an arbitrary pair $\{Z_i,Z_j\}$, where $Z$ is either $X$ or $Y$, depending on the case. When it came to setting the marginal frequencies, instead of allowing for two degrees of freedom on each marginal, a further restriction was introduced (apart from the sum being one): allele and genotype frequencies were constrained to be in Hardy--Weinberg equilibrium \citep{Hardy}, since all the SNPs in the schizophrenia database verify it (it is one of the quality controls that are used). So there is a single free parameter, which is the minor allele frequency, that is sampled from a uniform distribution on $[0.05,0.2]$. The lower limit mimics standard GWAS quality control filters (in settings with moderate sample size) and the upper one was set so that the resulting true interactions are not the easiest to detect.

\setcitestyle{square}There are a few options in literature for simulating epistasis between SNPs. Some models (like the ones by \citet{Marchini}) are overly simplistic, e.g. by not allowing to adjust the interaction intensity in order to assess the robustness against different alternatives. Some recent approaches (like the ones studied by \citet{Russ})\setcitestyle{round} make interpretability more difficult, in the sense that we are very interested in quantifying the intensity of interaction (i.e., deviation from the null hypothesis) when assessing the power of our test. In order to overcome such shortcomings, we introduce our own models for SNP-SNP interaction.

The most straightforward model is one in which the probability of each genotype is the product of the marginals (there is independence). For dependence, two kinds of models will be defined. On the one hand, model \texttt{qexp} conveys a dependence structure that becomes more intense as parameter $e\in[1,+\infty[$ increases, in the way that Table~\ref{qexp} describes. On the other hand, model \texttt{qmult} has $g\in[0,1]$ as its free parameter (Table~\ref{qmult}). Again, the closer the parameter is to $1$, the less notorious the association becomes.

\begin{table}%
	\centering
	\caption{Contingency table for model \texttt{qexp}.}
	\begin{tabular}{cccc|c}
		\multicolumn{1}{c|}{$Z_i$ \textbackslash $\:Z_j$} & 0     & 1     & 2     &  \\
		\cmidrule{1-4}    \multicolumn{1}{c|}{0} & $pr+q^es-qs$ & $ps-q^es+qs$ & $p(1-r-s)$ & $p$ \\
		\multicolumn{1}{c|}{1} & $qr-q^es+qs$ & $q^es$ & $q(1-r-s)$ & $q$ \\
		\multicolumn{1}{c|}{2} & $(1-p-q)r$ & $(1-p-q)s$ & $(1-p-q)(1-r-s)$ & $1-p-q$ \\
		\midrule
		& $r$   & $s$   & $1-r-s$ & $1$ \\
	\end{tabular}%
	\label{qexp}
\end{table}

\begin{table}%
	\centering
	\caption{Contingency table for model \texttt{qmult}.}
	\begin{tabular}{cccc|c}
		\multicolumn{1}{c|}{$Z_i$ \textbackslash $\:Z_j$} & 0     & 1     & 2     &  \\
		\cmidrule{1-4}    \multicolumn{1}{c|}{0} & $pr-(1-g)qs$ & $ps+(1-g)qs$ & $p(1-r-s)$ & $p$ \\
		\multicolumn{1}{c|}{1} & $qr+(1-g)qs$ & $gqs$ & $q(1-r-s)$ & $q$ \\
		\multicolumn{1}{c|}{2} & $(1-p-q)r$ & $(1-p-q)s$ & $(1-p-q)(1-r-s)$ & $1-p-q$ \\
		\midrule
		& $r$   & $s$   & $1-r-s$ & $1$ \\
	\end{tabular}%
	\label{qmult}%
\end{table}%

\subsection{Results of the simulation study}
\setcitestyle{square}%
Each simulation consisted in the study of one of the models for a SNP pair. This is an acceptable simplification because the current setting is a problem of multiple testing and not a single high-dimensional test (see \citet{AnnuRev:Cai} for a discussion of the methodological and conceptual differences), that is, there are no underlying asymptotic results when $L\to\infty$ that require a whole $n\times L$ matrix to be built and replicated.
\setcitestyle{round}

\begin{figure}
	\centering\includegraphics[width=.6\textwidth]{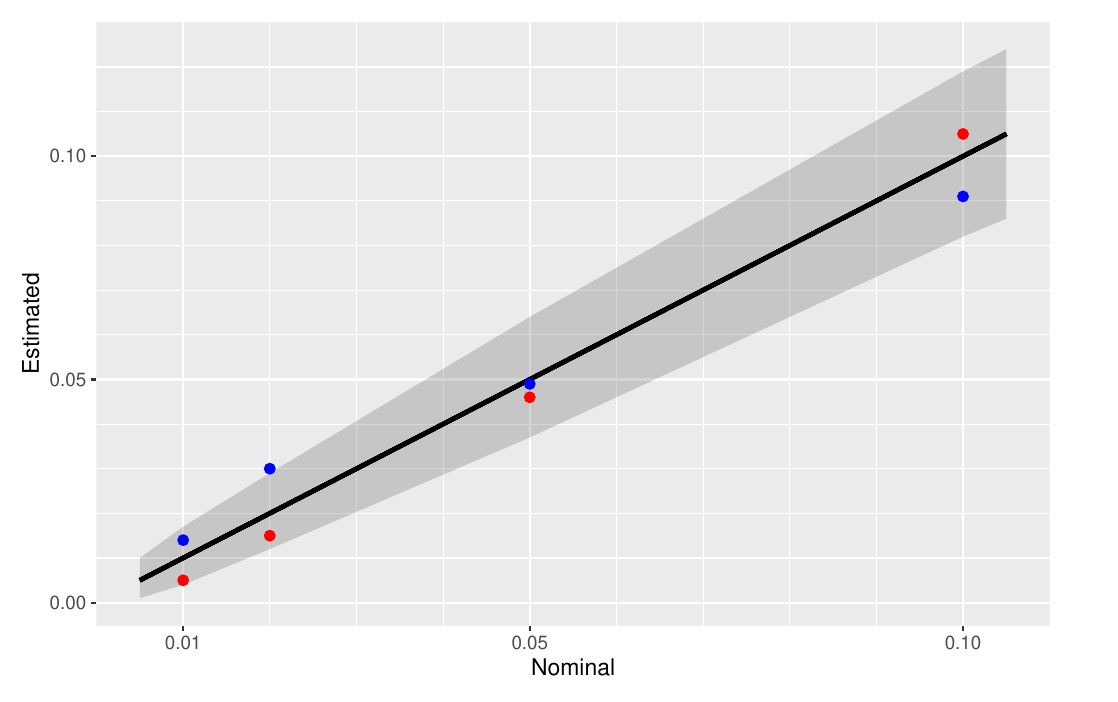}
	\caption{Nominal significance level ($\alpha$) versus empirical power under the null hypothesis ($\hat\alpha$), under model \texttt{indep} in cases and \texttt{indep} in controls. Blue dots correspond to \textit{dcov}; the red ones were generated with BOOST. The grey shadow is a 95 \% confidence band for $\hat\alpha$ given $\alpha$.}
	\label{Calibr:epistasis}
\end{figure}

\begin{figure}
	\centering\includegraphics[width=\textwidth]{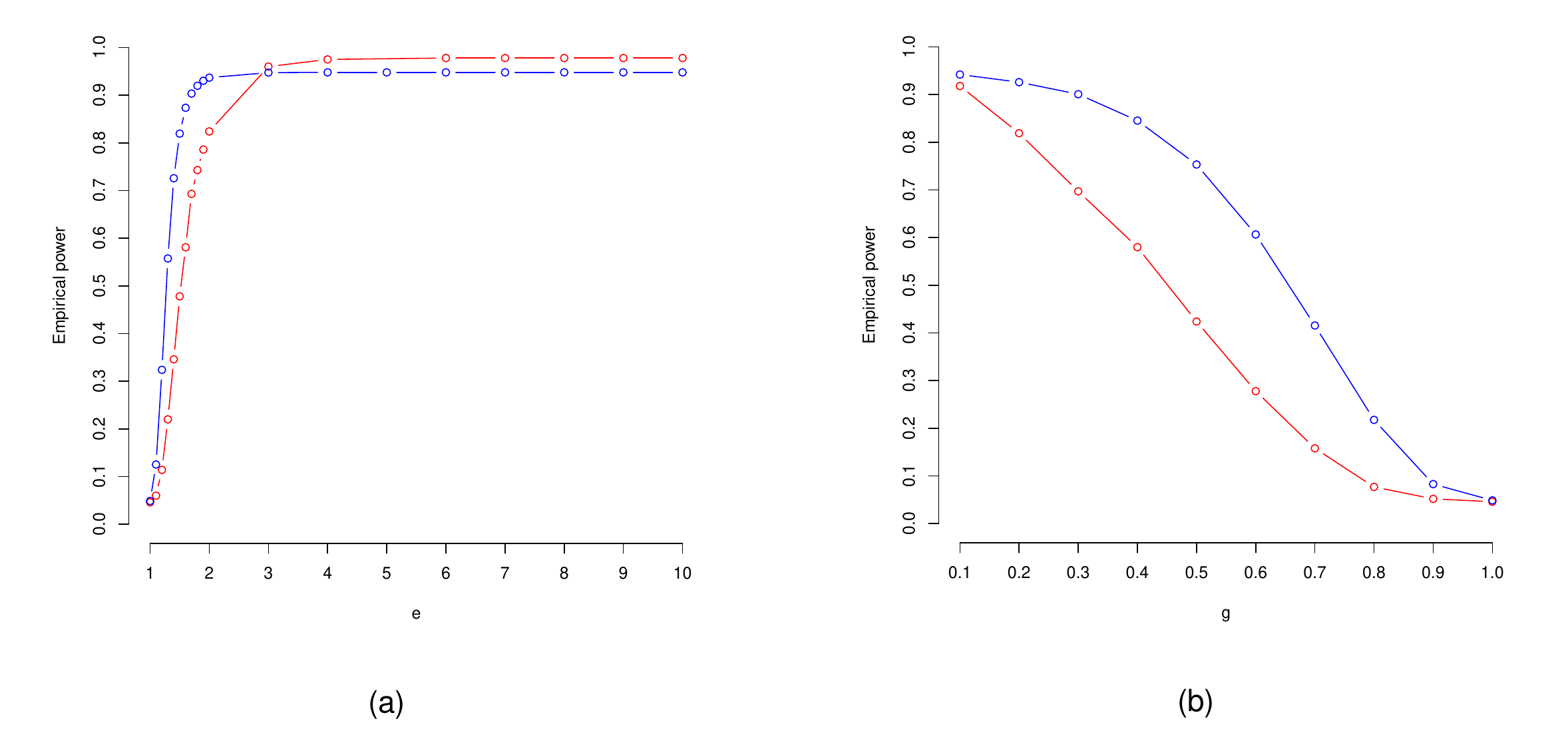}
	\caption{Empirical power when the SNP pair distributions for cases/controls are (a) \texttt{qexp} with parameter $e\in\mathbb{Z}^{+}$ and \texttt{indep}, and (b) \texttt{qmult} with parameter $g\in[0,1]$ and \texttt{indep}. Colour blue represents our distance covariance test; whereas red corresponds to BOOST.}
	\label{Power_indep_q}
\end{figure}

We now show some illustrative examples of the performance of our testing procedure. Firstly, Fig.~\ref{Calibr:epistasis} represents the calibration of significance for some usual nominal levels for the only scenario under the null hypothesis we expect to come across in practice, that is, independence in both cases and controls. On the other hand, empirical power is represented on Fig.~\ref{Power_indep_q}. In all cases, $R=1000$ replicates were carried out. For each plot, we also display the results we obtained with one of the most popular tools within the genomics community for the kind of epistasis we are studying --- it is called \textit{BOOST} \citep{BOOST} and is easily accessible from the widely used genetics software package PLINK \citep{PLINK}. As indicated in Appendix A, there is an extremely large number of options in the literature to perform this task and therefore it is not feasible to compare our technique with a representative fraction of them.

On the basis of the aforementioned tables, it can be concluded that the calibration of significance is acceptable or even good for the most usual levels of nominal $\alpha$. In addition, the plots on Fig.~\ref{Power_indep_q} show that the power is satisfactory (for the models under consideration) and that, as expected, it increases as one gets further away from the null hypothesis. In the scenarios we studied, we have either comparable or more power than BOOST.

\section{Application to a case-control study of schizophrenia}\label{SCZ}

The genomic database that we study in this chapter is described in detail in Section~\ref{gen:database}. It contains observations of $6\,371\,078$ SNPs across all the genome, from a case-control study of schizophrenia in Galicia \citep{Galicia}, with $n_1=585$ cases and $n_2=573$ controls.

For a better understanding of the nature of the dataset and the quality controls \citep{gwas:bimj} and downstream analysis it underwent, we refer the reader to Sections~\ref{gen:database} and~\ref{reprod:deets:epistasis}. Section~\ref{reprod:deets:epistasis} also contains further details on reproducibility.

We now present two experimental setups we carried out to better understand our methodology, by using it to interrogate the schizophrenia dataset. In them, we interpret our analyses of DNA data at ``higher'' levels on the biomolecular hierarchy (proteins and RNA), based on missense SNPs and genetically-regulated gene expression, respectively. As with the simulations, we restrict ourselves to the discrete metric, in order to be agnostic regarding the underlying genetic model.

In each of the two experiments, we will apply the methodology described in Section~\ref{sec:test} to SNP pairs across the human genome, to then interpret the results by performing several tests comparing proportions and ranks in cases versus controls. The key rationale is that the set of putative interactions detected with our testing procedure (i.e., the SNP pairs for which independence is rejected in cases and not in controls) will include both pairs in ``true'' epistasis and instances of population substructure, whereas the SNP pairs where independence is rejected in controls and not in cases only consists of spurious interactions.

\subsection{Genomic database}\label{gen:database}
The SNP data around which the whole Chapter~\ref{ch3} pivots comes from a case-control study of schizophrenia, which was performed on $n_1=585$ patients and $n_2=573$ control blood donors, all of them of Galician origin, as described by \citet{Galicia}.

\begin{sloppypar}
	Each individual's genome was sequenced using microarray \mbox{\textit{PsychArray-24 BeadChip}} (Illumina, San Diego, California). After genotyping, several conventional quality controls were performed. Namely, to avoid experiment-derived problems, it was decided to leave out from the database every SNP that verified any of the following conditions:
\end{sloppypar}
\begin{enumerate}
	\item The minor allele frequency (MAF) is less than 1 \% in our samples.
	\item The genotype proportions differ significantly from Hardy--Weinberg equilibrium in the control sample, for nominal $\alpha=0.001$.
	\item The \emph{call rate} (proportion of non-missing data after genotyping) is under 95 \%, or either it is significantly different between cases and controls ($p$-value of less than 0.001).
\end{enumerate}

Had not the previous conditions been imposed, many badly-behaved SNPs would remain in the database, that is, for many SNPs it would not be possible to clearly discriminate between the three possible genotypes.

Individuals were removed when, after the SNP quality control, the genotyping for more than 5 \% of their SNPs was missing. For assessing cryptic relatedness, we computed the identity by descent proportion (pi-hat statistic) for each pair of individuals and, for every pair in which $\hat\pi>0.15$ one of its members was removed.

All the aforementioned restrictions were applied with the default algorithms and implementations for GWAS quality controls on PLINK \citep{PLINK}.

Data recollection followed the guidelines of the Declaration of Helsinki, was approved by the Galician Ethical Committee for Clinical Research, and
participants signed an informed consent; as stated in \citet{Galicia}.

\subsection{Experiment I: Functional enrichment}

Taking into account the goal of this first experiment, it is sensible to restrict ourselves to a certain subset of the initial database, comprising $L=8030$ missense SNPs.

Firstly, we apply our test procedure separately to cases and controls (as previously discussed), with a \citet{BH} nominal FDR threshold of 0.05. We only consider SNP pairs consisting of two variants that lay on different chromosomes or that are more than 1 Mb apart (i.e., not physically close). This prevents evident cases of spurious findings due to linkage disequilibrium \citep{BOOST}.

We thus obtain 113 out of $\comb{L}{2}$ SNP pairs that show association in cases and not in controls (which we would consider putative interactions), versus 95 in controls and not in cases (which just reflect population substructure). The difference (in proportions) is not significant; with a $p$-value of $0.12$, which could be lower. These 113 and 95 pairs correspond, respectively, to 222 and 189 unique SNPs, a proportion difference with $p$ of 0.055. Those SNPs lay on 220 and 191 different genes. Removing the 13 that are common among both lists, we get 207 and 178 genes ($p$ of 0.07).

We hypothesise that the genes known to be involved in synapse ---which is the biological structure that allows for nervous impulses to be transmitted, and it is known to be closely related to schizophrenia--- will be overrepresented in our group of putative interactions with respect to the spurious ones. Intersecting the genes we had with the list of synaptic genes by \citet{SynGO}, we see that 13 of the 207 and 9 out of 178 genes are known to be related to synapse. The proportion difference has a $p$-value of 0.26, so our results for Experiment I are negative and we can show no (strong) evidence that we are detecting any signal related to synapse.

However, this is not to say that our method cannot offer interesting insight on this data. The current knowledge on complex disease genetics indicates that regulatory regions play a crucial role \citep{Sullivan:Geschwind}, so one should focus on genetically-regulated gene expression, rather than on missense SNPs. This motivates Experiment II.

\subsection{Experiment II: Gene expression}

With this second data example, we want to show that our results make sense at the level of genetically-regulated gene expression (i.e., mRNA). For this task, as explained in Section~\ref{reprod:deets:epistasis}, only some variables on our schizophrenia database can be used, comprising some $L=6456$ SNPs that regulate gene expression in the brain, but not in any other tissue of the human body, according to data from the \citet{GTEx}.

We now apply our procedure as in Experiment I, seeing that there are significantly more pairs in putative interaction than in spurious one: 1272 versus 1137 (with \textit{p} of 0.032), after applying the physical distance threshold of 1 Mb. These pairs represent 1539 and 1439 unique SNPs respectively, again a significant difference ($p$-value $\approx0.019$, which drops to 0.0024 by removing SNPs in both sets).

We finally order the $p$-values we obtained for each of the $\comb{L}{2}$ tests we performed on cases, and do the same for controls. We then take the absolute value of the difference of both ranks for each SNP pair. We hypothesise that those absolute rank differences will tend to be greater on the true positive list than on the false positives. We perform a Wilcoxon--Mann--Whitney $U$ test and we find that we can confirm that it is the case, with a $p$-value of less than $2.2\cdot10^{-16}$.

All in all, the results of Experiment II indicate that we are detecting some genuine signal, at the genetically-regulated gene expression level. This would be very unlikely if our method did not function correctly.

\subsection{Reproducibility details}\label{reprod:deets:epistasis}

We now explain some non-essential technicalities that were left out of the explanation of the previous section.

\subsubsection*{Experiment I: Functional enrichment}

For this experiment, among all the autosomal SNPs, only the missense SNPs (i.e., those that induce a change in the aminoacid sequence of a protein) are considered, as a way to detect interaction at the protein level.

To determine which SNPs are missense and which not, our reference was the  ENSEMBL Biomart database \citep{Biomart}. Since our schizophrenia data refers to the GRCh37.p13 (hg19) assembly of the human genome, we chose the Biomart version accordingly. For a comprehensive review on ENSEMBL and Biomart, we refer the reader to \citet{ENSEMBL:rev}.

As we want to have some ability to detect some signal, and it is remarkably difficult to detect any instance of epistasis in real data \citep{Russ}, we restrict ourselves to the SNPs for which the least frequent of the two alleles is observed with a frequency of at least $0.05$ in our samples and $0.01$ on \citet{Biomart}. The latter filter also ensures that the missense SNPs we are studying are well-known and annotated. This yields a SNP count of $19\,356$.

We then remove all the SNPs that lay on any of the 25 small regions of the human genome which are known to suffer from \emph{long-range linkage disequilibrium} (LD), a phenomenon that can cause confusion between true SNP-SNP interactions related to schizophrenia and associations due to the architecture of chromosomes \citep{HighLD:ajhg}. The list of those 25 regions we referred to can be found, for example, on \citet{Facal}.

Furthermore, once the high LD regions were removed, we followed standard practice among geneticists \citep{Abdellaoui:2013} and \emph{pruned} those SNPs from the remaining $18\,268$ showing evidence of short-range LD, setting $r^2<0.1$ in PLINK 1.9 \citep{PLINK}. We used windows of width 500 SNPs, shifting them $+1$ SNP on each step. Thus, we obtained the $L=8030$ SNPs that were analysed with our method on Experiment I.

Whenever we assign SNPs to genes, we once again follow the annotations on ENSEMBL Biomart.

\subsubsection*{Experiment II: Gene expression}

On Experiment II, we will only use SNPs that regulated gene expression (i.e., eQTLs) in any of the brain tissues, and nowhere else within the human body, according to data from the \citet{GTEx}. With this aim, we downloaded \texttt{GTEx\textunderscore Analysis\textunderscore v7\textunderscore}\texttt{eQTL.tar.gz} (single tissue cis-eQTL data for GTEx Analysis V7, dbGAP accession  phs000424.v7.p2 ) from
\begin{center}
	\url{https://www.gtexportal.org/home/datasets},
\end{center}
which again refers to the GRCh37.p13 (hg19) assembly of the human genome. There we simply removed anything that is not a SNP (e.g., insertions), we created lists of SNPs that regulate gene expression on brain and non-brain, and performed a set difference.

We therefore chose those SNPs present on our study which were also among the $97\,913$ SNPs that act as eQTLs only in brain (and not in other tissues). Removing data on sexual chromosomes, as well as the high LD regions (as in Experiment I), the SNP count drops to 56 395. After once again pruning the short-range LD, we get the final SNP list for this experiment, which has length $L=6456$.

\section{Discussion and conclusion}\label{epist:discu}

Distance correlation has been shown to characterise independence for 3-point marginal spaces. With this approach, a hypothesis test based on the general characterisation of independence that distance correlation offers has been designed, extending the idea of LCTs \citep{Cai:Liu} to ternary data.

We derive the explicit asymptotic null distribution of the distance-covariance statistics that arise. To our knowledge, the usage of distance correlation in discrete spaces (in genomics or elsewhere) ---and, in particular, its application to the search for SNP-SNP interactions--- has no precedents in literature. Moreover, no previously published research has attempted to perform large-scale multiple testing with any of the techniques derived from energy statistics \citep{TEOD}. However, what does exist in the literature is the usage of distance correlation for finding the association between genetic data (as observations of continuous random variables in Euclidean spaces) and a phenotype \citep{Hua:kernel:gwas}, which is another interesting problem, but completely different both regarding biological and mathematical factors.

Simulations show that the calibration of significance is adequate and that power is considerably high against various alternatives. We also show that we generally outperform one of the most popular epistasis detectors \citep{BOOST} in the scenarios we have studied.

The schizophrenia database has been interrogated with our methodology, obtaining biologically sound results at the level of genetically-regulated gene expression. Some very recent studies show evidence of epistasis between regulatory regions of the human genome \citep{Lin,Patel}, which supports our findings.

In order to frame our results, we would like to emphasise that all popular epistasis detectors find large amounts of false positives and do not have a really high power \citep{Russ}. Therefore, the main limitation of our method (as it is of any other for this task) is that it is very difficult to make any solid discoveries when working with real data. Epistasis detection is an extremely challenging biostatistical problem, in which there is much still progress to be made, given its key role in human complex genetics \citep{Van Steen:Moore}.


%% file: ch4_v14.tex
\chapter{Testing for gene-phenotype associations in human complex traits}
\thispagestyle{empty}
\label{ch4}
\graphicspath{{./fig_ch4/}}

Unraveling the relationship between genes and observable (phenotypic) features has been a central question to genetics since the inception of the discipline \citep{Zschocke}. For the last 15 years, the GWAS has been the most prominent design for human trait studies. A main goal is to detect SNPs that are significantly associated with the variability of the phenotypic feature of interest \citep{15y}.

It is often assumed by practitioners that every association between a genetic variant and a quantitative phenotype is linear and additive, a simplification that is not substantiated by biological knowledge, as indicated in previous chapter. In this context, we present generalised distance covariance (GDC) as a novel tool for approaching GWAS. As already explained in Chapter~\ref{ch2} and illustrated in Chapter~\ref{ch3}, GDC characterises any kind of dependency ---not only the linear one--- and, with a convenient choice of the distance that one uses on the SNPs, it is possible to select a priori the kind of genetic model that it is desired to test for. This allows for profound biological interpretations. The GDC theory is mathematically equivalent to the Hilbert--Schmidt independence criterion, which in turn is dual to a linear global test in the space of kernel features.

We propose a family of hypothesis tests for marginal effects of SNPs on the trait of interest, one per distance/kernel that we define. We firstly prove consistency against all functional alternatives. We then explicitly derive the asymptotic null distribution of the test statistic. This way, we avoid the resampling schemes that are the rule in the kernel and distance literature, which is key to perform quickly and precisely in simulations. With further theoretical developments, we showcase how each of our tests is the locally most powerful one for a certain underlying model. In addition, we adjust our testing for nuisance covariates, which is crucial in genomics. We finally show satisfactory performance in simulated datasets, and demonstrate applicability by studying the serum levels of liver enzymes.

The rest of the chapter is structured as follows. Section~\ref{GWAS} introduces the discipline of quantitative trait genetics and its state of the art. Section~\ref{Models:SNP} introduces some additional genetic concepts, as well as the three modern independence testing traditions that we will be focusing on. In Section~\ref{gwas:testing} we introduce our family of tests for marginally significant SNPs and delve into some of their theoretical properties, to then move towards their local optimality and interpretation in Section~\ref{GT}. Section~\ref{gwas:covariates} presents the theory that allows us to account for confounders when testing. We make some practical remarks in Section~\ref{gwas:practical}. We illustrate the performance of our technique, both with simulations (\S~\ref{gwas:simu}) and a real data example (\S~\ref{rdata:hepatic}). We finalise with a discussion of the results and the conclusions thereof (\S~\ref{gwas:discu}).

The contents of this chapter are collected in \citet{F:gwas}.

\section{Complex human traits and genome-wide association studies}\label{GWAS}

In humans, a vast majority of the phenotypic traits are multifactorial, that is, their variability is due to a large and complex combination of environmental and genetic factors, with each of them contributing with very small effects, as a general rule.

As already explained in Chapter~\ref{ch1} and reiterated in Chapter~\ref{ch3}, a GWAS dataset contains the information for a large number of sampled (human) individuals on an even larger number of SNPs. A prominent goal is to identify genotype-phenotype associations. Thus, the response variable corresponds to a phenotypic characteristic of interest, which in this chapter we will assume to be continuous. This means that we will not be considering the case-control experimental design of Chapter~\ref{ch3}, but rather a situation where there is a single large cohort of individuals for which the complex trait under study is quantified.

As we did in Section~\ref{sec:space3}, we remark that for a biallelic SNP there are three possible genotypes an individual can have. If we denote by $C$ and $T$ the two alleles, the support of the random element `Genotype of the SNP under consideration' is the 3-point set:
$$\{CC,CT,TT\}.$$

It is important to use computationally efficient and statistical powerful testing methodology, which can capture the particular structure of the data. For the testing for the association between a quantitative phenotype and individual SNPs, almost invariably, a standard linear regression model is applied \citep{Brandes}, in which practitioners code the three possible states by counting the number of minor alleles. In our example, if we assume that $f(T)<f(C)$ without loss of generality, we get:$$0:=CC;\;\;\;1:=CT;\;\;\;2:=TT.$$Then one would treat the possible values $\{0,1,2\}$ of each SNP as either categorical or continuous. However, it has been shown that these approaches often lead to suboptimal results \setcitestyle{square}(e.g., sometimes the maximum phenotypic effect is achieved in heterozygosity \citep{Costas:Heteroz_opt})\setcitestyle{round}%
and anyhow nothing ensures that two copies of the minor allele will have an effect of twice the size of that of a single copy (i.e., additivity may not hold), and in such scenarios the traditional test can have little to no power. Hence, it is sensible and necessary to consider different models of genotype-phenotype association \citep{SIM}. On top of that, given that sample sizes for human studies can only increase up to a certain upper bound, there is a need for new statistical techniques that can detect causal SNPs that are being overlooked by traditional GWA analyses.

In this chapter, we present a novel method for testing the association of a single SNP with a quantitative response, by assuming no particular structure on the marginal space $\{CC,CT,TT\}$ for each SNP. In order to work with this abstract type of data, we equip the $3$-point space with a premetric structure. Trying to find associations in a space where we can only work with distances naturally leads to basing our testing procedure generalised distance covariance \citep{SRB,Jakobsen,Lyons}, which we introduced in detail in Chapter~\ref{ch2}. Distance covariance vanishes if and only if there is independence, thus allowing for the detection of any kind of dependencies, and it is equivalent to its kernel counterpart, the HSIC \citep{Sejdinovic}, which we had already introduced in Section~\ref{hsic}. Both tests are tantamount to performing the locally most powerful test of significance of a certain regression model derived from the data, with this third tradition of independence testing being know as the ``Global Test'' \citep{goeman2006testing}, as we had explained in Section~\ref{gt}.


Our methodology yields a different hypothesis test each time one changes the distance/kernel with which the support of the SNPs is equipped. Only some distances/kernels make sense for that purpose, and we thus define a family of tests for large-scale detection of SNPs that are significantly associated with the phenotypic trait of interest. We will show how we have consistency against all (functional) alternatives, as well as high power against many alternatives regardless of the choice of the distance. Our techniques are based on approximating the true null distribution of the test statistic using theoretical developments, which proves to be very computationally efficient and precise (i.e., we demonstrate applicability). Each of our tests is the locally most powerful one under certain model assumptions, and we can know what model this is a priori, given that it is determined by the initial choice of the distance. This means that, in any real data application, we can very easily interpret our results. Finally, we highlight that our test can be adjusted for covariates, which is a fundamental requisite for any GWAS analysis tool.

\section{Models for the association between SNPs and quantitative traits}
\label{Models:SNP}

We now introduce some concepts of basic quantitative genetics modelling, which will allow us to interpret our testing framework and the results it offers from a biological perspective.

As previously mentioned, we will assume without loss of generality that SNPs are biallelic loci, i.e., they can manifest themselves as either major allele $A_1$ or minor allele $A_2$, with the latter having a lower frequency in the population by definition. With this notation, the three possible genotypes each individual can carry are: $A_1 A_1$ (major allele in homozygosity), $A_1 A_2$ (heterozygosity) and $A_2 A_2$ (minor allele in homozygosity). To be consistent with standard genetic notation, we will encode those three genotypes as the values of a random element $X$ with support $\{0,1,2\}$, which counts the occurrences of $A_2$.

\begin{table}[H]
	\begin{center}
		\caption{Association models between a SNP $X$ and an absolutely continuous quantitative trait $Y$.}
		\begin{tabular}{r|rrr}
			& $X=0 \, (A_1 A_1)$ & $X=1 \, (A_1 A_2)$ & $X=2 \, (A_2 A_2)$ \\
			\hline
			mean of $Y$ conditional to $X$ & $\mu_0$ & $\mu_1$ &  $\mu_2$\\
			standardised effect (for $\mu_0 \neq \mu_2)$ &$0$ & $h$  & $1$ \\
		\end{tabular}
		\label{Table:Models:SNP}
	\end{center}
\end{table}

For studying different models between the state $X \in \{0,1,2\}$ of a certain SNP and an absolutely continuous response $Y \in \R$, let us define the conditional mean of $Y$ given $X$: $$\mu_j = \E [Y|X=j],$$ where $j \in \{0,1,2\}$. In classical quantitative genetics \citep[][Section 3.2]{GI}, the association between $X$ and $Y$ is represented as on Table~\ref{Table:Models:SNP}, where one is generally assuming that the means of the two homozygous states are different: $\mu_2 \neq\mu_0$. The standardised effect for each state $j\in\{0,1,2\}$ is hereby calculated as $\frac{\mu_j-\mu_0}{\mu_2-\mu_0}$.

The association models are then classified based on the biological interpretation of the value of the parameter $h:=\frac{\mu_1-\mu_0}{\mu_2-\mu_0}\in\mathbb R$, known as {\em heterozygous effect}:

\begin{itemize}
	\item $h<0$: \emph{underdominant} model (or negative overdominant model)
	\item $h=0$: \emph{dominant-recessive} model; where $A_1$ is \emph{dominant}, $A_2$ is {\it recessive}.
	\item $h=1$: \emph{dominant-recessive} model; where $A_2$ is \emph{dominant}, $A_1$ is \textit{recessive}.
	\item $h \in ]0,1[$: \emph{codominant model} (or incomplete dominance model). A codominant model with $h=\tfrac12$ is called \emph{additive} model.
	\item $h>1$: \emph{overdominant} model.
\end{itemize}

In the course of this chapter, we will also consider models for which $\mu_0 = \mu_2$ and $\mu_1 \neq \mu_0$, which we will refer as {\em purely heterozygous}.

We say that the cases $h\in\{0,1\}$ correspond to the dominance of (the phenotype of) $A_1$ and $A_2$ respectively, as current nomenclature of medical genetics indicates that dominance refers to the fact of observing the exact same phenotype of homozygosity also under heterozygosity \citep{Zschocke}. And it is in that sense that we understand $h$ as a measure of \emph{dominance} and hence the names of the genetic models it defines have all something to do with that word.

\section{A distance-based test for gene-phenotype dependence and its kernel counterpart}\label{gwas:testing}

\subsection{Tailoring premetrics to SNP data}\label{tailoring}

In this section, we investigate generalised distance covariance $\V_{\rho_\mX,\rho_\mY}$ for testing independence between a SNP $X \in \mX := \{0,1,2\}$ and a quantitative response $Y \in \R$. As elucidated in Equation~\eqref{eq:gendcov2}, $\V_{\rho_\mX,\rho_\mY}$ is fully specified by choosing premetrics $\rho_\mX$ and $\rho_\mY$ on $\mX$ and $\mY$, respectively. While many distances on $\R$ appear sensible, we restrict ourselves to
\begin{equation}\label{rho_Y}
\rho_\mY(y,y') = \frac{1}{2} |y-y'|^2,
\end{equation}
since it leads to both tractable test statistics (Section~\ref{gwas:testing}) and illustrative interpretations (\S~\ref{GT}). 

For defining meaningful distances on the support space of the SNPs, we note that $0$ and $2$ correspond to homozygous states, while $1$ denotes the heterozygous state. The definition which homozygous state is $0$ and which is $2$ is typically given by the convention of using $0$ for the more frequent allele $A_1$. We argue that any reasonable testing procedure should be invariant to the arbitrary labeling of $A_1$ and $A_2$. Consequently we only consider distances for which $d(0,1) = d(1,2)= 1,$ where we set the unit scale by normalising these distances to one (note that the conclusions of the test would be the same under any scale transformations).

The resulting family of distances is characterised by the nonnegative real number $b:=d(0,2)$ and we will denote them as $d_b$ in the following. For a premetric to define a distance covariance in the sense of Section \ref{gdc:semimetric}, it must be of negative type and for this in turn, its square root must satisfy the triangle inequality. This holds if and only if $\sqrt{d_b(0,2)} \leq \sqrt{d_b(0,1)} + \sqrt{d_b(1,2)} = 2$, which is equivalent to $b \leq  4$. Proposition 3 in \citet{Sejdinovic} implies that $b \in ]0,4]$ indeed defines valid semimetrics of negative type. For $b=0$, $d_b$ obviously does not define a semimetric, since two distinct points have distance $0$. However, it is clear that the theory by \cite{Sejdinovic} easily extends to premetrics, assimilating points that are separated with distance zero (i.e., dropping the identity of indiscernibles).

We will hence study the family of premetrics $\{d_b\}_{b\in[0,4]}$, where $d_b:\spx\times\spx\longrightarrow\R$ is such that $d_b(0,1) = d_b(1,2) =1$ and $d_b(0,2)=b \in [0,4]$. Important special cases are:
\begin{itemize}
	\item The discrete metric $d_1(0,1) = d_1(1,2) = d_1(0,2) = 1$, as studied in Chapter~\ref{ch3}.
	\item The Euclidean distance $d_2(x,x') = |x-x'|$, connected to standard distance covariance on the ordered set $\{0,1,2\} \subset \mathbb{R}$.
	\item The squared distance $d_4(x,x') = (x-x')^2$, linked to linear regression on the ordered set $\{0,1,2\} \subset \R$.
\end{itemize}

We also note that any premetric $d_b$ with $b \in ]1,4[$ is related to the $\alpha-$distance covariance of \citep{SRB} for $\alpha = \log_2 b$.

Consider now the classical genotype-phenotype association models introduced in Section \ref{Models:SNP}. If the association model is known beforehand, it appears sensible to tailor the distance $d_b$ on the genotype level to the model under consideration. In particular, it is immediately clear that the distance $d_0$ reflects a purely heterozygous model, where $\mu_0 = \mu_2$; moreover it is easy to see that $d_4$ is a sensible choice for the additive model with heterozygous effect $h= \frac{\mu_1 - \mu_0}{\mu_2 - \mu_0} = \frac{1}{2}$.

However, the exact genotype-phenotype association model is typically unknown in practice, and we will see in the following that it is precisely for this situation that the GDC based on $d_b$ shows its strengths. In particular, we will see in Section \ref{GT} that each $d_b$ corresponds to the locally most powerful tests in a specific situation where the association model is uncertain.

For the rest of the chapter, we use the simplified notation
$$
{\V}_b := {\V}_{d_b,\rho_\mY}, \quad  \widehat{\V}_b := \widehat{\V}_{d_b,\rho_\mY}, 
$$
where $\rho_\mY$ is given in~\eqref{rho_Y}.

We now recall, from Section~\ref{hsic}, that the duality between the HSIC and distance covariance is based on duality between kernels and premetrics. Along those lines, the following proposition provides kernels induced by the family of distances $d_b$, as a particular case of Equation~\eqref{eq:smindkernel}.

\begin{proposition} \label{prop:kernel} 
	The distance $d_b$ induces the kernel $k_b$ with
	$$
	k_b(0,0) = k_b(2,2) = 1; \quad k_b(1,1) = k_b(0,1) = k_b(1,2) = 0; \quad k_b(0,2) = 2-b.
	$$
\end{proposition}

Once again echoing Section~\ref{hsic}, by virtue of Mercer's theorem, each (nonsymmetric, positive definite) kernel $k: \mZ \times \mZ \to \R$ can be decomposed into \emph{features}, that is, there is a map $\boldsymbol{\Phi}: \mZ \to \R^d$ (for $d\in\Zplus\cup\{\infty\}$) such that $$k(z,z') = \langle \boldsymbol{\Phi}(z), \boldsymbol{\Phi}(z') \rangle
\;\text{ for all }z,z'\in\mZ;$$
where $\langle \cdot, \cdot \rangle$ denotes the standard inner product in $\R^p$. And whenever we have a premetric, we can obtain a feature map of the kernel induced by that distance. The following result provides a feature map of $d_b$.

\begin{proposition}\label{prop:fm}
	A  feature map $\boldsymbol{\Phi}=(\phi_1,\phi_2)$ of $d_b$ is given by
	$$
	\phi_1(x) = \sqrt{\frac{b}{2}} (-1_{\{x=0\}} + 1_{\{x=2\}}) , \quad \phi_2(x) = \sqrt{\frac{4-b}{2}} 1_{\{x=1\}}
	$$
	or in vector notation (that we will use throughout the chapter),
	$$
	\phi_1 = \sqrt{\frac{b}{2}} \begin{pmatrix} -1 \\ 0 \\1 \end{pmatrix}, \, \phi_2 = \sqrt{\frac{4-b}{2}} \begin{pmatrix} 0 \\ 1 \\ 0 \end{pmatrix}.
	$$
\end{proposition}

\subsection{Characterisation of fluctuations in the conditional mean of the response}

Unlike classical distance covariance, ${\V}_b$ does not characterise independence because $\rho_\mY$ is not of strong negative type. However, ${\V}_b$ can detect all associations defined via the classical phenotype-genotype association models introduced in Section \ref{Models:SNP}. For this purpose, we again consider
$$
\mu_j = \E[Y|X=j]
$$
for $j\in\spx\equiv\{0,1,2\}.$ Moreover, we define
$$
p_j = \Prob(X=j)
$$
for $j\in\spx\equiv\{0,1,2\}.$ Then, if $b \in ]0,4[$ and the first moment of $Y$ exists, under some regularity conditions, we have that the distance covariance between $X$ and $Y$ vanishes if and only if the mean effects of $Y$ are homogeneous among the categories of $X$, i.e. if and only if: $\mu_0=\mu_1=\mu_2$.

\begin{theorem} \label{th:dcovzero}
	Let $(X,Y)$ be jointly distributed random variables in $ \{0,1,2\} \times \R$ with $\E[Y] < \infty$. If $\mu_0 = \mu_1 = \mu_2$, then
	$$
	\V_b^2(X,Y) = 0.
	$$
	Moreover, if $b \in ]0,4[$ and $p_j >0$ for $j \in \{0,1,2\}$, then $\mu_i \neq \mu_j$ for some $i\neq j$ implies that
	$$
	\V_b^2(X,Y) > 0.
	$$
\end{theorem}

The second part of Theorem \ref{th:dcovzero} does not hold true for the ``boundary cases'' of GDC with  $b \in \{0,4\}$; these are exactly the cases where $\widehat{\V}_b$ is tailored to one single genetic model (the purely heterozygous model for $b=0$ and the additive model for $b=4$).

\begin{proposition}\label{prop:nocondmeanfully}
	Let $b$ be either $0$ or $4$ and let $X$ be a random variable on $\{0,1,2\}$ with $p_j >0$ for $j \in \{0,1,2\}$. Then we can define a random variable $Y$ on the same underlying probability space such that $\mu_i \neq \mu_j$ for some $i\neq j$, but $\V^2_b(X,Y) = 0$.
\end{proposition}

Theorem \ref{th:dcovzero} implies that, for $b \in ]0,4[$, the empirical version $\widehat{\V}_b$ may be used to establish consistent tests for the null hypothesis
$$
H_0: \mu_0 = \mu_1 = \mu_2.
$$

In the following, we will introduce tests based on the asymptotic and the finite sample distribution of $\widehat{\V}_b$.

\subsection{Asymptotic and finite-sample distribution}

The asymptotic distribution of  distance covariance is known to follow an infinite weighted sum of chi-squared distributed random variables \citep{TEOD}, which is almost never exploited in the specialised literature when it comes to applying the test in practice. This is due to the difficulty of estimating the coefficients of the series and of deciding where to truncate. However, resampling is hardly feasible in the GWAS setting where a large number of tests have to be performed.

In the following, we will derive a closed-form expression for our version of generalised distance covariance $\widehat{\V}_b$, which enables testing at a reasonable speed.

\begin{theorem}\label{testasy}
	Let $\bX=(X_1,\ldots,X_n)$ and $\bY = (Y_1,\ldots,Y_n)$ denote IID samples of jointly distributed random variables $(X,Y) \in \{0,1,2\} \times \R$ with $\Var(Y) = \sigma_Y^2 <\infty$. If $X$ and $Y$ are independent, then, for $n \to \infty$,
	$$
	n \, \widehat{\V}_b^2 \stackrel{\mathcal{D}}{\longrightarrow}  
	\sigma_Y^2 (\lambda_1 Q_1^2 + \lambda_2 Q_2^2),
	$$
	where $Q_1^2$ and $Q_2^2$ are chi-squared distributed with one degree of freedom and $\lambda_1$ and $\lambda_2$ are the eigenvalues of matrix 
	$$
	K = \begin{pmatrix} \frac{b}{2} (p_0 + p_2 - (p_2 -p_0)^2) & \sqrt{\frac{b \, (4-b)}{4}} p_1 (p_0-p_2)\\
		\sqrt{\frac{b \, (4-b)}{4}} p_1 (p_0-p_2) & \frac{4-b}{2} (p_1-p_1^2)\end{pmatrix}.
	$$
\end{theorem}

Using the asymptotic distribution for testing is typically more problematic in GWAS than for standard settings, since the convergence is  slower in the tails of the distributions and we are often interested in approximating very small $p$-values.

Assuming that the phenotype for each of the three genetic states is normally distributed with homogeneous variance, we can derive the finite-sample distribution of $\widehat{\V}_b^2$.

\begin{theorem} \label{testfinite}
	For $n \in \Zplus$, let $\bX=(X_1,\ldots,X_n) \in \{0,1,2\}^n$ denote a fixed sample and let $\bY = (Y_1,\ldots,Y_n)$ be defined by
	$$
	Y_i = \mu_j \, 1_{\{X_i = j\}} + \varepsilon_i,
	$$
	where $\boldsymbol{\mu} = (\mu_0, \mu_1, \mu_2)^t \in \R^3$  and $(\varepsilon_1,\ldots,\varepsilon_n)$ is IID with $\varepsilon_i \sim \mathcal{N}(0,\sigma^2_Y)$. If $\mu_0 = \mu_1 = \mu_2$, then,
	\begin{equation}\label{eq:testfinite1}
		\Prob \left( \frac{n \, \widehat{\V}_b^2}{\widehat{\sigma}_Y^2} \geq k \right) = \Prob(T_n \geq 0),
	\end{equation}
	where $\widehat{\sigma}_Y^2 = \frac{1}{n} \sum_{j=1}^n (Y_j -\tfrac{1}{n} \sum_{i=1}^n Y_i)^2 $,
	$T_n$ is defined by
	\begin{equation} \label{eq:testfinite2}
		T_n = \Big(\widehat{\lambda}_1- \frac{k}{n} \Big) \, Q_1^2 + \big(\widehat{\lambda}_2- \frac{k}{n} \big) \, Q_2^2 - \frac{k}{n} Q_3^2 - \cdots - \frac{k}{n} Q_{n-1}^2
	\end{equation}
	and $Q_1^2,\ldots,Q_{n-1}^2$ are IID chi-squared with one degree of freedom each; $\widehat{\lambda}_1$ and $\widehat{\lambda}_2$ are the eigenvalues of matrix 
	$$
	K = \begin{pmatrix} \frac{b}{2} (\widehat{p}_0 + \widehat{p}_2 - (\widehat{p}_0 -\widehat{p}_2)^2) & \sqrt{\frac{b \, (4-b)}{4}} (-\widehat{p}_1 (\widehat{p}_0-\widehat{p}_2) \\
		\sqrt{\frac{b \, (4-b)}{4}} (-\widehat{p}_1 (\widehat{p}_0-\widehat{p}_2) & \frac{4-b}{2} (\widehat{p}_1-\widehat{p}_1^2)\end{pmatrix},
	$$
\end{theorem}

\subsection{Computing \textit{p}-values} \label{sec:pvalues}

In GWA studies, it is standard practice to make decisions on individual SNPs based on the ``genome-wide significance threshold'', which is defined as $\alpha = 5 \cdot 10^{-8}$ \citep{Tam}. In this setting, using the previous asymptotic results leads to some inflation of the type I error rate even for moderately large sample sizes. For this reason, we recommend to use the finite-sample distribution in Theorem~\ref{testfinite}, except for very large sample sizes, say $n > 30\,000$.

For calculating $p$-values, we first observe that by Theorem \ref{testfinite}, for $\widehat{\lambda}_2- \frac{k}{n} > 0$,
\begin{align*}
	\Prob \left( \frac{n \, \widehat{\V}_b^2}{\widehat{\sigma}_Y^2} \geq k \right)&= \Prob \left( \frac{\big(\widehat{\lambda}_1- \frac{k}{n} \big) \, Q_1^2 + \big(\widehat{\lambda}_2- \frac{k}{n} \big) \, Q_2^2}{\frac{1}{n-3} (Q_3^2 - \cdots - Q_{n-1}^2)} \geq \frac{k (n-3)}{n} \right) \\ &=1 - G_{F(2 (\widehat{\lambda}_1- \frac{k}{n}), 2 (\widehat{\lambda}_2- \frac{k}{n}); n-3)} \left(\frac{k (n-3)}{n} \right),
\end{align*}

where $G_{F(\alpha_1,\alpha_2; \nu)}$ is the cumulative distribution function of a generalised $F$ distribution in the terminology of \cite{Ramirez}. A closed-form expression for $G_{F(\alpha_1,\alpha_2; \nu)}$  can be derived from the general result in \cite{Dunkl}, yielding:
\begin{equation} \label{eq:appell}
	G_{F(\alpha_1,\alpha_2; \nu)}(x) =  \,\Big(\frac{\nu \alpha_2}{2 x+ \nu \alpha_2} \Big)^{\nu/2+1} \, \frac{x}{ \sqrt{\alpha_1 \alpha_2}} F_1 \left( \frac{\nu}{2} +1, \frac{1}{2}, 1; 2; \frac{(1-\frac{\alpha_2}{\alpha_1})x}{(x+ \frac{\nu \alpha_2}{2})} , \frac{x}{(x+ \frac{\nu \alpha_2}{2})} \right),
\end{equation}
where $F_1$ is the first Appell (hypergeometric) series \citep{Appell}.
The test described in Theorem \ref{testfinite} can be regarded as a generalisation of the classical $F$-test in linear regression. In particular, for $b\in\{0,4\}$, it follows that $\widehat{\lambda}_2 = 0$ and we obtain exactly the $F$-statistic for a simple linear regression model with predictors $1_{\{X=1\}}$ (corresponding to a purely heterozygous model) and $X$ (corresponding to an additive model), respectively.

For calculating the $p$-value, one can either numerically evaluate the closed form expression using efficient algorithms for the Appell $F_1$ hypergeometric series or use one of the many algorithms for the evaluation of the distribution function of quadratic forms of Gaussian variables \citep{Duchesne}  using Equation~\eqref{eq:testfinite1}. From our experience, the former option is both computationally more efficient and more precise, so it will be our choice any time we apply the finite-sample distribution throughout this chapter. The main part of the code is written in R, and from it we call the Python package \texttt{mpmath} \citep{mpmath} for a precise and computationally efficient calculation of the Appell $F_1$ hypergeometric series. To further speed up the calculation, we now derive upper and lower bounds for the $p$-values yielded by the finite-sample distribution (as per Theorem \ref{testfinite}).

\begin{proposition}\label{prop:mM}
	Let $G_{\chi^2(w_1,w_2)}$ denote the cumulative distribution function of random variable $w_1 Q_1^2 + w_2 Q_2^2$, where $Q_1^2$ and $Q_2^2$ are IID chi-squared distributed with one degree of freedom. Further, let $G_{F(d_1,d_2)}$  denote the cumulative distribution function of the classical $F$-distribution with $d_1$ and $d_2$ degrees of freedom. Then:
	$$
	p^*  \leq \Prob \left( \frac{n \, \widehat{\V}_b^2}{\widehat{\sigma}_Y^2} \geq k \right) \leq p^{**},
	$$    
	where for $\widehat{\lambda}_2 - \frac{k}{n} >0$, 
	\begin{align*}
	p^* = 1- \min\Bigg\lbrace &G_{\chi^2(\widehat{\lambda}_1- \frac{k}{n}, \widehat{\lambda}_2- \frac{k}{n} )} \left(\frac{k (n-3)}{n} \right) , \\
	&G_{F(1,n-3)} \left(\frac{k (n-3)}{\widehat{\lambda}_1 n - k} \right),\\
	& G_{F(2,n-3)} \Bigg(\frac{k (n-3)}{ \prod_{i=1}^2 (\widehat{\lambda}_i n - k)^{1/2}	\, } \Bigg)\Bigg\rbrace
	\end{align*}   
	and 
	$$
	p^{**} =  5 \, \left( 1- G_{F(1,n-2)} \left(\frac{k (n-2)}{\widehat{(\lambda}_1 + \widehat{\lambda}_2) n - 2 \, k } \right) \right).
	$$
	For $\widehat{\lambda}_2 - \frac{k}{n} \leq 0$, 
	$$
	p^* = 1-  G_{F(1,n-2)} \left(\frac{k (n-2)}{\widehat{\lambda}_1 n - k} \right),
	\, \text{and} \quad
	p^{**} = 1-  G_{F(1,n-3)} \left(\frac{k (n-3)}{\widehat{\lambda}_1 n - k} \right).
	$$  
\end{proposition}

When performing GWA studies in practice, if the goal is to detect genome-wide significant variants, it is usually not interesting to calculate precisely the largest $p$-values (say, for example, greater than $M =  10^{-4}$). On the other hand, it may also be not sensible to precisely evaluate extremely small $p$-values (say smaller than $m = 10^{-64}$). This is the fundamental idea under the computational trick we explain below. However, it should also be noted that there are tasks related to GWASs (like the evaluation of polygenic scores) where one may be interesting in also being accurate for larger $p$-values. In those cases, the value of $M$ should be chosen accordingly.

For a fast algorithm, we first calculate the approximations $p^*$ and $p^{**}$ for all SNPs. This can be done extremely efficiently, for example by using the algorithms for convolutions of gamma variables by \citet{coga}, which are conveniently available as package \texttt{coga} in \texttt{R} \citep{R}. Precise evaluation of the $p$-values in Theorem \ref{testfinite} is then only carried out for the SNPs satisfying $p^* < M$ and $p^{**} > m$. In Section~\ref{gwas:simu}, the computational efficiency of this fast algorithm is compared to that of the naive algorithm, which evaluates the precise $p$-value for all SNPs.

\section{Locally most powerful property and interpretation}\label{GT}
In Section \ref{gwas:testing}, we derived a computationally efficient test that can detect all  alternatives that can be expressed by the classical genetic associations in Section \ref{Models:SNP}. In the following, we show that for each $b \in [0,4]$, $\widehat{\V}_b^2$ features a valuable interpretation as the locally most powerful test statistic in certain models. This provides both a theoretical guarantee for the statistical efficiency of $\widehat{\V}_b^2$ and contributes to better understanding which choices of $b$ are the most suitable from a biological perspective.

The classical score test \citep{cox1979theoretical} for a model with likelihood $\ell^{*}(\theta; \bZ)$ where $\bZ \in \R^n$ is an observation and $\theta \in \Theta \subset \R$ is a univariate parameter, is a one-sided test of
$$H_0^* : \theta = \theta_0\text{ against }H_1^* : \theta > \theta_0$$ that rejects $H_0^*$ if
$$
S^* = \frac{d\log \ell^*(\theta_0; \bZ)}{d \theta} \geq c
$$
for some critical value $c$. The score test is also known as the {\it locally most powerful test} since it satisfies the following optimality property

\setcitestyle{square}
\begin{lemma}[\cite{goeman2006testing}, Lemma 2] \label{lem:goeman}
	For $\theta \in \Theta$, denote by $Z_\theta \in \R^n$ a random variable distributed corresponding to $\ell^{*}(\theta; \bZ)$ and denote its probability measure by $P_{\theta}$. Suppose that the derivative $\frac{d \ell^*(\theta; \bZ)}{d \theta}$ exists for all $Z \in \R^n$ and is bounded in a neighbourhood of $\theta_0$. Then, for any test of $H_0^*$ with critical region $A$ and power function $w(\theta) = P_{\theta}(A)$, the derivative $\frac{d w (\theta_0)}{d \theta}$ exists. Also, denote the power function of the score test statistic by $w^* (\theta) = P_{\theta} (S^* \geq c)$ for some $c \geq 0$. Then $$
	w(\theta_0) \leq w^*(\theta_0)
	$$
	implies
	$$
	\frac{d}{d \theta} w(\theta_0) \leq  \frac{d}{d \theta} w^*(\theta_0).
	$$
\end{lemma}
\setcitestyle{round}%

Since
$$
P_{\theta_0+h} (A) = w(\theta_0+h) 
= w(\theta_0) + h \, \frac{d}{d \theta} w(\theta_0) + o(h),
$$
Lemma \ref{lem:goeman} implies that no test of the same size can be more powerful for infinitesimally small deviations from $\theta_0$. This implies that the score test is the most powerful test for detecting local alternatives corresponding to infinitesimally small deviations from $\theta_0$ or short {\it locally most powerful test}.

\cite{DJ} have shown that, if the squared Euclidean distance is applied on the response, the generalised distance covariance arises from the score test statistic in certain Gaussian regression models.
This implies that $\widehat{\V}^2_b$ has an interpretation as locally most powerful test statistic, which we state in Theorem \ref{th:lomopo} and Remark \ref{rem:lomopo}.

Using its HSIC representation (cf. Equations \ref{eq:smindkernel} and \ref{eq:HSICemp}), $\widehat{\V}_b^2$ can alternatively be written as
\begin{equation} \label{eq:HSICstat}
	\widehat{\V}^2_b(\bX, \bY) = \frac{1}{2 \, n^2} \sum_{i,j=1}^n k_b(X_i,X_j) (Y_i - \widehat{\mu}_Y) \, (Y_j - \widehat{\mu}_Y)
\end{equation}
with 
$ \widehat{\mu}_Y = \frac{1}{n} \sum_{j=1}^n Y_j$.

Theorem \ref{th:lomopo} provides an interpretation of $\widehat{\V}^2_b$ as the locally most powerful test statistic in a Gaussian regression model.

\begin{theorem} \label{th:lomopo}
	Let $(\phi_1,\ldots,\phi_r)$ be a feature map induced by the distance $d_b$ with corresponding kernel $k_b$, as e.g. provided by Proposition \ref{prop:fm}.
	Consider the model
	\begin{equation} \label{eq:lomopo}
		Y_i = \sum_{j=1}^r \beta_j \phi_j (X_i) + \mu_Y + \varepsilon,
	\end{equation}
	where $\mu_Y$ is known, $\varepsilon \sim \mathcal{N}(0,\sigma^2)$ and, for $j \in \{1,\ldots,r\}$, $\beta_j = \tau B_j$ with $\tau \in \R$ and $B_1,\ldots,B_r$ are mutually uncorrelated random variables with $\E[B_j] = 0$ and $\E[B_j^2]=1$. Then the locally most powerful test statistic for testing 
	$$
	H_0 : \tau^2 = 0 \text{ against } H_1 : \tau^2 > 0
	$$
	is  given by
	\begin{equation}\label{eq:lomopo:ts}
		\widehat{\mathcal{U}}_b^2 =  \frac{1}{n^2}	\sum_{i,j=1}^n k_b(X_i,X_j) (Y_i- \mu_Y) (Y_j-\mu_Y).
	\end{equation}
\end{theorem}

\begin{remark} \label{rem:lomopo}   
	The population mean $\mu_Y$ is typically unknown in practice.
	By plugging in the sample mean $\widehat{\mu}_Y$ for $\mu_Y$ in \eqref{eq:lomopo:ts}, we see that a pivot statistic for $\widehat{\mathcal{U}}_b^2$ is given by the squared generalised distance covariance $\widehat{\V}_b^2$ in \eqref{eq:HSICstat}.
\end{remark}

In GWASs, it is usually conjectured that the effect of a single SNP on a quantitative trait is small, whence the assumption of a small $\tau$ appears sensible. Consequently, the locally most powerful property is particularly desirable for this setting. 
Theorem \ref{th:lomopo} does neither specify the marginal distribution of  $(B_1,\ldots,B_r)$ nor the feature map $(\phi_1,\ldots,\phi_r)$. In the following section, we elaborate on how different choices of  $(B_1,\ldots,B_r)$ and $(\phi_1,\ldots,\phi_r)$ lead to different interesting interpretations of Theorem \ref{th:lomopo}, providing insights into the nature of  $\widehat{\V}_b^2$.

For a first interpretation, we consider that the random vector $(B_1,\ldots,B_r)$ in Theorem \ref{th:lomopo} satisfies $P(B_i \neq 0, B_j \neq 0) =0$ for $i \neq j$. This implies that only one of the coefficients $\beta_1, \ldots, \beta_r$ is nonzero and hence only one of the features in Equation~\eqref{eq:lomopo} is involved for each realisation of the model.

\begin{corollary} \label{cor:mix}
	Let $(\phi_1,\ldots,\phi_r)$ be a feature map induced by the distance $d_b$ and, for $j \in \{1,\ldots,r\}$, let $c_j >0$; further denote $\psi_j(\cdot) = \phi_j (\cdot) /c_j$.
	Let $U$ be a discrete random variable with $P(U = j) = \frac{c_j^2}{\sum_{k=1}^n c_k^2}$ and consider the model
	$$
	y_i = \tau \,A \sum_{j=1}^r 1_{\{U=j\}} \psi_j (x_i) + \mu_Y + \varepsilon,
	$$
	where $\tau \in \R$, $\mu_Y$ is known, $\varepsilon \sim \mathcal{N}(0,\sigma^2)$ and $A$ is a random variable, independent of $U$ with $\E[A]=0$ and $0 < \E[A^2] < \infty$ (e.g. $P(A=1)=P(A=-1)=\tfrac12$). Then the locally most powerful test for testing 
	$
	H_0 : \tau^2 = 0 \text{ against } H_1 : \tau^2 > 0
	$ is given by \eqref{eq:lomopo:ts}.
\end{corollary}

For facilitating interpretation, the factors $c_j$ should be chosen in a way such that the standardised features $\psi_j$ are on a comparable scale. The variable $A$ balances positive and negative effects of a feature (guaranteeing $E[B] = 0$ in Theorem \ref{th:lomopo}).

Corollary \ref{cor:mix} states that $\widehat{\V}_b^2$ is nearly (cf. Remark \ref{rem:lomopo}) the locally most powerful test statistic in a model, where each of $r$ different association patterns (specified by the $r$ standardised features of the feature maps) between a SNP $X$ and a quantitative response $Y$  is present with a certain probability. We note that this is different from a mixture model, in the sense that the random parameters $U$ and $A$  does not depend on $i$, but are only drawn once and hence the same model is true for all samples $i$. 

Considering that we would typically apply the same test for each of a large number of SNPs, the corresponding test is optimal for situations in which the association patterns expressed by $\psi_j$ shows up for a fraction of $\frac{c_j^2}{\sum_{i=1}^n c_i^2}$ of the SNPs. 

We now introduce new feature maps leading to a particularly helpful interpretation of Corollary \ref{cor:mix} : 
For $b \in [2,4]$, we easily see that that a feature map of $d_b$ is given by,
$$
\phi_1 = \sqrt{{4-b}} \begin{pmatrix} 0 \\  0 \\ 1\end{pmatrix}, \, \quad \phi_2 = \sqrt{{4-b}} \begin{pmatrix} 0 \\  1 \\ 1\end{pmatrix}, \, \quad \phi_3 = 2\, \sqrt{b-2} \begin{pmatrix} 0 \\ \tfrac12 \\ 1 \end{pmatrix}
$$
Applying Corollary \ref{cor:mix} with $c_1 = c_2 = \sqrt{{4-b}}, \, c_3 = 2\,\sqrt{b-2}$ yields that for $b \in [2,4]$, $\V_b^2$ is in a certain sense optimal for a setting where the absolute difference between the two homozygous states is $|\tau|$ (with small $\tau$) and the heterozygous state takes the value of each of the homozygous states with probability $\frac{4-b}{2 \,b}$ and the average of the two values with probability $\frac{4\,(b-2)}{2b}$. This corresponds to the situation, where a dominant and recessive model hold for a fraction of $\frac{4-b}{2 \,b}$ of the SNPs each, and an additive model holds for a fraction $\frac{4\,(b-2)}{2b}$ of the SNPs.

In particular, $\V_2$ is optimal if all SNPs associated with $Y$ follow a dominant-recessive model and each of the homozygous states is dominant for one half of the SNPs. $\V_3$ on the other hand is optimal for a situation, where a dominant-recessive model is present with probability $\frac{1}{3}$ (for which each of the homozygous state is dominant with the same probability) and an additive model is present with probability $\frac{2}{3}$. The extreme case $\V_4$ corresponds to the locally most powerful test in a purely additive model and is equivalent to the test statistic obtained from linear regression with the SNP $X \in \{0,1,2\}$ as single predictor.

Similarly, for $b \in [0,2]$, we can derive the feature map
$$
\phi_1 = \sqrt{{b}} \begin{pmatrix} 0 \\  0 \\ 1\end{pmatrix}, \, \quad \phi_2 = \sqrt{{b}} \begin{pmatrix} 0 \\  1 \\ 1\end{pmatrix}, \, \quad \phi_3 = \sqrt{2-b} \begin{pmatrix} 0 \\ 1 \\ 0\end{pmatrix}.
$$
Applying Corollary \ref{cor:mix} with $c_1 = c_2 = \sqrt{b}, \, c_3 = \sqrt{2-b}$ yields, that, for $b \in [0,2]$, $\V_b^2$ is in a certain sense optimal for situations where a dominant-recessive model is present with probability $\frac{2 \,b}{2+b}$ (in which each of the homozygous states is dominant with equal probability) and a purely heterozygous model is present with probability $\frac{2-b}{2+b}$. 
For, $\V_2$, $c_3$ is zero and we obtain the same interpretation as above. $\V_1$ is optimal for a situation in which two means are equal and for each $j \in \{0,1,2\}$, $\mu_j$ differs from the other two means for $\tfrac13$ of the associated SNPs. This model is agnostic in the sense that it does not make any difference between the states $0, 1, 2$, which is also clear from $d_1(0,1)= d_1(0,2) = d_1(1,2) = 1$. For $b=0$, we obtain $c_1 = c_2 =  0$; hence $\widehat{\V}_0$ is optimal  for a purely heterozygous model --- the corresponding test statistic is equivalent to the one obtained from a linear regression with predictor $Z_i = 1_{\{X_i=1\}}$.

While it is common that the response values for the heterozygous state lie between the values of the two homozygous states, it seems rather unlikely that we encounter an exact additive model. Instead, the response values of the heterozygous state will typically lie closer to one of the homozygous states. A model which assumes that the response values for the heterozygous state lie somewhere between the response values  of the two homozygous states is referred to as a {\it partially dominant model}, as indicated in Section~\ref{Models:SNP}.

We will now show that, for $b \in ]2,4]$, $\V^2_b$ can be interpreted as the locally most powerful test statistic in certain random partially dominant models. For $b \in [0,2)$, we obtain a similar interpretation based on overdominant models. For this purpose, we first state the following alternative formulation of the locally most powerful property.

\begin{theorem} \label{th:lomopo2}
	Consider the distance $d_b$ and
	assume the model
	\begin{align*}
		Y_i = \begin{cases}  \mu_Y + \varepsilon, &\text{ if $x_i =0$}, \\
			\mu_Y + \beta_1 + \varepsilon,  &\text{ if $x_i =1$}\\
			\mu_Y + \beta_1 + \beta_2 + \varepsilon	&\text{ if $x_i =2$,}
		\end{cases}
	\end{align*}
	where $\mu_Y$ is known, $\varepsilon \sim \mathcal{N}(0,\sigma^2)$ and $(\beta_1, \beta_2) = \tau B$ with $\tau \in \R$ and $B$ is a random variable with $E[B] = 0$ and 
	$$
	E[B B^t] =  c \, \begin{pmatrix} 1 & \frac{b}{2}-1 \\ \frac{b}{2}-1 & 1  \end{pmatrix},
	$$
	where $c$ is some constant.
	Then the locally most powerful test for testing 
	$
	H_0 : \tau^2 = 0 \text{ against } H_1 : \tau^2 > 0
	$ is given by \eqref{eq:lomopo:ts}.
\end{theorem}

This yields the interpretation of $\widehat{\V}_b$ as locally most powerful test statistics in regression models with correlated regression parameters. For $b \in [0,2[$, the correlation between $\beta_1$ and $\beta_2$ is negative. In this case, we can choose $B$ in a way such that $\beta_1$ and $\beta_2$ always have opposing signs. For $b \in ]2,4]$ on the other hand, the correlation between $\beta_1$ and $\beta_2$ is positive and hence we can choose $B$ in a way such that $\beta_1$ and $\beta_2$ always have the same sign.

Remembering the association models introduced in Section 2.1, we can interpret $\V_b$ with $b \in ]0,2[$ as the locally most powerful test in an overdominant model with random heterozygous effect $H$. Analogously $\V_b$ with $b \in ]2,4[$ can be interpreted as the locally most powerful test in a partially dominant model with random heterozygous effect $H$. 

By choosing $\beta_1, \beta_2$ as two-sided gamma distributions with same sign, we obtain the following corollary, providing a particularly helpful interpretation of $\V_b$ for $b \in ]2,4[$.

\begin{corollary} \label{cor:beta}
	Consider the distance $d_b$ with $b \in ]2,4[$ and
	assume the model
	\begin{align*}
		Y_i = \begin{cases}  \mu_Y + \varepsilon, &\text{ if $x_i =0$}, \\
			\mu_Y + \tau H A + \varepsilon,  &\text{ if $x_i =1$}\\
			\mu_Y + \tau A + \varepsilon	&\text{ if $x_i =2$,}
		\end{cases}
	\end{align*}
	where $\mu_Y$ is known, $\tau \in \R$, $\varepsilon \sim \mathcal{N}(0,\sigma^2)$ and the heterozygous effect $H$ is beta-distributed with parameters $(\frac{b-2}{4-b}, \frac{b-2}{4-b})$. $A$ is a random variable, independent of $H$ with $\E[A]=0$ and $\E[A^2]=1$ (e.g. $P(A = 1) = P(A = - 1) = \frac{1}{2}$).
	Then the locally most powerful test for testing 
	$
	H_0 : \tau^2 = 0 \text{ against } H_1 : \tau^2 > 0
	$ is given by \eqref{eq:lomopo:ts}.
\end{corollary}

Corollary \ref{cor:beta} states that, for $b \in ]2,4[$, $\V_b$ arises from the locally most powerful test in a partially dominant model for which the heterozygous effect parameter $H$ is beta-distributed with parameters $(\frac{b-2}{4-b}, \frac{b-2}{4-b})$. An important special case is $\V_3$, which is most powerful if the effect parameter $H$ is uniformly distributed on $[0,1]$ --- i.e. $\V_3$ is optimal for a random Gaussian regression model where the mean $\mu_1$ is uniformly distributed on the interval $[\mu_0,\mu_2]$.

A similar result as in Corollary \ref{cor:beta} can be obtained for $b \in ]2,4[$, see Appendix~\ref{apA}. We conclude this section with an overview of helpful interpretations for $\widehat{\V}_b$ for different parameter values $b$ in the range $[0,4]$ (Table \ref{tab:interpretation})

\begin{table}[htbp] 
	\caption{Genetic model against which $\widehat{\V}_b$ provides the locally most powerful test, for different values of $b\in[0,4].$	}
	\label{tab:interpretation}
	\begin{tabular}{l | l}
		& Genetic model \\
		\hline
		$b=0$ & purely heterozygous model ($\mu_0 =\mu_2$)\\
		$b \in ]0,1[$ & overdominant model with large heterozygous effect ($h = \frac{G_1}{G_1 - G_2}$ with $G_i \sim \Gamma(\frac{2-b}{b},1)$) \\
		$b=1$ & agnostic model, treating the states $\{0,1,2\}$ indifferently
		\\
		$b \in ]1,2[$ & overdominant model with small heterozygous effect ($h = \frac{G_1}{G_1 - G_2}$ with $G_i \sim \Gamma(\frac{2-b}{b},1)$) \\
		$b = 2$ & dominant-recessive model with equal probability for dominance and recessiveness\\
		$b \in ]2,3[$ & partially dominant model where $h$ tends to be close to $0$ or $1$ ($h \sim \beta(\frac{b-2}{4-b}, \frac{b-2}{4-b})$)\\
		$b =3$ & partially dominant model, heterozygous effect $h$ is uniformly distributed on $[0,1]$ \\
		$b \in ]3,4[$ & partially dominant model where $h$ tends to be close to $\tfrac12$ ($h \sim \beta(\frac{b-2}{4-b}, \frac{b-2}{4-b})$) \\
		$b = 4$ & additive model, $h= \frac{1}{2}$
	\end{tabular}
\end{table}

\section{Adjusting for nuisance covariates}\label{gwas:covariates}

In GWA studies it is often necessary or beneficial to control for nuisance covariates. For an illustrative example, consider that we aim to test the association of a SNP $X$ with height $Y$ in adults including elderly individuals. Then it appears sensible to adjust for both sex and age, reducing variation in the response and leading to higher power. Moreover, the phenomenon known as \emph{population stratification} (i.e., the systematic difference in allele frequencies between subgroups of the population, accompanied by a difference in the distribution of the phenotypic trait under study) has been identified since the very beginning of the genomic era as a main cause of false positives in GWASs \citep{Cardon,Brandes}. Consequently it may be necessary to control for strata in the population, which may be done by using information on ethnic groups or by taking the first few principal components of the full genomic information.

We now derive adjusted versions of $\V_b^2$ and $\widehat{\V}_b^2$ for testing in the presence of nuisance covariates. Different from other approaches \citep{Partial,Conditional}, we will adjust for the influence of the covariates in a linear fashion, which allows to retain both a tractable test statistic and a meaningful interpretation; nonlinear influences of the covariates can still be taken into account by transformations, using e.g. splines. 

For defining the linearly adjusted version of our GDC, let $Z = (1,Z_1,\ldots,Z_q)^t \in \R^{(q+1)}$ be a random vector with $\E [Z_j^2] < \infty$. Then define,
$$
\V^2_b(X\, , \, Y; \,Z) = \V^2_b(X, \, Y- Z^t \tilde{\bgamma}),
$$
where $\tilde{\bgamma}$ is given by
$$
\tilde{\bgamma} = \argmin_{\bgamma \in \R^{q+1}} (Y- Z^t \bgamma)^2.
$$
Assuming that $\E[Y^2] < \infty$, we obtain the classical representation:
$$
\tilde{\bgamma} = \E[Z Z^t]^{-1} E[Z Y]  .
$$

The following corollary is an immediate consequence of Theorem \ref{th:dcovzero}.

\begin{corollary} \label{th:nuiscovpop}
	If $\E[Y-\tilde{\bgamma}^t Z\mid X=j] = 0$ for all $j \in \{0,1,2\}$, then 
	$$
	\V^2_b(X\, , \, Y; \,Z) = 0.
	$$
	On the other hand, if $\E[Y-\tilde{\bgamma}^t Z\mid X=j] \neq 0$ for some $j \in \{0,1,2\}$ and $p_j >0$ for all $j \in \{0,1,2\}$, then, if $b \in ]0,4[$,
	$$
	\V^2_b(X\, , \, Y; \,Z) > 0.
	$$
\end{corollary}

\medskip

\noindent In particular, assuming $b \in ]0,4[$, Corollary \ref{th:nuiscovpop} yields, that in the setting of a linear regression,
$$
Y = \widetilde{\bgamma}^t Z  +\mu_j 1_{\{X=j\}} + \varepsilon ,
$$
where 
$\V^2_b(X\, , \, Y; \,Z)$ ) equals $0$ if and only if $\mu_0 = \mu_1 = \mu_2$.

Given jointly distributed IID samples $\bX$, $\bY$, $\bZ$, we define our test statistic:
$$
\widehat{\V}^2_b(\bX\, , \, \bY; \,\bZ) = \widehat{\V}^2_b(\bX, \, \bY-\bZ \widehat{\bgamma} ),
$$
where $\widehat{\bgamma}$ is the ordinary least-square estimate:
$$    
\widehat{\bgamma} = (\bZ^t \bZ)^{-1} \bZ^t \bY.
$$

Hence the adjusted version  $
\widehat{\V}^2_b(\bX\, , \, \bY; \,\bZ)$ is defined as the regular GDC $\widehat{\V}^2_b$ between $\bX$ and the residuals of a linear regression of $\bY$ on $\bZ$.

We now state the asymptotic distribution of $
\widehat{\V}^2_b(\bX\, , \, \bY; \,\bZ)$. Different from the case without covariates, naive resampling methods are not valid here because the samples $(X_i, Y_i - Z_i^t \widehat{\bgamma})$ are non-exchangeable. Hence, the derivation of the test statistic distribution is crucial even for the case where we only consider a small number of SNPs.

\begin{theorem} \label{th:covasy} Let $Z = (1,Z_1,\ldots,Z_q)^t \in \R^{q+1}$ and $X \in \{0,1,2\}$ be random variables with $\E[Z^2] < \infty$. Assume the model
	$$
	Y =  \bgamma^t Z + \mu_j 1_{\{X=j\}} + \varepsilon,
	$$
	where $\varepsilon \in \R$ is independent of $(X,Z)$ with $\E[\varepsilon] = 0$, $\E[\varepsilon^2] = \sigma_\varepsilon^2 < \infty$ and $(\mu_0,\mu_1,\mu_2) \in \R^3$.  Further assume that $Z$ is non-singular. Consider now jointly distributed IID samples $\bX \in \{0,1,2\}^n$, $\bY \in \R^n $ and $\bZ \in \R^{n \times (q+1)}$ of $(X,Y,Z)$.
	
	If $\mu_0 = \mu_1 = \mu_2$, then, for $n \to \infty$,
	$$
	n \, \widehat{\V}_b^2 (\bX, \, \bY; \, \bZ) \stackrel{\mathcal{D}}{\longrightarrow}  
	\sigma_\varepsilon^2 (\lambda_1 Q_1^2 + \lambda_2 Q_2^2),
	$$
	where $Q_1^2$ and $Q_2^2$ are chi-squared random variables with one degree of freedom and $\lambda_1$ and $\lambda_2$ are the eigenvalues of matrix:
	$$
	\bK = E[\bPhi(X) \bPhi(X)^t] - E[\bPhi(X) Z^t] \, (E[Z Z^t])^{-1}  E[\bPhi(X) Z^t]^t ,  $$
	where $\bPhi = (\phi_1,\ldots, \phi_r)^t$ is an arbitrary feature map of $d_b$.
\end{theorem}

Under Gaussianity, we can again derive the exact finite-sample distribution.

\begin{theorem} \label{th:covfinite}
	For $n \in \Zplus$, let $\bX=(X_1,\ldots,X_n) \in \{0,1,2\}^n$ denote a fixed sample and let $\bY = (Y_1,\ldots,Y_n)$ be defined by
	$$
	Y_i = \bgamma^t Z_i + \mu_j \, 1_{\{X_i = j\}} + \varepsilon_i,
	$$
	where $\boldsymbol{\mu} = (\mu_0, \mu_1, \mu_2)^t \in \R^3$, $Z_i \in \R^p$  and $(\varepsilon_1,\ldots,\varepsilon_n)$ is IID, and independent of $(\bX,\bZ)$, with $\varepsilon_i \sim \mathcal{N}(0,\sigma^2_\varepsilon)$. If $\mu_0 = \mu_1 = \mu_2$, then,
	$$
	\Prob \left( \frac{n \, \widehat{\V}_b^2(\bX, \, \bY; \, \bZ) }{\widehat{\sigma}_\varepsilon^2} >k \right) = \Prob(T_n > 0),
	$$
	where $\widehat{\sigma}_\varepsilon^2 = \frac{1}{n} \sum_{j=1}^n (\widehat{\varepsilon}_j -\tfrac{1}{n} \sum_{i=1}^n \widehat{\varepsilon}_i)^2 $ with
	$$
	\widehat{\varepsilon}_i = Y_i - Z_i  (\bZ^t \bZ)^{-1} \bZ^t \bY,
	$$
	$T_n$ is defined by
	$$
	T_n = \left(\widehat{\lambda}_1- \frac{k}{n} \right) \, Q_1^2 + \left(\widehat{\lambda}_2- \frac{k}{n} \right) \, Q_2^2 - \frac{k}{n} Q_3^2 - \cdots - \frac{k}{n} Q_{n-p-1}^2
	$$
	and $Q_1^2,\ldots,Q_{n-p-1}^2$ are IID chi-squared with one degree of freedom each; $\widehat{\lambda}_1$ and $\widehat{\lambda}_2$ are the eigenvalues of matrix 
	$$
	\bK = \frac{1}{n} \bU^t (\bI-{\bZ} ({\bZ}^t {\bZ})^{-1} {\bZ}^t) \bU,
	$$
	where $\bU \in \R^{n \times r}$ is a matrix with entries
	$$
	(\bU)_{ij} = \phi_j(X_i),
	$$
	and $\phi_1,\ldots, \phi_r$ is an arbitrary feature map of $d_b$.
	
\end{theorem}

As per Proposition \ref{prop:fm}, a feature map with $2$ features exists for each $b \in [0,4]$. Hence $K$ can always be represented by a $2 \times 2$ matrix enabling rapid evaluation of the eigenvalues, as demonstrated by the real data example in Section \ref{rdata:hepatic}.
$p$-values based on Theorem \ref{th:covfinite} can be approximated analogously to the setting without covariates in Section \ref{sec:pvalues}. All results in Section \ref{GT} regarding  the interpretation of $\widehat{\V}_b^2(\bX, \, \bY)$ hold true for $\widehat{\V}_b^2(\bX, \, \bY; \, \bZ)$ with the modification of  adding $\bgamma^t Z$ to the right-hand side of the corresponding Gaussian regression models.

\section{Practical aspects}\label{gwas:practical}

\subsection{Imputed data}

In practice, GWAS are often performed on imputed genotype data \citep{Abecasis:imputation}. In this case, the SNP information for numerous loci is not directly measured. Instead, the corresponding SNPs are imputed using information from other SNPs and complete data from a reference population. For these imputed SNPs, we do not observe the allele count $X \in \{0,1,2\}$, but the expected allele count $X$ in the interval $[0,2]$. Hence, the methodology explained in this chapter is not directly applicable in this setting.

A simple but clearly inefficient way to deal with this issue is to round the allele count before performing the analysis. Another straightforward generalisation is to use the $\alpha$-distance covariance with $\alpha = \log_2 b$; however this approach leads to a substantially more complicated distribution of the test statistic and hence to increased computing time.

In order to retain a similar test statistic while using all information on the expected allele counts, we propose to generalise the methodology by linearly interpolating the features, i.e. we use as a feature map the following modification of the one in Proposition \ref{prop:fm},
$$
\widetilde{\phi}_1(x) = \sqrt{\frac{b}{2}} x, \quad \widetilde{\phi}_2(x) = \sqrt{\frac{4-b}{2}} |x-1|.
$$
A straightforward calculation yields that the corresponding distance is:
$$
d_b(x,y) =  \begin{cases} (x-y)^2 &\quad \text{if } x \geq 1, y \geq 1 \vee x< 1, y < 1, \\ \frac{b}{4} (x-y)^2 +\frac{4-b}{4} (x+y-2)^2 &\quad  \text{if } x \geq 1, y < 1 \vee x< 1, y \geq 1.\end{cases}
$$
It is easy to see that, Theorem \ref{th:covasy} and \ref{th:covfinite} (which imply Theorems \ref{testasy} and \ref{testfinite} respectively) hold analogously replacing the feature maps in the formulations of the theorems by $(\widetilde{\phi}_1,\widetilde{\phi}_2)^t$.

\subsection{Multiallelic single-nucleotide polymorphisms}

As in most methodological work on GWAS, we were assuming for simplification that all SNPs are biallelic. However while this assumption is true for the majority of SNPs, numerous SNPs with three or more alleles (``multi-allelic SNPs'') have been identified \citep{multi}. An advantage of our approach is that it can be straightforwardly generalised to multiallelic SNPs by defining distances on the space $\{0,1,2\}^m$, where $m$ is the number of alleles. We propose the distance:
$$
\widetilde{d}_b((x_1,\ldots,x_m),(y_1,\ldots,y_m)) = \frac{1}{2} \sum_{i=1}^m d_b(x_i,y_i),
$$
where $x_i$ counts the number of alleles of type $i$ and $d_b$ is the distance on $\{0,1,2\}$ used before; this is easily seen to generalise the biallelic case, even in the case of imputed data.

The distribution of the test statistic in the multiallelic setting can be derived similarly as in the biallelic setting, however since there are $m+1 \choose 2$ states the corresponding asymptotic distribution features $t = {m+1 \choose 2} -1$ eigenvalues $\widehat{\lambda_1},\ldots \widehat{\lambda}_t$. If only $2$ alleles are present, the test statistic and its distribution reduce to the biallelic case; very rare alleles have virtually no influence on the test statistic. Since the test is directed towards alternatives corresponding to differences in the most frequent alleles, we expect good power properties in the multi-allelic setting; the test focuses on the variants where there is potentially enough power to detect a possible effect.

\subsection{Choice of $b$}

We have obtained model-based interpretations of the test statistic that are summarised in Table \ref{tab:interpretation}. We emphasise again that all choices of  $b \in ]0,4[$ are consistent against all alternatives; only the degenerate cases $b \in \{0,4\}$ do not guarantee that. 

Interpreting $\V_b$ as a mixture of additive and dominant recessive models, we easily calculate that $b=\frac{12}{5} = 2.4$ gives the locally most powerful test statistic for the setting, where we assume a dominant, recessive and additive model with probability $\frac{1}{3}$ each. $b = \frac{8}{3} \approx 2.67$ relates to the situation of a dominant and recessive model with probability $\frac{1}{4}$ each and an additive model with probability $\frac{1}{2}$. Finally, $b=3$ is optimal for the setting where the heterozygous effect is uniformly distributed on the interval $[0,1]$.

Considering that both partially dominant models and dominant-recessive models frequently arise in practice, it appears that $b \in [2,3]$ is a good choice for most applications; for this reason the GDC tests with $b=2$ and $b=3$ are investigated in further detail in the following simulation study. As discussed in Section~\ref{gwas:simu}, our simulation studies suggest that $b=2$ tests has higher power in most practical situations. For this reason, we recommend $b=2$ as default.

\section{Simulation study}\label{gwas:simu}
To demonstrate the performance of our methods, we will present a series of simulation studies. We will compare the proposed distance covariance test based on $\widehat{\V}_2$ and $\widehat{\V}_3$ with the following three competitors:
\begin{itemize}
	\item The additive model, performing a linear regression of $y$ on $X$, treating $X \in \{0,1,2\}$ as continuous predictor; this is equivalent  to the test based on $\widehat{\V}_4$ .
	\item A linear model, treating the SNP $X \in \{0,1,2\}$ as categorical predictor; this model is commonly referred to as ANOVA.
	\item A test based on the \texttt{nmax3} statistic, calculated as the maximum of three nonparametric trend tests, based on the recessive, additive and dominant model respectively; as implemented in the R package \texttt{AssocTests} \citep{nmax3}.
\end{itemize}

\subsection{Computation time}

\begin{figure}
	\centering
	\includegraphics[width=\textwidth]{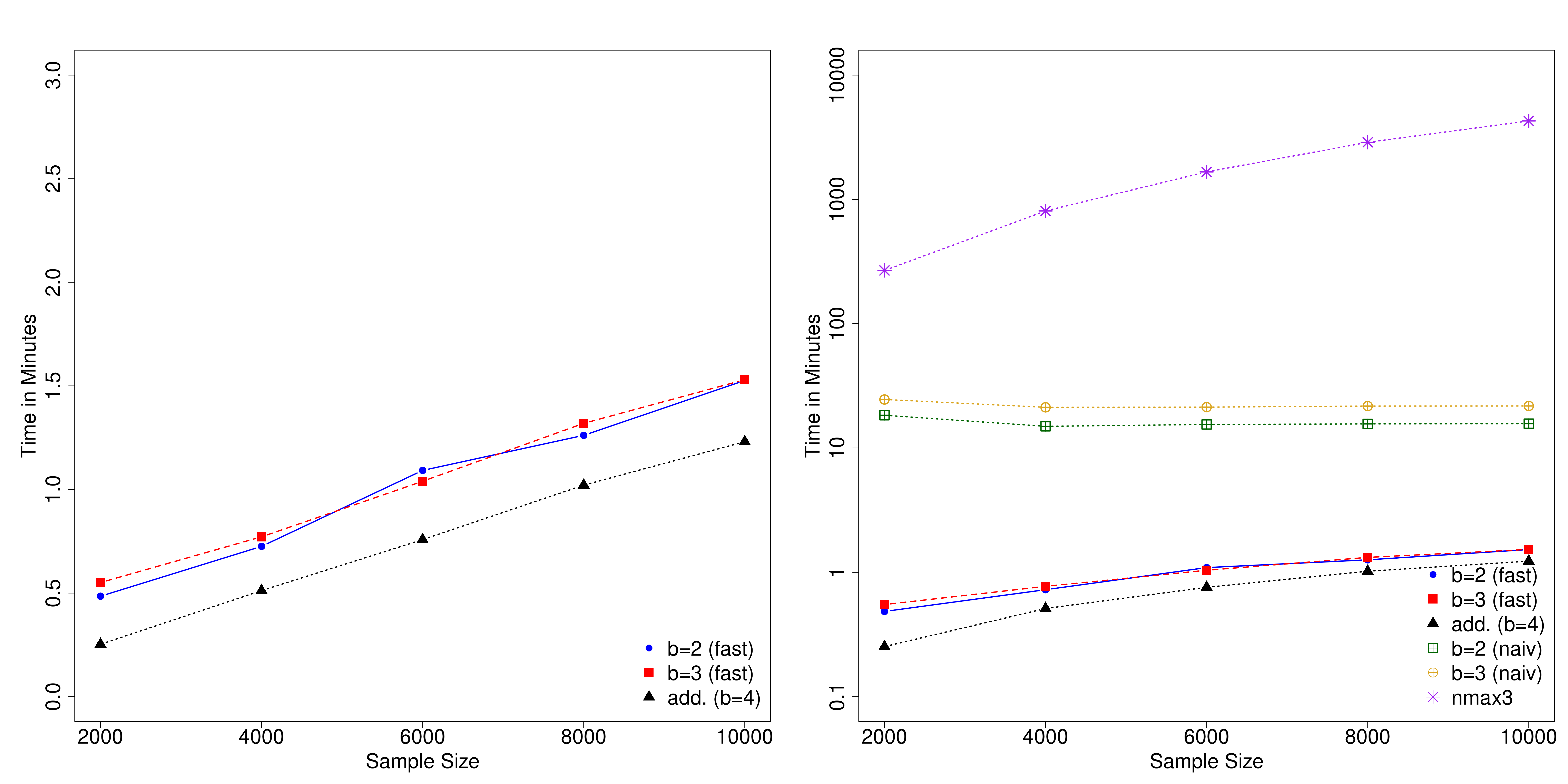}
	\caption{Computation time for different methods for SNP testing on $p=100\,000$ SNPs and sample size as indicated in the plot. The different methods are: GDC test with $b=2$ (blue), GDC test with $b=3$ (red), additive model (black), GDC test without $p$-value screening with $b=2$ (green), GDC test without $p$-value screening with $b=3$ (yellow) and the \texttt{nmax3} procedure (purple). The left-hand plot features the comparison on a linear scale; the one on the right, on a log-scale.}.
	\label{comptime_plot}
\end{figure}

With the aim of evaluating the computation time of the methods, let us consider $100\,000$ SNPs with \emph{minor allele frequency} (MAF) of $0.5$, and a varying sample size $n$. The response will be an Gaussian $Y$, independent of the SNPs. Each method was applied 50 times, with the minimum of the 50 computation times being displayed in Figure~\ref{comptime_plot}.

To allow for a fair comparison, the additive model was implemented by simply using the GDC algorithm with $b=4$. For the distance covariance with $b=2$ and $b=3$, we considered two different versions --- on the one hand, the recommended version (as described in Section \ref{sec:pvalues}), using a screening procedure filtering out SNPs with $p>10^{-3}$ in the first step using a guaranteed anticonservative approximation for our test distribution; and on the other hand, the naive implementation, which evaluates the precise $p$-value for each SNP. For this comparison, we also included a competing method found in the literature, namely the \texttt{nmax3} test, from the R package \texttt{AssocTests} \citep{nmax3}.

On the left-hand side of Figure~\ref{comptime_plot}, a comparison of the additive model and the recommended versions for $b=2$ and $b=3$ is provided, highlighting the excellent computational performance of these methods (in particular, $100\,000$ SNPs are evaluated in less than $2$ minutes for a sample size of $n=8000$). The right-hand side of Figure~\ref{comptime_plot} displays all $6$ methods (so that the subfigure on the left can be seen as a zoom-in of this more complete one), using a log-scale for the computation time. We note that the naive implementation of the GDC methods without screening leads to a substantially increased computation time, which is more than $10$ times higher than for the version with the pre-screening. Moreover, the GDC with no screening shows virtually no difference in computation time for the sample sizes under consideration; which is little surprising since the biggest part of the time is used to evaluate the $p$-values. Finally, we note that the given implementation of the \texttt{nmax3} procedure takes substantially longer computation time than all other methods, making it hard (but not impossible) to apply in practical situations.
All computations were run on a single core of an Intel Xeon E312xx system with 2.6 MHz.

As one can see from the computation times above, one can get precise $p$-values for $p=10^5$ SNPs in around 20 minutes. Since the algorithm is $O(p)$, a full conventional GWAS (where $p\approx5\cdot10^{6}$) will round in less than a day. This means that even the slower of the two algorithms we propose for $p$-value evaluation can be used in practice. For instance, one could realistically study polygenic scores for a GWAS with $5\cdot10^{6}$ SNPs, by first calculating the accurate $p$-values for each of those variants with the distance-covariance test developed in the present chapter.

\begin{figure}
	\centering
	\includegraphics[width=\textwidth]{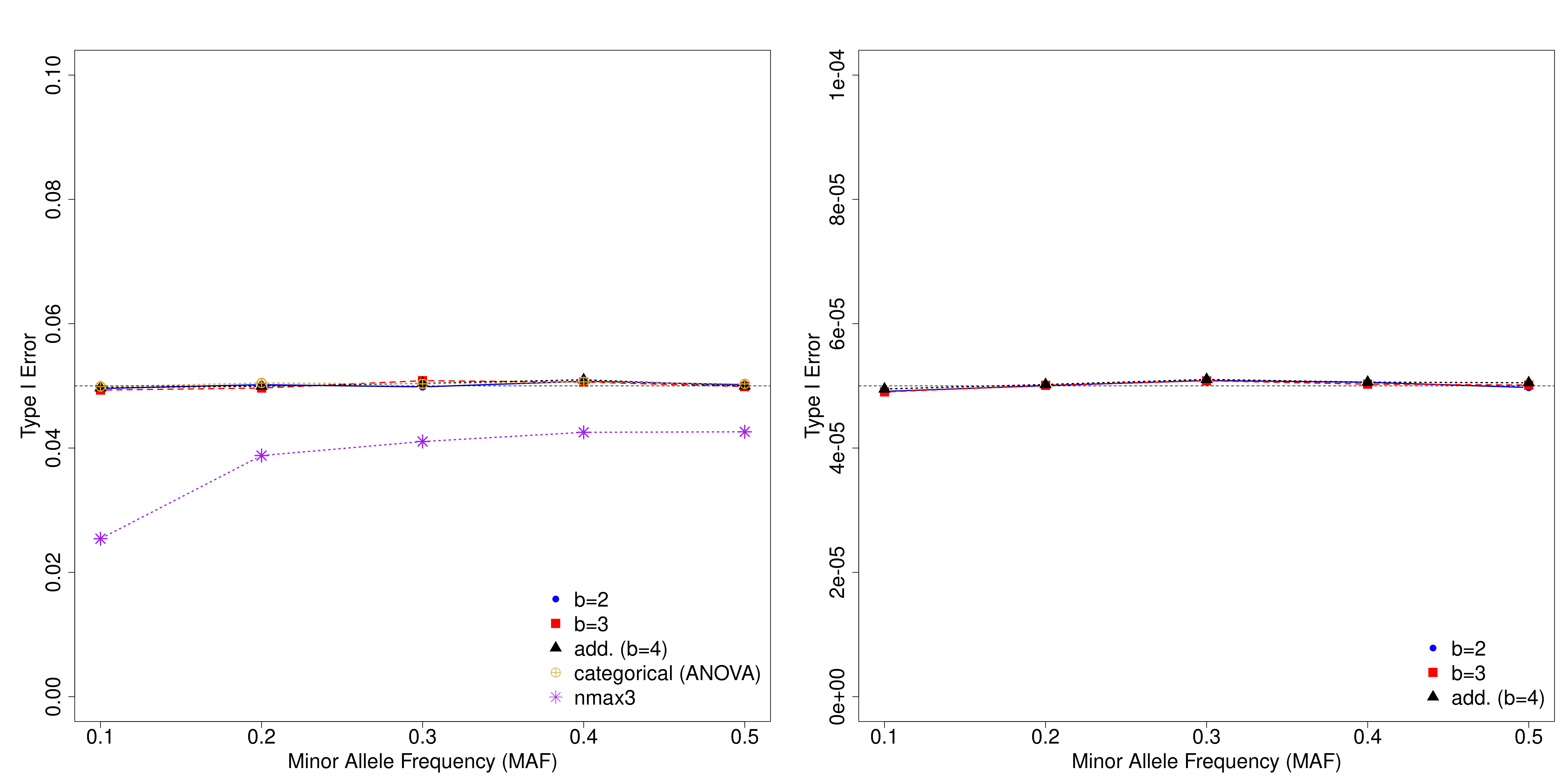}
	\caption{Empirical type I error for different SNP testing methods, for $n=300$ and normally distributed outcomes. The left hand-side corresponds to nominal $\alpha = 0.05$ ($10^4$ simulation runs); the right hand-side, to $\alpha = 5 \times 10^{-5}$ ($10^8$ replicates). Five testing procedures are displayed (see colour legend): GDC with $b=2$ (blue), GDC with $b=3$ (red), additive test (black), ANOVA (yellow), and \texttt{nmax3} (purple).}
	\label{fig:typeI}
\end{figure}

\subsection{Type I error}\label{gwas:simu:alpha}

For comparing type I error control of the different methods, we fix the sample size at $n=300$ and consider SNPs with MAF $0.1,0.2,\ldots,0.5$.

The data is simulated according to a null model under normality, i.e.
$
y_i = \varepsilon_i,
$
where $\varepsilon_1,\ldots,\varepsilon_n$ are IID standard Gaussian random variables.

We first fix the nominal level at $\alpha = 0.05$ and evaluate the empirical type I error for all methods under considerations using  $K=10\,000$ simulation runs; the results are provided in the left-hand side of Figure \ref{fig:typeI}. The empirical type I error of the tests based on the ANOVA model, the additive model, $\widehat{\V}_2$ and $\widehat{\V}_3$ are always very close to $0.05$, which is expected since the exact finite sample distribution is used for all four methods. Our simulations hence confirm Theorem~\ref{testfinite}. The \texttt{nmax3} procedure, on the other hand, is remarkably conservative, particularly for smaller MAFs.

To investigate if our methods suffer from numerical issues when approximating the Appell $F_1$ hypergeometric series in \eqref{eq:appell}, we further used $K=100$ million simulation runs to evaluate the empirical type I error for a nominal level of $\alpha = 5 \times 10^{-5}$. As can be seen from the right-hand side of Figure \ref{fig:typeI}, the empirical type I error of our methods is again very close to the nominal level.

\begin{figure}[!htbp]
	\centering
	\includegraphics[width=\textwidth]{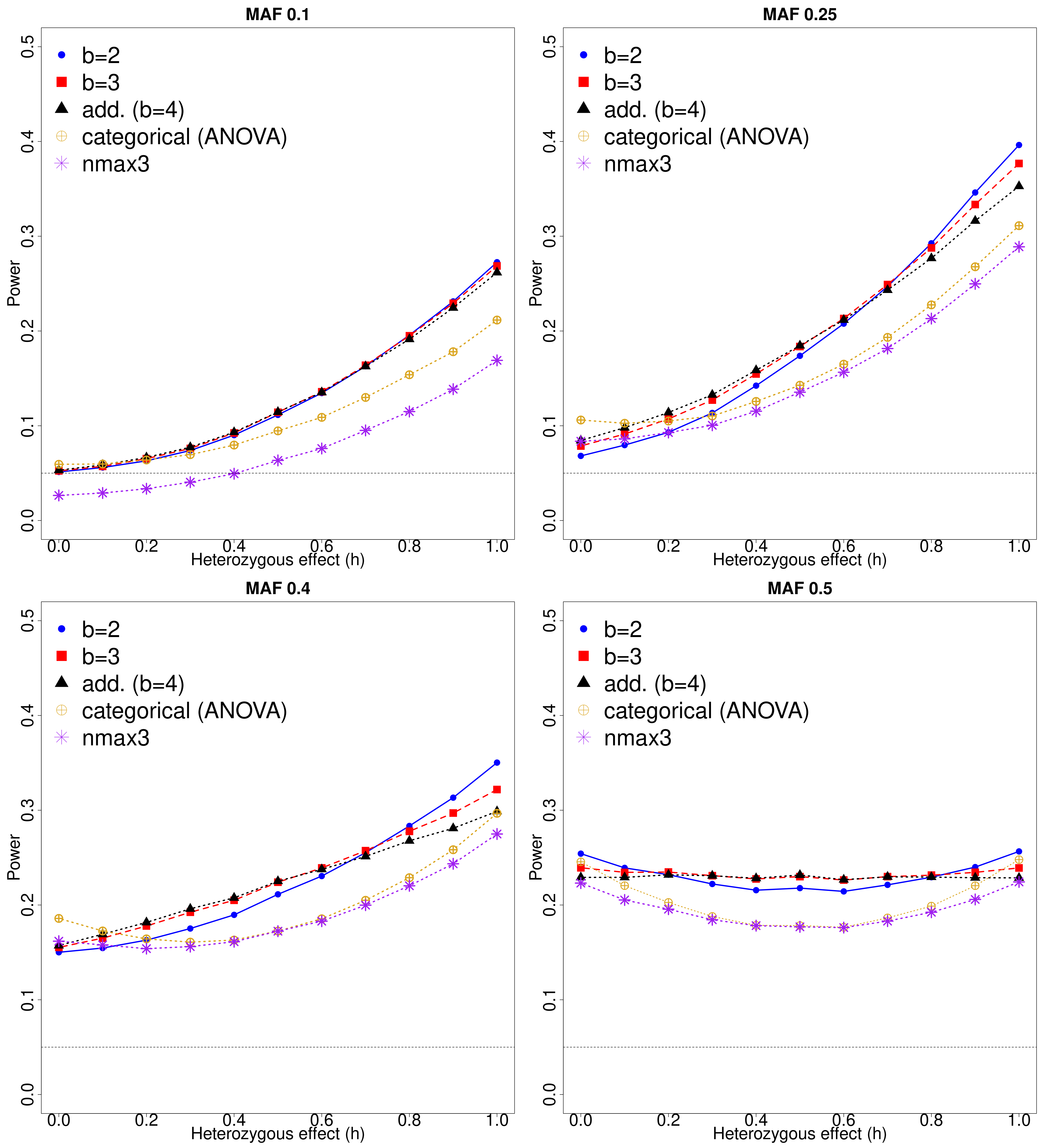}
	\caption{Power curves for different SNP testing methods, for $n=300$ and nominal $\alpha=0.05$, under model \eqref{eq:powermodel}. Each plot corresponds to a different value of MAF (left to right and top to bottom: $0.1,0.25,0.4,0.5$). The $X$-axis in each subfigure represents the heterozygous effect $h$ in its range $[0,1]$. Five testing procedures are represented (see colour legend): GDC with $b=2$ (blue), GDC with $b=3$ (red), additive test (black), ANOVA (yellow), and \texttt{nmax3} (purple).}
	\label{fig:power05}
\end{figure}

\begin{figure}[!htbp]
	\centering
	\includegraphics[width=\textwidth]{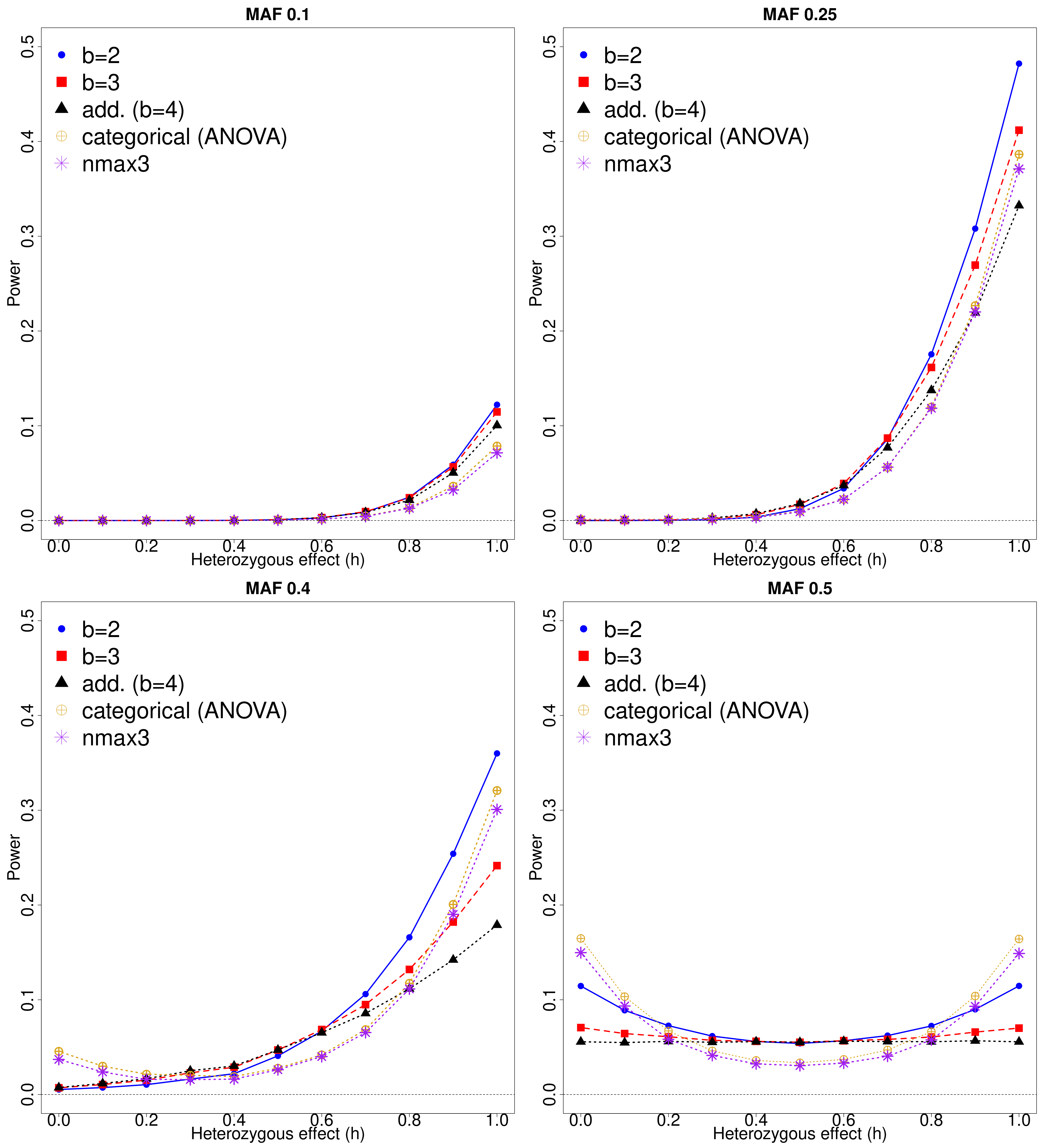}
	\caption{Power curves for different SNP testing methods, for $n=3000$ and nominal $\alpha=5\times\cdot10^{-8}$, under model \eqref{eq:powermodel}. Each plot corresponds to a different value of MAF (left to right and top to bottom: $0.1,0.25,0.4,0.5$). The $X$-axis in each subfigure represents the heterozygous effect $h$ in its range $[0,1]$. Five testing procedures are represented (see colour legend): GDC with $b=2$ (blue), GDC with $b=3$ (red), additive test (black), ANOVA (yellow), and \texttt{nmax3} (purple).}
	\label{fig:power5e8}
\end{figure}

\subsection{Power}

For comparing the power of the different testing procedures considered in Section~\ref{gwas:simu:alpha}, we assume the following population model:
\begin{equation} \label{eq:powermodel}
	y = h \beta \, 1_{\{X=1\}} +  \beta \, 1_{\{X=2\}} + \varepsilon;
\end{equation}
where $\varepsilon$ follows a normal distribution with mean $0$ and variance $25$. We let the heterozygous effect $h$ vary in $\{0,0.1,\ldots,1\}$. This can be considered an approximation of a situation where the quantitative trait is the sum of $26$ independent terms of the same size and we consider the power for detecting one of the terms.

We consider two situations; in the first the sample size is $n=300$ and the nominal level is $\alpha = 0.05$, for the second the sample size is $n=3000$ and $\alpha = 5 \times 10^{-8}$. For the MAF, we choose $0.1$, $0.3$ and $0.5$; for the $n=3000$ setting, we additionally consider a scenario with a rare allele with MAF $0.01$.

\section{Real data analysis}\label{rdata:hepatic}

So far, we have devised a novel approach to GWAS within the framework of distances, kernels and global tests; studied its theoretical properties extensively and demonstrated reasonable performance throughout our simulations. We will now examine how or methodology can be applied to a real dataset. One of the interests of psychiatric genetics is the study of addictions \citep{Hatoum}, with special focus in substance use disorders. Alcoholism, one of the most prominent examples due to its disease burden and wide spread across the globe \citep{Shield}, has been studied from geneticists since the pre-omic era, and it is ---together with related conditions--- one of the few examples of the survival into the GWAS era of large-effect loci identified by candidate gene studies \citep{Walters}.

Large-scale GWA studies are starting to reveal the polygenic architecture of several alcohol-related traits \citep{Gelernter}. In our case we will focus on one of the main causes of the high burden of alcohol-use disorders --- hepatic damage, which has as its biomarkers some well-known liver enzymes such as the aspartate aminotransferase (AST), the alanine aminotransferase (ALT) and the $\gamma$-glutamyltransferase (GGT). In 2024, new loci related to the variability of serum concentration of these enzymes are being identified and the search for them is a topic of current research interest, after a number of very large GWASs in populations of different ancestries \citep{Ghouse,Pazoki}.

Nowadays, for many research purposes in complex trait genetics (e.g., polygenic risk studies, Mendelian randomisation, meta-analyses), it suffices to use existing data from GWA studies that are publicly available in the form of summary statistics. However, when it comes to identifying loci related to the phenotype of interest (which is the goal of the methodology we are presenting in this chapter), it is necessary to access individual-level data, which are in general not free to use, both in an economical sense and in terms of privacy. After careful consideration of a number of databases and repositories, we found out that a database in the Database of Genotypes and Phenotypes (dbGaP) of the National Library of Medicine of the United States of America \citep{dbGaP} matched the scientific needs of this study.

The dataset was produced as part of the Trinity Student Study (dbGaP accession number: \href{https://www.ncbi.nlm.nih.gov/projects/gap/cgi-bin/study.cgi?study_id=phs000789.v1.p1}{phs000789.v1.p1}) and has been described in several bibliographic references \citep{Mills:Molloy,Molloy:Brody:2016,Desch}. The cohort was sampled during the academic year 2003--2004 in the Trinity College of the University of Dublin, with the goal of researching the genetics of quantitative traits. Only students with no serious medical condition, and of Irish ethnicity (based on the geographic origin of their grandparents), were included. Thus, the sample comprised $2407$ individuals (1409 of them, females), with age range $[19,28]$ (in years) and 94.4 \% of the subjects in $[20,25]$ .

We consider a total of $p=757\,577$ SNPs, which is the exact number of variants in the PLINK \citep{PLINK} files available through dbGaP. As indicated by \citet{Desch}, the array used for genotyping was the Illumina HumanOmni1-Quad Beadchip. In that article they also speak roughly of $758\,000$ SNPs, although not the exact same number that we have, which we attribute to small differences in quality controls across the various research articles among which the information on the Trinity dataset is spread. There is a similar situation with the sample size, where \citet{Desch} also describes approximately $2400$ individuals and then leaves a few out, but not in the exact same figures as we have. Checking the other literature on this database did not help clarify the situation either.

\begin{figure}[!htbp]
	\centering\includegraphics[width=0.8\textwidth]{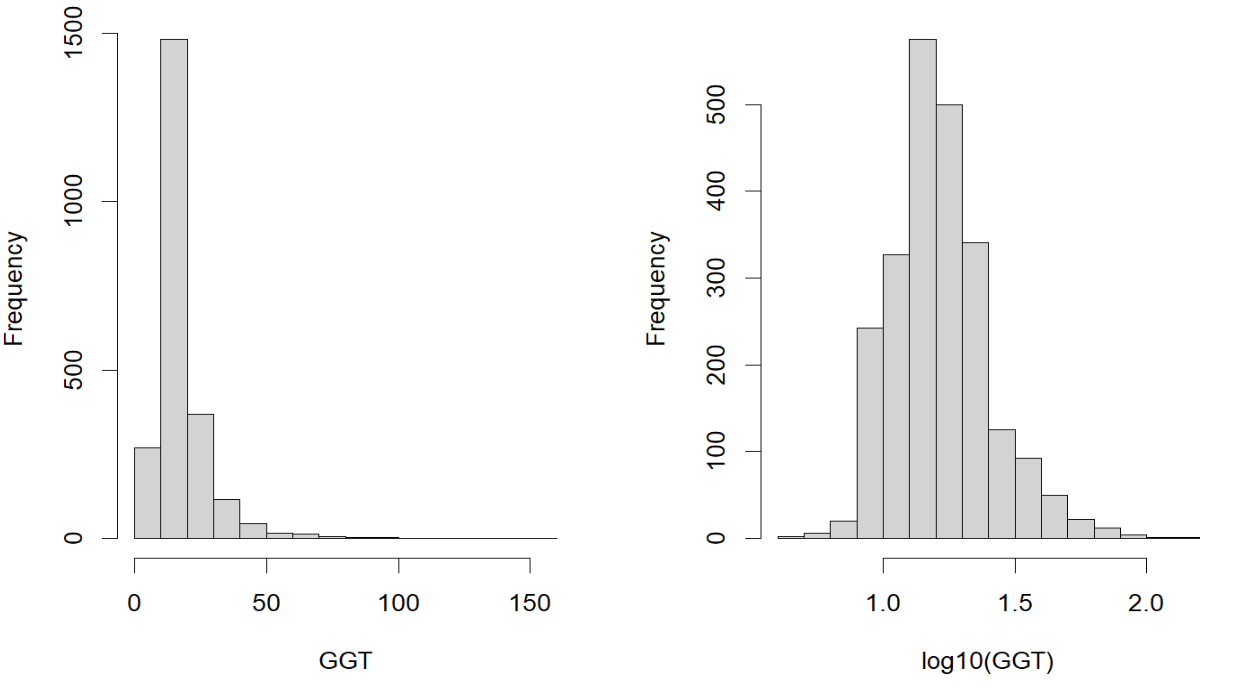}
	\caption{Histograms of the raw values of the GGT serum concentration for the Trinity data (left) and its logarithm to base $10$.}
	\label{hist:ggt}
\end{figure}

In order not to complicate the interpretation of our results, and taking into account that our goal is to demonstrate applicability of our method, we will focus in only one of the more than 50 phenotypic variables available in the dataset, namely in the GGT serum concentration. The empirical distribution of $Y$ has a median of 15 (units per liter), with an interquartile range of 8. As seen in Figure~\ref{hist:ggt}, it is justified to consider as our $Y$ the logarithm (to base $10$) of the GGT concentration \citep[][page 59]{Gelman:Hill}, which we will do for our analyses. The data for GGT is missing for 87 of the individuals, which lowers the sample size to 2320. When intersecting those 2320 individuals with the 2232 for which there is genotype data available in dbGaP, we get a final $n$ of 2152.

\begin{figure}[!htbp]
	\centering\includegraphics[width=0.98\textwidth]{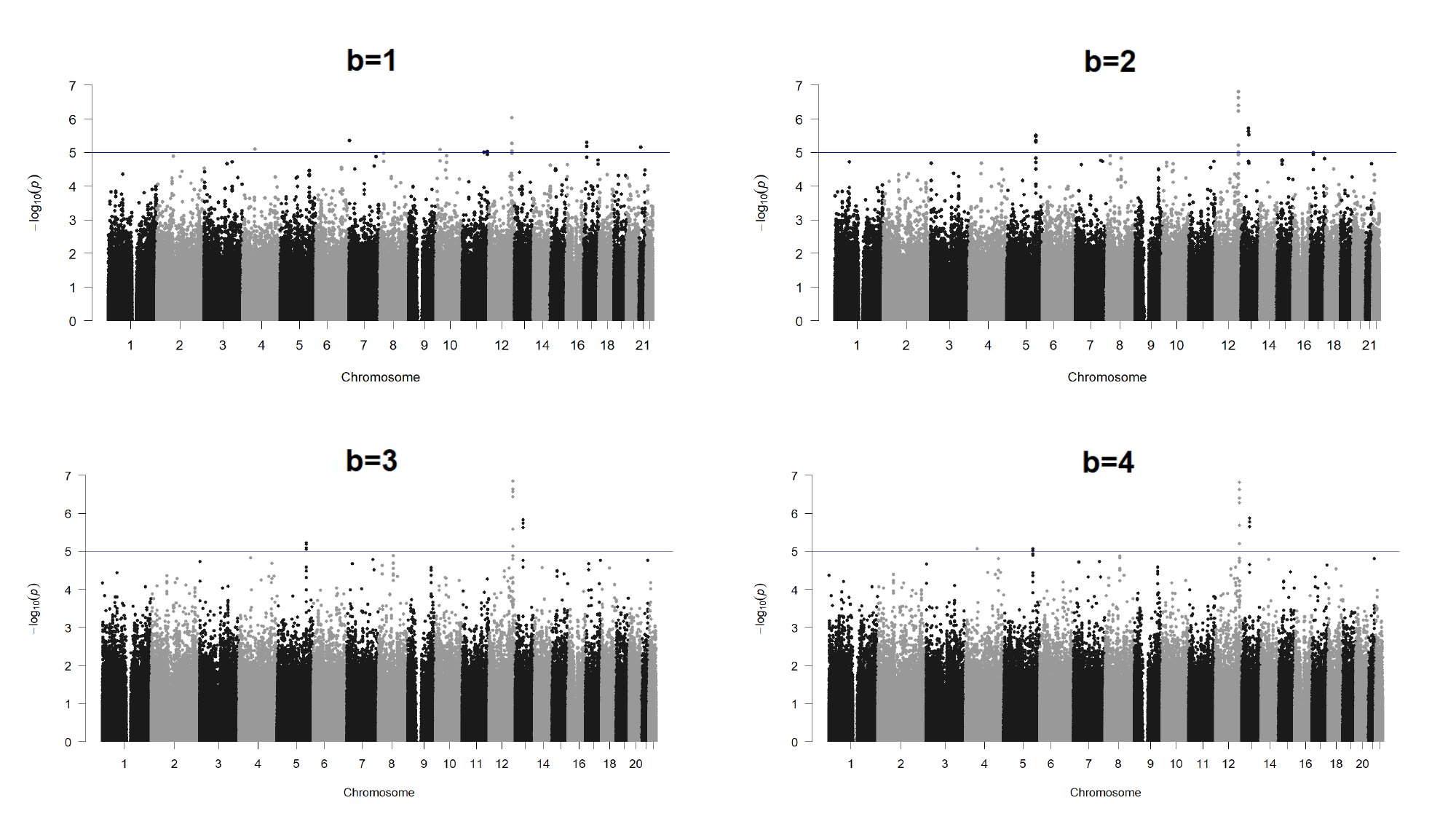}
	\caption{Manhattan plots for the Trinity dataset analysis with the distance covariance test, for $b\in\{1,2,3,4\}$ (left to right, and top to bottom). Blue horizontal lines indicate a significance threshold of $10^{-5}$. The test is corrected for age, sex and population structure.}
	\label{manh_dc_agesx}
\end{figure}

We have applied our method for $b\in\{1,2,3,4\}$ to the Trinity dataset and, with the help of R package \texttt{qqman} \citep{qqman}, we display the results in Figure~\ref{manh_dc_agesx} as Manhattan plots \citep{Manhattan}. These graphics are a standard visualisation in GWA studies which represents the minus $\log_{10}$ of the $p$-value versus the physical location of each SNP considered in the genome (left to right, chromosomes 1 to 22; and then ordering within them according to the nucleotide position). Therefore, the highest `skyscrappers' indicate where the most significant SNPs are located. We see that the signal is as sparse as one would expect in this setting. Note that there are small portions in the $X$-axis with no observed SNPs --- these correspond to pericentromeric regions, for which existing technologies cannot genotype common variation very well (due to very repetitive nucleotidic patterns).

We also use this example to demonstrate how our method works when accounting for covariates, since we have considered two of the ones present in the original dataset (namely, Age and Sex) that made sense for our analyses. Even though the experimental design tends to ensure ethnic and sociodemographic homogeneity, this is not enough to completely neglect the role that population stratification may play, as both \citet{Desch} and \citet{Carter:Mills} noted. Hence, we also include as covariates in our analysis the first 3 principal components of ancestry, as generated by flag \texttt{$-\,-$pca} in PLINK \citep{PLINK}.

\begin{figure}[!htbp]
	\centering\includegraphics[width=0.7\textwidth]{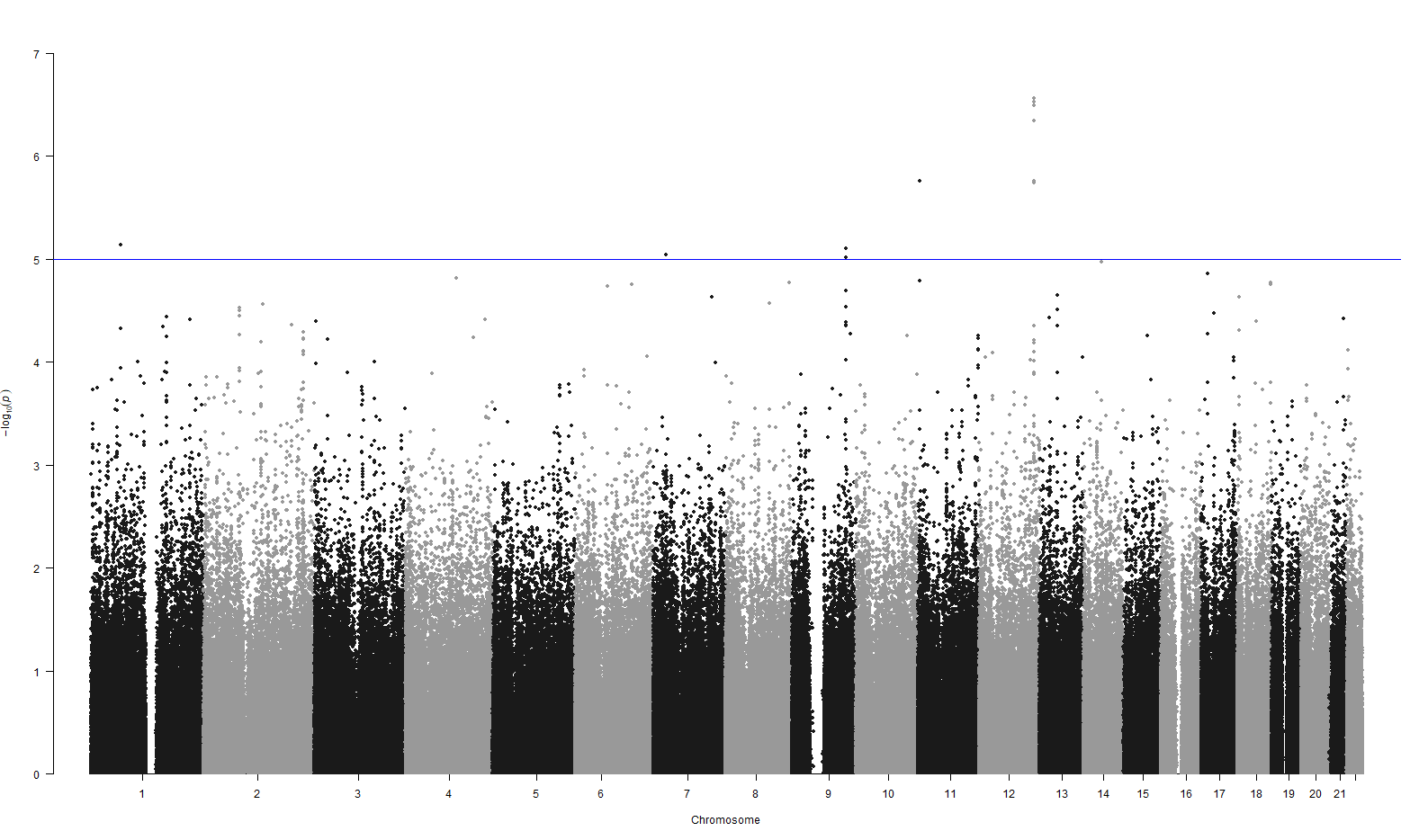}
	\caption{Manhattan plot for the Trinity dataset analysis with PLINK's linear test, correcting for Age, Sex and population structure. The blue horizontal line indicates a significance threshold of $10^{-5}$.}
	\label{manh_plink}
\end{figure}

A fully-fledged post-GWAS functional validation of the specific results obtained \citep{Tam} goes beyond the scope of this dissertation, but we will attempt to interpret the results from a biological point of view, to some degree. With that aim, we repeated the real data analysis with conventional statistical methodology. We have chosen the \texttt{$-\,-$linear} default test of PLINK \citep{PLINK} for this purpose, where we once more consider sex and age as covariates, and correct for the first 3 principal components of the genetic information. Figure~\ref{manh_plink} displays the corresponding Manhattan plot.

For all 5 methods under consideration, we have studied each SNP with a $p$-value under $10^{-5}$, which amounts to a total of 10 to 20 SNPs per method (with some overlap between them), as it can displayed in Tables~\ref{tab:ggt:plink}--\ref{tab:ggt:b4}. We have used the \href{https://www.ebi.ac.uk/gwas/studies/GCST001307}{NHGRI-EBI GWAS Catalog} \citep{GWAS:Catalog} to firstly look if any of our positives had previously been described to be associated to GGT serum levels in independent samples of European ancestry. Our search has revealed that SNP \emph{rs1169288} was genome-wide significant in the study by \citet{Middelberg}. This SNP has $p<10^{-5}$ for the GDC test with $b\in\{1,2,3,4\}$ and also for PLINK's linear test, and in addition each of the 5 methods detected 3 to 6 SNPs with a chromosome position that would tend to indicate linkage disequilibrium with \emph{rs1169288} (all in less than 20 kbp, within chromosome 12). Note that these hits correspond to the highest `skyscrapper' in each of the Manhattan plots (Figures~\ref{manh_dc_agesx}--\ref{manh_plink}).

Finally, we used once more the GWAS Catalog to look for the largest published GWAS for GGT serum levels in population of European ancestry, to compare our results with those for independent samples. We chose the study by \citet{Pazoki}, which has a sample size of $437\,194$. As they tested with a linear model, it is not possible to use their results as a benchmark of what the `truth' is, but we can use it to show that we do not perform worse than the linear model in our samples. Namely, using $\alpha=0.05$ as a reference and excluding the hits in chromosome 12 we already mentioned, PLINK found in the Trinity dataset 2 loci with low $p$-values in \citet{Pazoki} and 3 with high; $b=1$ detected 2 low, 2 of approximately 0.05, and 2 high; and $b\in\{2,3,4\}$ all found only 2 loci, both with high $p$-values in \citep{Pazoki}. All in all, the only signal with strong bibliographical evidence of being genuine is that of chromosome 12, which is found with every method. The other hits may or may not correspond to relevant biological discoveries.

\begin{table}[!htbp]
	\centering
	\caption{SNPs with a $p$-value of less than $10^{-5}$ with PLINK's linear test. The columns represent: dbSNP ID, chromosome, position in base pairs, reference allele, the aforementioned $p$-value, and that of \citet{Pazoki} for the same SNP.}
	\begin{adjustbox}{width=.95\textwidth}
			\begin{tabular}{@{}llllll@{}}
				\toprule
				SNP       & Chromosome & Position (bp) & Ref. allele & $p$-value PLINK & $p$-value Pazoki \\
				\cmidrule(l{-1pt}r{-1pt}){1-6}
				rs1169300 & 12         & 119915608     & A           & 2.76E-07        & 0                \\
				rs2464196 & 12         & 119919810     & T           & 2.97E-07        & 0                \\
				rs2259820 & 12         & 119919725     & A           & 3.23E-07        & 0                \\
				rs1182933 & 12         & 119939005     & A           & 4.55E-07        & 0                \\
				rs1863514 & 11         & 4416924       & C           & 1.74E-06        & 0.61             \\
				rs3213545 & 12         & 119955720     & T           & 1.76E-06        & 0                \\
				rs1169302 & 12         & 119916685     & G           & 1.80E-06        & 8.90E-262        \\
				rs1169288 & 12         & 119901033     & G           & 1.82E-06        & 0                \\
				rs2375754 & 1          & 65384221      & G           & 7.38E-06        & 0.55             \\
				rs915281  & 9          & 119007784     & C           & 7.99E-06        & 0.67             \\
				rs7801967 & 7          & 28211820      & T           & 9.19E-06        & 3.10E-08         \\
				rs6478298 & 9          & 119034512     & C           & 9.72E-06        & 0.97            \\
				\bottomrule
			\end{tabular}
	\end{adjustbox}
	\label{tab:ggt:plink}
\end{table}

\begin{table}[!htbp]
	\centering
	\caption{SNPs with a $p$-value of less than $10^{-5}$ with the distance covariance test for $b=1$. The columns represent: dbSNP ID, chromosome, position in base pairs, reference allele, the aforementioned $p$-value, and that of \citet{Pazoki} for the same SNP.}
	\begin{adjustbox}{width=.95\textwidth}
			\begin{tabular}{@{}llllll@{}}
				\toprule
				SNP       & Chromosome & Position (bp) & Ref. allele & $p$-value DC $b=1$ & $p$-value Pazoki \\
				\cmidrule(l{-1pt}r{-1pt}){1-6}
				rs1169288  & 12 & 119901033 & G & 9.20E-07 & 0      \\
				rs7794763  & 7  & 3501031   & T & 4.37E-06 & 0.057  \\
				rs12601826 & 17 & 14436360  & T & 5.01E-06 & 0.49   \\
				rs1169300  & 12 & 119915608 & A & 5.45E-06 & 0      \\
				rs2464196  & 12 & 119919810 & T & 5.45E-06 & 0      \\
				rs4299187  & 17 & 14426169  & A & 6.66E-06 & 0.61   \\
				rs2825610  & 21 & 19793455  & C & 7.10E-06 & 0.053  \\
				rs1588514  & 4  & 61054173  & T & 7.88E-06 & 0.85   \\
				rs12766994 & 10 & 20180814  & C & 8.15E-06 & 0.019  \\
				rs2259820  & 12 & 119919725 & A & 9.02E-06 & 0      \\
				rs11220787 & 11 & 126493283 & G & 9.26E-06 & 0.12   \\
				rs7120599  & 11 & 110602153 & C & 9.88E-06 & 0.0068 \\
				\bottomrule
			\end{tabular}
	\end{adjustbox}
	\label{tab:ggt:b1}
\end{table}

\begin{table}[!htbp]
	\centering
	\caption{SNPs with a $p$-value of less than $10^{-5}$ with the distance covariance test for $b=2$. The columns represent: dbSNP ID, chromosome, position in base pairs, reference allele, the aforementioned $p$-value, and that of \citet{Pazoki} for the same SNP.}
	\begin{adjustbox}{width=.95\textwidth}
			\begin{tabular}{@{}llllll@{}}
				\toprule
				SNP       & Chromosome & Position (bp) & Ref. allele & $p$-value DC $b=2$ & $p$-value Pazoki \\
				\cmidrule(l{-1pt}r{-1pt}){1-6}
				rs1169288 & 12 & 119901033 & G & 1.59E-07 & 0    \\
				rs1169300 & 12 & 119915608 & A & 2.36E-07 & 0    \\
				rs2464196 & 12 & 119919810 & T & 2.36E-07 & 0    \\
				rs2259820 & 12 & 119919725 & A & 4.04E-07 & 0    \\
				rs1182933 & 12 & 119939005 & A & 5.84E-07 & 0    \\
				rs9527666 & 13 & 56968400  & T & 1.93E-06 & 0.19 \\
				rs1409244 & 13 & 56961368  & A & 2.43E-06 & 0.26 \\
				rs354786  & 13 & 57026834  & T & 3.00E-06 & 0.26 \\
				rs2303071 & 5  & 147468560 & G & 3.16E-06 & 0.21 \\
				rs880687  & 5  & 147466870 & G & 3.37E-06 & 0.22 \\
				rs2303062 & 5  & 147460200 & A & 4.41E-06 & 0.22 \\
				rs2303063 & 5  & 147460220 & G & 4.41E-06 & 0.2  \\
				rs2303065 & 5  & 147460305 & T & 4.41E-06 & 0.21 \\
				rs2303067 & 5  & 147461148 & A & 4.85E-06 & 0.21 \\
				rs3213545 & 12 & 119955720 & T & 6.16E-06 & 0 \\
				\bottomrule
			\end{tabular}
	\end{adjustbox}
	\label{tab:ggt:b2}
\end{table}

\begin{table}[!htbp]
	\centering
	\caption{SNPs with a $p$-value of less than $10^{-5}$ with the distance covariance test for $b=3$. The columns represent: dbSNP ID, chromosome, position in base pairs, reference allele, the aforementioned $p$-value, and that of \citet{Pazoki} for the same SNP.}
	\begin{adjustbox}{width=.95\textwidth}
			\begin{tabular}{@{}llllll@{}}
				\toprule
				SNP       & Chromosome & Position (bp) & Ref. allele & $p$-value DC $b=3$ & $p$-value Pazoki \\
				\cmidrule(l{-1pt}r{-1pt}){1-6}
				rs1169300 & 12 & 119915608 & A & 1.42E-07 & 0         \\
				rs2464196 & 12 & 119919810 & T & 1.42E-07 & 0         \\
				rs2259820 & 12 & 119919725 & A & 2.33E-07 & 0         \\
				rs1169288 & 12 & 119901033 & G & 2.73E-07 & 0         \\
				rs1182933 & 12 & 119939005 & A & 3.69E-07 & 0         \\
				rs9527666 & 13 & 56968400  & T & 1.47E-06 & 0.19      \\
				rs1409244 & 13 & 56961368  & A & 1.84E-06 & 0.26      \\
				rs354786  & 13 & 57026834  & T & 2.38E-06 & 0.26      \\
				rs3213545 & 12 & 119955720 & T & 2.60E-06 & 0         \\
				rs2303071 & 5  & 147468560 & G & 5.93E-06 & 0.21      \\
				rs880687  & 5  & 147466870 & G & 6.52E-06 & 0.22      \\
				rs1169302 & 12 & 119916685 & G & 7.39E-06 & 8.90E-262 \\
				rs2303062 & 5  & 147460200 & A & 8.11E-06 & 0.22      \\
				rs2303063 & 5  & 147460220 & G & 8.11E-06 & 0.2       \\
				rs2303065 & 5  & 147460305 & T & 8.11E-06 & 0.21      \\
				rs2303067 & 5  & 147461148 & A & 8.84E-06 & 0.21   \\
				\bottomrule
			\end{tabular}
	\end{adjustbox}
	\label{tab:ggt:b3}
\end{table}

\begin{table}[!htbp]
	\centering
	\caption{SNPs with a $p$-value of less than $10^{-5}$ with the distance covariance test for $b=4$. The columns represent: dbSNP ID, chromosome, position in base pairs, reference allele, the aforementioned $p$-value, and that of \citet{Pazoki} for the same SNP.}
	\begin{adjustbox}{width=.95\textwidth}
			\begin{tabular}{@{}llllll@{}}
				\toprule
				SNP       & Chromosome & Position (bp) & Ref. allele & $p$-value DC $b=4$ & $p$-value Pazoki \\
				\cmidrule(l{-1pt}r{-1pt}){1-6}
				rs1169300 & 12 & 119915608 & A & 1.52E-07 & 0         \\
				rs2464196 & 12 & 119919810 & T & 1.52E-07 & 0         \\
				rs2259820 & 12 & 119919725 & A & 2.41E-07 & 0         \\
				rs1182933 & 12 & 119939005 & A & 3.96E-07 & 0         \\
				rs1169288 & 12 & 119901033 & G & 5.31E-07 & 0         \\
				rs9527666 & 13 & 56968400  & T & 1.35E-06 & 0.19      \\
				rs1409244 & 13 & 56961368  & A & 1.68E-06 & 0.26      \\
				rs3213545 & 12 & 119955720 & T & 2.09E-06 & 0         \\
				rs354786  & 13 & 57026834  & T & 2.24E-06 & 0.26      \\
				rs1169302 & 12 & 119916685 & G & 6.35E-06 & 8.90E-262 \\
				rs2303071 & 5  & 147468560 & G & 8.65E-06 & 0.21      \\
				rs6830854 & 4  & 57316600  & G & 8.65E-06 & 1         \\
				rs880687  & 5  & 147466870 & G & 9.65E-06 & 0.22      \\
				\bottomrule
			\end{tabular}
	\end{adjustbox}
	\label{tab:ggt:b4}
\end{table}

\section{Discussion and conclusion}\label{gwas:discu}

In this chapter, we have derived novel methodology for testing the association of SNPs with a quantitative response based on the generalised distance covariance $\V_b$. We have further provided a model-based interpretation for the method and investigated different choices of parameter $b$. Our tests are consistent against functional alternatives and have high power against many alternatives, with each of our tests being the locally most powerful one under some model assumptions. We demonstrate good performance in simulations and sound results in a real data example. Moreover, we show in theory and practice that we can satisfactory adjust for nuisance covariates.

Our literature review provides no direct competing methods from the distance and kernel communities, but we did find a couple of works that handle related problems. \citet{Fischer:kernel} focus on case-parent trios (a different and simpler setting) and do not give any particular structure to the space of genotypes (instead, they define some notion of similarity matrix). \citet{Hua:mult:comp:gwas} had the idea of applying distance covariance to some kind of GWAS. Even though they restrict themselves to Euclidean spaces, by the virtue of using the energy association measure, they are able to capture more signal than traditional methodology. They lack the interpretation that we have and also the ability of detecting non-additive effects, but they do investigate how to treat missing data (a central problem in genomics) and the effects different schemes for FDR control have, both of which could be future lines of work for us too. A year later than Hua's paper, \citet{Carlsen} suggested another approach for GWASs with distance covariance as a first filter to then detect marginally significant SNPs for a binary trait using ordinary ridge regression with an FDR control mechanism. Their usage of distance correlation as a sure independence screening mechanism \citep[see][]{Li:SIS} is based on using the rough genotype values $\{0,1,2\}$ as such, on the real line and with the Euclidean metric. As extensively discussed throughout this chapter, we deem the latter assumption too stringent and, from that point on, our path completely diverges from that of \citet{Carlsen}. In yet another article on the topic, \citet{Rongling} followed a similar approach to Carlsen's, but in their case for quantitative phenotypes and with a two-stage variable selection procedure \citep[each of them with the DC-SIS by][]{Li:SIS}, to then finalise with LASSO or similar methods. It is worth mentioning that \citet{Rongling} open the door in their regression models to consider non-additive effects of SNPs, albeit they restrict themselves to discussing full dominant/recessive scenarios as the only alternative. Again, our approach differs from the very beginning, but we considered it of interest to highlight the main pieces of literature that in any way use distance or kernel methodology for GWA studies. Distance correlation has also proven to be useful in other ``omic'' scenarios, e.g. to study expression \citep{Guo,SysBio}, which again reinforces the potential of these statistical techniques for such kind of data.

When using ours or any other method to detect marginally significant SNPs, one should take into account that the positives one finds may occur due to three main reasons \citep{Cardon}: their being genuine causative agents of phenotypic variation (i.e., true positives), chance or an artifact (e.g., selection bias or presence of confounders), or the SNP being in linkage disequilibrium with the truly causative SNP (and therefore truly associated with the response). Our method tackles confounders by design, and selection bias is in principle something the study design should take care of. On the other hand, determining which exact SNP of a small region (or locus) is the causative agent of the observed phenotype-genotype relationship is more a biological question to which one should apply domain knowledge from that field \citep{Brandes}. It is interesting to note that the methodology presented in Chapter~\ref{ch3} detected as a by-product pairs of SNPs physically near each other, which means that distances are a good way of studying this problem too, as an alternative way to the standard techniques for the \emph{pruning} and \emph{clumping} procedures in GWAS settings.

All in all, as of 2024, there is still great interest in finding new SNPs that marginally influence a given phenotype remains a central question to genetics as of today, while the discoveries that have already been made keep improving clinical practice and basic understanding of human biology \citep{15y}. The need for finding even more trait-associated loci is justified by the fact that GWAS generally discover genetic variants with small effect sizes and that therefore explain a modest proportion of the overall heritability \citep{Tam}. Traditional GWAS analyses will keep yielding new \textit{bona fide} associations as long as sample size will keep being increased (which has lead to the large biobank era), but there are natural and pragmatical limits to how many human beings one can sample, so the need for new statistical avenues to this problem is clear. We argue that ideas like ours (i.e., applying modern statistics to modern genomics) have the potential to transform the field.


%% file: ch5_v14.tex
\fancyhead[LO]{\rightmark}
\fancyhead[RE]{\leftmark}
\renewcommand{\headerright}{\thechapter}

\chapter{Comparison of distance-based tests with classical methodology for categorical data}
\thispagestyle{empty}
\label{ch6} 
\graphicspath{{./fig_ch6/}} 

Categorical variables are of uttermost importance in biomedical research. When two of them are considered, it is often the case that one wants to test whether or not they are statistically dependent. This can be achieved by extending the distance-covariance philosophy of Chapter~\ref{ch3} to two-dimensional contingency tables of arbitrary (finite) size. We show weaknesses of classical methods and we propose testing strategies based on distances that lack those drawbacks.

We then apply the same fundamental ideas to one-dimensional tables, namely to the testing for goodness of fit to a discrete distribution, for which we resort to an analogous statistic called \emph{energy distance}, which had already been mentioned in Chapter~\ref{ch2}.

We prove that, in both settings, our methodology has desirable theoretical properties, and we show how we can calibrate the null distribution of our test statistics without resorting to any resampling technique. We illustrate all this in simulations, as well as with some real data examples, demonstrating the adequate performance of our approach in practice.

The scope of this chapter will be to address the testing for independence and goodness of fit with categorical data, using the aforementioned techniques, collectively known as \emph{energy statistics} \citep{TEOD}. We first use Section~\ref{intro:ct} to introduce the statistical methodology that is conventionally used for theses problems. Section~\ref{test:dc:categorical} contains our novel approach to the testing for independence between two categorical variables. In Section~\ref{test:ed:categorical}, we develop the testing for goodness of fit to a discrete distribution using the same basic notions, but with different theoretical tools. Some illustrative simulations are reported in Section~\ref{simu:categorical}. In Section~\ref{rdata:categorical}, we apply the method to real data, to show applicability. In the former case, we study the ability of polygenic risk scores to capture the chronicity of schizophrenia, and in the latter, we look for departures from the Hardy--Weinberg equilibrium. Concluding remarks are given in Section~\ref{discu:categorical}. Proofs of the theoretical results are given in Section~\ref{ap:proofs:ch6} of the appendix.

The contents of this chapter are also publicly available as a separate article \citep{F:categorical}.

\section{Classical tests for categorical data}\label{intro:ct}

In Chapter~\ref{ch3}, an interesting dataset from complex disease genomics motivated us to define distances on discrete spaces of cardinality 3 and test independence among variables whose support lies on such spaces. Since the times of Karl Pearson (more than a century ago), the corresponding test for categorical variables with an arbitrary finite number of categories has been of paramount interest to manifold applications. As a matter of fact, independence of categorical variables ranks among the most often tested hypotheses in biomedical practice \citep{BS}. Discrete data arise in health sciences in a variety of contexts \citep{Agresti,Preisser} --- for measuring responses to treatments, signposting the stage of a disease (or whether the disease is present), establishing subgroups after a diagnosis, and so forth.

In this chapter, we present the distance and kernel counterpart of what \citet{Pearson} did. We derive some theory for independence testing and extend it to the problem of goodness of fit. We finally illustrate the performance of our methodology with synthetic and real data examples, including the comparison with competing methods.

Let us first consider the testing for independence, between two categorical variables: $X\in\{1,\ldots,I\}$ and $Y\in\{1,\ldots,J\}$. Given an IID sample $\{(X_m,Y_m)\}_{m=1}^n$, one can construct the $I\times J$ contingency table $(n_{ij})_{i,j}$ by counting the observations per pair of categories $(X,Y)$:
$$
n_{ij} = \sum_{m=1}^n 1_{\{X_m = i, Y_m = j \}}.
$$
Under the null hypothesis, we expect to observe, in each cell:
$$
n_{ij} ^*:=\frac{1}{n} \sum_{k=1}^J n_{ik} \sum_{k=1}^I n_{kj}\;\:.
$$
One of the most common test statistics is Pearson's:
$$
\chi^2=\sum_{i=1}^{I} \sum_{j=1}^{J}  \frac{(n_{ij} - n^*_{ij})^2}{n^*_{ij}},
$$
for which the $p$-values are either computed using a chi-squared distribution with $(I-1)(J-1)$ degrees of freedom, or with permutations. The same holds for the null distribution of the $G$-test:
$$
G=2\sum_{i=1}^{I} \sum_{j=1}^{J}  n_{ij}\log\left( \frac{n_{ij}}{n^*_{ij}}\right),
$$
which is essentially the likelihood ratio test for this problem \citep[][\S~2.4.1]{Agresti}. Other available methods include Fisher's exact test \citep{FET} and the $U$-statistic permutation test (USP) by \citet{BS}. The authors of this last work very illustratively show how classical methods have important limitations related to imbalanced cell counts, which justifies the need for new techniques for such a relevant problem.

For the problem of goodness of fit, it is customary to resort to Pearson's (chi-squared) test, for which the philosophy is, once more ``the squared difference of the observed and the expected, divided by the expected;'' now with the difference that the table is $1\times I$ and the expected cell counts will be: $$n_i^*=np_i\;;$$ for $i=1,\ldots,I$; with $p_i=\Prob_{H_0}\{X=i\}$ being the probability of $X$ being observed as $i$ under the distribution for which goodness of fit is being tested for.

\section{The distance covariance test of independence between two categorical variables}\label{test:dc:categorical}

Given an IID sample $\{(X_m,Y_m)\}_{m=1}^n$ of $(X,Y)$, a consistent (but biased) estimator for the generalised distance covariance \citep{TEOD} between our jointly distributed two random variables is given by
$$
\widehat{V} = \widehat{T}_1 - 2 \widehat{T}_2 + \widehat{T}_3,
$$
where
\begin{align*}
	\widehat{T}_1 &= \frac{1}{n^2} \sum_{l,m =1}^n \dx(X_l,X_m) \, \dy(Y_l,Y_m),\\
	\widehat{T}_2 &= \frac{1}{n^3} \sum_{l =1}^n  \left[ \sum_{m =1}^n \dx(X_l,X_m) \right] \, \left[ \sum_{m =1}^n \dy(Y_l,Y_m) \right] , \\
	\widehat{T}_3 &= \frac{1}{n^4} \left[\sum_{l,m =1}^n \dx(X_l,X_m) \right] \, \left[\sum_{l,m =1}^n \dy(Y_l,Y_m) \right].
\end{align*}

We assume that the supports $\spx$ and $\spy$ of $X$ and $Y$ respectively are finite, with cardinality $I\in\Zplus$ and $J\in\Zplus$. When it comes to deciding which distances $\dx$ and $\dy$ to equip them with, the only restriction we have for distance covariance and associated techniques to work out is that we need to be in a premetric structure of strong negative type, as seen in Chapter~\ref{ch2}. Now the question would be which of those feasible distances is the most convenient to use. Since we are working with categorical data and we want to be as agnostic as possible in terms of the underlying relationships among categories, in the following we will restrict ourselves to the case in which the metric structure on both marginal spaces reflects this agnosticism. In other words, we will equip both $\spx$ and $\spy$ with the discrete metric (which we will henceforward denote simply as $d$ for both spaces), already defined in Equation~\eqref{disc:metric}.

Alternatively, we could obtain the same test statistic by identifying the $I$ categories of $X$ with an orthonormal basis of $\mathbb{R}^I$ and then using the Euclidean distance and classical distance covariance \citep{SRB}, instead of its extension to metric spaces \citep{Jakobsen,Lyons}.

We now construct the $I \times J$ contingency table for the IID sample $\{(X_m,Y_m)\}_{m=1}^n$ of $(X,Y)$. Its $(i,j)$-th cell will be denoted by $n_{ij}$:
$$
n_{ij} = \sum_{m=1}^n 1_{\{X_m = i, Y_m = j \}}.
$$
We call the $n_{ij}$'s \emph{observed} cell counts, whereas their \emph{expected} counterparts are their expected values under the null hypothesis (i.e., independence of $X,Y$).

We now introduce the notation $n_{i \cdot}$ and $n_{ \cdot j}$ for the row and column sums of the contingency table:
$$
n_{i \cdot} := \sum_{j=1}^J n_{ij} = \sum_{m=1}^n 1_{\{X_m = i\}};
$$
$$
n_{\cdot j} := \sum_{i=1}^I n_{ij} = \sum_{m=1}^n 1_{\{Y_m = j\}}.
$$	
These allow us to define the expected cell counts (under independence):
$$
n^*_{i j} = \frac{1}{n} n_{i \cdot} n_{\cdot j}
$$
By performing some algebraic manipulations, one can see that our test statistic can compactly be written as:
\begin{equation}\label{ts:indep}
	\widehat{V} =\frac{1}{n^2}  \sum_{i=1}^{I} \sum_{j=1}^{J}  (n_{ij} - n^*_{ij})^2
\end{equation}

On the other hand, Pearson's (chi-squared) test for independence is based on the statistic 
$$
\chi^2=\sum_{i=1}^{I} \sum_{j=1}^{J}  \frac{(n_{ij} - n^*_{ij})^2}{n^*_{ij}},
$$
which only differs in a ``normalising'' denominator in each term of the sum.

We now state the following result on the null distribution of our test statistic~(\ref{ts:indep}), which is proven in Appendix~\ref{ap:proofs:ch6}. 

\begin{theorem} \label{th:indep}
	Let $(X_1,\ldots,X_n)$ and $(Y_1,\ldots,Y_n)$ be IID samples of jointly distributed random variables $(X,Y) \in \{1,2,\ldots,I\} \times\{1,2,\ldots,J\}$, with $q_i := P(X=i)$ and $r_j := P(Y=j)$.
	
	Consider $\mathcal X$ and $\mathcal{Y}$ equipped with the discrete metric. Then the empirical distance covariance between the two random variables can be written as:
	$$
	\dCovh_{\text{discrete}}^2(X,Y) = \frac{1}{n^2}  \sum_{i=1}^{I} \sum_{j=1}^{J}  (n_{ij} - n^*_{ij})^2
	$$
	
	In addition, whenever $X$ and $Y$ are independent, for $n \to \infty$,
	$$
	n \, \dCovh_{\text{discrete}}^2(X,Y)
	\distrilim 
	\sum_{i=1}^{I-1} \sum_{j=1}^{J-1} \lambda_i \mu_j Z_{ij}^2
	$$
	
	where $Z_{ij}^2$ are independent chi-squared variables with one degree of freedom each. $\lambda_1,\ldots,\lambda_I$ are the eigenvalues of matrix $\mathbf A=(a_{ij})_{I\times I}$, whose entries are:
	$$
	a_{ij} = q_i \delta_{ij} - q_i q_j,
	$$
	where $\delta_{ij}$ is the Kronecker delta.
	Similarly, $\{\mu_1,\ldots, \mu_J\}$ is the spectrum of  $\mathbf B=(b_{ij})_{J\times J}$, with
	$$
	b_{ij} = r_i \delta_{ij} - r_i r_j.
	$$
\end{theorem}

It should be noted that $\mathbf A$ and $\bB$ are the covariance matrices of a multinomial distribution multiplied by a factor (actually, of a ``multi-Bernoulli'' distribution).

In practice, when it comes to using the distribution above, we will take the empirical estimators $\hat q_i$ and $\hat r_j$, then construct estimators of $\bA$ and $\bB$ from them, to finally use the products of their eigenvalues as the coefficients in the linear combination of IID $\chi^2_1$'s.

Hence, obtaining the $p$-values of our test boils down to evaluating the distribution function of weighted sums of chi-squared variables. The approximation of quadratic forms of Gaussian variables has been very well studied historically and it arises fairly often in statistical practice \citep{Duchesne}. The algorithm by \citet{Imhof} is arguably one of the best known ones, but its speed can come at the price of precision \citep{Jelle:Biomet}. We have instead chosen to resort to \citet{Farebrother} for our approximations, in the implementation by \citet{Duchesne}.

\section{The energy test for goodness of fit to a discrete distribution}\label{test:ed:categorical}
Let us once again consider a categorical variable $X$ with support $\spx$ of cardinality $I\in\mathbb Z^+$, which we will assume to be $\{1,\ldots,I\}$ without loss of generality. We observe a sample $X_1,\ldots,X_n$ IID $X$ and we will use it to test for $X\sim F$ having been drawn from a certain distribution $F_0$:
$$
H_0:F= F_0
$$

The distance-based statistic for this kind of test would be the adaptation of the one by \citet{N} to our setting. Let $d$ denote once more the discrete distance on the support of $X$. Then, the energy distance between the empirical distribution and $F$ (which equals $F_0$ under the null hypothesis) is:
$$
\mathcal E_n=n\left[ \frac{2}{n}\sum_{l=1}^n\E d(x_l,X)-\E d(X,X') -\frac{1}{n^2}\sum_{l,m=1}^n d(x_l,x_m) \right] ;
$$
where $\{x_l\}_{l=1}^n$ is a sample realisation of $\{X_l\}_{l=1}^n$ and $X'$ is an IID copy of $X$. We refer the reader to \citet{ED} for a more comprehensive review on this kind of statistics.

We recall from Section~\ref{intro:ct} that the expected cell count for each category is $n_i^*=n p_i$, whereas the observed cell count is simply:
$$
n_i:=\sum_{l=1}^n 1_{ \{X_l=i\} }
.$$
With this notation, and after some algebra, we can write our test statistic for $H_0:F= F_0$ as:
$$
\mathcal E_n=\frac{1}{n}\sum_{i=1}^I(n_i-n_i^*)^2,
$$
which again resembles Pearson's without its denominator. As of its null distribution, we present the following result.

\begin{theorem} \label{th:gof}
	Let $(X_1,\ldots,X_n)$ be an IID sample of random variable $X \in \mathcal X=\{1,2,\ldots,I\}$.
	
	Consider $\mathcal X$ equipped with the discrete metric. Then the energy distance test statistic for goodness of fit to a fixed distribution $\bp=(p_i)_{i=1}^I$ on $\{1,\ldots,I\}$ is:
	$$
	\mathcal E_n=\frac{1}{n}\sum_{i=1}^I(n_i-n_i^*)^2,
	$$
	with the observed counts being $n_i:=\sum_{l=1}^n 1_ { \{X_l=i\} } $ and the expected ones: $n_i^*=np_i$.
	
	Then, whenever $X$ is distributed according to $\bp$, for $n \to \infty$,
	$$
	\mathcal E_n
	\distrilim 
	\sum_{i=1}^{I-1} \lambda_i Z_{i}^2
	$$
	where $Z_{i}^2$ are independent chi-squared variables with one degree of freedom each. $\lambda_1,\ldots,\lambda_I$ are the eigenvalues of matrix $\bC=(c_{ij})_{I\times I}$ with
	$$
	c_{ij} = p_i \delta_{ij} - p_i p_j,
	$$
	where $\delta_{ij}$ is the Kronecker delta.
\end{theorem}

Note that, matrix $\bC$ here is, once again, a covariance matrix of a multinomial, and therefore has zero as one of its eigenvalues and $I-1$ as its rank.

For the proof of the preceding theorem, we forward the reader to Appendix~\ref{ap:proofs:ch6}.

\section{Simulation study}\label{simu:categorical}

We will now show how the tests proposed in Sections~\ref{test:dc:categorical} and~\ref{test:ed:categorical} perform numerically, by simulating some population models that we consider illustrative. Subsection~\ref{simu:dc} is devoted to the distance-covariance test and Subsection~\ref{simu:ed}, to the one based on the energy distance.

\subsection{Distance-covariance test of independence}\label{simu:dc}

As previously mentioned, the test statistic we present in Section~\ref{test:dc:categorical} is (almost) the same as the USP test statistic by \citet{BS}, with the substantial ---albeit not fundamental--- difference being that theirs is the $U$-statistic counterpart of our $V$-statistic. The approach for the testing, however, is completely different, since they use permutations, whereas we derive the (asymptotic) null distribution of the test statistic (Theorem~\ref{th:indep}). We will therefore use the family of models for contingency tables with exponentially decaying marginals described by \citet{BS}, as it provides a good framework for both assessing the calibration of significance and the performance in terms of power. We will compare our method with theirs, as well as with Pearson's chi-squared test, Pearson's test with permutations, Fisher's exact test and the $G$-test.

Let us first define the model. For given $I$ and $J$, we define the cell probabilities of our contingency table under independence as:
$$
p_{ij}^{(0)}:=\frac{2^{-(i+j)}}{(1-2^{-I})(1-2^{-J})}\text{; for }i=1,\ldots,I; j=1,\ldots, J.
$$
The above expression is clearly the product of the marginal probabilities. It is also easy to see that the probability mass is maximised in the top-left corner of the contingency table and it decreases rightwards and downwards.

Now, for each $\eps\in\Rplus$ small enough so that no probabilities are out of $[0,1]$, we define $p_{ij}^{(\eps)}$ as the following perturbation of $p_{ij}^{(0)}$:
$$
p_{ij}^{(\eps)}:=\begin{cases}
	p_{ij}^{(0)}+\eps&\text{if }(i,j)\in\{(1,1),(2,2)\}\\
	p_{ij}^{(0)}-\eps&\text{if }(i,j)\in\{(1,2),(2,1)\}\\
	p_{ij}^{(0)}&\text{otherwise }
\end{cases};
$$
where $\eps\leq\min\brc{ \left[ 8(1-2^{-I})(1-2^{-J}) \right]^{-1}, 1-\left[ 4(1-2^{-I})(1-2^{-J}) \right]^{-1} }$. The larger $\eps$ is (within its range), the further the contingency table is from the null hypothesis. The upper bound for $\eps$ can be arbitrarily close to $0.125$ (as both $I$ and $J$ tend to infinity), but for us it will be approximately $\frac{1024}{7905}\approx 0.1295$, as we will be restricting our simulated contingency tables to the dimensions we state below.

To follow exactly the footprints of \citet{BS}, we consider $M=10^4$ replicates of contingency tables with $I=5$ rows and $J=8$ columns, containing $n=100$ observations. For each of the methods based on permutations, we chose $B=999$ as the number of resamples and we use the algorithm by \citet{Patefield} to uniformly draw the contingency tables with given marginals.

\begin{figure}
	\centering\includegraphics[width=.7\textwidth]{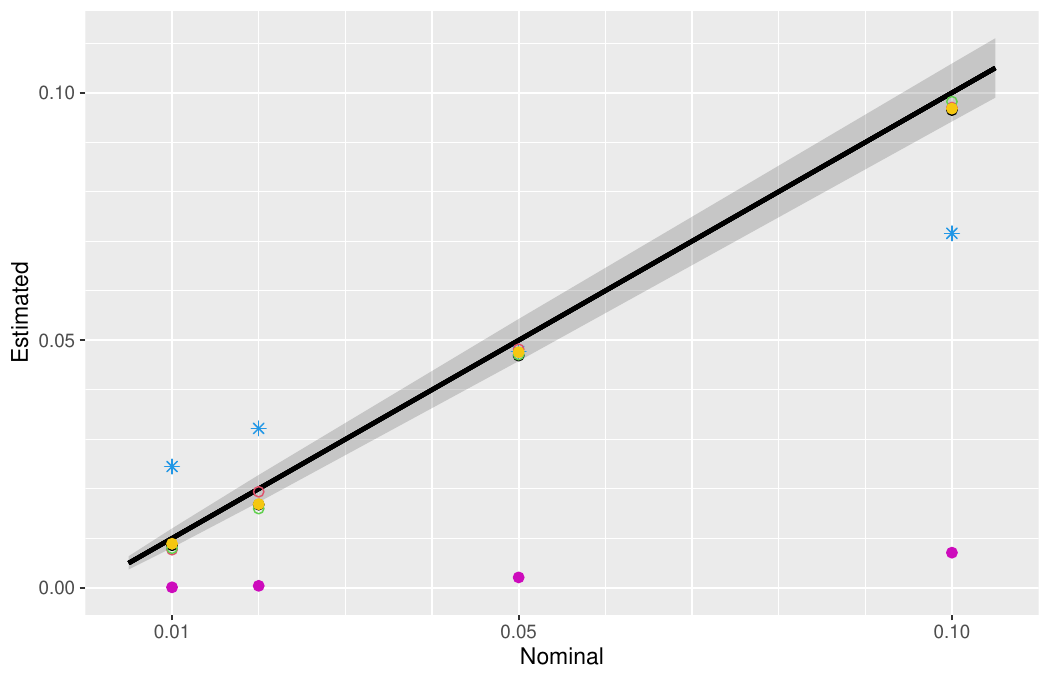}
	\caption{Empirical power under the null hypothesis ($\hat\alpha$) versus nominal significance level ($\alpha$), for the decaying marginals model, comparing our distance covariance method (golden points), Pearson's chi-squared test (pale blue), Pearson's test with permutations (dark red), the USP (black), Fisher's exact test (green) and the $G$-test (purple). The grey shadow is a 95 \% confidence band for $\hat\alpha$ given $\alpha$.}
	\label{Calibr:categorical}
\end{figure}

For $\eps=0$ we can see how we calibrate significance. Figure~\ref{Calibr:categorical} shows the results with our method for some reference values of nominal $\alpha$, and allows for a comparison with competing techniques. We see that we control type I error very satisfactorily, both when considering our results only and when comparing them with Pearson's test with permutations, the USP and Fisher's exact test. All the aforementioned tests perform satisfactorily in terms of calibration of $\alpha$. The $G$-test, however, proves to be far too conservative. Pearson's chi-squared fails, too, when it comes to controlling the type I error, but does so in a less dramatic fashion (and it actually produces a good result for nominal $\alpha$ of 0.05). To find an explanation to this phenomenon, one should note that the model we are using features very small expected cell counts, which will tend to break down the heuristic rules as to when to use the chi-squared distribution with $(I-1)(J-1)$ degrees of freedom to compute $p$-values or not.

\begin{figure}
	\centering\includegraphics[width=.7\textwidth]{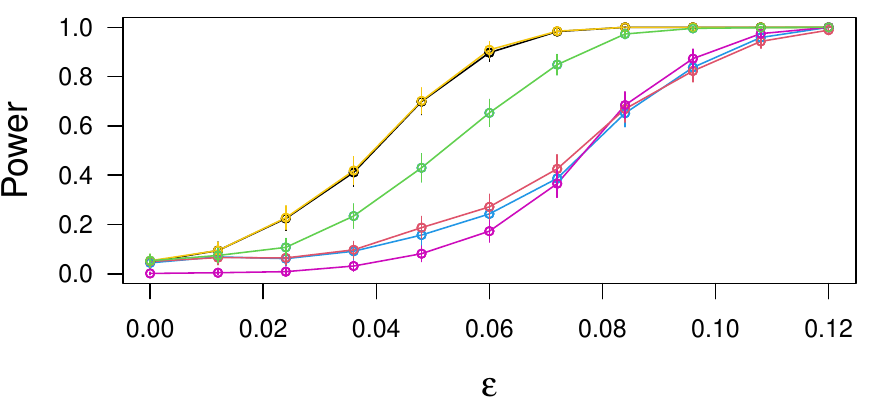}
	\caption{Power curve comparison for the decaying marginals model, displaying our distance covariance method (golden curve), Pearson's chi-squared test (pale blue), Pearson's test with permutations (dark red), the USP (black), Fisher's exact test (green) and the $G$-test (purple). The $5\times 8$ cells of each contingency table were filled with $n=100$ observations. $M=10^4$ replicates were considered. Error bars span from $-3$ to $+3$ standard deviations for each value of parameter $\eps$, which indicates the distance from the null hypothesis.}
	\label{Power_comparison_plot}
\end{figure}

In terms of power, Figure~\ref{Power_comparison_plot} shows that we perform very similarly to the USP (which shows how our derivation of the null distribution is correct and that the asymptotic approximation is not very far off when $n=100$). The power curve of Fisher's exact test is clearly under ours, whereas the one for the remaining classical methods is quite low for most values of $\eps$.

Other than the theoretical insight that using distance covariance provides (i.e., characterising general independence, the relationship to kernels and global tests, and so forth), we provide a relevant practical improvement with respect to the USP --- running time. Our experiments show that we are 3 orders of magnitude faster in testing than the USP. This remarkable difference in speed is not due to anything being intrinsically slow about computing the USP statistic, but it is simply a consequence of comparing a testing approach that uses a closed-form null distribution with another one that requires almost a thousand permutations in its default settings \citep{BS}.

\subsection{Energy-distance test of goodness of fit}\label{simu:ed}

We will firstly summarise the notion of Hardy--Weinberg equilibrium (HWE), an important genetic concept that was independently introduced in 1908 by the eponymous authors \citep{Hardy,Weinberg}. Let us consider a biallelic locus, whose alleles we will denote as $A_1$ and $A_2$. Under panmixia and in the absence of evolutionary influences, the frequencies of both alleles and of each possible genotype ($A_1A_1$, $A_1A_2$ and $A_2A_2$) remain constant from generation to generation. If we use the following notation for the allele frequencies:
$$
\theta_1:=f(A_1);\;\;\; \theta_2:=f(A_2);
$$
the genotype frequencies that are to be maintained under the HWE are:
$$
f(A_1A_1)=\theta_1^2;\;\;\; f(A_1A_2)=2\theta_1\theta_2;\;\;\; f(A_2A_2)=\theta_2^2;
$$
where $\theta_1+\theta_2=1$. We point out that the \emph{frequencies} that geneticists denote by $f$ are what a statistician would call \emph{proportions} in the population. It is also noteworthy that those frequencies that the HWE predicts are the terms of the expansion of
$$\left( \theta_1+\theta_2\right)^2 $$
as a sum.

\begin{figure}[!htbp]
	\centering\includegraphics[width=.9\textwidth]{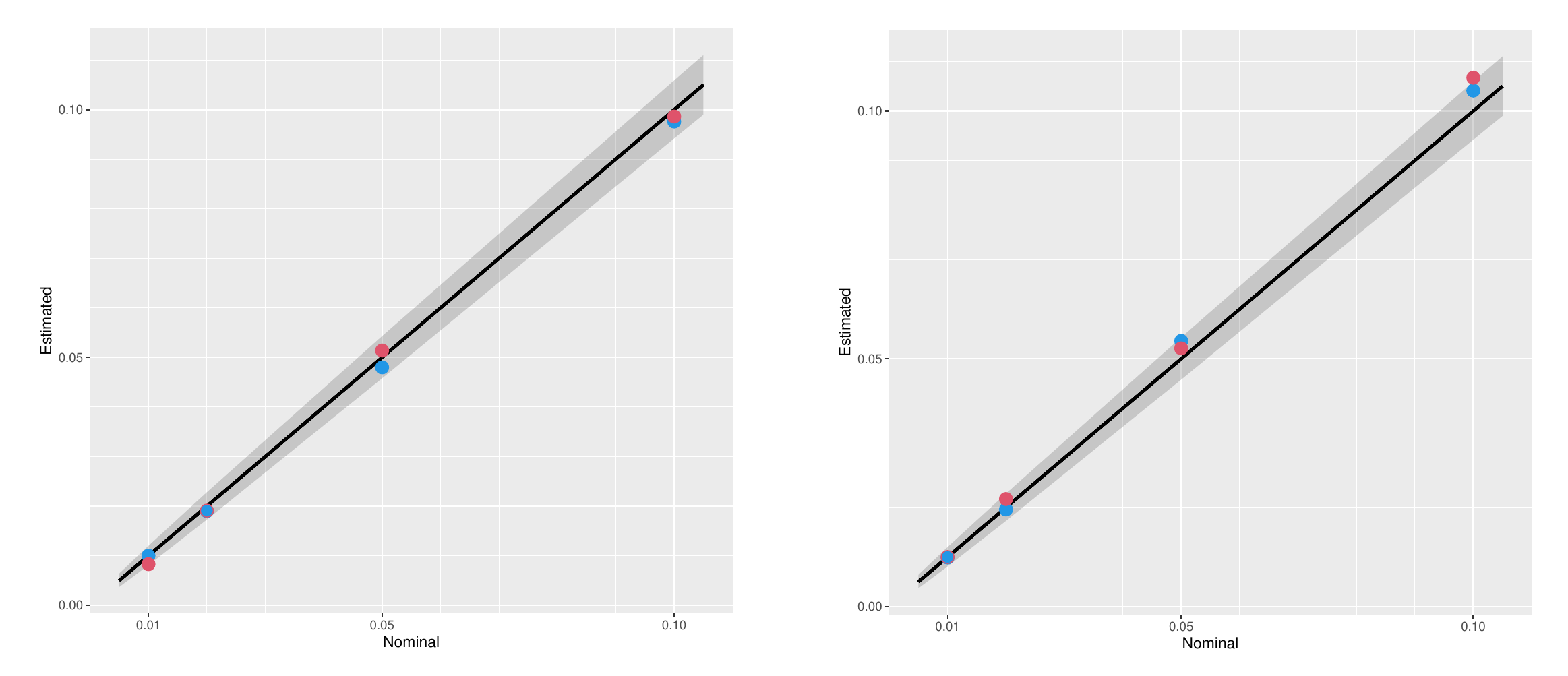}
	\caption{Empirical power under the null hypothesis ($\hat\alpha$) versus nominal significance level ($\alpha$), for the goodness-of-fit test of the biallelic Hardy--Weinberg equilibrium, when $\theta_1=\frac{2}{3}$ (left-hand plot) and $\theta_1=\frac{1}{2}$ (right). Red dots correspond to our energy distance method; blue are those for Pearson's chi-squared test. The grey shadow is a 95 \% confidence band for $\hat\alpha$ given $\alpha$.}
	\label{calibr:gof2}
\end{figure}

We will now start the simulations by showing the calibration of significance for some reference values of nominal $\alpha$ for our energy-distance test and the chi-squared test of goodness of fit. Based on the values for the allele frequencies we have encountered in the real data examples that we will be presenting in Subsection~\ref{rdata:ed}, we have chosen $\frac{2}{3}$ and $\frac{1}{2}$ as representative values of $\theta_1$ for our simulations. Figure~\ref{calibr:gof2} shows that both our method and the $\chi^2$ test perform well in terms of type I error. Every simulation in this subsection will take $n=500$ observations for each of the $M=10^4$ replicates. The sample size is a rounding of the one we have in Section~\ref{rdata:categorical}, but our numerical experiments show qualitatively similar conclusions for other values of $n$.

We now introduce two models that depart from the null hypothesis. For model 2S, we first consider the HWE genotype frequencies for the case where $\theta_1=\frac{2}{3}$:

\begin{center}
	\begin{tabular}{ccc}
		$A_1A_1$& $A_1A_2$ & $A_2A_2$  \\
		\hline
		$\frac{4}{9}$& $\frac{4}{9}$ &$\frac{1}{9}$  \\
	\end{tabular}
\end{center}

And we introduce a parameter $s\in[0,1]$ which is zero under the null hypothesis and it increases as so does the distance from $H_0$:
\begin{center}
	\begin{tabular}{ccc}
		$A_1A_1$& $A_1A_2$ & $A_2A_2$  \\
		\hline
		$\frac{4(1-s)}{9}$& $\frac{4(1-s)}{9}$ &$\frac{1+8s}{9}$  \\
	\end{tabular}
\end{center}

On the other hand, model 2K introduces parameter $k\in[-1,1]$, whose absolute value is an indicator of divergence from the HWE with $\theta_1=\theta_2=\frac{1}{2}$:
\begin{center}
	\begin{tabular}{ccc}
		$A_1A_1$& $A_1A_2$ & $A_2A_2$  \\
		\hline
		$\frac{1-k}{4}$& $\frac{k+1}{2}$ &$\frac{1-k}{4}$  \\
	\end{tabular}
\end{center}

\begin{figure}[!htbp]
	\centering\includegraphics[width=.9\textwidth]{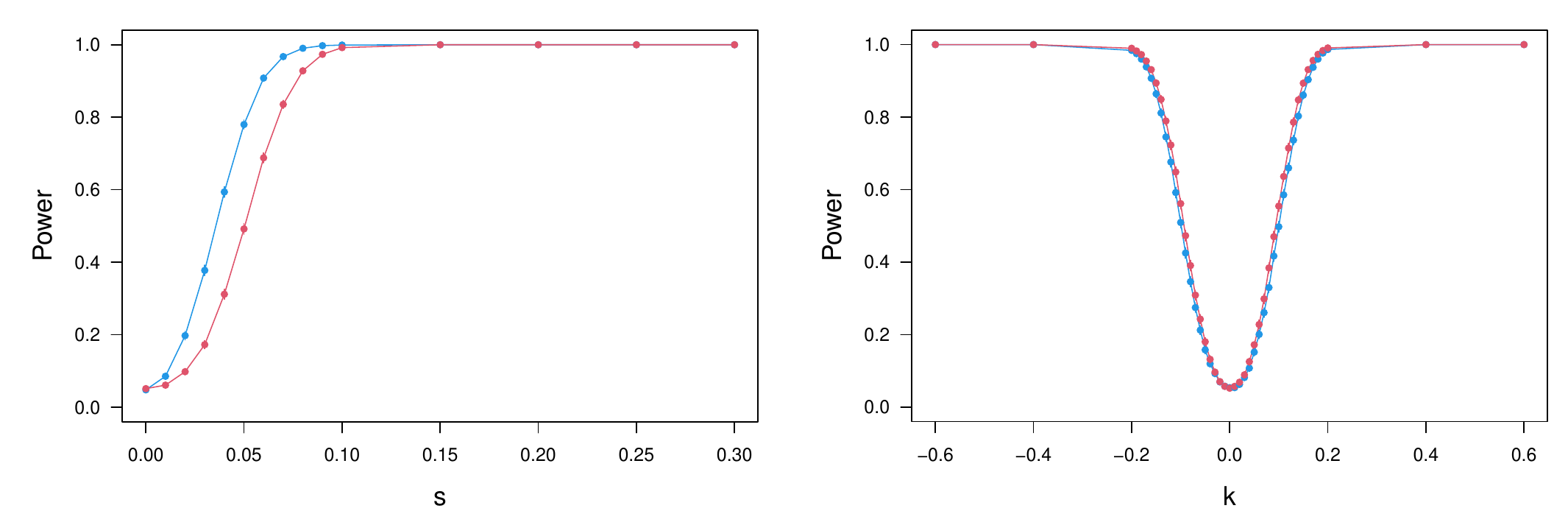}
	\caption{Power curve comparison for models 2S (left) and 2K (right), displaying our energy distance method (red lines and dots) and Pearson's chi-squared test (blue). $M=10^4$ replicates with sample size $n=500$ were considered. Error bars are barely visible in this case, but they span from $-3$ to $+3$ standard deviations for each value of parameters $s$ and $k$, which indicates the distance from the null hypothesis.}
	\label{power:2allele}
\end{figure}

We present power curves for models 2S and 2K for both dCov and the $\chi^2$ test in Figure~\ref{power:2allele}. We observe that both tests perform very satisfactorily, even for divergences from the null hypothesis that are not the highest in magnitude.

In order not to restrict ourselves to the case where the number of categories is only $3$, we will now generalise the notion of HWE. One way of doing so would be to increase the ploidy, which would yield as genotype frequencies the terms of the binomial expansion of
$$\left( \theta_1+\theta_2\right)^c $$
for $c>2$. We will however opt for a generalisation that one can encounter in humans, that is, increasing the number of possible alleles. Let us consider a triallelic locus with allele frequencies
$$
\theta_1:=f(A_1);\;\;\; \theta_2:=f(A_2);\;\;\; \theta_3:=f(A_3);
$$
where $\theta_1+\theta_2+\theta_3=1$. Then the Hardy--Weinberg genotype frequencies are:
\begin{center}
	\begin{tabular}{cccccc}
		$A_1A_1$& $A_2A_2$ & $A_3A_3$& $A_1A_2$& $A_1A_3$ & $A_2A_3$  \\
		\hline
		$\theta_1^2$&$\theta_2^2$&$\theta_3^2$& $2\theta_1\theta_2$ &$2\theta_1\theta_3$ &$2\theta_2\theta_3$   \\
	\end{tabular}
\end{center}

As with the biallelic case, we first consider a scenario where the allele frequencies are unbalanced: $\theta_1=0.70$, $\theta_2=0.25$ and $\theta_3=0.05$. Model 3S departs from the HWE for those values as parameter $s\in[0,1]$ increases:
\begin{center}
	\begin{tabular}{cccccc}
		$A_1A_1$& $A_2A_2$ & $A_3A_3$& $A_1A_2$& $A_1A_3$ & $A_2A_3$  \\
		\hline
		$0.49(1-s)$&$\frac{1+15s}{16}$&$0.0025(1-s)$& $0.35(1-s)$ &$0.07(1-s)$ &$0.025(1-s)$   \\
	\end{tabular}
\end{center}

And we also consider the case where $\theta_1=\theta_2=\theta_3=\frac{1}{3}$. By introducing parameter $k\in[0,1]$ to tune the intensity of the departure from the null, we define model 3K:
\begin{center}
	\begin{tabular}{cccccc}
		$A_1A_1$& $A_2A_2$ & $A_3A_3$& $A_1A_2$& $A_1A_3$ & $A_2A_3$  \\
		\hline
		$\frac{2k+1}{9}$& $\frac{2k+1}{9}$& $\frac{2k+1}{9}$ &$\frac{2-2k}{9}$ &$\frac{2-2k}{9}$ &$\frac{2-2k}{9}$   \\
	\end{tabular}
\end{center}

\begin{figure}[!htbp]
	\centering\includegraphics[width=.9\textwidth]{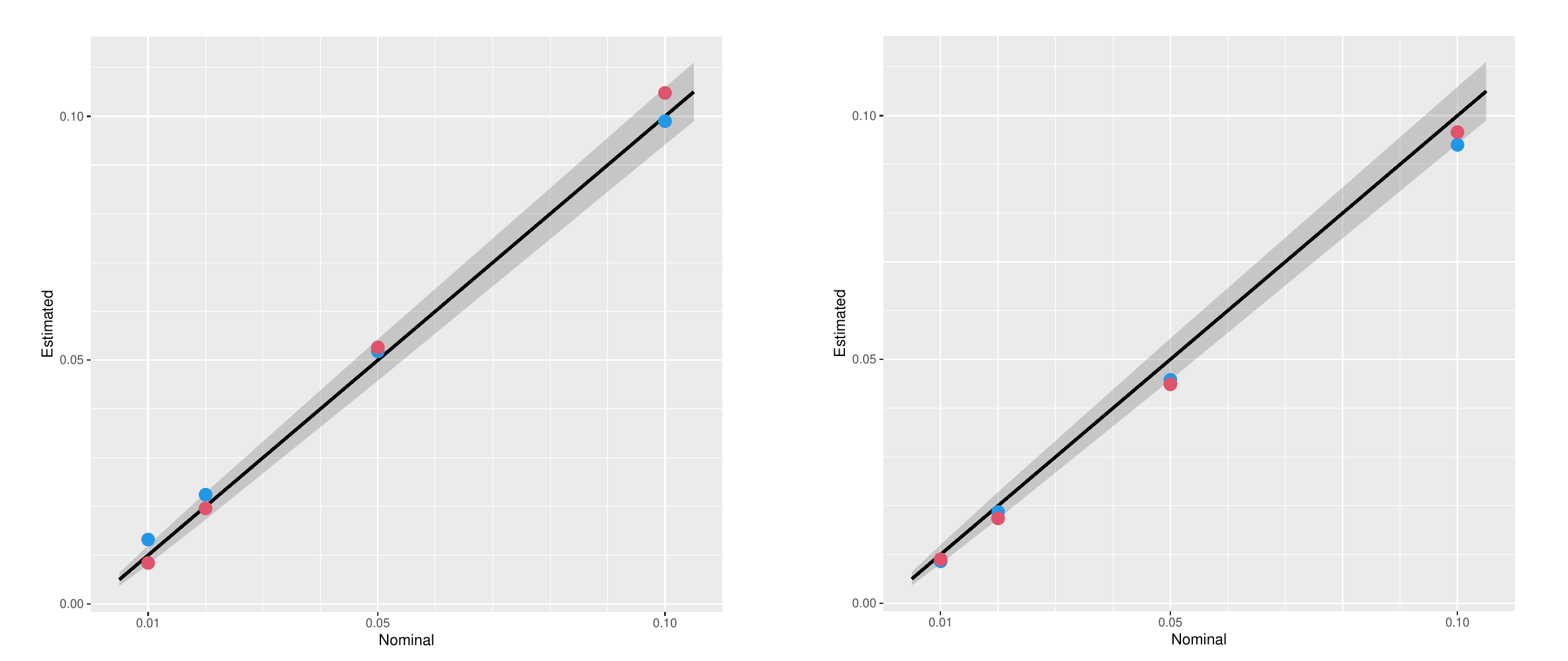}
	\caption{Empirical power under the null hypothesis ($\hat\alpha$) versus nominal significance level ($\alpha$), for the goodness-of-fit test of the triallelic Hardy--Weinberg equilibrium, when $(\theta_1,\theta_2, \theta_3)=(0.70, 0.25, 0.05)$ (left-hand plot) and $\theta_1=\theta_2=\theta_3=\frac{1}{3}$ (right). Red dots correspond to our energy distance method; blue are those for Pearson's chi-squared test. The grey shadow is a 95 \% confidence band for $\hat\alpha$ given $\alpha$.}
	\label{calibr:gof3}
\end{figure}

\begin{figure}[!htbp]
	\centering\includegraphics[width=.9\textwidth]{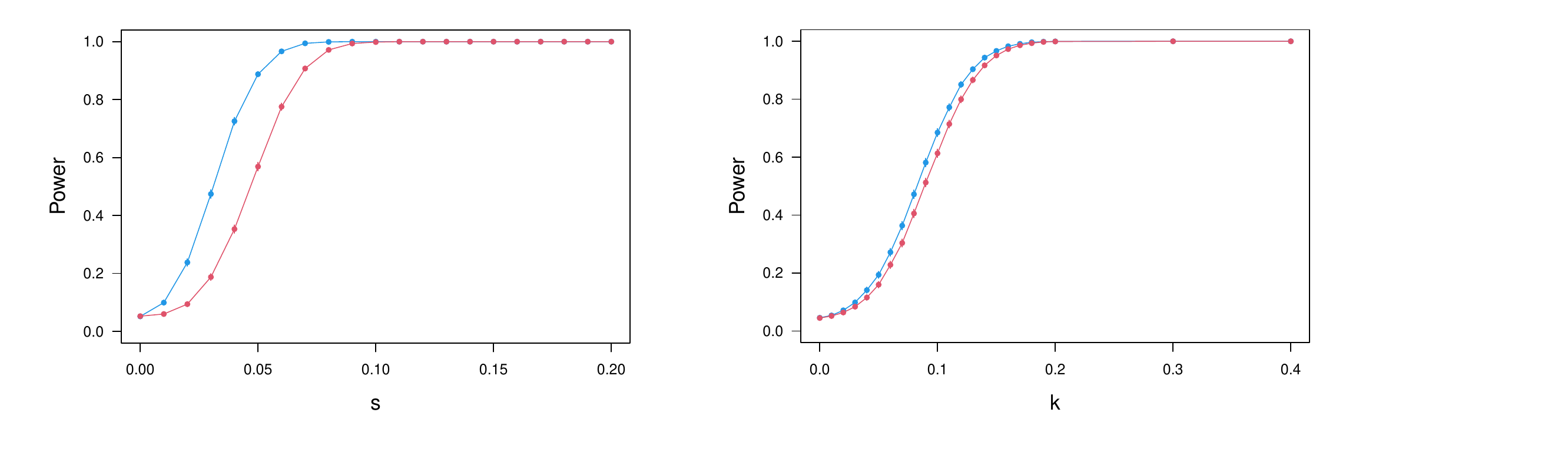}
	\caption{Power curve comparison for models 3S (left) and 3K (right), displaying our energy distance method (red lines and dots) and Pearson's chi-squared test (blue). $M=10^4$ replicates with sample size $n=500$ were considered. Error bars are barely visible in this case, but they span from $-3$ to $+3$ standard deviations for each value of parameters $s$ and $k$, which indicates the distance from the null hypothesis.}
	\label{power:3allele}
\end{figure}

Figure~\ref{calibr:gof3} shows that, once again, both the energy distance and Pearson's chi-squared control type I error. The power curves in Figure~\ref{power:3allele} show $\mathcal E$ a bit below the $\chi^2$, but we do not perform a great deal worse.

\section{Real data analyses}\label{rdata:categorical}

To complete the numerical analyses in Section~\ref{simu:categorical}, we now demonstrate the applicability of the methodology introduced in this chapter. We introduce two examples of interest to biomedical practice that arise from a dataset produced by us \citep{Facal:Scand}. Subsection~\ref{rdata:dc} explores the potential of our distance-covariance independence test for interpreting the clinical significance of polygenic scores, whereas Subsection~\ref{rdata:ed} presents real-life examples of the Hardy--Weinberg models introduced in Subsection~\ref{simu:ed}.

\subsection{Distance-covariance test of independence}\label{rdata:dc}

We begin by showing with a real biomedical example how our test for dependence can be used in practice. We consider data from \citet{Facal:Scand}, where $6\,007\,158$ SNPs were genotyped for $n=427$ patients of schizophrenia. For each of them, we consider a categorical variable $X$ indicating how chronic the psychiatric disorder is in that person (an index with four possible values, based on the admission history in health facilities), and another categorical variable $Y$ which indicates the PRS tercile (i.e., whether the \emph{polygenic risk score} for schizophrenia of the patient is low, medium or high).

Although the clinical utility of PRSs is very limited at the individual level, they may be useful for the identification of specific quantiles of risk for stratification of a population to apply specific interventions \citep{PRS:Topol}. This is why it makes the most sense to consider PRS as a categorical variable (and not one with many categories) instead of working with its raw individual scores. The data for our example can be seen in Table~\ref{tab:chronic}.

\begin{table}
	\centering
		\caption{Contingency table for the chronicity dataset, which observations of $n=427$ individuals. Rows of the table correspond to the categories of random variable `Chronicity of schizophrenia', whereas columns represent the terciles of the polygenic risk score.}
		\begin{tabular}{cccc|c}
			\multicolumn{1}{c|}{Chr. \textbackslash $\:$PRS} & $\mathrm{T}_1$     & $\mathrm{T}_2$     & $\mathrm{T}_3$     &  \\
			\cmidrule{1-4}    \multicolumn{1}{c|}{Low} & $12$ & $9$ & $4$ & $25$ \\
			\multicolumn{1}{c|}{Middle-Low} & $37$ & $20$ & $29$ & $86$ \\
			\multicolumn{1}{c|}{Middle-high} & $40$ & $58$ & $44$ & $142$ \\
			\multicolumn{1}{c|}{High} & $53$ & $55$ & $66$ & $174$ \\
			\midrule
			& $142$   & $142$   & $143$ & $427$  \\
		\end{tabular}%
	\label{tab:chronic}
\end{table}

We can now apply the different methods of Section~\ref{simu:categorical} to our dataset. Pearson's test yields similar results with and without permutations, due to the lack of low (expected) cell counts. In both cases, the $p$-value is around 0.025 and one would reject independence for a nominal $\alpha$ of 0.05. The $G$-test offers a $p$ of 0.022, in line with Pearson's. Fisher's exact test also does not diverge much, with 0.024. Finally, the USP and the distance covariance yield $p$-values of 0.047 and 0.044. All things considered, in this case one would tend to reject the null hypothesis of independence (when $\alpha=0.05$), which is consistent with the hypothesis that the PRS can measure how ``sick'' a patient is (or, more generally, how intense the trait of interest is).

\subsection{Energy-distance test of goodness of fit}\label{rdata:ed}
We will now see two examples of how one can test for goodness of fit with our methodology. Let us consider again the cohort of $n=427$ individuals by \citet{Facal:Scand}. As previously mentioned, a frequent quality control for GWAS data is whether or not each SNP is in HWE. Let us consider, for example, the biallelic SNP rs9545047 because it is one of the variants in the most current list of loci known to influence gene expression in relationship with schizophrenia, as per Extended Data Table 1 in \citet{PGC3}. This SNP has also the peculiarity of not being in a protein-coding gene, but near one, whose expression it regulates by getting transcribed into the so-called \emph{long intergenic non-protein coding RNA} (lincRNA). For this locus, we observe genotype \textit{AA} 139 times; \textit{CA}, 232 times and \textit{CC}, 56 times. Using the online tool UCSC Genome Browser \citep{UCSC}, we can retrieve several useful information about this SNP, including the allele frequencies according to the GnomAD database, which gives us: $$f(C)\approx0.41\,.$$

GnomAD v4.1.0 offers allele frequencies for different ancestries, and we have chosen the value for European (non-Finnish) population, since it is the best match for the geographical origin of our 427 individuals, which are from the northwestern Iberian peninsula. We have opted for GnomAD because it is the online resource for human population genetics with the largest sample size that we are aware of.

Therefore, the expected cell counts are:
\begin{center}
	\begin{tabular}{ccc}
		$AA$& $CA$ & $CC$  \\
		\hline
		$148.6$& $206.6$ &$71.8$  \\
	\end{tabular}
\end{center}

On the other hand, in our data we observe:
\begin{center}
	\begin{tabular}{ccc}
		$AA$& $CA$ & $CC$  \\
		\hline
		$139$& $232$ &$56$  \\
	\end{tabular}
\end{center}

When applying our energy testing procedure, it yields a $p$-value of $0.027$, which coincides with the one obtained with Pearson's. This means that both tests would reject the null hypothesis for nominal $\alpha$ of $0.05$. This is a perfectly logical result for a SNP linked to schizophrenia, which is expected to have the frequency of one of its haplotypes at a frequency that departs from the one that would be encountered under the HWE. One should also note that SNPs like this one are not left out during the quality control phase of the GWAS (described in Section~\ref{gen:database}) because the Hardy--Weinberg filter only applies to the controls.

Given that not many triallelic SNPs exist, we will just be considering one of them for illustrative purposes, without giving much profound interpretation to the results. We choose SNP rs2594292, for which the observed genotypes are:
\begin{center}
	\begin{tabular}{cccccc}
		$AA$& $GG$ & $TT$ & $AG$ & $AT$ & $TG$  \\
		\hline
		$214$& $34$ &$0$ & $148$ & $16$ & $15$  \\
	\end{tabular}
\end{center}

Once again resorting to GnomAD, we get the following population allele frequencies:
$$f(A)\approx0.69;\;\;f(G)\approx0.26;\;\;f(T)\approx0.05\,.$$
Using them to calculate the expected cell counts, we get a $p$-value of $0.24$ with our method and of $0.07$ with Pearson's. In this case we observe more dissimilar results, but with none of the tests finding significant evidence of divergence from the HWE with nominal $\alpha$ of $0.05$, which is a logical result for any SNP not known to be linked to schizophrenia.

\section{Discussion and conclusion}\label{discu:categorical}

We have proposed a new test for the independence of categorical variables (one of the most often tested hypotheses in biomedical research) by using distance covariance, an association measure that characterises general statistical independence. As we allow for arbitrary dimensions of the contingency table, this extends the possibilities we showed in Chapter~\ref{ch2} for the $3\times3$ case. We have as well developed a novel testing strategy for the goodness of fit to a discrete distribution. For both methods, we demonstrate good performance and applicability, with simulations and analyses of relevant biomedical examples.

The test statistic we derive for independence happens to have a simple algebraic expression similar in spirit to that of Pearson's $\chi^2$ test. We are not the first to see the connection between the two tests, as it was already mentioned in Remark 3.12 of \citet{Lyons} and explored in some detail in the final section of \cite{DJ}. Nevertheless, the proofs we provide are original and we are the first ones (to our knowledge) to analyse the matter in detail. On top of that, we are not aware of any previous instance in the literature where a test for goodness of fit to a discrete distribution is built based on energy statistics.

Another test for independence that is related to ours is the one in \citet{BS}, initially introduced in \citet{BKS}. The main conceptual difference in our approaches is that we derive the asymptotic null distribution of our $V$-statistic and are able to satisfactorily use it in practice, whereas their testing is based on permutations (of a $U$-statistic). It is also noteworthy that, in that article, no mention is made of distance--based association measures, a relationship that we thoroughly explore. In return, we obtain from their results the conclusion that our test statistic is very close to being the minimum-variance unbiased estimator of the population USP-divergence statistic. As they indicate, if one assumes that the population quantity is meaningful (and we now know it is, given its connection to distance covariance), then the test statistic is a very good estimator of it.

A remarkable pragmatical difference between our goodness-of-fit test and the one for independence is that the former does not require to plug in any frequencies to then estimate the multinomial covariance matrix and get the coefficients of the linear combination of chi-squared's. In this case, the $p_i$'s are fixed and known, since they are given by the null hypothesis. However, when testing whether or not the population distribution belongs to a certain family of distributions, one would need to plug in the parameters in which the family is indexed. The effect that the estimation of such parameters has in $U-$ and $V-$statistics has been studied by authors such as \citet{Wet:Randles} and \citet{Gamero}.

All in all, we have presented new methodology to address important problems of practitioners, proven solid theoretical properties, explored connections with well-known methods, and illustrated all of it in simulated and real datasets.


%% file: ch6_v14.tex
\fancyhead[LO]{\rightmark}
\fancyhead[RE]{\leftmark}
\renewcommand{\headerright}{\thechapter}

\chapter{Conclusions}
\thispagestyle{empty}
\label{ch7} 
\graphicspath{{./fig_ch7/}} 

For the sake of giving a sense of closure to the body of work we have presented throughout these pages, and to help the readership get an overview of our research, we summarise our results in Section~\ref{discu:all}, accompanied by a brief discussion (more in-depth comments on our results can be found in Sections~\ref{epist:discu}, \ref{gwas:discu} and \ref{discu:categorical}). We conclude this chapter by laying out some lines of future work in Section~\ref{fut:work}.

Any reader interested in our research output can find a list of contributions from page~\pageref{further:info} on.

\section{Results and discussion}\label{discu:all}

The topic of this dissertation is the testing for association between random elements with support in spaces whose structure represents settings of interest to the genetics of human complex traits. To that purpose, we used Chapter~\ref{ch1} to introduce both the mathematical and biological sides of our field of interest. In one word, today we are living unprecedented development in the ways we produce, store and process data; and the \emph{science} within \emph{data science} has a strong computational component, but its methodology is governed by statistics. In parallel to that transformation, the landscape of (human) biology has also undergone a deep change, evolving from a discipline that use to produce few observations of a small number of variables of similar nature, to a true high-throughput science that produces ultra-large, very heterogeneous datasets, with the advent of the `omic' era.

Many problems of current interest in human genetics boil down to looking for dependencies between variables that have a particular structure. When that is the goal, the toolbox of classical statistics falls short of providing robust and versatile techniques for testing general independence. Hence, in Chapter~\ref{ch2} we introduced the abstract theory that allows to define a general association measure, called \emph{distance correlation}, that characterises independence in most metric, semimetric and premetric spaces that one may encounter in practice. This is part of the broader topic of \emph{energy statistics} \citep{SR:book}, currently quite popular among mathematical statisticians. It turns out that all this theory is equivalent to the testing derived from the `kernel trick' \citep{Sejdinovic}, ubiquitous in the machine learning community. A further third school of independence testing, that of the so-called Global Tests (i.e., locally most powerful tests in Gaussian regression) is shown to be dual to the preceding two, when one simply transforms the data with the feature maps of the kernels in question and carries out conventional linear regression there \citep{DJ}.

Those state-of-the-art approaches to independence testing are the basis of the contributions presented in the remainder of the dissertation. In addition to the non-trivial literature review in Chapter~\ref{ch2}, our research has developed statistical methodology that allows to test for relevant biological hypotheses, including:
\begin{itemize}
	\item genetic interaction (Chapter~\ref{ch3});
	\item gene-phenotype association (Chapter~\ref{ch4});
	\item general dependencies between clinical variables (Chapter~\ref{ch6}); and
	\item Hardy--Weinberg equilibrium (also Chapter~\ref{ch6}).
\end{itemize}

In most of those settings, we first identified a problem of interest in complex disease genetics, to then propose abstract spaces whose structure best fits the data type and what is known about it, to finally develop testing procedures and other theoretical results. The `creative' process followed the opposite direction in the case of Chapter~\ref{ch6}, where it was the extension of a statistical approach what cross-fertilised new domains of application within genetics, and not the other way around.

We have shown that our methodology performs quite satisfactorily in simulations, including comparisons with preexisting competing testing procedures. On top of that, we have thoroughly studied real datasets to close the circle, bringing to practical utility the techniques we developed thinking about those very examples. The biological conclusions we are able to draw vary in each case, but they generally convey the idea of reasonable performance.

A crucial point for each of those chapters is that the statistical methods that are conventionally applied in GWA settings are based on the additivity of the allele effects in each SNP, an assumption that is known to be too restrictive in practice and to hamper the finding of signal that follows other inheritance patterns \citep{Mammals,Costas:Heteroz_opt}. With that aim, we first explored some general premetrics that can model the structure of the support space of possible genotypes better than the Euclidean one, to then consider all the possible ones in Chapter~\ref{ch4}, as well as their interpretations. For the rest of our contributions, we opted for an agnostic approach to the underlying inheritance model, for different reasons --- in Chapter~\ref{ch3} because transcending additivity already means a contribution to knowledge with respect to the literature on epistasis we are aware of (and because considering many metrics would complicate interpretation in that case); and in Chapter~\ref{ch6}, due to the fact that it is the discrete distance the one that provides interesting connections to very well-known classical methods \citep{Pearson} and the state of the art \citep{BKS}.

The test statistics that arise from distance covariance and associated methodology are, for the most part, $V$- and $U$-statistics. As a general rule, they asymptotically follow a weighted sum of chi-squared distributed random variables with one degree of freedom each \citep{TEOD}. While some approaches for approximating this distribution via moment-matching \citep{Berschneider,Huang:Huo} have been proposed, the predominant procedure for testing is still to resort to resampling methods, which is so computationally inefficient that it is not a reasonable approach in high-throughput sciences like genomics, as demonstrated in Section~\ref{comput:challenge}.

The beauty of the genetic problems we study is not only their real-life meaning, but it is also a mathematical one --- by making us work in very simple, finite support spaces, not only can we design the structure of those spaces to account for any biological reality we have in mind; but also the mathematical statistics behind them becomes slightly simpler. Namely, the finitude of the marginal spaces implies the finitude of the quadratic form that the empirical distance covariance (times the sample size) converges to. This means that, when combining the different strategies shown in Appendix~\ref{apA} for deriving the coefficients of the quadratic for with replacing them with their empirical counterparts, one can compute $p$-values in a very fast and precise way. We also do so for a slightly different problem, the testing of goodness of fit to a discrete distribution, where we resort to energy distance (a close relative of distance covariance) as a test statistic. Its asymptotic distribution has the remarkable feature of being completely specified under the null hypothesis, thus not requiring the estimation of any parameter for the computation of $p$-values.

As far as the comparison with preexisting methodology is concerned, in Chapter~\ref{ch3}, our simulations show that distance-based testing calibrates significance as well as the very popular alternative by \citet{BOOST}, and that power is better in our case (for the models considered). In Chapter~\ref{ch4}, when comparing the performance of distance covariance against that of the direct competitor by \citet{nmax3}, we prove to be superior both in terms of type I error control and of power. When comparing our method for various values of $b$, we see that the highest power is achieved with different $b$'s, depending on the value of the heterozygous effect $h$, that is, we confirm that we are able to specify \textit{a priori} against which inheritance model we want to be (the locally most) powerful. Finally, in Chapter~\ref{ch6}, our independence test outperforms classical methods such as Pearson's and the $G$-test, and is on par with the USP \citep{BS}; whereas the energy-distance goodness-of-fit test has a power curve that is slightly under that of Pearson's $\chi^2$. Also in that chapter, we accompany the empirical results with meaningful theoretical insight --- in this case, the connection between testing for all kinds of independence, the traditional Pearson's test and the cutting-edge USP.

When analysing the applied part of our work, Chapter~\ref{ch3} can be summarised as pointing towards epistasis taking place at the level of genetically-regulated gene expression, which is consistent with the findings of recent literature \citep{Lin,Patel}. Chapter~\ref{ch4} finds signal that is as sparse as expected in a GWAS, whose $p$-values are in the expected order of magnitude for the sample size in consideration, and that includes some positives that had already been found with independent samples of similar ancestry \citep{Middelberg}. Finally, the results of Chapter~\ref{ch6} are consistent with the ability of polygenic scores to measure the severity of a disorder \citep{PRS:Topol} and with the very basic conceptual notion that SNPs associated with schizophrenia will not be in Hardy--Weinberg equilibrium in the subpopulation of patients of schizophrenia.

All things considered, we have presented novel developments in mathematical statistics, orientated towards relevant applications in genetics, with very important computational demand. As a result, we have learned a great deal in the fields of mathematical statistics, biology and computer science over the last few years, and in the following section we present a road map for future learning.

\section{Future work}\label{fut:work}

We now sketch some promising lines for future research. A first interesting task would be to try to design a procedure to infer from the sample which distance is optimal in some way, for problems in which the knowledge of the domain of application does not clearly point towards any specific premetric.

Also from the point of view of mathematical statistics, it intrigues us the research question of exploring the connections between distance covariance and random forests --- if something meaningful could be worked out from it, theoretical and empirical insight would be gained. 

We also wonder how our methodology in Chapter~\ref{ch6} would adapt to the study of independence between binary and ternary variables. And by this we do not mean simply taking $I=2$ and $J=3$, but rather performing a study of the interactions between the mitochondrial (of which each individual only carries one copy) and nuclear genome (that manifests three possible genotypes, as previously indicated), and interpreting the results in the same way we did in Chapter~\ref{ch3}.

On top of that, there are currently several open questions on GWAS data that are fundamental, including: heritability estimation, testing for causality, or the prediction of phenotypes from genotypes \citep{Brandes}. Those goals go beyond the scope of this dissertation, but we believe that distance and kernel methods can allow to better conceptual approaches to any GWAS-related task, transcending simple additive and linear models, with approaches similar to the ones in this dissertation.

We have restricted ourselves to the study of humans, but our techniques have the potential to be used for other organisms. One challenge would arise when dealing with species of higher ploidy than humans (i.e., to those where each individual carries $c>2$ copies of their genome in each cell), owing to the fact that the cardinality of the finite support space for the $X$'s would differ from $3$ and then the adaptation of our methodology would not be straightforward. Recent research confirms that, at least for mammals (for which $c=2$), it is advantageous to not only consider additivity of effects, but to also consider dominant effects, in order to improve the power of GWA studies and uncover causality \citep{Mammals}.

Likewise, we focus on SNPs due to them occurring very frequently and being used often in genetic practice, but one can adapt our statistical techniques to any other kind of variant. For any of them, we would be using a finite support space, to then proceed as we did for SNPs. It would also be of interest to consider other response variables in Chapter~\ref{ch4} that are not of continuous nature. For example, one could extend the methodology to binary or survival outcomes.

On the other hand, biological knowledge indicates that genetic interactions may be, in practice, of order 3 and higher \citep{Russ}, which means that distance multivariance \citep{Boettcher1}, as already hinted in Chapter~\ref{ch3} and Appendix~\ref{apA} could be of great interest in practice, once the proper methodological developments have been carried out.

Finally, we once more emphasise that it is not only the genotype that explains the variability of phenotypes across individuals and cohorts, but rather the genotype `plus' the environment. It would hence be a promising line of future research to explore the \emph{conditional distance covariance} \citep{Conditional} as a way of incorporating environmental variables to the paradigm of the problems studied in this dissertation, which may in turn lead to better understanding the molecular basis of complex human disease.


%% file: apA_v14.tex
\fancyhead[LO]{\rightmark}
\fancyhead[RE]{\leftmark}
\renewcommand{\headerright}{\thechapter}

\chapter{Some theoretical results}
\thispagestyle{empty}
\label{apA}
\graphicspath{{./fig_apA/}}

In this appendix we present theoretical details of the mathematical statistics in Chapters~\ref{ch2} to~\ref{ch6}, which we did not include in the main body of the dissertation in order to make it easier to read. This includes proofs of theorems and propositions for the most part, with some additional results and observations.

\section{Proofs of Chapter~\ref{ch2}}\label{proofs:ch2}

\par\noindent\rm\textbf{\textsc{Proof of Proposition~\ref{prop:cr_ineq}}} ($c_r-$inequality).

\noindent(1) Let \boxx{r<1}. The case where $\beta$ vanishes is trivial, so one can assume $\beta\neq0$. The goal is to show that 
\[(t+1)^r\leq t^r+1,\;t:=\frac{\alpha}{\beta}\]
or, equivalently, that
\[f(t):=t^r+1-(t+1)^r\geq0\text{.}\]
And the latter inequality holds because $r-1<0$:
\[\forall t\in\Rplus,\;f'(t)=r(t^{r-1}-(t+1)^{r-1})>0\implica\forall t\in\Rplus,\;f(t)\geq f(0)=0\text{.}\]
\noindent(2) For \boxx{$r\geq1$}, the function $g(x):=x^r$ is convex in every $x\in\Rplus$. When $r>1$:
\[g''(x)=r(r-1)x^{r-2}>0,\;x\in\Rplus\text{.}\]
Geometrically, convexity implies that:
\[\pushQED{\qed} 
g\left(\frac{\alpha+\beta}{2}\right)\leq\frac{g(\alpha)+g(\beta)}{2}\eqv(\alpha+\beta)^r\leq2^{r-1}(\alpha^r+\beta^r)\text{.}\qedhere
\popQED
\]

\bigskip

\par\noindent\rm\textsc{\textbf{Proof of Proposition~\ref{prop:ineq_amu}}}.
\begin{sloppypar}
	(1) $D(\mu)=\int\dx(x',x'')\dmu^2(x',x'')\leq$
	
	$\leq\mu(\spx)\int\dx(x',x)\dmu(x')+\mu(\spx)\int\dx(x,x'')\dmu(x'')=2\amu(x)$.
	
	(2) Applying \emph{(1)} to $x$ and $y$ and adding side-by-side the resulting equations, one gets: $2D(\mu)\leq2\amu(x)+2\amu(y)$.
	
	(3) Integrate with respect to $\mu(z)$ both sides of: $\dx(x,y)\leq\dx(x,z)+\dx(y,z)$.
	
	(4) Idem to \emph{(3)}: $\dx(x,z)\leq\dx(x,y)+\dx(y,z)$.\qed
\end{sloppypar}

\bigskip

\par\noindent\rm\textsc{\textbf{Proof of Theorem~\ref{th:dmu_L2}}}.

It is convenient to firstly justify that, for any $(x,y)\in\spx^2$,
\[|d_\mu(x,y)|\leq2\amu(y)\text{.}\]
To see this, there are two cases to be considered:
\begin{itemize}
	\item{If \boxx{$d_\mu(x,y)\geq0$}, it suffices to apply the inequalities in Proposition~\ref{prop:ineq_amu}:
		\[|d_\mu(x,y)|=d_\mu(x,y)\overleq{(3)}D(\mu)\overleq{(1)}2\amu(y)\text{.}\]}
	\item{For \boxx{$d_\mu(x,y)<0$}, the arguments of \citet[page 10]{Jakobsen} make use of unnecessarily strong hypotheses. Instead, the following rationale:
		\[\forall z,t\in\spx:\:\dx(x,z)\leq\dx(x,y)+\dx(y,t)+\dx(t,z)\implica
		\]
		\[\implica
		\amu(x)\leq\dx(x,y)+\amu(y)+D(\mu)\text{;}\]
		yields $|d_\mu(x,y)|\leq2\amu(y)$.}
\end{itemize}
Now, using the aforementioned inequality, proving that $d_\mu\in\mathcal L^2(\mu_1\times\mu_2)$ turns out to be quite straightforward:
\[\int d_\mu(x,y)^2\dmu_1\times\mu_2(x,y)\leq4\int\amu(x)\amu(y)\dmu_1\times\mu_2(x,y)\overeq{Fubini}\]
\[\pushQED{\qed}
=4\int\dx(x,z)\dmu_1\times\mu(x,z)\:\int\dx(y,z)\dmu_2\times\mu(y,z)\overless{\boxx{$\dx\in\mathcal L^1$}}+\infty\text{.}\qedhere
\popQED\]

\bigskip

\par\noindent\rm\textsc{\textbf{Proof of Theorem~\ref{th:dcov:well_def}}}.
In order to check that \emph{dcov} is well-defined, it suffices to note that the integral of the product of two functions with respect to a (nonnegative) measure is always a scalar product (i.e., bilinear and semidefinite positive) and, as a result, it satisfies the Cauchy--Bunyakovsky--Schwarz inequality. It is also possible to prove this particular case of H\"older's inequality more directly:
\[0\leq\int[d_\mu(u)d_\nu(v)-d_\mu(v)d_\nu(u)]^2\dtheta^2(u,v)=2\int d_\mu^2\dtheta^2\:\int d_\nu^2\dtheta^2-2\left(\int d_\mu d_\nu\dtheta^2\right)^2\implica\]\[\overthen{\boxx{$d_\mu,d_\nu\in\mathcal L^2$}}|\dcov(\theta)|\leq\sqrt{\int d_\mu^2\dtheta^2\:\int d_\nu^2\dtheta^2}<+\infty\text{.}\]
A third approach is to derive a particular case of the AM-GM inequality (and also of Young's):
\[(d_\mu\pm d_\nu)^2\geq0\iff
\mp d_\mu d_\nu\leq\frac{d_\mu^2+d_\nu^2}{2}\iff
|d_\mu d_\nu|\leq\frac{d_\mu^2+d_\nu^2}{2}\text{,}\]
Anyhow, the key step is to show that the integrals on the right-hand side are finite. For instance, in the case of $d_\mu$: 
\[\int d_\mu(x,x')^2\dtheta^2((x,y)(x',y'))\overeq{Fubini}\iint d_\mu(x,x')^2\dtheta(x,y)\dtheta(x',y')\overeq{ACOV}\]
\[=\int d_\mu(x,x')^2\dmu^2(x,x')\overless{$d_\mu\in\mathcal L^2(\mu\times\mu)$}+\infty\text{;}\]
where the acronym ``ACOV'' stands for \emph{abstract change of variables}, which in this case takes a projection as the change of variables function. More formally, let $f$ be a measurable function in the following diagram:
\[\left(\spx\times\spy,\Borel{\spx}\tensor\Borel{\spy},\theta\right)\overarrow{$\pi_1$}\left(\spx,\Borel{\spx}\right)\overarrow{$f$}\left(\mathbb{R},\Borel{\mathbb{R}}\right)\text{.}\]
When $f\in\mathcal L^1(\theta\comp\pi_1^{-1})$, the aforementioned ACOV theorem ensures that:
\[\int_{\pi_1(\spx\times\spy)}f\wrt(\theta\comp\pi_1^{-1})=\int_{\spx\times\spy}(f\comp\pi_1)\dtheta\]
or, recalling that $\mu\overeq{def.}\theta\comp\pi_1^{-1}$:
\[\pushQED{\qed}
\int_{\spx}f(x)\dmu(x)=\int_{\spx\times\spy}f(x)\dtheta(x,y)\text{.}\qedhere
\popQED\]

\section{Technical notes on Chapter~\ref{ch3}}\label{proofs:ch3}

\subsection{A lemma for the discrete distance}\label{lemmas}

We now state a result by \citet{DJ} that is instrumental in the proof of the central result of Chapter~\ref{ch3}. For the theory in this chapter that we present and not use in practice, we make use of work by \citet{Huang:Huo} and \citet{Boettcher2}, but we refer the reader to those bibliographical references ---instead of reproducing their content here---, in order to stay on-topic.

\setcitestyle{square}
\begin{lemma}[\cite{DJ}, Theorem 7] \label{lem:discrete:dist}
Let $X$ and $Y$ be random variables with supports $\{1,2,\ldots,I\}$ and $\{1,2,\ldots,J\}$, respectively; where $I,J\in\Zplus$.

We now construct the entries of matrix $\bL^X=(L^X_{rs})_{I\times I}\in\R^{I\times I}$ as follows:
$$
L^X_{rs}=p_s\left( \delta_{rs}-p_r-p_s+\sum_{i=1}^I p_i^2\right) ,
$$
where $\delta_{\cdot\cdot}$ is the Kronecker delta and $p_i:=\Prob(X=i)$ is the probability mass of $X$ in $i\in\{1,\ldots,I\}$. Furthermore, let us denote by
$$\lambda^X_1,\ldots,\lambda^X_{I-1}$$
the nonzero eigenvalues of $\bL^X$. If $\{\lambda^Y_j\}_{j=1}^{J-1}$ are defined analogously, then the following limit in distribution holds under independence of $X$ and $Y$, as $n\to\infty$:
$$
n \, \dCovh_{\text{discrete}}^2(X,Y) \stackrel{\mathcal{D}}{\longrightarrow}
\sum_{i=1}^{I-1}\sum_{j=1}^{J-1}\lambda_i^X\lambda_j^Y Z_{ij}^2 ;
$$
with $\{Z_{ij}\}_{i,j}$ being IID standard Gaussian.

\end{lemma}
\setcitestyle{round}%

\subsection{Proof of Theorems~\ref{th:discrete} and~\ref{th:Euclidean}}

\par\noindent\rm\textbf{\textsc{Proof of Theorem~\ref{th:discrete}}} (discrete metric).

Applying Lemma~\ref{lem:discrete:dist}, one gets that, as $n\to\infty$, 	
$$
n \, \dCovh_{\text{discrete}}^2(X,Y) \stackrel{\mathcal{D}}{\longrightarrow}  
\lambda_1 \, \mu_1 Z_{11}^2 + \lambda_1 \mu_2 Z_{12}^2 + \lambda_2 \mu_1 Z_{21}^2 + \lambda_1 \mu_2 Z_{22}^2,
$$
where $(Z_{ij}^2)_{i,j =1}^3$ are IID $\chi^2_1$; whereas $\{\lambda_j\}_{j=1}^2$ and  $\{\mu_j\}_{j=1}^2$ are the non-zero eigenvalues of certain matrices. Namely, $\lambda_1$ and $\lambda_2$ are the non-null eigenvalues of the following matrix:
$$
\bA=\begin{pmatrix} (1-2 p_0 + \sum p_j^2) \, p_0 & ( -p_0 - p_1 + \sum p_j^2) \, p_1 & ( -p_0 - p_2 + \sum p_j^2) \, p_2 \\ (-p_0 - p_1 + \sum p_j^2) p_0 & (1- 2 p_1 + \sum p_j^2) \, p_1 & ( -p_1 - p_2 + \sum p_j^2) \, p_2 \\ (-p_0 - p_2 + \sum p_j^2) p_0 & (-p_1 - p_2+ \sum p_j^2) \, p_1 & ( 1 - 2 \, p_2 + \sum p_j^2) \, p_2\end{pmatrix}.
$$
By multiplying each of the rows of the matrix by $p_0$, $p_1$ and $p_2$ respectively, and adding up the rows, one easily sees that this matrix is singular. Using the relation $p_0+p_1+p_2=1$, it is easy to see that the characteristic polynomial of $\bA$ is:
$$
P(\lambda) = \lambda \, \left( \lambda^2 - \left( 1-\sum_{j=1}^3 p_j^2\right)  \lambda + 3 \prod_{j=1}^3 p_j\right), 
$$
where $P(\lambda)$ is defined using the sign convention that makes it monic: $\det(\lambda\,\bI_3-\bA)$.

Calculating the roots of $P(\lambda)$ yields $\lambda_1$ and $\lambda_2$. The derivation of $\mu_1$ and $\mu_2$ follows the same strategy.

\bigskip

\par\noindent\rm\textbf{\textsc{Proof of Theorem~\ref{th:Euclidean}}} (Euclidean metric).

Throughout the proof, we will use the notation:
$$
M = 2 p_0 p_1 + 2 p_1 p_2 + 4 p_0 p_2 = 2 p_0 (1-p_0) + 2 p_2 (1-p_2).
$$
Applying \citet[Theorem 4.12]{Huang:Huo}, we obtain that, when $n\to\infty$,
$$
n \, \dCovh_{\text{Euclidean}}^2(X,Y) \stackrel{\mathcal{D}}{\longrightarrow} \sum_{i,j=1}^3 \lambda_i \mu_j Z_{ij}^2;
$$
where $(Z_{ij}^2)_{i,j =1}^3$ are IID $\chi^2_1$ and $\{\lambda_j\}_{j=1}^3$, $\{\mu_j\}_{j=1}^3$ are non-negative real numbers.
By \citet[Lemma 4.14]{Huang:Huo}, $\lambda_1,\lambda_2,\lambda_3$ are the eigenvalues of
$$
\begin{pmatrix} (2 p_1 + 4 p_2 - M) \, p_0 & (-1 +  p_1 + 3 p_2 - p_0 +M) \, p_1 & M \, p_2 \\(-1 +  p_1 + 3 p_2 - p_0 +M) p_0 & (2 p_2 + 2 p_0 - M) \, p_1 & ( -1 + 3 p_0 + p_1 + p_2 -M) \, p_2 \\ M p_0 & ( -1 + 3 p_0 + p_1 + p_2 -M) \, p_1 & (2 p_1 + 4 p_0 - M) \, p_2\end{pmatrix};
$$
with $\mu_1,\mu_2, \mu_3$ being defined analogously.\\
Computing the characteristic polynomial of this matrix and proceeding as we did for Theorem~\ref{th:discrete} completes the current proof.

\subsection{Extensions to more than two SNPs}\label{ap:mvar}

\begin{theorem} \label{th:discretemult}
	Let $(X_1,\ldots,X_n)$, $(Y_1,\ldots,Y_n)$, $(U_1,\ldots,U_n)$ be IID samples of jointly distributed random variables $(X,Y,U) \in \{0,1,2\}^3$, with $p_j = P(X=j), q_j = P(Y=j), r_j = P(U=j), j=0,1,2$.
	
	Consider $\{0,1,2\}$ equipped with the discrete metric.
	
	Then, whenever $X, Y, U$ are jointly independent, for $n \to \infty$,
	$$
	n \, \widehat{\dMv}_{\text{discrete}}^2(X,Y,U) \stackrel{\mathcal{D}}{\longrightarrow}  
	\sum_{k,l,m=1}^2 \lambda_k \mu_l \gamma_m Z_{klm}^2;
	$$
	
	where $Z_{klm}^2$ with  $k,l,m \in \{1,2\}$ are IID chi-squared with one degree of freedom. The coefficients of their linear combination are given by:
	$$
	\lambda_{1,2} = \frac{1-\sum p_j^2}{2} \pm \sqrt{\frac{(1-\sum p_j^2)^2}{4} - 3 \prod p_j}; 
	$$
	$$
	\mu_{1,2} = \frac{1-\sum q_j^2}{2} \pm \sqrt{\frac{(1-\sum q_j^2)^2}{4} - 3 \prod q_j};
	$$
	$$
	\gamma_{1,2} = \frac{1-\sum r_j^2}{2} \pm \sqrt{\frac{(1-\sum r_j^2)^2}{4} - 3 \prod r_j}.
	$$
\end{theorem}

\bigskip

\begin{theorem}\label{th:Euclideanmult}
	Let $(X_1,\ldots,X_n)$, $(Y_1,\ldots,Y_n)$, $(U_1,\ldots,U_n)$ be IID samples of jointly distributed random variables $(X,Y,U) \in \{0,1,2\}^3$, with $p_j = P(X=j), q_j = P(Y=j), r_j = P(U=j), j=0,1,2$.
	
	Consider $\{0,1,2\}$ equipped with the Euclidean metric.
	
	Then, whenever $X, Y, U$ are jointly independent, for $n \to \infty$,
	$$
	n \, \widehat{\dMv}_{\text{Euclidean}}^2(X,Y,U) \stackrel{\mathcal{D}}{\longrightarrow}  
	\sum_{k,l,m=1}^2 \lambda_k \mu_l \gamma_m Z_{klm}^2;
	$$
	
	where $Z_{klm}^2$ with  $k,l,m \in \{1,2\}$ are IID chi-squared with one degree of freedom. The coefficients of their linear combination are given by:
	$$
	\lambda_{1,2} = p_0(1-p_0)+p_2(1-p_2) \pm \sqrt{\Big(p_0(1-p_0)+p_2(1-p_2)\Big)^2 - 4 \prod p_j};
	$$
	$$
	\mu_{1,2} = q_0(1-q_0)+q_2(1-q_2) \pm \sqrt{\Big(q_0(1-q_0)+q_2(1-q_2)\Big)^2 - 4 \prod q_j};
	$$
	$$
	\gamma_{1,2} = r_0(1-r_0)+r_2(1-r_2) \pm \sqrt{\Big(r_0(1-r_0)+r_2(1-r_2)\Big)^2 - 4 \prod r_j}.
	$$
\end{theorem}

\bigskip

\par\noindent\rm\textsc{\textbf{Proof of Theorems \ref{th:discretemult} and \ref{th:Euclideanmult}}}. By Equation (A16) in \cite{Boettcher2}, the asymptotic distribution of $n \, \widehat{\dMv}^2$ is the same as of the statistic $n \, \widetilde{\dMv}^2$ with
$$
\widetilde{\dMv}^2(\bX, \bY, \bZ) = \frac{1}{n^2}\sum_{i,j=1}^n \tilde{A}_{ij} \tilde{B}_{ij} \tilde{C}_{ij},
$$
where 
$$
\tilde{A}_{ij}:= -a_{ij} + \E[|X-X_j|] + \E[|X_i-X|] - E[|X-X'|].
$$
$X'$ denotes an IID copy of the random variable $X$. $\tilde{B}_{ij}, \tilde{C}_{ij}$ are defined analogously to $\tilde{B}_{ij}$. 

$\widetilde{\dMv}^2(\bX, \bY, \bZ)$ on the other hand is a degenerate V-statistic of order 2 and its distribution can be derived via classical results \citep{Serfling}. The closed-form expressions of the coefficients  $\lambda_1$, $\lambda_2$, $\mu_1$, $\mu_2$, $\gamma_1$, $\gamma_2$ can be in a totally analogous way as for distance covariance in the preceding proofs.
\qed.

\section{Theoretical notes on Chapter~\ref{ch4}}

\com{Think if I want to bold any matrices or other symbols in this section.}

\subsection{A lemma on locally most powerful tests}\label{lem:apA:ch4}
We first state a lemma that will be of use when proving results of Chapter~\ref{ch4}.

\setcitestyle{square}
\begin{lemma}[\cite{DJ}, Theorem 3] \label{lem:gt}
    Let $V:\mX \to \R$ be a stochastic process with $E[V(s)] \equiv 0$ and $\E[V(s)\,V(t)] = k(s,t)$ for some kernel $k$ (i.e., we assume it to be symmetric and positive definite).
	For $i=1,\ldots,n$, consider the univariate regression model
	$$
	y_i \sim \mathcal{N}(\mu + r_i, \sigma^2),
	$$
	where $r_i = \tau V(X_i)$, and $\mu,\tau \in \R$. Furthermore, we denote its likelihood by $g(r_i)$ . Then the locally most powerful test statistic for testing $$H_0:\tau^2 =0 \text{ against } H_1:\tau^2 > 0$$ in the marginal model
	\begin{equation} \label{eq:marglh}
		\overline{\ell}(\tau^2) = \E_{V(\cdot)|\tau^2} \Bigg[\prod_{i=1}^n g(r_i) \Bigg],
	\end{equation}
	is  (up to translation and multiplication by constants),
	\begin{equation} 
		\frac{1}{n^2} \sum_{i,j = 1}^n k(X_i,X_j) (y_i - \mu) (y_j - \mu).
	\end{equation}
\end{lemma}
\setcitestyle{round}%

\subsection{Proofs of theoretical results}

\par\noindent\rm\textsc{\textbf{Proof of Proposition~\ref{prop:kernel}}}. The kernel $k_b$ follows directly from taking $z_0 = 1$ in Equation~\eqref{eq:smindkernel} in the main body of the dissertation. For any $x,x'\in\spx\equiv\{0,1,2\}$, we have:
$$
k_b(x,x')=d_b(x,1)+d_b(x',1)-d_b(x,x').
$$
The evaluation of $k_b$ at each point of $\spx\times\spx$ concludes the proof.
\qed

\bigskip

\par\noindent\rm\textsc{\textbf{Proof of Proposition~\ref{prop:fm}}}. By direct evaluation, we see that the feature map $\boldsymbol{\Psi}(x) = (\psi_1(x), \psi_2(x))$ 
$$
\psi_1(x) = \sqrt{\frac{b}{2}} (-1_{\{x=0\}} + 1_{\{x=2\}}) , \quad \psi_2(x) = \sqrt{\frac{4-b}{2}}(1_{\{x=1\}} - 1)
$$
satisfies
$$
k_b(x,x') = \langle \boldsymbol{\Psi}(x), \boldsymbol{\Psi}(x') \rangle.
$$
It is easy to see that any translation of a feature map of $d_b$ is a feature map for $d_b$, which completes the proof.
\qed

\bigskip

\par\noindent\rm\textsc{\textbf{Proof of Theorem~\ref{th:dcovzero}}}. Applying Equations~\eqref{eq:hsic:feat:cov} and~\eqref{equiv:hsic:dc} in the main body of the dissertation, the generalised distance covariance $\V_{\rho_\mX, \rho_\mY}$ can be written as 
\begin{equation} \label{eq:dcovcov2}
	\V^2_{\rho_\mX, \rho_\mY}(X,Y) = \sum_{l=1}^{d_\mX} \sum_{m=1}^{d_\mY} \mbox{Cov}^2 (\bPhi_l^{\rho_\mX} (X), \bPhi_m^{\rho_\mY} (Y)),
\end{equation}
where $\bPhi^{\rho_\mX}$ and $\bPhi^{\rho_\mY}$ are feature maps of the (kernels induced by the) premetrics $\rho_\mX$ and $\rho_\mY$, respectively.

On the one hand, the premetric $\rho_\mY (y,y') = \tfrac12 |y-y'|^2$ induces the linear kernel $l(y,y,) = y y'$ with trivial feature map $\phi^{\rho_\mY}  = id_{\R}$. On the other hand, it is straightforward to see that a feature map of $d_b$ is given by:
$$
\phi_1^{d_b}(x) = \sqrt{\frac{b}{2}} (-1_{\{x=0\}} + 1_{\{x=2\}}) , \quad \phi^{d_b}_2(x) = \sqrt{\frac{4-b}{2}} 1_{\{x=1\}}.
$$
Inserting the these feature maps into Equation \eqref{eq:dcovcov2}, we obtain:
$$
\V_b(X,Y) = \frac{b}{2} (\mbox{Cov}(-1_{\{X=0\}} + 1_{\{X=2\}}, \, Y))^2 \, +  \, \frac{4-b}{2} (\mbox{Cov}(1_{\{X=1\}}, \, Y))^2.
$$
Expanding the covariances above and applying the law of total probability, we obtain
$$
\V_b(X,Y) =  \frac{b}{2} \, (- p_0 \, (\mu_0 - \mu_Y) +  p_2 \, (\mu_2 - \mu_Y))^2 + \frac{4-b}{2} \, (p_1 \, (\mu_1 - \mu_Y))^2,
$$
where $\mu_Y = E[Y]$.

If $\mu_0 = \mu_1 = \mu_2 = \mu_Y$, it  follows that $\V_b(X,Y) = 0$, completing the proof of the first part.

For the second part, assume that $\mu_i \neq \mu_j$ for some $i \neq j$. We first take care of the case $\mu_1 \neq \mu_Y$. Then 
$$
\frac{4-b}{2} \, (p_1 \, (\mu_1 - \mu_Y))^2 > 0,
$$
and hence $\V_b(X,Y)>0$. 

Now consider the remaining case $\mu_1 = \mu_Y$. In this case, either $\mu_0 < \mu_Y < \mu_2$ or 
$\mu_2 < \mu_Y < \mu_0$. For either possibility, it follows that
$$
\frac{b}{2} \, (- p_0 \, (\mu_0 - \mu_Y) +  p_2 \, (\mu_2 - \mu_Y))^2 > 0,
$$
and hence 
$\V_b(X,Y)>0$.
\qed

\bigskip

\par\noindent\rm\textsc{\textbf{Proof of Proposition~\ref{prop:nocondmeanfully}}}.

We consider first $b=0$. For any $(X,Y) \in \{0,1,2\} \times \R$, by the law of total probability,
$$
\mu_Y = p_0 \mu_0 + p_1 \mu_1 + p_2 \mu_2.
$$
Let $Y$ be such $\mu_0 = p_2, \mu_1 = 0$ and $\mu_2 = - p_0$.
Then $\mu_0 \neq \mu_2$, but $\mu_1 = \mu_Y$ and hence,
$$
\V_0(X,Y) = 2 \, (p_1 \, (\mu_1 - \mu_Y))^2 = 0.
$$

Now consider $b=4$.
Let $Y$ be such $\mu_0 = p_1 p_2$, $\mu_1 = - 2 p_0 p_2$ and $\mu_2 = p_0 p_1$. Then $\mu_0 \neq \mu_1$, however $\mu_Y = 0$ by the law of total probability and hence

$$
\V_4(X,Y) = 2 \, (- p_0 \, (\mu_0 - \mu_Y) +  p_2 \, (\mu_2 - \mu_Y))^2 = 0.
$$
\qed

\bigskip

\par\noindent\rm\textsc{\textbf{Proof of Theorem~\ref{testasy}}}. As in the proof of Theorem~\ref{th:dcovzero}, we consider the feature map corresponding to $d_b$ given by the vector notation,
$$
\phi_1 = \sqrt{\frac{b}{2}} \begin{pmatrix} -1 \\ 0 \\1 \end{pmatrix}, \, \phi_2 = \sqrt{\frac{4-b}{2}} \begin{pmatrix} 0 \\ 1 \\ 0 \end{pmatrix},
$$
i.e.
$$
\phi_1(x) = \sqrt{\frac{b}{2}} (-1_{\{x=0\}} + 1_{\{x=2\}}) , \quad \phi_2(x) = \sqrt{\frac{4-b}{2}} 1_{\{x=1\}}.
$$
We will further denote by $U_1 = \phi_1(X)$, $U_2 = \phi_2(X)$, $\bmu_U = (E[U_1], E[U_2])^t$ and $\mu_Y = E[Y]$. For a sample of size $n$, for each $i \in \{1,\ldots,n$\}, we define:
$$U_{1i} = \phi_1(X_i) \text{ and } U_{2i} = \phi_2(X_i)$$
We construct $\bU$ as the corresponding data matrix in $\mathbb{R}^{n \times 2}$:
$$
(\bU)_{kl} = U_{lk}, \quad k \in \{1,\ldots,n\},\, l \in \{1,2\}.
$$
Now, let $\bI$ denote the $n \times n$ identity matrix, $\bone = (1,\ldots,1)^t \in \mathbb{R}^n$ and $\bH=\frac{1}{n} \bone \bone^t$.
From \cite[Equation 3]{DJ}, it follows that $\widehat{\V}_b^2(\bX,\bY)$ can be written as
\begin{align*}
	n \, \V_b^2(\bX,\bY) = \frac{1}{n} \bY^t (\bI-\bH) \bU \bU^t (\bI-\bH) \bY = \bv^t \bv,
\end{align*}
with 
$$
\bv = \frac{1}{\sqrt{n}} \bU^t (\bI-\bH) \bY.
$$
Since $(\bI-\bH) \ba = \bzero$ for any vector with constant components $\ba=a\mathbf{1}$, $\bv$ can alternatively be written as:
\begin{align}
	\bv &= \frac{1}{\sqrt{n}} (\bU- \bone \bmu_U^t)^t (\bI-\bH) (\bY-\bone \mu_Y) \\
	&= \frac{1}{\sqrt{n}} (\bU- \bone \bmu_U^t)^t (\bY-\bone \mu_Y) - \frac{1}{\sqrt{n}} (\bU- 1 \bmu_U^t)^t \bH (\bY-\bone \bmu_Y) \label{eq:vdecomp} .
\end{align}
We first consider the second term in Equation \eqref{eq:vdecomp},
\begin{align*}
	\frac{1}{\sqrt{n}} (\bU- \bone \bmu_U^t)^t \bH (\bY-\bone \mu_Y) &= 
	\frac{1}{n^{3/2}} (\bU- \bone \bmu_U^t)^t \bone \bone^t (\bY-\bone \mu_Y) \\ &= \frac{1}{n^{3/2}} \begin{pmatrix}
		\sum_{i=1}^n \left(U_{1i} - E[U_1]\right) \\
		\sum_{i=1}^n \left(U_{2i} - E[U_2]\right)
	\end{pmatrix} \left(\sum_{i=1}^n \left(Y_{i} - E[Y]\right)\right). 
\end{align*}
Since $\frac{1}{\sqrt{n}} \begin{pmatrix}
	\sum_{i=1}^n \left(U_{1i} - E[U_1]\right) \\
	\sum_{i=1}^n \left(U_{2i} - E[U_2]\right)
\end{pmatrix} $ and $\frac{1}{\sqrt{n}} \sum_{i=1}^n \left(Y_{i} - E[Y]\right)$ both converge in probability to normal distributions due to the multivariate \emph{central limit theorem} (CLT), this term converges in probability to zero.
We now consider the first term in Equation \eqref{eq:vdecomp}. We now recall that, under the null hypothesis, $U$ and $Y$ are independent. Therefore the multivariate CLT yields the following asymptotic result (for $n \to \infty$):
$$
\frac{1}{\sqrt{n}} (\bU- \bone \bmu_U^t)^t (\bY-\bone \mu_Y) \stackrel{\mathcal{D}}{\longrightarrow} \Normal_2(\bzero,\bGamma),
$$
where 
$$
\bGamma = \begin{pmatrix} \sigma_Y^2 \, \Var(\phi_1(X)) & \sigma_Y^2\, \Cov(\phi_1(X), \phi_2(X)) \\ \sigma_Y^2 \, \Cov(\phi_1(X), \phi_2(X)) &  \sigma_Y^2 \, \Var(\phi_2(X)) \end{pmatrix}    
$$
We will assume in the following that $\bGamma$ has full rank; in cases where the rank of $\bGamma$ equals one, the proof can be carried out similarly. Then:
\begin{equation} \label{eq:wconv}
	\bw := \bGamma^{-1/2} \bv \stackrel{\mathcal{D}}{\longrightarrow} \Normal_2(\bzero, \bI_2)
\end{equation}
and 
\begin{align}
	n \V_b^2(\bX,\bY) &= \bw^t \,  \bGamma \, \bw  \nonumber \\ &= \sigma_Y^2 \, \bw^t \, \bQ \, \bLambda \, \bQ^t \, \bw, \label{eq:dcovfinalform}
\end{align}
where $\bQ$ is an orthogonal $2 \times 2$ matrix, $\bLambda$ is a diagonal matrix of the form 
$$
\bLambda = \begin{pmatrix} \lambda_1 & 0  \\ 0 &  \lambda_2 \end{pmatrix},    
$$
and $\lambda_1, \lambda_2$ are the eigenvalues of matrix: 
$$
\bK = \begin{pmatrix}  \Var(\phi_1(X)) &  \Cov(\phi_1(X), \phi_2(X)) \\ \Cov(\phi_1(X), \phi_2(X)) &   \Var(\phi_2(X)) \end{pmatrix} .
$$
Evaluation of the entries of $\bK$ is straightforward and yields the form given in the main body of the dissertation.
Since the standard normal is invariant under orthogonal transformations, combining Equations \eqref{eq:wconv} and \eqref{eq:dcovfinalform} yields
$$
n \widehat\V_b^2(\bX,\bY)  \stackrel{\mathcal{D}}{\longrightarrow} \sigma_Y^2 (\lambda_1 Q_1^2 + \lambda_2 Q_2^2),
$$
where $Q_1^2$ and $Q_2^2$ are chi-squared distributed, with one degree of freedom each. This completes the proof.
\qed

\bigskip

\par\noindent\rm\textsc{\textbf{Proof of Theorem~\ref{testfinite}}}. We use the same notation as in the proof of Theorem~\ref{testasy}. Hence
$\widehat{V}_b^2(\bX,\bY)$ can again be written as
\begin{align*}
	n \, \widehat\V_b^2(\bX,\bY) &= \frac{1}{n} \bY^t (\bI_n-\bH) \bU \bU^t (\bI_n-\bH) \bY \\ &= \frac{1}{n}  (\bY-\bone \mu_Y)^t (\bI_n-\bH) \bU \bU^t (\bI_n-\bH) (\bY-\bone \mu_Y), 
\end{align*}
where the second line follows from $\bone \mu_Y = \bH \bone \mu_Y$.
Consequently,
$$
\frac{n \, \widehat{\V}_b^2(\bX,\bY)}{\widehat{\sigma}_Y^2}  =  \frac{(\bY-\bone \mu_Y)^t (\bI_n-\bH) \bU \bU^t (\bI_n-\bH) (\bY-\bone \mu_Y)}{(\bY-\bone \mu_Y)^t (\bI_n-\bH) I_n (\bI_n-\bH) (\bY-\bone \mu_Y)}.
$$
Hence
$$
\left\{ \frac{n \, \widehat{\V}_b^2(\bX,\bY)}{\widehat{\sigma}_Y^2} \geq k \right\}
$$
is obviously equivalent to
$$
\left\lbrace  (\bY-\bone \mu_Y)^t (\bI_n-\bH) \frac{1}{n} (\bU \bU^t - k I_n) (\bI_n-\bH) (\bY-\bone \mu_Y) \geq 0 \right\rbrace   .
$$
Now consider the following matrix:
\begin{align*}
	\frac{1}{n} (\bI_n-\bH) (\bU \bU^t  - k \bI_n) (\bI_n-\bH) =  \frac{1}{n} (\bI_n-\bH) \bU \bU^t   (\bI_n-\bH) - \frac{k}{n} \, (I_n - H).
\end{align*}
The constant vector $\bo_n = \big(\sqrt{\frac{1}{n}}, \ldots,\sqrt{\frac{1}{n}}\big)^t$ is an eigenvector to eigenvalue $0$ for both matrices $(\bI_n-\bH) \bU \bU^t   (\bI_n-\bH)$ and $(I_n - H)$. Augmenting $\bo_n$ to an orthogonal basis (represented by matrix $\bO$) of $(\bI_n-\bH) \bU \bU^t   (\bI_n-\bH)$, we obtain:
\begin{align*}
	\frac{1}{n}  (\bI_n-\bH) \bU \bU^t   (\bI_n-\bH) - \frac{k}{n} \, (\bI_n - \bH) 
	&= \bO \widehat{\bLambda} \bO^t -  \bO  \bD_{n-1} \bO^t,
\end{align*}
where $\bD_{n-1}$ is a diagonal matrix with diagonal $(k/n,k/n,\ldots,k/n,0)$. Since the standard normal distribution is invariant under orthogonal transformations, we obtain that
\begin{align*}
	&(\bY-\bone \mu_Y)^t (\bI_n-\bH) (\bU \bU^t - k I_n) (\bI_n-\bH) (\bY-\bone \mu_Y) \\ &\stackrel{\mathcal{D}}{=} (\widehat{\lambda}_1 - k/n) Q_1^2 + (\widehat{\lambda}_2 - k/n) Q_2^2 - k/n Q_3^2 - \cdots - k/n Q_{n-1}^2,
\end{align*}
where $Q_1^2, \ldots, Q_{n-1}^2$ are chi-squared with one degree of freedom and $\widehat{\lambda}_1$ and $\widehat{\lambda}_2$ are the eigenvalues of $\widehat{K} = \frac{1}{n}  \bU^t (\bI_n-\bH) \bU $. The evaluation of the entries in $\widehat{K}$ is straightforward and completes the proof.
\qed

\bigskip

\par\noindent\rm\textsc{\textbf{Proof of Proposition~\ref{prop:mM}}}. 
We start with the case where  $\widehat{\lambda}_2- \frac{k}{n} > 0$ showing
$$
p^* \leq \Prob \left( \frac{n \, \widehat{\V}_b^2}{\widehat{\sigma}_Y^2} \geq k \right) = \Prob\left(\frac{(\widehat{\lambda}_1- \frac{k}{n})Q_1^2+(\widehat{\lambda}_2- \frac{k}{n})Q_2^2}{\frac{1}{n}(Q_3^2+\cdots+Q_{n-1}^2)} \geq k \right),
$$ 
separately for the tree terms over which the minimum is taken.
For the first term, we need to show that $V \leq_{st} U$, where
$$
U = \frac{(\widehat{\lambda}_1- \frac{k}{n})Q_1^2+(\widehat{\lambda}_2- \frac{k}{n})Q_2^2}{\frac{1}{n-3}(Q_3^2+\cdots+Q_{n-1}^2)}, \quad V= (\widehat{\lambda}_1- \frac{k}{n})Q_1^2+(\widehat{\lambda}_2- \frac{k}{n})Q_2^2.
$$
Also define,
$$
W = \frac{(\widehat{\lambda}_1- \frac{k}{n})Q_1^2+(\widehat{\lambda}_2- \frac{k}{n})Q_2^2}{\frac{1}{m-3}(Q_3^2+\cdots+X_{m-1}^2)},
$$
for some $m > n$ (where all $X_{j}^2$ are chi-squared distributed random variables).

Let $H_U$, $H_V$ and $H_W$ denote the cumulative distribution functions of the random variables $U$, $V$ and $W$, respectively and let $h_U$, $h_V$ and $h_w$ denote their corresponding densities. Using the series representation in Equation 97 of \citet{Kotz}, it follows that the family $$\{H_V(ax), a > 0\}$$ satisfies the monotone likelihood ratio property. Applying Proposition 2 of \citet{Rivest} now yields that $W$ is smaller than $U$ in the star-shaped order (cf. also Example 1 in \cite{Rivest}).

Using Theorem 1 by \citet{Dunkl} it is straightforward to show that
$$
h_U(0) \leq h_W(0).
$$
Applying \cite[Theorem 3]{Jeon} shows that  
$
W \leq_{st} U
$
from which $V \leq_{st} U$ follows with a simple limit argument.

For the second term, we first observe that
$$
\frac{(\widehat{\lambda}_1- \frac{k}{n})Q_1^2}{\frac{1}{n-3}(Q_3^2+\cdots+Q_{n-1}^2)} \leq \frac{(\widehat{\lambda}_1- \frac{k}{n})Q_1^2+(\widehat{\lambda}_2- \frac{k}{n})Q_2^2}{\frac{1}{n-3}(Q_3^2+\cdots+Q_{n-1}^2)} 
$$
and hence
$$\Prob \left( \frac{n \, \widehat{\V}_b^2}{\widehat{\sigma}_Y^2} \leq k \right) \leq G_{F(1,n-3)} \Big(\frac{k (n-3)}{\widehat{\lambda}_1 n - k} \Big).
$$
The inequality for the third term is a direct consequence of Equation (32) in \citet{Dunkl}.

For $p^{**}$, define the random variable
$$
Q = \frac{(\widehat{\lambda}_1 + \widehat{\lambda}_2- \frac{ 2\, k}{n})Q_1^2}{\frac{1}{n-2}(Q_3^2+\cdots+Q_{n-1}^2)}.
$$

By \cite{Extremal}, the denominators of $Q$ and $U$ satisfy

$$
\Prob\left( \Big(\widehat{\lambda}_1- \frac{k}{n})Q_1^2+(\widehat{\lambda}_2- \frac{k}{n})Q_2^2\Big) \geq x\right) \leq \Prob\left(\Big(\widehat{\lambda}_1 + \widehat{\lambda}_2- \frac{ 2\, k}{n}\Big) Q_1^2 \geq x\right),
$$
whenever one of the expressions is smaller than $0.215$. It follows by a simple combinatorial argument that, for all $x$,
$$
\Prob (U \geq x) \leq \frac{1}{0.215} \Prob (Q \geq x).
$$

Finally consider the case $\widehat{\lambda}_2- \frac{ k}{n} \leq 0$. Then
\begin{align*}
	\big(\widehat{\lambda}_1- \frac{k}{n} \big) \, Q_1^2  - \frac{k}{n} \, Q_2^2 - \frac{k}{n} Q_3^2 - &\cdots - \frac{k}{n} Q_{n-1}^2 \leq T_n \\ \leq & \big(\widehat{\lambda}_1- \frac{k}{n} \big) \, Q_1^2   - \frac{k}{n} Q_3^2 - \cdots - \frac{k}{n} Q_{n-1}^2.
\end{align*}
The proof now follows by elementary transformations.
\qed

\bigskip

\par\noindent\rm\textsc{\textbf{Proof of Theorem~\ref{th:lomopo}}}. For $t \in \{0,1,2\}$, let $V(t) = \sum_{j=1}^r B_j \phi_j (t)$. Since $\boldsymbol{\phi}(\cdot)$ is a feature map of $k_b$, we obtain 
$$
\E[V(s) V(t)]  = \sum_{j=1}^r \E [B_j^2] \, \phi_j(s) \, \phi_j(t) = k_b (s,t).
$$
The Theorem now follows from Lemma~\ref{lem:gt}.

\qed.

\bigskip

\par\noindent\rm\textsc{\textbf{Proof of Corollary~\ref{cor:mix}}}. For $j \in \{1,\ldots,r\}$, define $D_j = A \, 1_{\{U=j\}}/c_j$. Then $\E[D_j] = 0$, 

$$
\E[D_j^2] = \frac{\E[A^2]  P(U=j)}{c_j^2} = \frac{\E[A^2]}{\sum_{k=1}^n c_k^2}
$$
and, for $i \neq j$

$$
\E[D_i D_j] = \frac{\E[A^2]  P(U=i, U = j)}{c_i c_j} = 0.
$$

The result now follows from applying Theorem~\ref{th:lomopo} with $B_j = D_j / \sqrt{\frac{\E[A^2]}{\sum_{k=1}^n c_k^2}}$ and $\tau$ replaced by $\tau \, \sqrt{\frac{\E[A^2]}{\sum_{k=1}^n c_k^2}}$.

\qed

\bigskip

\par\noindent\rm\textsc{\textbf{Proof of Theorem~\ref{th:lomopo2}}}.

Define the stochastic process $V: \{0,1,2\} \to \R$ by,
$$
V(0) = 0, \quad V(1) =  B_1, \quad V(2) = (B_1 + B_2).
$$
Then $\E[V(t)] = 0$, $\E[V(0)^2] = \E[V(0) V(1)] = \E[V(0) V(2)] = 0$, $\E[V(1)^2] = 1$,
$$
\E[V(1) V(2)] = \E[B_1^2] + \E[B_1 B_2] = \frac{b}{2},
$$
and
$$
\E[V(2)^2 ] = \E[B_1^2] + 2\, \E[B_1 B_2] + E[B_2^2] = b.
$$
Moreover, by choosing $z_0 = 0$ in Equation~\eqref{eq:smindkernel}, we see that an alternative kernel induced by $d_b$ is $\widetilde{k}_b$ with
\begin{align*}
	&\widetilde{k}_b(0,0) = \widetilde{k}_b(0,1) = \widetilde{k}_b(0,2) = 0, \\ &\widetilde{k}_b(1,1) = 2, \quad\widetilde{k}_b(1,2) = b , \quad \widetilde{k}_b(2,2) = 2b.
\end{align*}
Hence $E[V(s) V(t)] = \frac{1}{2} \widetilde{k}_b(s,t)$.
The result now follows by applying Lemma~\ref{lem:gt}.
\qed

\bigskip

\par\noindent\rm\textsc{\textbf{Proof of Corollary~\ref{cor:beta}}}.

Although the proof is more instructive by constructing gamma-distributed variables and using Theorem~\ref{th:lomopo2}, it leads to some technicalities. To avoid these, we prove Corollary~\ref{cor:beta} directly, by first defining:
$$
V(0) = 0 ; \quad V(1) = H A ; \quad V(2) = A.
$$
It is easy to see that $V(0,0) = V(0,1) = V(0,2) = 0$. Moreover, by inserting the known first and second moments of the beta distribution, we obtain:
$$
\E[ V(1) V(1)] = \frac{1}{b} = \frac{1}{2 b } \widetilde{k}_b(1,1),
$$
$$
\E[ V(1) V(2)] = \frac{1}{2} = \frac{1}{2 b } \widetilde{k}_b(1,2),
$$
$$
\E[ V(2) V(2)] = 1 = \frac{1}{2 b } \widetilde{k}_b(2,2),
$$
where $\widetilde{k}_b$ is defined in the proof of Theorem~\ref{th:lomopo2}.
Applying Lemma~\ref{lem:gt} completes the proof.
\qed

\bigskip

\par\noindent\rm\textsc{\textbf{Proof of Theorem~\ref{th:covasy}}}. We will use the same notation as in the proof of Theorem~\ref{testasy}.
Moreover, let $\bH_Z$ denote the projection matrix  
$$
\bH_Z =  {\bZ} ({\bZ}^t {\bZ})^{-1} {\bZ}^t .
$$
Then,  $\widehat{\V}_b^2(\bX,\bY; \bZ)$ can be written as
\begin{align*}
	n \, \V_b^2(\bX,\bY; \bZ) &= \frac{1}{n} \bY^t (\bI-\bH_Z) \bU \bU^t (\bI-\bH_Z) \bY \\ &= \bv^t \bv,
\end{align*}
with 
$$
\bv = \frac{1}{\sqrt{n}} \bU^t (\bI-\bH_Z) \bY.
$$
Since $(\bI-\bH_Z) {\bZ} = 0$ , $\bv$ can alternatively be written as,
\begin{align}
	\bv &= \frac{1}{\sqrt{n}} (\bU- {\bZ} \balpha)^t (\bI-\bH_Z) (\bY- {\bZ} \bgamma) \\
	&= \frac{1}{\sqrt{n}} (\bU- {\bZ} \balpha)^t (\bY- {\bZ} \bgamma) - \frac{1}{\sqrt{n}} (\bU- {\bZ} \balpha)^t H_Z (\bY-{\bZ} \bgamma) \label{eq:vdecomp2} ,
\end{align}
where we denote $\balpha=\E[\bZ \bZ^t]^{-1} E[\bZ \bU]$.

We first consider the second term in Equation \eqref{eq:vdecomp2},
\begin{align*}
	\frac{1}{\sqrt{n}} (\bU- {\bZ} \balpha)^t H_Z (\bY- {\bZ} \bgamma) &= 
	\frac{1}{n^{3/2}} (\bU- {\bZ} \balpha)^t {\bZ} \,  (n^{-1} {\bZ}^t {\bZ})^{-1} {\bZ}^t (\bY- {\bZ} \bgamma). \\
\end{align*}
With similar arguments as in the proof of Theorem~\ref{testasy}, it follows that $\vecop((\bU- {\bZ} \balpha)^t {\bZ})$ and ${\bZ}^t (\bY- {\bZ} \bgamma)$ converge to normal distributions with mean $0$, whereas $n^{-1} {\bZ}^t {\bZ}$ converges to the (augmented) covariance matrix of $Z$. Hence, this term converges to $0$.

We now consider the first term in Equation \eqref{eq:vdecomp}. Applying the multivariate CLT and remembering that $U$ and $Y$ are independent under the null hypothesis, we observe that, for $n \to \infty$
$$
\frac{1}{\sqrt{n}} (\bU- {\bZ} \balpha)^t (\bY- {\bZ} \bgamma)  \stackrel{\mathcal{D}}{\longrightarrow} \Normal(\bzero,\bGamma),
$$
where 
$$
\bGamma =\sigma_\varepsilon^2 \bK 
$$

The rest of the proof is analogous to that of Theorem~\ref{testasy}.
\qed

\bigskip

\par\noindent\rm\textsc{\textbf{Proof of Theorem~\ref{th:covfinite}}}. We use the same notation as in the proofs of Theorems~\ref{testasy},~\ref{testfinite} and~\ref{th:covasy}. We first write:
\begin{align*}
	n \, \V_b^2(\bX,\bY;\bZ) &= \frac{1}{n} \bY^t (\bI_n-\bH_Z) \bU \bU^t (\bI_n-\bH_Z) \bY \\ &= \frac{1}{n}  (\bY-{\bZ} \bgamma)^t (\bI_n-\bH_Z) \bU \bU^t (\bI_n-\bH_Z) (\bY- {\bZ} \bgamma).
\end{align*}
Consequently,
$$
\frac{n \, \widehat{\V}_b^2(\bX,\bY;\bZ)}{\widehat{\sigma}_\varepsilon^2}  =  \frac{(\bY-{\bZ} \bgamma)^t (\bI_n-\bH_Z) \bU \bU^t (\bI_n-\bH_Z) (\bY- {\bZ} \bgamma)}{(\bY-{\bZ} \bgamma)^t (\bI_n-\bH_Z) I_n (\bI_n-\bH_Z) (\bY- {\bZ} \bgamma)}.
$$
Hence
$$
\left\{ \frac{n \, \widehat{\V}_b^2(\bX,\bY;\bZ)}{\widehat{\sigma}_\varepsilon^2} \geq k \right\}
$$
is obviously equivalent to
$$
\{ (\bY-{\bZ} \bgamma)^t (\bI_n-\bH_Z) \frac{1}{n} (\bU \bU^t - k I_n) (\bI_n-\bH_Z)(\bY-{\bZ} \bgamma)\geq 0\}.
$$
Now consider matrix
\begin{align*}
	\frac{1}{n} (\bI_n-\bH_Z) (\bU \bU^t  - k I_n) (\bI_n-\bH_Z) =  \frac{1}{n} (\bI_n-\bH_Z) \bU \bU^t   (\bI_n-\bH_Z) - \frac{k}{n} \, (\bI_n - \bH_Z).
\end{align*}
Take $p+1$ orthogonal eigenvectors $\bo_{n-p-1},\ldots \bo_n$ to eigenvalue $0$ of $(I_n - H_Z)$. Then $\bo_{n-p-1},\ldots \bo_n$  are obviously also eigenvectors to eigenvalue $0$ of matrix $(\bI_n-\bH_Z) \bU \bU^t   (\bI_n-\bH)$. Augmenting $\bo_{n-p-1},\ldots \bo_n$ to an orthogonal basis (represented by matrix $\bO$) of $(\bI_n-\bH_Z) \bU \bU^t   (\bI_n-\bH_Z)$, we obtain,
\begin{align*}
	\frac{1}{n}  (\bI_n-\bH_Z) \bU \bU^t   (\bI_n-\bH_Z) - \frac{k}{n} \, (\bI_n - \bH_Z) 
	&= \bO \widehat{\Lambda} \bO^t -  \bO  \bD_{n-p-1} \bO^t,
\end{align*}
where $\bD_{n-p-1}$ is a diagonal matrix with $n-p-1$ times $k/n$ and $p+1$ zeros in the diagonal and $\widehat{\Lambda}$ is a diagonal matrix with diagonal $(\widehat{\lambda_1},\widehat{\lambda_2},0,\ldots,0)$. Since the standard normal distribution is invariant under orthogonal transformations we obtain that
\begin{align*}
	&(\bY-{\bZ} \bgamma)^t (\bI_n-\bH_Z) \frac{1}{n} (\bU \bU^t - k I_n) (\bI_n-\bH_Z)(\bY-{\bZ} \bgamma) \\ &\stackrel{\mathcal{D}}{=} (\widehat{\lambda}_1 - k/n) Q_1^2 + (\widehat{\lambda}_2 - k/n) Q_2^2 - k/n Q_3^2 - \cdots - k/n Q_{n-p-1}^2.
\end{align*}
\qed

\subsection{Comments to Theorem~\ref{th:lomopo2} and extension of Corollary~\ref{cor:beta}}

For constructing bivariate random vectors with zero mean for which the marginals have equal or opposite sign, we note that for any pair of non-negative random variables $G =(G_1,G_2)^t$ we can use a random variable $A$ with $P(A=1) = P(A=-1) = \tfrac12$, independent of $G$ to construct mean zero random variables
$$
B = \begin{pmatrix} B_1 \\ B_2 \end{pmatrix} = \begin{pmatrix} A G_1 \\ A G_2 \end{pmatrix} \quad 
\widetilde{B} = \begin{pmatrix} \widetilde{B}_1 \\ \widetilde{B}_2 \end{pmatrix} = \begin{pmatrix} A G_1 \\ - A G_2 \end{pmatrix}
$$
Then the marginals of $B$ have equal sign, the ones of $\widetilde{B}$ have opposing signs, and 
$$
\Cor(B_1,B_2) = \frac{\E[A_1 A_2]}{\sqrt{\E[A_1^2] \E[A_2^2]}} \quad \quad \Cor(\widetilde{B_1},\widetilde{B_2}) = -\frac{- \E[A_1 A_2]}{\sqrt{\E[A_1^2] \E[A_2^2]}}
$$
Choosing $G_1$ and $G_2$  gamma-distributed with equal rate parameter and shape parameter $\frac{a-2}{4-a}$ ($a \in ]2,4[$) leads to (cf. Corollary~\ref{cor:beta}),
$$
\Cor(B_1,B_2) = \frac{a}{2}-1
$$
and consequently,
$$
\Cor(\widetilde{B_1},\widetilde{B_2}) = 1- \frac{a}{2}.
$$
Now let $b \in ]0,2[$ and choose $a= 4-b$. Then
$$
\Cor(\widetilde{B_1},\widetilde{B_2}) = \frac{b}{2} - 1.
$$
One directly obtains the following corollary of Theorem~\ref{th:lomopo2}, which extends Corollary~\ref{cor:beta} for $b \in ]0,2[$.

\begin{corollary} \label{cor:beta:ext}
	Consider the distance $d_b$ with $b \in ]0,2[$ and
	assume the model
	\begin{align*}
		Y_i = \begin{cases}  \mu_Y + \varepsilon, &\text{ if $x_i =0$}, \\
			\mu_Y + \tau G_1 A + \varepsilon,  &\text{ if $x_i =1$}\\
			\mu_Y + \tau (G_1-G_2) A + \varepsilon	&\text{ if $x_i =2$,}
		\end{cases}
	\end{align*}
	where $\mu_Y$ is known, $\tau \in \R$, $\varepsilon \sim \mathcal{N}(0,\sigma^2)$ and $G =(G_1, G_2)^t$ are independently gamma-distributed with shape parameters  $\frac{2-b}{b}$ and equal rate parameters; $A$ is a random variable, independent of $G$ with $\E[A]=0$ and $\E[A^2]=1$ (e.g. $P(A = 1) = P(A = - 1) = \frac{1}{2}$).
	Then the locally most powerful test for testing 
	$
	H_0 : \tau^2 = 0 \text{ against } H_1 : \tau^2 > 0
	$ is given by Equation~\eqref{eq:HSICstat} in the main body of the dissertation.
\end{corollary}

Hence, for $b \in ]0,2[$, the distance covariance test can be interpreted as the locally most powerful one in the case where the heterozygous effect is distributed as $\frac{G_1}{G_1 - G_2}$, where $G_1,G_2$ are independently gamma-distributed with shape parameters  $\frac{2-b}{b}$.

\section{Theoretical notes on Chapter~\ref{ch6}}\label{ap:proofs:ch6}

\subsection{Proof of Theorem~\ref{th:indep}}\label{proof:dist}

We will firstly show that the distance covariance test statistic has the compact form similar to Pearson's that we stated in the main body of the dissertation, to then prove the asymptotic null distribution.

We will investigate the terms $\widehat{T}_1, \widehat{T}_2, \widehat{T}_3$ one by one, to then see how $\widehat V$ can be written as a simple expression.
\begin{align*}
	\widehat{T}_1 &= \frac{1}{n^2} \sum_{l,m =1}^n d(X_l,X_m) \, d(Y_l,Y_m) \\
	&= \frac{1}{n^2} \sum_{l,m =1}^n 1_{\{X_l \neq X_m, Y_l \neq Y_m\}} \\
	&= \frac{1}{n^2} \sum_{l,m =1}^n \left( 1 - 1_{ \{X_l = X_m\} } - 1_{ \{Y_l = Y_m\} } + 1_{\{X_l = X_m, Y_l = Y_m\}}\right)  \\
	&= 1 - \frac{1}{n^2} \sum_{i=1}^{I} n_{i \cdot}^2 - \frac{1}{n^2} \sum_{j=1}^{J} n_{\cdot j}^2 +  \frac{1}{n^2}  \sum_{i=1}^{I} \sum_{j=1}^{J} n_{ij}^2.
\end{align*}  
For $\widehat{T}_2$, we first observe that 
$$
\sum_{m =1}^n d(X_l,X_m)
= \sum_{m =1}^n \left( 1 - 1_{ \{X_l = X_m\} } \right) 
= n - n_{X_l \cdot}
$$
and hence 
\begin{align*}
	\widehat{T}_2 &= \frac{1}{n^3} \sum_{l=1}^n (n- n_{X_l \cdot}) (n- n_{\cdot Y_l}) \\
	&= \frac{1}{n^3} \sum_{i=1}^{I} \sum_{j=1}^{J} (n - n_{i \cdot}) (n - n_{\cdot j }) n_{ij} \\
	&= 1 - \frac{1}{n^2} \sum_{i=1}^{I} n_{i \cdot}^2 - \frac{1}{n^2} \sum_{j=1}^{J} n_{\cdot j}^2 +   \frac{1}{n^3} \sum_{i=1}^{I} \sum_{j=1}^{J} n_{i \cdot} n_{\cdot j} n_{ij}.
\end{align*}	
Finally,
$$
\sum_{l,m =1}^n d(X_l,X_m) = \sum_{l =1}^n \left( n - n_{X_l \cdot} \right)   = n^2 - \sum_{i=1}^I n_{i \cdot}^2
$$
and hence
\begin{align*}
	\widehat{T}_3 &= \frac{1}{n^4} \left( n^2 - \sum_{i=1}^I n_{i \cdot}^2\right)  \left( n^2 - \sum_{j=1}^J n_{\cdot j}^2\right) \\
	&= 1 - \frac{1}{n^2} \sum_{i=1}^{I} n_{i \cdot}^2 - \frac{1}{n^2} \sum_{j=1}^{J} n_{\cdot j}^2 +  \frac{1}{n^4} \sum_{i=1}^{I} \sum_{j=1}^{J} n_{i \cdot}^2 n_{\cdot j}^2.
\end{align*}

When adding up the terms to obtain $\widehat{V}$, the terms $1$, $\frac{1}{n^2} \sum_{i=1}^{I} n_{i \cdot}^2$ and $\frac{1}{n^2} \sum_{j=1}^{J} n_{\cdot j}^2$ all cancel out and we obtain
\begin{align*}
	\widehat{V}  &= \frac{1}{n^2}  \sum_{i=1}^{I} \sum_{j=1}^{J} n_{ij}^2 - \frac{2}{n^3} \sum_{i=1}^{I} \sum_{j=1}^{J} n_{i \cdot} n_{\cdot j} n_{ij} + \frac{1}{n^4} \sum_{i=1}^{I} \sum_{j=1}^{J} n_{i \cdot}^2 n_{\cdot j}^2 \\
	&= \frac{1}{n^2}  \sum_{i=1}^{I} \sum_{j=1}^{J} \left( n_{ij} - \frac{1}{n} n_{i \cdot} n_{\cdot j}\right) ^2 \\
	&= \frac{1}{n^2}  \sum_{i=1}^{I} \sum_{j=1}^{J}  (n_{ij} - n^*_{ij})^2,
\end{align*} 	
which is what we wanted to achieve.

Now, to start the way towards the asymptotic null distribution, let $\spz$ be either $\{1,\ldots,I\}$ or $\{1,\ldots,J\}$. Then the discrete metric
\[
d(z,z')=1-\delta_{zz'},
\]
is dual to the following kernel in the sense of \citet{Sejdinovic}:
\[
k(z,z')=\delta_{zz'},
\]
which is known as the \emph{discrete kernel}. Then clearly one can take the dummy function on each of $\spx$ and $\spy$ as a feature map of the corresponding kernel/distance. We will denote them by $\phi:\spx\longrightarrow\R^I$ and $\psi:\spy\longrightarrow\R^J$, where:
$$
\phi_i(X) = 1_{\{X=i\}} , \quad \psi_j(Y) = 1_{\{Y=j\}}.
$$
Now we construct matrices $\bU=(U_{ij})_{n\times I}$ and $\bV=(V_{ij})_{n\times J}$ by transforming the $X$ and $Y$ samples with the feature maps: 
$$
U_{ki} = \phi_i(X_k) \quad V_{kj} = \psi_j(Y_k) .
$$
Note that each of row of the previous matrices contains an observation of $$\phi(X)\sim\Multib(\bq) \text{ or } \psi(Y)\sim\Multib(\br)$$ (respectively). Therefore:
$$
\bone\transp\bU\sim\Multin_I(n,\bq)
$$
$$
\bone\transp\bV\sim\Multin_J(n,\br)
$$
Now, applying Equation (3) in \citet{DJ} to our feature maps, we get:
$$
n \, \dCovh_{\text{discrete}}^2(X,Y)=\frac{1}{n}\sum_{i=1}^I\sum_{j=1}^J[\bU\transp(\bI_n-\bH)\bV]^2_{ij} ,
$$
where $\bI_n$ is the $n\times n$ identity matrix and $\bH=\frac{1}{n}\bone\bone\transp$ has constant entries equal to $\frac{1}{n}$. If we now define $\bC\equiv(C_{ij})_{I\times J}:=\frac{1}{\sqrt n}\bU\transp(\bI_n-\bH)\bV$, we can compactly write our test statistic as a trace:
$$
n \, \dCovh_{\text{discrete}}^2(X,Y)=\tr[\bC\bC\transp]=\tr[\bC\transp\bC]=\sum_{i=1}^I\sum_{j=1}^JC_{ij}^2 .
$$
Expressing an empirical distance covariance as a trace of a matrix product, as we did above, is not unusual \citep{TEOD} and indeed it is a very computationally efficient way of evaluating it. Nonetheless, for continuing the proof we are going to write:
$$
n \, \dCovh_{\text{discrete}}^2(X,Y)=\bc\transp \bc ;
$$
where $\bc:=\vecop(\bC)\in\R^{IJ}$ is the vectorisation of matrix $\bC$ (i.e., its image by the linear isomorphism $\R^{I\times J}\cong \R^{IJ}$).

If one adds a vector with constant components $\ba=a\bone$ to a column or row of a matrix, the result of centring it with matrix $\bI-\bH$ will be the same. Therefore, we can expand $\bC$ as:
$$
\bC=
\frac{1}{\sqrt{n}} (\bU\transp-\bq\bone\transp)(\bI-\bH)(\bV-\bone\br\transp)=
$$
$$
=\frac{1}{\sqrt{n}} (\bU\transp-\bq\bone\transp)(\bV-\bone\br\transp)-
\frac{1}{n^{3/2}}(\bU\transp-\bq\bone\transp)\bone\bone\transp(\bV-\bone\br\transp).
$$
The second term of the previous sum is:
$$
\bD:=\frac{1}{\sqrt n} \:
\left[ \frac{1}{\sqrt n}\begin{pmatrix}
	\sum_{m=1}^n \left( \phi_1(X_m)-q_1\right)  \\
	\ldots \\
	\sum_{m=1}^n \left( \phi_I(X_m)-q_I\right)
\end{pmatrix}\right] \:
\left[ \frac{1}{\sqrt n}
\left( 
\sum_{m=1}^n \left( \psi_1(Y_m)-r_1\right),
\ldots,
\sum_{m=1}^n \left( \psi_J(Y_m)-q_J\right)
\right) 
\right]
$$
By the central limit theorem, it is easy to see that each entry $D_{ij}$ of $\bD$ converges in probability to zero, owing to the fact that:
$$
\frac{1}{\sqrt n}\sum_{m=1}^{n}\left( \phi(X_m)-\bq\right) \distrilim\Normal_I(\bzero,\bA)
$$
$$
\frac{1}{\sqrt n}\sum_{m=1}^{n}\left( \psi(Y_m)-\br\right) \distrilim\Normal_J(\bzero,\bB) .
$$
Hence, $\vecop(\bD)$ converges in probability to the $IJ-$dimensional null vector, and the limit in distribution of $\bc$ will be that of the vectorisation of:
$$\bE:=\frac{1}{\sqrt{n}} (\bU\transp-\bq\bone\transp)(\bV-\bone\br\transp).$$
We can write the $(i,j)$th entry of the previous matrix as: $E_{ij}=\frac{1}{\sqrt n}\sum_{m=1}^{n} G_{mij}$, where
$$
G_{mij}=\left( \phi_i(X_m)-q_i\right) \left( \psi_j(Y_m)-r_j\right).
$$
Now, we see that we can apply the CLT to
$$
\vecop(\bE)=\frac{1}{\sqrt{n}}\sum_{m=1}^{n}\vecop(\bG_m).
$$
For a fixed $m\in\{1,\ldots,n\}$, let us see how the first and second moments of $\vecop(\bG)\equiv\vecop(\bG_m)$ look like. For $i\in\{1,\ldots,IJ\}$, the $i$th component of $\E[\vecop(\bG)]$ vanishes under the null hypothesis (i.e., independence of $X$ and $Y$):
$$
\E[G_{(i-1)\%I+1,\ceil{i/I}}]
=\E[\left( \phi_{(i-1)\%I+1}(X)-q_{(i-1)\%I+1}\right)  ]
\E[\left( \psi_{\ceil{i/I}}(Y)-r_{\ceil{i/I}}\right)  ]=0\cdot0=0.
$$
We have used the notation $\%$ to indicate the remainder of an integer division, and $\ceil{\cdot}$ for the ceiling.

The $(i,j)$th entry of the variance-covariance matrix of $\vecop(\bG)$ is:
$$
\Cov(G_{(i-1)\%I+1,\ceil{i/I}},G_{(j-1)\%J+1,\ceil{j/J}})
=
$$
$$
=
\E[\left( \phi_{(i-1)\%I+1}(X)-q_{(i-1)\%I+1}\right) 
\left( \phi_{(j-1)\%J+1}(X)-q_{(j-1)\%J+1}\right)
]$$
$$\times
\E[\left( \psi_{\ceil{i/I}}(Y)-r_{\ceil{i/I}}\right)
\left( \psi_{\ceil{j/J}}(Y)-r_{\ceil{j/J}}\right)
]=
$$
$$
=
a_{(i-1)\%I+1,(j-1)\%J+1}\,b_{\ceil{i/I},\ceil{j/J}}
=
[\bB\tensor\bA]_{ij}\; ,
$$
with $\tensor$ denoting the Kronecker product.

Applying the central limit theorem once more, we get the limiting distribution of $\bc$:
$$
\bc\distrilim
\Normal_{IJ}(\bzero,\bGamma);\quad \bGamma=\bB\tensor\bA
$$
Now, one would be tempted to take $\bGamma$ to the $-\frac{1}{2}$ and standardise $\bc$, but the reality is that $\bGamma$ is never of full rank because $\bA$ and $\bB$ never are. So we are going to first take some sort of matrix root and then consider its inverse, instead of the other way round.

Let us write $\bGamma=\bM \bM\transp$, where $\bM\in\R^{IJ\times r}$ has rank $r:=\rank(\bGamma)\leq IJ$. If $\bM^+$ denotes the Moore--Penrose (pseudo)inverse of $\bM$, we can easily conclude that:
$$
\bw:=\bM^{+}\bc\distrilim
\Normal_{r}(\bzero,\bI)
$$
by taking into account that
$$\bM^+\bGamma(\bM^+)\transp=\bM^+\bM(\bM^+\bM)\transp=\bM^+\bM\bM^+\bM=\bM^+\bM=\bI_r,$$
with the last equality owing to the fact of $\bM$ having full column rank.

We can finally go back to the expression of the empirical distance covariance:
$$
n \, \dCovh_{\text{discrete}}^2(X,Y)=
\bw\transp\bGamma \bw .
$$
As $\bGamma$ is symmetric, we can diagonalise it with an orthogonal modal matrix $\bQ\in\R^{IJ\times IJ}$:
$$
\bGamma=\bQ\transp\bLambda\bQ,
$$
where $\bLambda\in\R^{IJ\times IJ}$ is a diagonal matrix and has the eigenvalues of $\bB\tensor\bA$ in its diagonal (which are the $IJ$ products of the eigenvalues $\{\lambda_i\}_i$ and $\{\mu_j\}_j$ of $\bA$ and $\bB$, respectively).
This allows us to conclude:
$$
n \, \dCovh_{\text{discrete}}^2(X,Y)\distrilim\sum_{i,j}\lambda_i\mu_{j} Z_{ij}^2 ,
$$
where $\brc{Z_{ij}}_{i,j}$ are IID standard Gaussian.\qed

\subsection{Proof of Theorem~\ref{th:gof}}\label{proof:gof}

We will first derive the compact expression of $\mathcal E_n$. To that purpose, we firstly recall the definition of energy distance:
\begin{equation}\label{def:E:app}
	\mathcal E_n=n\left[ \frac{2}{n}\sum_{l=1}^n\E d(x_l,X)-\E d(X,X') -\frac{1}{n^2}\sum_{l,m=1}^n d(x_l,x_m) \right];
\end{equation}
where all the notation so far is the same as in the main body of the dissertation.

We firstly note that, for the discrete metric, we have: $$\E d(x_l,X)=\Prob\{X\neq x_l\}.$$ Summing over $l$ and multiplying by $\frac{2}{n}$:
$$
\frac{2}{n}\sum_{l=1}^n \E d(x_l,X)=\frac{2}{n}\sum_{l=1}^n (1-\Prob\{X= x_l\})=\sum_{i=1}^I \frac{n_i}{n} (1-p_i)=\sum_{i=1}^I \hat p_i (1-p_i);
$$
where $\hat p_i:=\frac{n_i}{n}$ is the estimated probability of category $i\in\{1,\ldots,I\}$ given the sample.

Secondly, we write the straightforward identity $$\E d(X,X')=1-\sum_{i=1}^I p_i^2.$$

And finally, for the remaining term of $\mathcal E_n/n$, we apply similar arguments to conclude:
$$
\frac{1}{n^2}\sum_{l,m=1}^n d(x_l,x_m)=1-\sum_{i=1}^I \hat p_i^2 .
$$

Now, adding up the three expressions:
$$
\frac{\mathcal E_n}{n}=2\sum_{i=1}^I \hat p_i(1-p_i)-\left[ 1-\sum_{i=1}^Ip_i^2\right] -\left[ 1-\sum_{i=1}^I\hat p_i^2\right] =
$$
$$
-2\sum_{i=1}^I\hat p_i p_i + \sum_{i=1}^I p_i^2+ \sum_{i=1}^I \hat p_i^2 =
\sum_{i=1}^I (\hat  p_i - p_i)^2=
\frac{1}{n^2}\sum_{i=1}^I (n_i-n_i^*)^2.
$$
We will now derive the asymptotic null distribution of $V$-statistic $\mathcal E_n$ from classical $U-$statistic theory (our $V$-statistic is a $U$-statistic plus an asymptotically constant term). By conveniently working out expression~\eqref{def:E:app}, we get:
$$
\mathcal E_n/n=\frac{1}{n^2}\sum_{l,m=1}^n\left[ -d(x_l,x_m) +\E d(x_l,X) +\E d(x_m,X)-\E d(X,X') \right] \equiv \frac{1}{n^2}\sum_{l,m=1}^n h(x_l,x_m);
$$
where we define $h$ as the symmetric function: $$h(y,z):=-d(y,z) +\E d(y,X) +\E d(z,X)-\E d(X,X').$$

By grouping the terms:
$$\mathcal E_n/n=\frac{1}{n^2}\sum_{l\neq m}h(x_l,x_m)+\frac{1}{n^2}\sum_{l=1}^n\E d(x_l,X)-\frac{1}{n}\E d(X,X').
$$
Now multiplying both sides by $n$, the following expression for the energy distance arises:
\begin{equation}\label{E:for:asym}
	\mathcal E_n=\frac{n(n-1)}{n^2}\:n\,\mathcal U + \frac{1}{n}\sum_{i=1}^I\hat p_i(1-p_i)-\E d(X,X').
\end{equation}
Applying the unnumbered theorem on Section 5.5.2 of \citet{Serfling}, we see that
$$ n\,\mathcal U\distrilim \sum_{i=1}^I  \lambda_i (Z_i^2-1) $$
as $n\to\infty$, where we note that $\mathcal U=\frac{1}{n(n-1)}\sum_{l\neq m}h(x_l,x_m)$ is a $U$-statistic and $\{\lambda_i\}_i$ is the spectrum of matrix
$$
\bC=(p_i \delta_{ij} - p_i p_j)_{I\times I}.
$$
Summing the elements of its diagonal yields its trace:
$$
\tr(\bC)=\sum_{i=1}^I (p_i-p_i^2)=1-\sum_{i=1}^I p_i^2=\E d(X,X') .
$$
We finally see that the middle term in~(\ref{E:for:asym}) converges in distribution to $0$ under the null, owing to the fact that $\hat p_i\aslim p_i$ by the strong law of large numbers. In conclusion:
$$
\mathcal E_n \distrilim \sum_{i=1}^I  \lambda_i (Z_i^2-1) + \sum_{i=1}^I  \lambda_i=
\sum_{i=1}^I  \lambda_i Z_i^2,
$$
where $\{Z_i^2\}_{i=1}^I$ are IID chi-squared variables with one degree of freedom each.\qed


%% file: apB_v14.tex
\fancyhead[LO]{\rightmark}
\fancyhead[RE]{\leftmark}
\renewcommand{\headerright}{\thechapter}

\chapter{Software and instructions for reproducibility}
\thispagestyle{empty}
\label{apB}
\graphicspath{{./fig_apB/}}

In line with the commitment of the broader scientific community with making empirical research reproducible, in this appendix we provide instructions for reproducing the numerical examples in the dissertation, which correspond to Chapters~\ref{ch3},~\ref{ch4} and~\ref{ch6}. Please note that Chapters~\ref{ch1},~\ref{ch2} and~\ref{ch7} contain no numerical examples, and they are therefore left out from the current appendix.

All the relevant reproducible research materials are publicly available in the repository named
\begin{center}
\texttt{rr\_phd\_dissertation},
\end{center}
which is publicly available at:
\begin{center}
\url{https://github.com/fer-cp/rr_phd_dissertation} .
\end{center}

Should the previous URL stop working at any point in the coming years, please search online the present email address of the author of this dissertation, who will do their best to fulfill any request for reproducibility materials.

We now provide an overview of the documentation of the repository, presenting that same information in a way that is easier to read in one viewing than the tree structure of the repository.

\section{General system requirements}

The software in the repository mostly relies on R \citep{R}, and on R packages developed by various authors. We recommend R version 4.3.1+, in Windows 10+, for running our scripts.

The applications to genetics depend on PLINK v1.9 \citep{PLINK}, by summoning \emph{plink.exe} from the R scripts. The \emph{*.exe} file is expected to have been downloaded from the PLINK website into the current working directory of R. Users of operating systems other than Windows should adapt the command line for calling PLINK to the requirements of their system, by manually editing the R scripts in the same way they would run any other command-line instruction from R in their system. The same holds for well-known platform-specific R commands, of which we only use the ones related to exporting graphics.

\section{Numerical examples of Chapter~\ref{ch3}}

In the subfolder \href{https://github.com/fer-cp/rr_phd_dissertation/tree/main/epistasis_dc}{epistasis\textunderscore dc} of the repository, all the reproducibility materials for Chapter~\ref{ch3} are available. We now describe them very briefly.

\subsection{Simulations}

For reproducing our simulation study (calibration of the type I error and power comparison with preexisting methodology), the reader should run first the R script \texttt{masterscript\textunderscore power.R}. This generates the data tables (as \emph{*.dat} files) necessary to produce the plots that we display as a result of our simulation study.

The power plots (which include the comparison with competing method BOOST, in a different colour) are directly generated when running the masterscript.

The code for the calibration plots is a bit more cumbersome, due to the confidence band, so we split it to a separate script. Please run \texttt{plotting\textunderscore calibration.R} to obtain those figures.

In order to generate plots or numerical results for other models, one should either perform small manual edits in the scripts, or run the simulation functions with different values of the parameters.

\subsection{Real data analyses}

We made two different experiments, as indicated in the main body of the manuscript, both with the full schizophrenia database by \citet{Galicia}.

\subsubsection{Experiment I}
We assume that we have a triplet of PLINK files (\emph{*.bed, *.bim, *.fam}) within the \texttt{experiment\_i} folder, which must be set as our current working directory.

We are not allowed to share our original PLINK files (due to ethical issues pertaining informed consent), but one can run the script

\begin{center}
\texttt{filtering\_snps\_experiment\_i.R}
\end{center}

to obtain the matrices with the observations for cases and controls with the same filters that we describe in the supplement. One can do so, for example, with the \texttt{toy.ped} example supplied at the PLINK demo.

One should run \texttt{masterscript\_experiment\_i.R} to reproduce Experiment I. It uses as input the data from the 8030 SNPs for cases and controls (\texttt{Matrix\_X.dat} and \texttt{Matrix\_Y.dat}), as well as the SNP IDs in the chromosome-position format (with some alterations for the sake of anonymity of sampled individuals, in order to make this data shareable; such modifications do not influence the results we present). The latter can be found in \texttt{chr\_pos.dat}.

Every relevant result has been written down as a comment in the \emph{*.R} file. We recommend using the search function with the query “result present in the manuscript” to find the exact lines of code that replicate every numerical result for Experiment I that is cited in the main manuscript.

At many points of the script, we generate intermediate result files, in order to ease running only parts of it. We do this in light of the moderately long running times of some segments, but it is also feasible to run the entire script within reasonable time in any modern desktop computer. Please note that it is necessary to set as the working directory the location of the masterscript R file before running it.

\subsubsection{Experiment II}
As in Experiment I, we begin with the full GWAS database. We assume that we have a triplet of PLINK files (\texttt{*.bed, *.bim, *.fam}) within the \texttt{experiment\_ii} folder. We are not allowed to share ours, but one can run the script

\begin{center}
	\texttt{filtering\_snps\_experiment\_ii.R}
\end{center}

to obtain the matrices with the observations for cases and controls with the same filters that we describe in the supplement. We indicate in the comments of the \emph{*.R} file the results of the relevant steps.

An important caveat is that we do not attach the GTEx files necessary for running this script. For obtaining them, one should visit the GTEx Portal at \url{https://www.gtexportal.org/home/datasets} , select ``Adult GTEx'' and ``QTL'' from the drop-down menu, download the file

\begin{center}
	\emph{GTEx\_Analysis\_v7\_eQTL.tar.gz}
\end{center}
(single tissue cis-eQTL data for GTEx Analysis V7, dbGAP accession \emph{phs000424.v7.p2} ), unzip it and place all the \emph{Brain\_*.signifpairs.txt} files within the folder

\begin{center}
	\texttt{experiment\_ii/gtex\_v7\_signifpairs/brain} ,
\end{center}

and all the remaining \emph{*.signifpairs.txt} files (the ones not beginning with \emph{Brain\_*}) in the analogous \texttt{nonbrain} folder.

To replicate Experiment II, the reader is kindly asked to run

\begin{center}
\texttt{masterscript\_experiment\_ii.R}.
\end{center}

All the observations we made for the masterscript of Experiment I also apply to this one.

\section{Numerical examples of Chapter~\ref{ch4}}

In the subfolder \href{https://github.com/fer-cp/rr_phd_dissertation/tree/main/gwas_dc}{gwas\textunderscore dc} of the repository, all the reproducibility materials for Chapter~\ref{ch4} are available.

We use the R package \texttt{reticulate} \citep{reticulate} to call the Python package \texttt{mpmath} \citep{mpmath} for a precise and computationally efficient calculation of the Appell $F_1$ hypergeometric series.

We will now give some details on the simulations and real data application.

\subsection{Simulations}

The numerical results for simulations of type I error and power can be re-run by sourcing the R files with self-explanatory names in subfolder \texttt{simulations} in the repository. The naming of such files is of the form \emph{typeI*.R} and \emph{powersimu*.R}). The computation times are estimated in the script \emph{comptime.R}.

All the graphics in the main body of the dissertation can be reproduced by first running the numerical results and then using the plotting configuration in \emph{plots.R} .

There is a script with testing functions, which is used every time that a numerical result for our methodology is generated. It is named \emph{sim\_functions\_snp\_pheno.R} and it calls the Python script \emph{pvalue\_python.py} for the evaluation of $p$-values with the library \emph{mpmath} \citep{mpmath}.

The following R packages are used:
\begin{itemize}
\item\texttt{AssocTests} (for comparing with preexisting competing tests);
\item\texttt{parallel} (allows for multi-thread or multi-core computations, whenever the hardware meets these needs);
\item\texttt{microbenchmark} (measuring times).
\end{itemize}

\subsection{Real data analyses}

The scripts in subfolder \texttt{liver\_enzymes} correspond to the example of hepatic enzymes we study in the main body of the dissertation. This data is available through dbGaP to anyone who fulfills their strict requirements on information security, and the agreement we have signed does not allow for sharing the data with third parties. Therefore, the software we here share can potentially be used with that or other similar data, but we do not provide any specific files for it. Once more, it is an option to use the toy example that comes with every release of PLINK (the genetic software that we again use in this application).

The individual numerical results for each SNP are obtained by running the script \emph{enzymes.R}, which again depends on R functions that call Python for the computation of $p$-values. Once this has been run, Manhattan plots can be created by means of \emph{manh\_plots.R}.

R packages used:
\begin{itemize}
	\item\texttt{coga} (for the generalised $F$ distribution);
	\item\texttt{qqman} (Manhattan plots);
	\item\texttt{data.table} (fast and efficient reading and writing of external files).
\end{itemize}

\section{Numerical examples of Chapter~\ref{ch6}}

In the subfolder \href{https://github.com/fer-cp/rr_phd_dissertation/tree/main/categorical_es}{categorical\textunderscore es} of the repository, all the reproducibility materials for Chapter~\ref{ch6} are available.

We outline the reproducing instructions in the following subsections.

\subsection{Simulations}

Chapter~\ref{ch6} proposes testing procedures for two separate problems with categorical data: independence of two variables, and goodness of fit of one variable to a given distribution.

There is an R script called \emph{test\_functions\_ct\_dcov.R} which provides the testing functions necessary for both problems, and then the simulations for each of the two are organised in different folders.

For independence, the numerical results are generated with \emph{simu\_indep\_with\_plots.R} , which also provides plots for the power curve comparison of our methodology with competitors. The figures related to the type I error control can be generated with

\begin{center}
\emph{plotting\_calibration\_methods\_indep.R} .
\end{center}
	
For goodness of fit, the numerical results are crunched in \emph{simu\_gof.R} . Power plots are created by sourcing \emph{plotting\_power\_gof.R} . The figures related to the type I error control can be generated with \emph{plotting\_calibration\_gof.R} .

R packages used:

\begin{itemize}
	\item\texttt{CompQuadForm} (evaluation of the distribution function of quadratic forms of Gaussian variables);
	\item\texttt{ggplot2} (advanced graphic functions that expand those in R by default).
\end{itemize}

\subsection{Real data analyses}

As with the simulations, here we do everything twice, once for the independence test and another time for that of goodness of fit. We will be referring to subfolder \texttt{real\_data} of the repository.

Unlike in other chapters of the dissertation, the real data applications that we present in Chapter~\ref{ch6} do not involve individual-level genotype data, so there is no privacy concerns. Therefore, the real data examples here can be run fully.

For independence, we provide the dataset for the example on admission history of schizophrenia patients in \emph{admission\_data.txt} . It can be then analysed in \emph{admission.R} . Relevant results and intermediate steps are marked as comments in that script.

For goodness of fit, the data of the allelic frequencies is typed out inside the corresponding R scripts, and the external data file \emph{pgc3\_snps.txt} contains a list of SNPs known to be associated with schizophrenia, against which we check the variants that we consider in each example. The testing for HWE in a biallelic locus is carried out in \emph{hwe\_2allele.R} , whereas the triallelic setting is dealt with in \emph{hwe\_3allele.R} .


%% file: res_gl_v14.tex
\renewcommand{\headerright}{Resumo}

\fancyhead[LO]{{Resumo en galego}} 
\fancyhead[RE]{{Resumo en galego}} 
\fancyhead[RO,LE]{\thepage}  
\chapter{Resumo en galego}
\thispagestyle{empty}
\label{resumo:gl}


Esta tese, intitulada \textit{Contrastes non paramétricos de independencia en alta dimensión, con aplicacións á xenética de doenzas complexas}, reflicte o traballo de investigación realizado pola persoa candidata ao título de doutor Fernando Castro Prado, durante a súa permanencia no Programa de Doutoramento en Estatística e Investigación Operativa da Universidade de Santiago de Compostela. Os contidos da tese foron elaborados coa colaboración e apoio das dúas persoas directoras da tese, Wenceslao González Manteiga (Universidade de Santiago de Compostela) e Javier Costas (Instituto de Investigación Sanitaria de Santiago de Compostela); así como dos coautores Dominic Edelmann (Centro Alemán de Investigacións Oncolóxicas, en Heidelberg), Fernando Facal (Servizo Galego de Saúde), Jelle J. Goeman (Centro Médico da Universidade de Leiden, nos Países Baixos) e David R. Penas (Misión Biolóxica de Galicia, do Consello Superior de Investigacións Científicas, en Pontevedra).

A continuación presentamos de forma compendiada os contidos da tese en galego, lingua oficial da universidade en que se cursaron os estudos de doutoramento. Estruturaremos esta presentación por bloques temáticos que se corresponden cos capítulos da tese:
\begin{enumerate}
	\item Introdución ao campo de coñecemento.
	\item Contrastes de independencia en espazos métricos e alén.
	\item Tests de interacción xene-xene en doenzas complexas.
	\item Tests de asociacións xenotipo-fenotipo en trazos complexos humanos.
	\item Comparación de contrastes baseados en distancias con metodoloxía clásica para datos categóricos.
	\item Discusión, conclusións e futuras liñas de traballo.
\end{enumerate}

\section*{Capítulo~\ref{ch1}. Introdución ao campo de coñecemento}

Nos últimos anos produciuse un desenvolvemento sen precedentes na maneira en que producimos, almacenamos e procesamos a información, na mesma maneira en que a primeira revolución industrial consistiu na transformación na maneira de producir, almacenar e procesar a enerxía \citep{Schoelkopf}. Esta revolución, como aquela do século XVIII, só foi posible grazas a enormes avances na ciencia relacionada co recurso na cerna da revolución: hoxe en día, os datos. Falamos dunha \emph{ciencia de datos}, a cal se fundamenta na estatística matemática, acompañada dunha forte compoñente computacional e do coñecemento do dominio de aplicación de interese.

En paralelo á revolución dos datos, a bioloxía (humana) tamén experimentou a súa propia transformación, pasando de ser unha disciplina que historicamente producía poucas observacións dun reducido número de variables de similar natureza entre si, a converterse nunha disciplina xeradora de \textit{big data}, na que a heteroxeneidade é un dos maiores desafíos (\textcolor{gray}{Holmes e Huber}, \citeyear{MSMB}). Tanto é así que a xenética estúdase xa ao nivel de toda a información hereditaria nun individuo (falamos xa de ciencia da \emph{xenómica} e de moitas outras disciplinas \textit{-ómicas}) ou mesmo toda a información xenética nunha cohorte de milleiros de individuos (estamos na era dos biobancos).

Con todo, en 2024, dispoñendo de datos de millóns de persoas tomados en miles de estudos, aínda queda unha moi grande marxe para o progreso, con moitos descubrimentos que facer, algúns dos cales poderán ser trasladados á práctica clínica mediante a medicina personalizada. A xenética, como todas as ciencias biomédicas, teñen moito traballo por diante, o de responder preguntas moi complexas en base a datos moi complexos. E a mellor ciencia baseada en datos biomédicos combinará metodoloxía estatística, habilidades informáticas e coñecemento do eido de aplicación. Por iso falamos dunha \emph{ciencia de datos biomédicos} \citep{Levitt}.

A independencia estatística é un tipo de relación entre dúas características das unidades experimentais que son obxecto de estudos que se corresponde co concepto informal de que unha variable non estea asociada coa outra de ningún xeito. A dependencia totalmente determinista é o contrario da independencia estatística, existindo un continuo de intensidade da asociación entre eses dous extremos. Matematicamente, dúas variables aleatorias son independentes se, e só se, a súa distribución de probabilidade conxunta é o produto das marxinais.

O principal obxectivo desta tese de doutoramento é o uso de técnicas non paramétricas para a obtención de contrastes de independencia en espazos métricos, semimétricos e premétricos xerais; en diferentes escenarios de alta dimensionalidade que son de interese para a xenómica de doenzas complexas. Isto dará lugar a varias aplicacións relevantes, dado que moitos dos problemas de interese en xenética (como en moitas das ciencias empíricas) redúcense á procura de asociacións entre variables.

Na bibliografía xenética, asúmese de xeito case universal que as variantes xenéticas actúan dun xeito linear, aditivo. Esta simplificación non ten por que cumprirse na práctica. Polo tanto, para nós é de interese un certo tipo de metodoloxía estatística para a detección de asociacións de toda índole (non unicamente as lineares), a cal presentamos no Capítulo~\ref{ch2}. Estas técnicas permitiranos presentar contribucións estatísticas de interese para a xenética de doenzas complexas nos Capítulos~\ref{ch3},~\ref{ch4}~e~\ref{ch6}. Finalmente, no Capítulo~\ref{ch7} faise unha discusión global do noso traballo de investigación, presentando así mesmo algunhas conclusións e futuras liñas de traballo.

\section*{Capítulo~\ref{ch2}. Contrastes de independencia en espazos métricos e alén}

Cando dúas variables (ou vectores ) $X$ e $Y$ toman valores en espazos euclidianos, é posible definir unha medida que caracteriza a súa independencia, chamada \emph{covarianza de distancias} \citep{SRB}, que se define como unha certa distancia $L^2$ ponderada entre a función característica conxunta e o produto das marxinais. A covarianza de distancias ten unha moi importante propiedade que a distingue doutros parámetros poboacionais máis convencionais: vale cero sé ---e só se--- hai independencia:
$$
\dCov(X,Y)=0\iff X, Y \;\text{independentes.}
$$

Este enfoque é popular entre a comunidade da estatística matemática nos últimos anos, mentres que neste período os científicos máis algorítmicos que traballan con datos tiveron como un dos seus principais focos de atención o chamado ``truco \textit{kernel}''. En lugar de transformar os seus grandes, complexos e heteroxéneos datos cunha distancia, utilizan funcións chamadas \textit{kernels}, que se definen con distintas propiedades pero que dan lugar a tests que son duais aos baseados en distancias \citep{Sejdinovic}. Estas dúas escolas de contrastes de independencia non só converxen entre si, senón que tamén o fan cos \emph{Global Tests} de \citet{goeman2006testing}, os cales veñen sendo os contrastes localmente máis potentes en certos modelos gaussianos de regresión.

Presentamos a covarianza de distancias partindo de espazos euclidianos, para logo estender este paradigma a espazos métricos, semimétricos e premétricos \citep{Jakobsen,Lyons,Sejdinovic}. A exploración pormenorizada dos aspectos matemáticos relativos a esta técnica e a aquelas que son duais a ela conclúe o capítulo.

\section*{Capítulo~\ref{ch3}. Tests de interacción xene-xene en doenzas complexas}

Malia os moitos esforzos da comunidade científica desde comezos do século XXI, a herdanza de trazos relativos ás enfermidades comúns dos humanos aínda non se comprende plenamente a nivel molecular. A este respecto, crese que unha das claves poden ser as interaccións xenéticas, en cuxa detección non se teñen realizado grandes progresos.

Unha limitación da metodoloxía existente para esta tarefa é a antedita hipótese de que os efectos son lineares. Non hai ningunha razón biolóxica para isto, polo que decidimos empregar a covarianza de distancias (que caracteriza a independencia estatística xeral, non só a linear) neste problema.

O gran tamaño das bases de datos xenómicas fai escasamente factible a nivel computacional a aplicación de tests de hipóteses baseados en distancias da maneira que é predominante na bibliografía, é dicir, mediante permutacións. Por este motivo, desenvolvemos a distribución nula asintótica do estatístico de contraste. Á parte desta contribución teórica, realizamos simulacións nas que obtivemos unha calibración do erro de tipo I satisfactoria, así como potencia que é comparable ou mellor que a de metodoloxía preexistente \citep{BOOST}. Concluímos cunha aplicación a datos de esquizofrenia (unha doenza de grande interese, pola súa elevada carga socioeconómica), obtendo resultados que son compatibles coa hipótese biolóxica de que a interacción a nivel de expresión xenética en cerebro regulada xeneticamente xoga un papel relevante na base molecular deste trastorno psiquiátrico \citep{Lin,Patel}.

\section*{Capítulo~\ref{ch4}. Tests de asociacións xenotipo-fenotipo en trazos complexos humanos}

Un dos obxectivos fundamentais dos estudos xenómicos é a detección de variantes no ADN humano que están significativamente asociadas coa variabilidade dun trazo (fenotípico) cuantitativo de interese. De igual maneira que o capítulo anterior centrábase ma detección de interaccións xenotipo-xenotipo, este céntrase nas asociacións fenotipo-xenotipo.

Argumentando novamente que o efecto das variantes xenéticas non segue necesariamente un patrón aditivo nin linear, desenvolvemos metodoloxía estatística baseada en distancias. Tras caracterizar todas aquelas que teñen sentido, vimos que a escolla dunha ou doutra permite seleccionar a priori a clase de modelo xenético que se está buscando, o cal resulta de grande interese biolóxico.

Demostramos que o noso procedemento de contraste de hipóteses é consistente contra todas as alternativas funcionais. Logo obtivemos unha forma pechada para a distribución nula asintótica do estatístico de contraste, o cal novamente permite evitar os inconvenientes computacionais da remostraxe. Botando man da equivalencia cos \textit{Global Tests}, demostramos que cada un dos nosos contrastes é o localmente máis potente baixo un determinado modelo. Ademais, presentamos a maneira de axustar para o caso no que hai que axustar por covariables, unha tarefa fundamental en xenómica.

O noso estudo de simulación amosou unha calibración axeitada do erro do tipo I, así como unha potencia satisfactoria. Na parte aplicada deste capítulo, estudamos unha base de datos de niveis en soro de encimas hepáticos, que actúan de biomarcadores de cirrose, unha doenza que asociada ao alcoholismo (manténdonos así dentro da temática da xenética psiquiátrica). Como resultado, atopáronse asociacións que son compatibles coas evidencias bibliográficas máis recentes \citep{Pazoki}.

\section*{Capítulo~\ref{ch6}. Comparación de contrastes baseados en distancias con metodoloxía clásica para datos categóricos}
Os datos categóricos son omnipresentes na investigación biomédica e xorden en moitos contextos de especial relevancia na investigación e na clínica. Polo tanto, resulta de interese ---tanto a nivel teórico como aplicado--- ver que sucede coa metodoloxía do Capítulo~\ref{ch3} cando os soportes marxinais teñen un número arbitrario de puntos (dentro da finitude).

O estatístico de contraste da independencia neste contexto ten unha forma moi semellante á de procedementos de contraste clásicos e moi coñecidos como o de Pearson e a razón de verosimilitudes (coñecida como test $G$). Estes son débiles en situacións nas que algunhas das celas da táboa de continxencia están case baleiras, mentres que o noso procedemento é insensible a este fenómeno. Á parte diso, amosamos boa calibración do erro de tipo I e potencia, comparando cos anteditos métodos clásicos. Así mesmo, exploramos a nivel teórico e aplicado as conexións da nosa metodoloxía coa de \citet{BKS}. Todo isto aplicámolo a un exemplo que ilustra que o xenoma ten capacidade preditiva do risco de esquizofrenia.

Por outra banda, outro contraste que adoita resultar de interese para datos categóricos en soportes arbitrarios é o de bondade de axuste a unha distribución (discreta). Unha vez máis usando procedementos baseados en distancias, obtemos unha distribución nula asintótica explícita que funciona de maneira satisfactoria en simulacións, mesmo para tamaños mostrais non excesivamente grandes. Aplicamos a nova metodoloxía proposta ao contraste de bondade de axuste ás proporcións preditas polo equilibrio de \citet{Hardy} e \citet{Weinberg}, cuns resultados que son consistentes co coñecemento biolóxico existente sobre os SNPs considerados.

\section*{Capítulo~\ref{ch7}. Resultados, conclusións e futuras liñas de traballo}

Imos proporcionar agora algunhas conclusións xerais sobre os resultados da tese, os cales produciron unha serie de manuscritos que se atopan en diverso grao de progreso cara á publicación en revistas da área de estatística. As persoas lectoras desta tese poden atopar unha listaxe destas contribucións desde a páxina~\pageref{further:info} en adiante.

O tema desta tese é o contraste de asociación entre elementos aleatorios con soporte en espazos cuxa estrutura representa escenarios de interese na xenética dos trazos humanos complexos. Con este obxectivo, empregamos o Capítulo~\ref{ch1} para introducir o campo do coñecemento e algunhas nocións fundamentais relativas á nosa metodoloxía e obxectivos.

Moitos problemas de interese en xenética humana redúcense á busca de dependencias entre variables que teñen unha certa estrutura. Neste contexto, vimos como a estatística clásica non proporciona as mellores ferramentas para deseñar os procedementos de contraste desexados. Isto motivou que, no Capítulo~\ref{ch2} introduciramos a teoría abstracta que permite definir unha medida xeral da asociación chamada \emph{covarianza de distancias}, que caracteriza a independencia na maioría de espazos que un pode atopar na práctica. Este enfoque baseado en distancias é equivalente a aquel baseado en \textit{kernels} e tamén aos \textit{Global Tests}.

A nosa investigación permitiu o desenvolvemento de metodoloxía estatística que permite contrastar hipóteses biolóxicas de relevancia, incluíndo:
\begin{itemize}
	\item interacción xenética (Capítulo~\ref{ch3});
	\item asociación xene-fenotipo (Capítulo~\ref{ch4});
	\item dependencias xerais entre variables clínicas (Capítulo~\ref{ch6}); and
	\item equilibrio de Hardy--Weinberg (tamén no Capítulo~\ref{ch6}).
\end{itemize}

En cada un deses casos, propuxemos espazos abstractos cuxa estrutura reflicte o tipo de dato e o que se sabe sobre el, para así desenvolver procedementos de contraste e outros resultados teóricos. As nosas simulacións amosan un comportamento satisfactorio da nosa metodoloxía, tanto en termos absolutos coma en termos relativos á metodoloxía estatística preexistente para cada tarefa. Ademais, empregamos datos reais para ilustrar as achegas teóricas, obtendo conclusións biolóxicas que, no seu conxunto, dan a idea dun funcionamento correcto das nosas técnicas.

Un punto crucial en cada un deses capítulos é que os métodos estatísticos que se adoitan aplicar na práctica biomédica están baseados en asumir a aditividade dos efectos das variantes xenéticas, o cal pode resultar demasiado restritivo ou directamente falso \citep{Mammals,Costas:Heteroz_opt}. Para isto, exploramos as premétricas que poden dar lugar a estruturas dos soportes marxinais de maneira máis axeitada que a euclidiana, dando interpretacións de cada unha delas.

Os estatísticos de contraste que xorden a partir da covarianza de distancias e mais da metodoloxía asociada son, en xeral, $V$- e $U$-estatísticos. A súa distribución nula asintótica é a miúdo unha suma ponderada de variables independentes, distribuídas todas elas consonte unha khi-cadrado cun grao de liberade (\textcolor{gray}{Székely e Rizzo}, \citeyear{TEOD}). Aínda que existen uns poucos exemplos na literatura en que se realiza algún tipo de aproximación desta distribución límite \citep{Berschneider,Huang:Huo}, o enfoque predominante para o contraste segue a consistir no uso de técnicas de remostraxe, o cal é tan ineficiente computacionalmente que non é razoable aplicalo na práctica xenómica.

A beleza do tipo de problemas xenéticos que estudamos non só pasa pola súa utilidade na vida real, senón que tamén se manifesta no plano matemático: ao esixirmos os nosos problemas aplicados o uso de espazos simples e finitos, non só podemos deseñar a estrutura deses espazos para reflectir unha ampla diversidade de realidades biolóxicas, senón que ao mesmo tempo a estatística matemática subxacente simplifícase. En concreto, a finitude dos espazos marxinais implica a finitude da forma cuadrática á que converxe a covarianza de distancias empírica (multiplicada polo tamaño da mostra) baixo independencia. Iso significa que, ao combinar as distintas estratexias expostas no Apéndice~\ref{apA} para a obtención dos coeficientes coa estimación dos parámetros mediante os seus análogos empíricos, é posible obter $p$-valores con rapidez e precisión. 

Tamén aplicamos a mesma filosofía a un problema un tanto diferente, mais relacionado: o contraste de bondade de axuste a unha distribución discreta, onde utilizamos a \emph{distancia de enerxía} (un estatístico semellante á covarianza de distancias). A distribución asintótica do estatístico de contraste ten a peculiaridade de estar totalmente especificada baixo a hipótese nula (que é simple), co cal non é preciso estimar ningún parámetro á hora de obter $p$-valores.

No tocante á comparación coa metodoloxía preexistente, no Capítulo~\ref{ch3}, as nosas simulacións indican que o noso contraste baseado en distancias calibra o nivel de significación tan ben como o moi popular competidor \citet{BOOST}, e que a potencia é mellor no noso caso (para os modelos considerados). No Capítulo~\ref{ch4}, ao comparar a covarianza de distancias co seu rival \texttt{nmax3} \citep{nmax3}, a metodoloxía por nós proposta sae vencedora, tanto en termos de erro tipo I coma de potencia. Ademais, o noso contraste ten a vantaxe adicional de que permite seleccionar a priori o modelo fronte ao cal se desexa que o test sexa o (localmente) máis potente.

Finalmente, no Capítulo~\ref{ch6}, por unha banda o noso contraste de independencia demostra ser mellor que métodos clásicos como o de Pearson, o test $G$ e mais o exacto de Fisher; e móstrase á par do USP de \textcolor{gray}{Berrett e Samworth} (\citeyear{BS}). E por outra banda, o test de bondade de axuste baseado na distancia de enerxía ten unha curva de potencia que se sitúa un pouco por debaixo da do test $\chi^2$ de Pearson. A comparativa con metodoloxía preexistente deste capítulo tamén a efectuamos a nivel teórico, xa que demostramos as conexións entre contrastar a independencia con xeneralidade, o tradicional test de Pearson e o moderno USP.

Ao facer balance da parte aplicada do noso traballo, vemos que o Capítulo~\ref{ch3} indica que a interacción xene-xene podería estar tendo lugar ao nivel da expresión xenética regulada xeneticamente, o cal é consistente con descubrimentos publicados recentemente \citep{Lin,Patel}. No Capítulo~\ref{ch4} atópase sinal que é tan disperso como se esperaba, cuxos $p$-valores están nunha orde de magnitude razoable en relación ao tamaño mostral, e que inclúe algúns positivos que xa se atoparan en mostras independentes da mesma procedencia étnica que a da nosa mostra \citep{Middelberg}. Finalmente, os resultados do Capítulo~\ref{ch6} son consistentes coa capacidade dos índices de risco polixénico para medir a severidade dun trastorno \citep{PRS:Topol} e coa noción conceptual básica de que os xenotipos correspondentes a variantes xenéticas asociadas á esquizofrenia non se van observar a igual frecuencia na subpoboación de pacientes de esquizofrenia que na poboación xeral.

En síntese, o traballo presentado nesta tese contén desenvolvementos relevantes no eido da estatística matemática, orientados cara a aplicacións xenéticas de interese, onde os recursos computacionais xogan un papel fundamental. Secasí, quedan liñas de traballo que un podería seguir neste campo, que detallamos a continuación.

Unha tarefa interesante sería a de deseñar un procedemento que permita inferir, a partir da mostra, que distancia é óptima nalgún sentido. Tamén é natural preguntarse que resultados se obterían na práctica ao adaptar a metodoloxía dos Capítulos~\ref{ch3} e~\ref{ch6} á busca de dependencias entre variables binarias e ternarias, o cal permitiría a aplicación á busca de interaccións entre variantes xenéticas no xenoma nuclear e no mitocondrial.

Ademais, hai moitos obxectivos fundamentais da xenómica, que non se abordaron nesta tese, como por exemplo: a estimación da herdabilidade, os contrastes de causalidade, ou a predición de fenotipos a partir de xenotipos \citep{Brandes}. Unha idea de futuro sería a aplicación de métodos baseados en distancias e \textit{kernels} a estes problemas, co obxectivo de crear ferramentas estatísticas cun maior sentido conceptual e unha mellor rendemento empírico que aquelas existentes na actualidade.

O noso foco é o estudo da xenética humana, pero as nosas técnicas poderían usarse para outros organismos. Mentres estes sexan diplontes, o soporte dos $X$'s seguirá a ser de cardinal $3$, co cal a metodoloxía non requiriría ningunha adaptación. O coñecemento actual apunta a que, polo menos en mamíferos, ten sentido transcender a aditividade dos efectos á hora de estudar a causalidade das variantes xenéticas na variabilidade dos trazos fenotípicos \citep{Mammals}.

Tamén podería resultar de interese a adaptación da metodoloxía do Capítulo~\ref{ch4} a variables resposta que non sexan de natureza continua, como poderían ser os indicadores de presenza-ausencia dunha enfermidade (variables binarias) ou a supervivencia (datos censurados). Por outra banda, o coñecemento biolóxico apunta a que as interaccións xenéticas son, na práctica de orde 3 e superior \citep{Russ}, co cal o uso da multivarianza de distancias \citep{Boettcher1} do que se deu unha idea superficial no Capítulo~\ref{ch3} podería ser unha idea de enorme interese práctico. Finalmente, unha vía de investigación extremadamente prometedora para o estudo do efecto de variables ambientais no fenotipo é a chamada \emph{covarianza de distancias condicional} \citep{Conditional}, o cal contribuiría á comprensión das causas da variabilidade entre individuos e subpoboacións de caracteres relacionados coas doenzas complexas humanas.

%% file: fi_v14.tex
\renewcommand{\headerright}{Further information}

\fancyhead[LO]{{Further information}} 
\fancyhead[RE]{{Further information}} 
\fancyhead[RO,LE]{\thepage}  
\chapter{Further information}
\thispagestyle{empty}
\label{further:info}
In compliance with the regulations for PhD studies at the University of Santiago de Compostela (namely, the \textit{Regulamento dos estudos de doutoramento na USC, DOG de 16 de setembro de 2020}), we hereby provide the information that is required from us regarding the research output of this dissertation. We will be referring to \emph{arXiv} e-prints, since none of our manuscripts have been accepted in a journal at the moment of handing in this dissertation (a situation that may change from now to the point of defending our PhD work). The public repository \emph{arXiv} (Cornell University Library) hosts a large proportion of current research in fields like mathematics and statistics ---including preprints, postprints and technical reports---, making them openly available for free.

Given that Chapter~\ref{ch1} is the introduction and that the last one (i.e., Chapter~\ref{ch7}) discusses the results and serves as the conclusion of the main body of the dissertation, we will restrict ourselves to Chapters~\ref{ch2}--\ref{ch6} for the description of the research output below.

\section*{ Research output of Chapter~\ref{ch2} }

The highly non-trivial reviewing effort carried out for Chapter~\ref{ch2} helped in the writing of the introductory sections of the papers that we list as contributions for the remaining chapters, but it also directly produced the following technical report:

\vspace*{.6cm}

{Castro-Prado, F.$^{1,2,3}$ and Gonz\'alez-Manteiga, W.$^{1,2}$} (2020). Nonparametric independence tests in metric spaces: What is known and what is not. Available at \url{https://arxiv.org/abs/2009.14150}.

$^1$ Department of Statistics, Mathematical Analysis and Optimisation; Faculty of Mathematics, University of Santiago de Compostela (USC). R\'ua Lope G\'omez de Marzoa s/n, 15782 Santiago de Compostela, Spain.

$^2$ Galician Centre for Mathematical Research and Technology (CITMAga). R\'ua Constantino Candeira s/n, 15782 Santiago de Compostela, Spain.

$^3$ Psychiatric Genetics Laboratory, Santiago Health Research Institute (IDIS). University Hospital, Travesía da Choupana s/n, 15706 Santiago de Compostela, Spain.

\vspace*{.6cm}

The PhD candidate contributed to the conceptualisation of the paper, bibliographical review, development of small mathematical results, discovery and correction of mistakes in published research by other authors, writing of the original manuscript, revision and editing.

This e-print is licensed under an \href{https://creativecommons.org/licenses/by-nc-sa/4.0/}{Attribution-NonCommercial-ShareAlike 4.0 International (CC BY-NC-SA 4.0) license}, meaning that anyone is free to copy, redistribute, mix and transform its content; as long as the purposes are non-commercial, the original work is appropriately cited, and any derivatives are shared under the same terms. This license cannot be revoked.

\section*{ Research output of Chapter~\ref{ch3} }

The contents of Chapter~\ref{ch3} correspond to those of the following preprint, which is as of June 2024 is undergoing the third round of revision in a journal of the area of statistics.

\vspace*{.6cm}

{Castro-Prado, F.$^{1,2,3}$, Costas, J.$^3$, Edelmann, D.$^4$, Gonz\'alez-Manteiga, W.$^{1,2}$ and Penas, D. R.$^5$} (2023). Testing for genetic interaction with distance correlation. Available at \url{https://arxiv.org/abs/2012.05285}.

$^1$ Department of Statistics, Mathematical Analysis and Optimisation; Faculty of Mathematics, University of Santiago de Compostela (USC). R\'ua Lope G\'omez de Marzoa s/n, 15782 Santiago de Compostela, Spain.

$^2$ Galician Centre for Mathematical Research and Technology (CITMAga). R\'ua Constantino Candeira s/n, 15782 Santiago de Compostela, Spain.

$^3$ Psychiatric Genetics Laboratory, Santiago Health Research Institute (IDIS). University Hospital, Travesía da Choupana s/n, 15706 Santiago de Compostela, Spain.

$^4$ Biostatistics Department, German Cancer Research Center (DKFZ), Im Neuenheimer Feld 280, 69120 Heidelberg, Germany.

$^5$ Computational Biology Laboratory, Spanish National Research Council (MBG-CSIC), Pazo de Salcedo, 36143 Pontevedra, Spain.

\vspace*{.6cm}

The PhD candidate contributed to the conceptualisation of the paper, bibliographical review, creation of new statistical methodology, software development, simulation study, search for appropriate datasets, real data application, writing of the original manuscript, revision and editing.

This preprint is licensed under an \href{https://creativecommons.org/licenses/by-nc-sa/4.0/}{Attribution-NonCommercial-ShareAlike 4.0 International (CC BY-NC-SA 4.0) license}, meaning that anyone is free to copy, redistribute, mix and transform its content; as long as the purposes are non-commercial, the original work is appropriately cited, and any derivatives are shared under the same terms. This license cannot be revoked.

\section*{ Research output of Chapter~\ref{ch4} }

The contents of Chapter~\ref{ch4} are mostly the same as those of the following manuscript, which we are preparing to submit to a journal of the area of statistics at the time of handing in this dissertation. This means that the final version may differ to some extent in the title, authorship, affiliations or content. However, we consider it more informative to include it as a research output `as is' than not doing so.

\vspace*{.6cm}

{Castro-Prado, F.$^{1,2,3}$, Edelmann, D.$^4$ and Goeman, J. J.} (2024a). A generalized distance covariance framework for genome-wide association studies. [Preprint.]

$^1$ Department of Statistics, Mathematical Analysis and Optimisation; Faculty of Mathematics, University of Santiago de Compostela (USC). R\'ua Lope G\'omez de Marzoa s/n, 15782 Santiago de Compostela, Spain.

$^2$ Galician Centre for Mathematical Research and Technology (CITMAga). R\'ua Constantino Candeira s/n, 15782 Santiago de Compostela, Spain.

$^3$ Psychiatric Genetics Laboratory, Santiago Health Research Institute (IDIS). University Hospital, Travesía da Choupana s/n, 15706 Santiago de Compostela, Spain.

$^4$ Biostatistics Department, German Cancer Research Center (DKFZ), Im Neuenheimer Feld 280, 69120 Heidelberg, Germany.

$^5$ Department of Biomedical Data Sciences, Leiden University Medical Center. Albinusdreef 2,
2333 ZA Leiden, the Netherlands.

\vspace*{.6cm}

The PhD candidate contributed to the conceptualisation of the paper, bibliographical review, creation of new statistical methodology, software development, simulation study, search for appropriate datasets, real data application, writing of the original manuscript, revision and editing.

\section*{ Research output of Chapter~\ref{ch6} }\label{res:output}

The contributions of Chapter~\ref{ch6} are to be found in the latest of our preprints, which is undergoing its second round of peer reviewing in a journal of the area of statistics, as of June 2024.

\vspace*{.6cm}

{Castro-Prado, F.$^{1,2,3}$, González-Manteiga, W.$^{1,2}$, Costas, J.$^3$, Facal, F.$^3$ and Edelmann. D.$^4$} (2024b). Tests for categorical data beyond Pearson: A distance covariance and energy distance approach. Available at \url{https://arxiv.org/abs/2403.12711}.

$^1$ Department of Statistics, Mathematical Analysis and Optimisation; Faculty of Mathematics, University of Santiago de Compostela (USC). R\'ua Lope G\'omez de Marzoa s/n, 15782 Santiago de Compostela, Spain.

$^2$ Galician Centre for Mathematical Research and Technology (CITMAga). R\'ua Constantino Candeira s/n, 15782 Santiago de Compostela, Spain.

$^3$ Psychiatric Genetics Laboratory, Santiago Health Research Institute (IDIS). University Hospital, Travesía da Choupana s/n, 15706 Santiago de Compostela, Spain.

$^4$ Biostatistics Department, German Cancer Research Center (DKFZ), Im Neuenheimer Feld 280, 69120 Heidelberg, Germany.

\vspace*{.6cm}

The PhD candidate contributed to the conceptualisation of the paper, bibliographical review, creation of new statistical methodology, software development, simulation study, search for appropriate datasets, real data application, writing of the original manuscript, revision and editing.

This preprint is licensed under an \href{https://creativecommons.org/licenses/by-nc-sa/4.0/}{Attribution-NonCommercial-ShareAlike 4.0 International (CC BY-NC-SA 4.0) license}, meaning that anyone is free to copy, redistribute, mix and transform its content; as long as the purposes are non-commercial, the original work is appropriately cited, and any derivatives are shared under the same terms. This license cannot be revoked.

%% file: bib_v14.tex
\renewcommand{\headerright}{Bibliography}
\fancyhead[LO]{{Bibliography}} 
\fancyhead[RE]{{Bibliography}} 
\fancyhead[RO,LE]{\thepage}